\documentclass[a4paper,11pt]{article}
\pdfoutput=1 

\usepackage{jheppub} 

\usepackage{multirow,mathrsfs}
\usepackage{amsxtra,amstext,latexsym,dsfont,amsfonts}
\usepackage{color,eucal,bbm}
\usepackage{xcolor}

\usepackage{float}   
\usepackage{subfig}
\usepackage{booktabs} 
\usepackage{array} 

\usepackage{ytableau}
\usepackage{algorithm}

\def\ben{\begin{equation}}
\def\een{\end{equation}}
\def\bea{\begin{eqnarray}}
\def\eea{\end{eqnarray}}

\newcommand{\bra}[1]{\mbox{$\langle #1 |$}}
\newcommand{\ket}[1]{\mbox{$| #1 \rangle$}}

\title{\huge Complexity of PXP scars revisited}

\author[a,b,c]{Pawel Caputa,}
\author[d,e]{Xuhao Jiang,}
\author[f]{Sinong Liu.}

\emailAdd{pawel.caputa@fysik.su.se}
\emailAdd{xuhao.jiang@tum.de}
\emailAdd{sinongliu@mail.tsinghua.edu.cn}

\preprint{YITP-25-94}

\affiliation[a]{The Oscar Klein Centre and Department of Physics, Stockholm University, AlbaNova, 106 91 Stockholm, Sweden}
\affiliation[b]{Yukawa Institute for Theoretical Physics, Kyoto University, Kitashirakawa Oiwakecho, Sakyo-ku, Kyoto 606-8502, Japan}
\affiliation[c]{Faculty of Physics, University of Warsaw, Pasteura 5, 02-093 Warsaw, Poland}

\affiliation[d]{Technical University of Munich, TUM School of Natural Sciences, Physics Department, 85748 Garching, Germany}
\affiliation[e]{Faculty for Physik, Ludwig-Maximilians-Universit\"at M\"unchen, 
Schellingstra\ss e 4, 80799, Munich, Germany}

\affiliation[f]{Yau Mathematical Sciences Center, Tsinghua University, Haidian District, Beijing 100084, China}

\abstract{
We revisit a quantum quench scenario in which either a scarring or thermalizing initial state evolves under the PXP Hamiltonian. Within this framework, we study the time evolution of spread complexity and related quantities in the Krylov basis. We find that the Lanczos coefficients $b_n$, as functions of the iteration number $n$, exhibit a characteristic arched growth and decay, followed by erratic oscillations which we refer to as buttress. The arched profile predominantly arises from contributions within the quantum many-body scar subspace, while the buttress is linked to thermalization dynamics.
To explain this behavior, we utilize the representation theory of $\mathfrak{s}l_3(\mathbb{C})$, allowing us to decompose the PXP Hamiltonian into a linear component and a residual part. The linear term governs the formation and width of the arch, and we observe that that there exists a threshold of arch width which determines whether a given initial state exhibits scarring. Meanwhile, the residual term accounts qualitatively for the emergence of the buttress. We estimate an upper bound for the extent of the buttress using Lucas numbers.
Finally, we demonstrate that spread complexity oscillates periodically over time for scarred initial states, whereas such oscillations are suppressed in thermalizing cases.
}
\begin{document}

\maketitle

\section{Introduction and summary} \label{intro}
The notion of complexity is closely tied to emergence, where intricate behavior at large scales arises from interactions among simple microscopic constituents. A wide range of physical phenomena, including turbulence \cite{Frisch_1995}, topological order \cite{wen1990topological}, and spacetime geometry in holography \cite{Takayanagi:2025ula}, are understood as emergent. Despite its importance, a precise and general definition of complexity in physics remains elusive, and substantial efforts are focused on formulating a universal and physically motivated framework \cite{Baiguera:2025dkc}.

Recent works have proposed a promising approach to quantifying the complexity of growth of quantum operators \cite{Parker:2018yvk} and spread of states \cite{Balasubramanian:2022tpr} using the Krylov basis \cite{Krylov1931} and the recursion method \cite{LanczosBook}. Studies in many-body systems, quantum field theories, and holographic models indicate that this definition is sensitive to topological phases \cite{Caputa:2022eye,Caputa:2022yju}, signatures of quantum chaos \cite{Rabinovici:2022beu,Erdmenger:2023wjg,Balasubramanian:2022dnj,Balasubramanian:2023kwd}, and the late-time behavior of dynamical systems (see review \cite{Nandy:2024htc}). Interestingly, fulfilling expectations from black hole physics \cite{Susskind:2014rva,Magan:2018nmu}, spread complexity has been shown to match the volume of the Einstein Rosen bridge in JT gravity \cite{Rabinovici:2023yex,Balasubramanian:2024lqk,Heller:2024ldz}, and its growth rate agrees with the proper momentum of probes falling into anti-de Sitter spacetimes \cite{Caputa:2024sux,Caputa:2025dep,Aguilar-Gutierrez:2025kmw}. These results suggest that Krylov-based complexity measures offer valuable insight into quantum dynamics across physics. Nevertheless, their full range of applicability and fundamental limitations remain to be understood. Therefore, testing their sensitivity is extremely important and valuable at this current, development stage.

One of the most intriguing nonequilibrium problems for testing complexity measures is thermalization, a phenomenon that draws the attention of many areas in physics, from quantum many-body systems to quantum gravity. For decades, it has been widely believed that the Eigenstate Thermalization Hypothesis (ETH) \cite{Deutsch:1991msp, Srednicki:1994mfb} is sufficient to predict the evolution of generic, non-integrable, isolated quantum many body systems when driven out of equilibrium. According to ETH, evolution of an initial state in such a systems, regardless of its specific features, will relax to thermal equilibrium with a well-defined effective temperature. However, as our understanding of nonequilibrium dynamics deepens, the universal validity of ETH across the entire Hilbert space has come under scrutiny.
A major breakthrough in this area came from experiments in 2018. Using a new class of Rydberg atom quantum simulators \cite{Bernien:2017ubn}, they observed both periodic revivals and quantum thermalization, depending on the choice of initial states. This striking behavior, which departs significantly from ETH predictions, has been identified as quantum many-body scars (QMBS) \cite{Turner:2018}. These phenomena cannot be accounted for by integrability \cite{Caux:2010by} or by disorder \cite{Abanin:2018yrt}, and are now regarded as manifestations of a weak form of ETH, sparking significant interest among theoretical physicists.

Recent studies of non-integrable quantum models \cite{Wildeboer:2022vjh,Lerose:2023akj,Wang:2024tso} have led to the conjecture that Hamiltonians hosting quantum many-body scars can be effectively decomposed into a direct sum of a scarred and a thermal sector: $H \approx H_{\text{scar}} \oplus H_{\text{thermal}}$ \cite{Moudgalya:2021ixk}. In such systems, a subspace of the Hilbert space is spanned by scarred eigenstates that remain dynamically decoupled from the thermalizing sector. How can this decoupled scarred subspace be identified or constructed in practice? Broadly, two distinct mechanisms have been proposed to generate or identify the scarred subspace\footnote{We do not consider here the third mechanism of projector embedding \cite{2017PhRvL.119c0601S}, which addresses how to embed a given scarred subspace into a thermal spectrum, rather than how to construct or identify it from first principles.}: the spectrum-generating algebra (SGA) \cite{Bohm1971DynamicalGA} and Krylov-restricted thermalization (KRT) \cite{Moudgalya:2019vlp}. These approaches are to some extent complementary. The SGA mechanism seeks an operator $Q^{\dagger}$, whose commutation relation with the Hamiltonian is proportional to itself within the Hilbert subspace of scars, so that the scarred subspace can be generated by acting $Q^{\dagger}$ on some eigenstate of the Hamiltonian. Clearly, $Q^{\dagger}$, interpreted as creation operator of quasi-particle excitations, and its conjugate transpose $Q$ play the role of a pair of ladder operators. However, finding such an operators is generally model-dependent and often analytically intractable, limiting the method’s applicability.
In contrast, the KRT framework takes a more general and operational perspective. Starting from an initial state, it constructs the Krylov subspace generated by repeated action of the Hamiltonian, effectively capturing the dynamics of that state. This approach is universally applicable to any model, regardless of whether an algebraic structure like the SGA exists. However, its generality comes with a trade-off: the Krylov subspace reflects the nature of the chosen initial state. Without prior knowledge of whether the initial state is scarred or thermal, the resulting subspace may not isolate the scarred sector.

Arguably, to understand quantum scars, we need a strategy for obtaining scarred subspaces that can combine the advantages of the both approaches above. Given the requirement on the structure of the models of SGA, it is more natural to achieve this goal by improving KRT. In particular, we can distinguish scarred from thermal subspaces by studying the by-products of KRT that behave significantly differently when generating Krylov subspaces from scarred or thermal initial states. This leads to two natural questions: {\it What are the characteristic by-products of KRT, and how do they evolve when an initial state, scarred or thermal, is driven out of equilibrium by the dynamics of the Hamiltonian?}

Fortunately, the first question can be naturally connected with the framework of the {\it spread complexity} \cite{Balasubramanian:2022tpr}, since it is closely related to the Krylov methods mentioned above. Indeed, the evaluation of the spread complexity consists of three steps: first we generate the Krylov basis applying {\it Lanczos algorithm} \cite{Lanczos:1950zz} to the Krylov subspace that contains the initial state and all the powers of the Hamiltonian acting on it (see next section). In the Krylov basis the Hamiltonian is a tri-diagonal, with {\it Lanczos coefficients} $a_n$s (on the diagonal) and $b_n$s (sub-/supra-diagonal) -- this the first by-product. Then we project the time evolution of the initial state onto each Krylov basis vector to obtain probability amplitudes -- the second by-product. These amplitudes satisfy a discrete Sch\"odinger equation with Lanczos coefficients and completely characterize the dynamics by mapping it to a particle hopping on a 1D chain. Last but not least, spread complexity is computed as average position of this particle on the chain -- the third by-product. In addition, since there is more information about the dynamics in the probability than in the average position on the chain, we can compute additional information-theoretic quantities such as e.g. the Shannon entropy of this probability distribution, called {\it Krylov entropy} (or K-entropy for short) \cite{Barbon:2019wsy}.

In this work, we will employ these Krylov basis tools in the {\it PXP model} \cite{Lesanovsky:2012}; the idealized model of atoms in the Rydberg blockade. This famous setup and its quantum scars have already been analyzed from various perspectives, including methods and tools of quantum information such as fidelity for a class of initial, product states of the form
\begin{equation}
    |{Z}_k\rangle = |\dots {1}\underbrace{{0}\dots {0}}_{k-1}\,{1}\dots\rangle,\qquad  k-1 \in \mathbb{Z}^+~.
    \tag{\ref{Zkqubit}}
\end{equation}
Moreover, \cite{Bhattacharjee:2022qjw} and \cite{Nandy:2023brt}, already applied the Krylov basis analysis and computed spread complexity in this model for the evolution of scar states with {\it PXP Hamiltonian}. Both found that a typical distribution of $b_n$ for a scarred state contains $L$ Lanczos coefficients ($L$ being the size of the lattice) lying on an arch, which is similar to the plots of Hamiltonian or Liouvillian belonging to $\mathfrak{su}(2)$ algebra \cite{Caputa:2021sib}. However, unlike for $\mathfrak{su}(2)$, $b_n$s do \uppercase{not} stop at $n=L$, and start increasing afterward. In \cite{Bhattacharjee:2022qjw} this behavior was modelled by an ad-hoc $q$-deformed $\mathfrak{su}(2)$ algebra. On the other hand, in \cite{Nandy:2023brt} the forward scattering approximation (FSA) was used instead of the original Lanczos algorithm to limit the number of coefficients to the $\mathfrak{su}(2)$ form. As such, we find both works, to some extent, unsatisfactory, and, since understanding the precise definition and characteristic features of quantum scars is so important, we decided to reconsider this problem. In particular, in a more systematic way, we carefully analyze the spread complexity in the PXP setup exploring the intrinsic symmetry (and algebraic structure) of the model.

Let us summarize our general strategy for the following: we consider some typical product initial states, scarring or not, unitarily evolved with the PXP Hamiltonian. By applying the Lanczos algorithm, we construct the Krylov basis, find Lanczos coefficients $b_n$ and calculate the amplitudes and the spread complexity. We then carefully examine their properties and identify features that can distinguish between the initial state being the scar or not.

With these in mind, the paper is organized as follows: 
\autoref{spreadcomplex} provides a brief review of spread complexity and related concepts such as the Krylov basis, and the Lanczos coefficients. 
\autoref{qmbs} reviews quantum many-body scars (QMBS) and two possible mechanisms, Krylov restricted thermalization (KRT) and spectrum-generating algebra (SGA), behind QMBS. A brief introduction of ETH, its range of validity and the resulting Hilbert space decomposition, follows in section \ref{eth2hilbert}. The logic of \autoref{qmbs} benefits heavily from \cite{Serbyn:2020wys, Moudgalya:2021xlu}.
Starting from \autoref{pxpmodel}, we concentrate on the PXP model. The early and late-time behaviors of the fidelity, showing quantum revivals for specific initial states, are reviewed in section \ref{revival}. 
Section \ref{lanczos-alg} investigates the properties and their causes of the Lanczos coefficients resulting from the PXP Hamiltonian acting on specific initial states. More specifically, section \ref{lanczos-numeric} exhibits the typical shapes of the growth of the Lanczos coefficients, which can be divided into two parts, an arch and a buttress. In section \ref{lanczos-analytic} we review the representation theory of $\mathfrak{s}l_3(\mathbb{C})$ and determine the physical meaning and the origin of the arch and the buttress. Section \ref{lanczos-numberic-random} provides supporting numerical results. 
In section \ref{pxpkcomplex}, we present plots of evolution of the spread complexity and Krylov entropy, and analyze the time evolution of the components of the target state in the Krylov basis (probabilities) used for the calculations of these two quantities. Finally, in \autoref{conclusion} we conclude and include some mathematical background and details of calculations related to \autoref{lanczos-alg} in appendices.
\section{Basics of Spread Complexity} \label{spreadcomplex}
In this section, we briefly summarize the Krylov basis approach to quantifying quantum complexity, and the definition of the spread complexity of quantum states \cite{Balasubramanian:2022tpr}. We closely follow \cite{Caputa:2021sib,Caputa:2022eye,Caputa:2022yju} but readers can find more pedagogical introduction and recent review of this technology in \cite{Nandy:2024htc}.

Consider a quantum state $|\Psi (s) \rangle$ evolved from an initial state $|\Psi_0 \rangle$ by a time-independent Hamiltonian $H$
\begin{equation}
|\Psi (s) \rangle = e^{-i H s} |\Psi_0 \rangle ~.
\label{evolve}
\end{equation}
We may interpret \eqref{evolve} as a computational task represented by a unitary circuit $e^{- i H s}$ from a ``reference state" $|\Psi_0 \rangle$ to a ``target state" $|\Psi(s) \rangle$. The circuit is parametrized by the circuit time $s$, that typically varies from 0 to 1, corresponding to the reference and the target state, respectively. 
Intuitively, the evolution of a ``simple" initial state with generic Hamiltonian will spread the state over the system's Hilbert space, and the state at time $s>0$ will become more ``complex". 

This can be made precise by following the spread of the state $|\Psi(s) \rangle$ over the basis in Hilbert space. Namely, we first pick a numbered basis $\mathcal{B} = \{ |B_n \rangle, n=0,1,2, \cdots \}$, with the initial state coinciding with our starting state $|B_0 \rangle = |\Psi_0 \rangle$. Projection of $|\Psi(s) \rangle$ on the basis elements will provide the probability amplitudes for the support of the state on each vector. Then, using the jargon of complexity/computation, we define the cost function associated with these probabilities
\begin{equation}
    \mathscr{C}_{\mathcal{B}}(s) =  
    \sum_n n | \langle \Psi (s) | B_n \rangle |^2
   ~,\label{SpreadB}
\end{equation}
where each contribution is weighted by the non-decreasing index (here taken to be simply $n$).
Finally, we define {\it spread complexity} as a minimum of this cost function over all choices of basis
\begin{equation}
    \mathscr{C}_K(s) = \min_{\mathcal{B}} \mathscr{C}_{\mathcal{B}}(s)\,.
\end{equation}
In general, such an abstract definition can be hard to compute in physical settings, especially with infinite-dimensional Hilbert spaces. Fortunately, \cite{Balasubramanian:2022tpr} proved (for general discrete time settings as well as for continuous, but finite times) that the above minimization is achieved by the so-called {\it Krylov basis} that we now define. 

The Krylov basis $|K_n \rangle$ is obtained by first forming the Krylov subspace \cite{Krylov1931} by successively applying to the initial state $|\Psi_0 \rangle$ all the powers of the Hamiltonian that drives the evolution $\mathscr{Kr}=\operatorname{span} \{ |\Psi_0 \rangle, H | \Psi_0 \rangle, H^2 |\Psi_0 \rangle , \cdots \}$. Next, we can iteratively construct an orthonormal basis in this subspace using the Gram-Schmidt procedure, referred to as the {\it Lanczos algorithm} \cite{Lanczos:1950zz} in this context:
\begin{equation}
 |A_{n+1} \rangle =(H-a_n) |K_n \rangle-b_n|K_{n-1} \rangle\,,\qquad\ket{K_n}=b^{-1}_n\ket{A_n}\,,\label{LanczosAlg}
\end{equation}
starting from $n=0$ and $|K_0\rangle=|\Psi_0 \rangle$, $b_0=0$.
The two sets of coefficients $a_n$s and $b_n$s, are the so-called {\it Lanczos coefficients} and are given by
\begin{equation}
    a_n=\langle K_n| H|K_n\rangle\,,\qquad b^2_n=\langle A_n|A_n\rangle\,.
\end{equation}
The procedure should be continued until $n=n^*$ for which $b_{n^*}=0$ (sometimes $n^*$ can be infinite), and the dimension of the Krylov basis, in general bounded by the total dimension of the Hilbert space, is $n^*+1$. 

In practice, the Lanczos coefficients can be extracted from the moments of the {\it return-amplitude} (also known as auto-correlator, survival amplitude or Loschmidt amplitude)
\begin{equation}
    S(s) = \langle \Psi(s) | \Psi_0 \rangle= \langle \Psi_0|e^{iHs} | \Psi_0 \rangle = \sum_{k=0}^{\infty} \frac{s^k}{k!}  \langle \Psi_0 | (i H)^k |\Psi_0 \rangle ~,
    \label{returnamp}
\end{equation}
and the relation between Lanczos coefficients and moments follows from the representations of Laplace transform of $S(s)$ as a continued fraction (see \cite{Balasubramanian:2022tpr,LanczosBook}). In particular, for return amplitudes with time reflection symmetry $s\to-s$, that we will also encounter in this work, we only get even moments implying $a_n=0$.

Clearly, the Lanczos algorithm \eqref{LanczosAlg} is equivalent to a tri-diagonal action of the Hamiltonian on the Krylov basis
\begin{equation}
    H |K_n \rangle = b_n |K_{n-1} \rangle + a_n |K_n \rangle + b_{n+1} |K_{n+1} \rangle~,
    \label{recurrel}
\end{equation}
and this allows us to extract the probability amplitudes that we need for the spread complexity \eqref{SpreadB}, i.e., the coefficients of the expansion of $|\Psi(s) \rangle$ in the Krylov basis
\begin{equation}
    |\Psi (s) \rangle = \sum_n \varphi_n (s) |K_n \rangle~, \qquad \varphi_n (s) = \langle K_n | \Psi (s) \rangle\,,
    \label{krylovexpansion}
\end{equation}
with the $0$-th amplitude related to the return amplitude $S (s)= \varphi_0^*(s)$. Indeed, the amplitudes satisfy the following discrete Schr\"odinger equation
\begin{equation}
    i \partial_s \varphi_n (s) = a_n \varphi_n (s)+b_n \varphi_{n-1}(s) + b_{n+1} \varphi_{n+1} (s)~,
    \label{dschrodinger}
\end{equation}
with initial condition $\varphi_n(0)=\delta_{n,0}$. These way, we map the evolution to a particle hopping on a 1D chain with sites corresponding to the Krylov basis vectors labelled by $n$. Moreover, the amplitudes provide a probability distribution on this chain
\begin{equation}
    p_n(s)=|\varphi_n(s)|^2,\qquad \sum_n |\varphi_n (s)|^2 =1\,.
\end{equation}
This way, we can intuitively understand the spread complexity as the average position of this particle on the 1D chain 
\begin{eqnarray}
    \mathscr{C}_K(s)=\langle n\rangle = \sum_n n|\varphi_n (s) |^2 ~,
\end{eqnarray}
and this definition naturally generalizes operator size \cite{Roberts:2018mnp,Qi:2018bje,Lin:2019qwu} and Krylov complexity for the Heisenberg evolution of quantum operators \cite{Parker:2018yvk}.

Even though conceptually elegant, in many practical applications we can only evaluate Krylov and spread complexities numerically. Nevertheless, there is a handful of analytical examples that will also play an important roles in the PXP model. Such analytical progress is possible by noticing that, for some return amplitudes, and their Lanczos coefficients, effective dynamics on the Krylov chain can be represented in terms of a dynamical algebra \cite{Caputa:2021sib}. Namely, in these examples, we can represent the Hamiltonian in the Krylov basis as
\begin{equation}
    H \doteqdot \gamma L_0+\alpha ( L_+ + L_- )\,,
\end{equation}
where $L_0, L_{\pm}$ are the generators (ladder operators) of some specific Lie algebra. Then, if the Krylov basis is identified with the Lie algebra basis, the Lanczos coefficients naturally follow the algebraic data
\begin{equation}
 \gamma L_0 |K_n \rangle = a_n |K_n \rangle\,,\quad \alpha L_- |K_n \rangle = b_{n} |K_{n-1} \rangle\,,\quad   \alpha L_+ |K_n \rangle = b_{n+1} |K_{n+1} \rangle\,.
    \label{lanczoscoe}
\end{equation}
We will come back to this story for the $\mathfrak{su}(2)$ algebra in later sections.
\section{General Profiling of Quantum Many-Body Scars} \label{qmbs}
\subsection{Weak ETH and Hilbert Space Decomposition} \label{eth2hilbert}
One of the most common protocols for studying thermalization in quantum systems (in both experiment and theory) is {\it quantum quench}. In this scenario, we usually start (at $t=0$) with a ground state $|\Psi_0\rangle$ of a quantum system governed by a time-independent Hamiltonian  $H_0$ (initial Hamiltonian). Then, we perform a unitary time evolution with a different Hamiltonian $H$ (evolving Hamiltonian) for $t>0$. This defines a non-equilibrium state 
\begin{equation}
    |\Psi(t) \rangle = e^{-i H t} |\Psi_0 \rangle~,
    \label{quench-setup}
\end{equation}
identical to \eqref{evolve}, if we identify the circuit time $s$ with physical time $t$.

Usually we choose the initial Hamiltonian $H_0$ from some free  theory, such that the initial state $|\Psi_0 \rangle$ is a product state, and the evolving Hamiltonian $H$ can be a deformation of $H_0$ some operator. The fortune of $|\Psi_0 \rangle$ during the quench can be observed by measuring physical quantities such as 2-point correlators of some local observables, entanglement entropy \cite{Caputa:2017ixa}, quantum complexity \cite{Camargo:2018eof}, etc. 
It is expected that during quantum quenches in generic, non-integrable systems, those probes asymptotically approach their values predicted by the micro-canonical ensemble, which is called quantum {\it thermalization}. A rigorous statement of this hypothesis is the ETH  \cite{Srednicki:1994mfb, Deutsch:1991msp}. The mathematical form of this conjecture \cite{Dalessio:2016} can be stated as follows
\begin{equation}
    \langle E_m | \hat{A} | E_n \rangle = \bar{A} \left( E \right) \delta_{m,n} + R_{m,n} \Omega \left(E  \right)^{-1/2} f_A \left( E, \omega \right) ~,
    \label{eth}
\end{equation}
where $\hat{A}$ is some local observable/operator, and $|E_n \rangle $ labels the eigenstates of the Hamiltonian $H$ with energy $E_n$. Hence, the LHS of \eqref{eth} represents the matrix elements of $\hat{A}$ in the basis of eigenstates. Denoting the mean energy by $E = (E_m+E_n)/2$ and energy differences by $\omega=E_m-E_n$, on the RHS, we have a smooth function $\bar{A} (E)$ that represents the thermal expectation value of $\hat{A}$ at energy $E$, and another smooth, (intensive) function $f_A (E,\omega)$. The $R_{m,n}$ stands for a pseudo-random variable with zero mean and unit variance, and the density of states is denoted by $\Omega(E)$. The case where $m=n$ in \eqref{eth} is called the {\it diagonal} ETH in the literature \cite{Shiraishi:2018}. One may have realized that the initial state is absent in the diagonal ETH, which implies that information about {\it any} initial condition will be scrambled and ``forgotten" as the quantum system evolves.  

The ``strong diagonal ETH" hypothesis provides a scenario where the entire set of eigenstates $\{ |E_n\rangle \}$ spanning the Hilbert space obeys \eqref{eth}. On the other hand, we might ``settle for the next best thing" and require that only a (large) subset of eigenstates obeys the diagonal ETH. This is the weak diagonal ETH hypothesis that permits the existence of different time evolution depending on the initial states, i.e., we can effectively decompose the Hilbert space into subspaces
\begin{equation}
    \mathscr{H} \approx \mathscr{H}_{\text{thermal}} \oplus \mathscr{H}_{\text{non-thermal}}~.
    \label{hilbertdecomp}
\end{equation}

When the dimension of the non-thermal subspace $\mathscr{H}_{\text{non-thermal}}$ is not exponentially small, relatively to the dimension of the thermal subspace $\mathscr{H}_{\text{thermal}}$, it is possible to observe the non-thermal phenomena in experiments. According to the structure of $\mathscr{H}_{\text{non-thermal}}$, we can roughly divide the quantum systems that break the strong ETH while obey the weak ETH into two types: 
One is called {\it quantum many-body scars} (QMBS) \cite{Turner:2018} if $\mathscr{H}_{\text{non-thermal}}$ is spanned by a few highly excited eigenstates of the Hamiltonian $H$. Usually, the number of these highly excited eigenstates grows exponentially slower than the size of the entire Hilbert space.
The other is called {\it Hilbert space fragmentation} \cite{Sala:2020}, where the Hilbert space splits into exponentially many dynamically disconnected parts, of which some are non-integrable and expected to show quantum thermalization, while others are not. Hence the Hilbert space is further decomposed into
\[ \mathscr{H} \approx \left( \bigoplus_j^{K_{\text{thermal}}} \mathscr{H}_{{\text{thermal}},j} \right) \oplus \left( \bigoplus_{j'}^{K_{\text{non-thermal}}} \mathscr{H}_{{\text{non-thermal}},j'} \right)~, \]
where $K_{\text{thermal}} + K_{\text{non-thermal}} \sim e^L$ for system size $L$. One can further divide Hilbert space fragmentation into weakly fragmented and strongly fragmented, depending on whether there is a dominant\footnote{By dominant we mean that the growth of the subspace dimension is comparable with that of the entire Hilbert space.} non-integrable Hilbert subspace \cite{Sala:2020}. 

This short review should be sufficient for setting up the terminology regarding the structure of Hilbert spaces and ETH. Clearly, ETH is a relatively restrictive scenario and it is hard to verify it in generic quantum systems, except a handful of setups \cite{Kim:2014,Garrison:2018}\footnote{Although it is  believed to be an important feature of holographic CFTs. See e.g. recent discussion in \cite{Kawamoto:2024vzd}.}. Before we move on, we should also mention that there are certainly quantum systems that break weak diagonal ETH, of which one class is the well-known integrable systems and the other class is called {\it many-body localized} (MBL) systems (see e.g. \cite{Nandkishore:2014kca}).
\subsection{Krylov Restricted Thermalization and Spectrum-Generating Algebra}\label{krtvssga}
Let us now focus on the structure of the Hilbert space. Suppose we perform the Lanczos algorithm mentioned in \autoref{spreadcomplex} starting from a known, thermal or non-thermal, initial quantum state\footnote{Here ``a known thermal or non-thermal quantum state" means the state lies entirely in one of the two Hilbert subspaces $\mathscr{H}_{\text{thermal}}$ and $\mathscr{H}_{\text{non-thermal}}$, since arbitrary thermal or non-thermal $|\Psi_0 \rangle$ may contain components from both subspaces, leading to the Krylov basis generated spanning entire Hilbert space. In conclusion, due to the arbitrariness of the initial state, it remains unclear whether the Krylov subspace generated is a thermal one or a non-thermal one  unless we already know whether the initial state is thermal or non-thermal. This is a shortcoming of KRT, as we mentioned in \autoref{intro}.}. Because the two Hilbert subspaces $\mathscr{H}_{\text{thermal}}$ and $\mathscr{H}_{\text{non-thermal}}$ are disconnected, the Lanczos algorithm only generates a basis within one of the two Hilbert subspaces. Whether a thermal or non-thermal basis is generated depends on the Hilbert subspace where $|\Psi_0 \rangle$ belongs to. A direct corollary from this is that the time evolution of $|\Psi_0 \rangle$ under system's Hamiltonian $H$, $|\Psi (t) \rangle$ remains within the same subspace, since the Krylov basis is exact. 
This mechanism for producing exactly embedded subspaces of the Hilbert space is named {\it Krylov restricted thermalization} (KRT) \cite{Moudgalya:2019vlp}. However, despite its power in searching exactly embedded subspaces, KRT does not guarantee a specific dynamics of $|\Psi(t) \rangle$. As we reviewed before, there are still not-trivial steps from the Krylov basis to $|\Psi(t) \rangle$ that require the amplitudes $\varphi_n(t)$ solving the discrete Shr\"odinger equation \eqref{dschrodinger}, hence a good understanding of the Lanczos coefficients.

Instead, assume that we start from an eigenstate of the Hamiltonian $H$ with energy $E$, $|\mathcal{E} \rangle$, rather than some arbitrary quantum state $|\Psi_0 \rangle$. 
 If we are able to define a local operator $\hat{Q}^{\dagger}$ that satisfies the following commutation relation with the Hamiltonian
\begin{equation}
    [ H, \hat{Q}^{\dagger} ] = \omega \hat{Q}^{\dagger}~,
    \label{eq:SGA}
\end{equation}
then, within the subspace $|\mathcal{E} \rangle$, we can generate a tower of eigenstates with energy $\{ E+ n\omega, n=0,1,2,\cdots \}$ by repeated action of $\hat{Q}^{\dagger}$ on $|\mathcal{E} \rangle$, i.e., $\left\{\left(\hat{Q}^{\dagger} \right)^n |\mathcal{E} \rangle, n=0,1,2,\cdots \right\}$. These states also span the Hilbert subspace. Moreover, the presence of such tower has implications on the dynamics of $|\Psi(t) \rangle$ in the same subspace. This is usually studied using fidelity i.e., the norm square of the return-amplitude $S(t)$ \eqref{returnamp}
\begin{equation}
    | S(t) |^2 = \sum_{m,n} |c_n c_m |^2 e^{i (E_m -E_n) t}~.
    \label{fidelitySGA}
\end{equation}
Indeed, it is not difficult to see that $| S(t) |^2$ shows revivals \cite{Eberly:1980} with period $T = \frac{2\pi}{\omega}$, significantly differing from quantum thermalization. Such a mechanism that constructs a family of highly excited non-thermal eigenstates is called {\it spectrum-generating algebra} (SGA) \cite{Bohm1971DynamicalGA}. The advantage of SGA is exactly the disadvantage of KRT -- the former unveils the non-thermal behaviors well. However, there is always a trade-off -- it is not always possible to find $\hat{Q}^{\dagger}$ in practice\footnote{Sometimes the degree of difficulty in applying spectrum-generating algebra to a model is up to whether the Hamiltonian contains a certain symmetry. Successful cases include the Hubbard model \cite{Yang:1989,Zhang:1990}, the Spin-1 XY magnets \cite{Schecter:2019}, and the AKLT model \cite{Moudgalya:2018nuz}.}.

The contrast between the two mechanisms illustrated in this section helps us to sketch a feasible strategy for generating non-thermalizing Hilbert subspaces, or to be more concrete, quantum many-body scars. That is, we first produce the embedded subspaces of the Hilbert space by KRT. This, of course, leads to complicated discrete Schr\"odinger equations for $\varphi_n(t)$. However, as pointed out in \autoref{spreadcomplex}, the Lanczos coefficients $a_n$s and $b_n$s play a crucial rule in determining the dynamics of $\varphi_n(t)$s. Thus, for dynamics of $\varphi_n(t)$s to result in the quantum revival of $|\Psi(t)\rangle$, Lanczos coefficients must exhibit some characteristic features that we should determine. To achieve this, we will need a specific model that we discuss next.
\section{Quantum Many-Body Scarring from Experiment: the PXP Model} 
\label{pxpmodel}
The PXP model \cite{Lesanovsky:2012} has attracted significant attention in quantum simulations, particularly with experimental realizations in Rydberg quantum simulators using rubidium and strontium atoms \cite{Bernien:2017ubn}. In the limit where the nearest neighbour interaction is much larger than the detuning and the Rabi frequency, the system can be effectively described by a spin-$\frac{1}{2}$ Hamiltonian on a chain of size $L$
\begin{equation}
    H_{\text{PXP}} = \sum_{i=1}^LP_{i-1}X_{i}P_{i+1}~,
    \label{pxp-hamiltonian}
\end{equation}
where $P_i = \ket{0}\bra{0}$ is the Projector to $\ket{0}$ state, and $X_i=\ket{0}\bra{1}+\ket{1}\bra{0}$ is the Pauli X operator, with the subscript $i$ indicating that the operator acts on the $i^{\text{th}}$ site. The presence of the projector $P$ arises from a mechanism called Rydberg Blockade, subject to which adjacent atoms cannot simultaneously occupy $\ket{1}$ states, leading to a non-trivial local constraint.
The most striking experimental observation is the non-thermalizing behavior associated with quantum many-body scars. When initialized in specific states (such as e.g. the Néel state $\ket{{Z}_2}=\ket{1010...}$), the system exhibits persistent oscillations with amplitudes that decay slowly over time \cite{Bernien:2017ubn}.

The PXP model has the spatial inversion symmetry $\mathcal{I}:i\to L-i+1$ that commutes with the Hamiltonian $[\mathcal{I},H_{\text{PXP}}]=0$. Its two eigenvalues are $+1$ (for symmetric eigenstates, $\mathcal{I}|\psi\rangle=+1|\psi\rangle$) and $-1$ (for antisymmetric eigenstates, $\mathcal{I}|\psi\rangle=-1|\psi\rangle$) \cite{Buijsman:2022bkj}:
\begin{equation}
    \mathcal{I}\ket{\sigma_1,\sigma_2,...,\sigma_L} = \ket{\sigma_L,...,\sigma_2,\sigma_1}~,
\end{equation}
where $\ket{\sigma_i}$ stands for $\ket{0}$ or $\ket{1}$.
The model also shows a time-reversal symmetry $\mathcal{C}=\prod_iZ_i$, with $Z_i$ the Pauli Z operator acting on the $i^{\text{th}}$ site. $\mathcal{C}$ anti-commutes with the PXP Hamiltonian $\{\mathcal{C},H_{\text{PXP}}\}=0$
, making the spectrum of $H_{\text{PXP}}$ symmetric about 0, that is, for every eigenstate $\ket{E}$ with energy $E$, there exists another state $\mathcal{C}\ket{E} $ with energy $-E$. Furthermore, if the {\it periodic boundary condition} (PBC) $i \to i \bmod{L}$ has been applied, which will be the case in the rest of our paper, the PXP model acquires an additional translation symmetry $\mathcal{T}:i \to i+1$ with eigenvalue $e^{ip}$, where $p$ represents the momentum of the eigenstates. When considering the PBC, the PXP Hamiltonian \eqref{pxp-hamiltonian} becomes
\begin{equation}
    H_{\text{PXP}} = \sum_{i=1}^L P_{(i-1) \bmod L}X_{i}P_{(i+1) \bmod L}~,
    \label{pxp-hamiltonian-pbc}
\end{equation}
and this will be the form that we restrict to in all of our work.

Despite these symmetries, the results of \cite{Turner:2018} indicate that for relatively small lattices, the energy level spacing of the PXP model closely follows a semi-Poisson distribution. However, as the lattice size increases, the level statistics gradually converges towards a Wigner-Dyson distribution. In both cases, the behavior deviates from the Poisson level statistics typically associated with integrable systems, suggesting that the PXP model exhibits more non-integrable features.
\subsection{Quantum Revivals in the PXP Model} \label{revival}
In the Hilbert space of the PXP model, one can single out a class of special states $| {Z}_k \rangle$ that exhibit intriguing dynamical behavior under Hamiltonian evolution. Formally, these states can be written as
\begin{equation}
    |{Z}_k\rangle = |\dots {1}\underbrace{{0}\dots {0}}_{k-1}\,{1}\dots\rangle,\qquad  k-1 \in \mathbb{Z}^+~.
    \label{Zkqubit}
\end{equation}
Particularly noteworthy is the N\'eel state $| {Z}_2 \rangle$, which is prominently associated with the phenomenon of quantum many-body scars, demonstrating persistent coherent oscillations and revivals. In contrast, states with higher periodicity (such as $|{Z}_4\rangle$, etc.) show weaker and faster-decaying revivals, indicating significantly reduced dynamical stability.

\begin{figure}[H]
    \centering
    \subfloat[$|O\rangle$]{
        \includegraphics[width=0.48\textwidth,height=0.25\textheight, keepaspectratio]{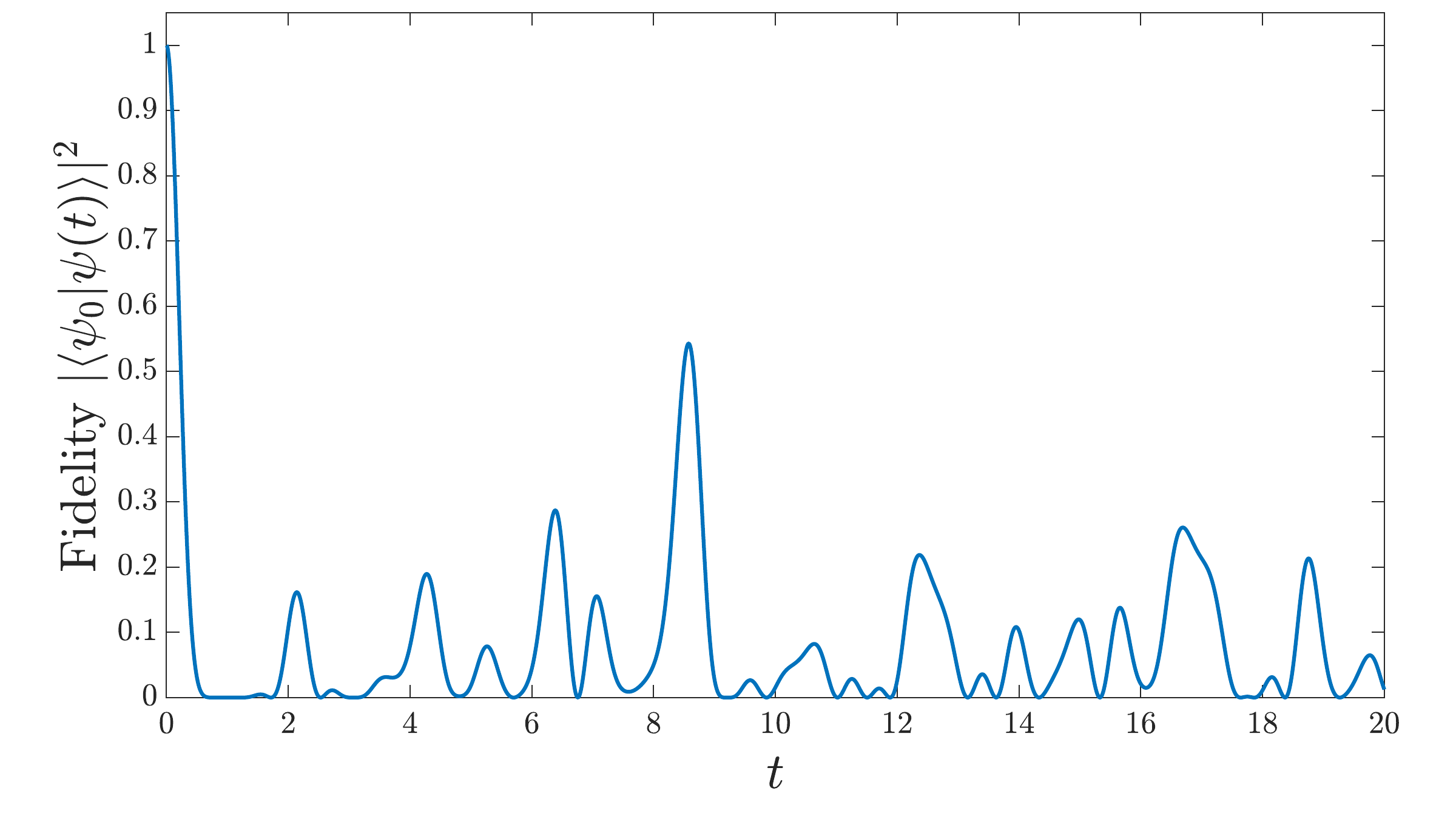}
        \label{fig:fiZ020}
    }
    \subfloat[$| {Z}_2 \rangle$]{
        \includegraphics[width=0.48\textwidth,height=0.25\textheight, keepaspectratio]{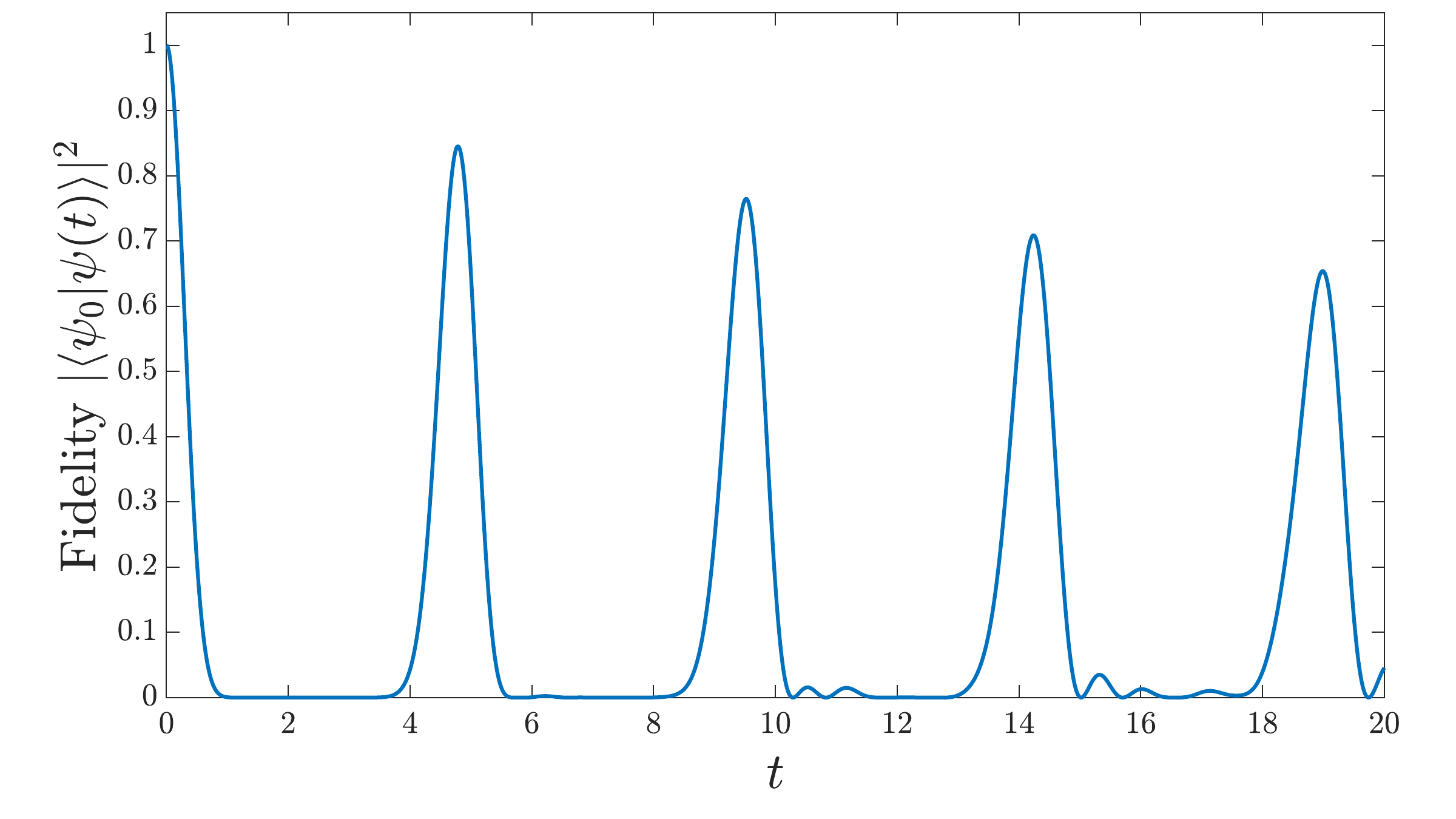}
        \label{fig:fiZ220}
    }\\
    \subfloat[$| {Z}_3 \rangle$]{
        \includegraphics[width=0.48\textwidth,height=0.25\textheight, keepaspectratio]{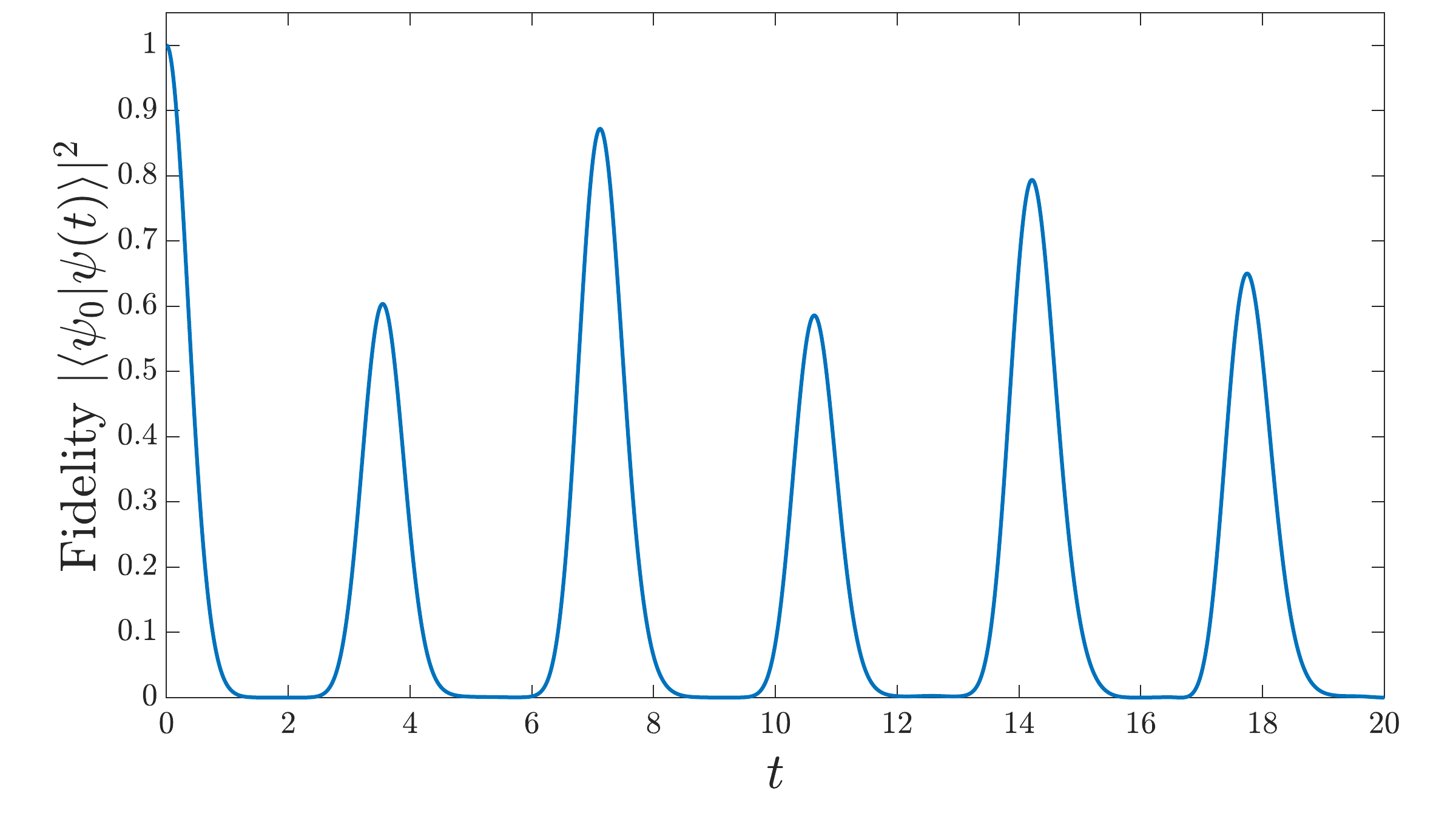}
        \label{fig:fiZ320}
    }    
    \subfloat[$| {Z}_4 \rangle$]{
        \includegraphics[width=0.48\textwidth,height=0.25\textheight, keepaspectratio]{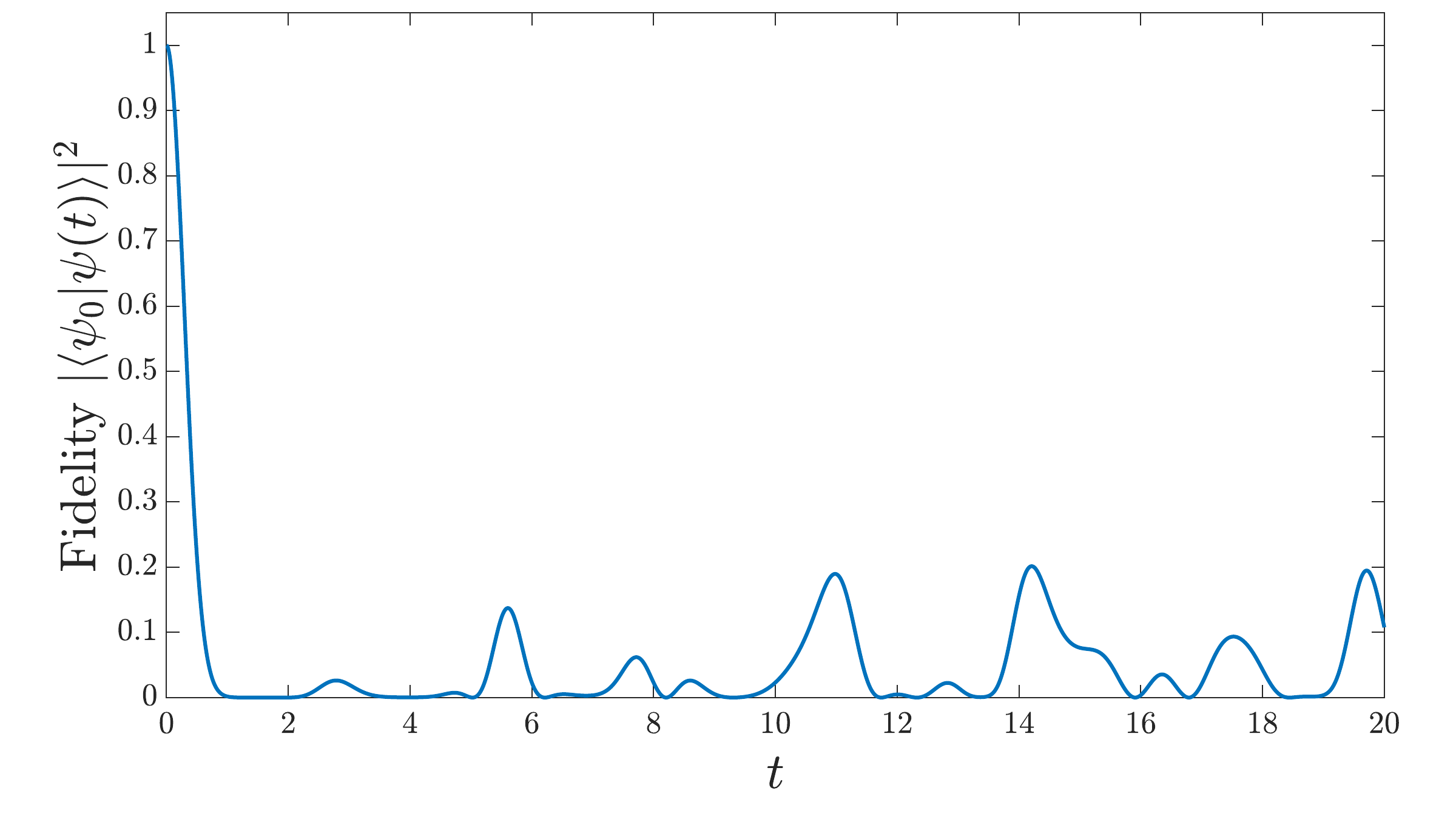}
        \label{fig:fiZ420}
    }   
    \caption{Fidelity of the time evolution of different initial states $| {Z}_k \rangle$, in particular, $|O\rangle$ (\ref{fig:fiZ020}), $|Z_2\rangle$ (\ref{fig:fiZ220}), $|Z_3 \rangle$ (\ref{fig:fiZ320}) and $|Z_4\rangle$ (\ref{fig:fiZ420}), at early times, subject to the PXP Hamiltonian in \eqref{pxp-hamiltonian-pbc} with lattice size $L=12$. Note that $| {Z}_2 \rangle$ and $| {Z}_3 \rangle$ show strong revivals.}
    \label{fig:fi20}
\end{figure}

In \autoref{fig:fi20} we plot early time evolution of fidelity $\mathcal{F}=|\langle \psi(t=0)| \psi(t) \rangle|^2 $ for different  $ |{Z}_k \rangle$ states, as well as the state $|O\rangle \equiv |0 \rangle^{\otimes L}$ (also denoted as $|Z_{\infty}\rangle$ in the rest of our paper), where a definition of $|\psi(t) \rangle$ is given in \eqref{quench-setup}.

It can be clearly seen that the $| {Z}_2 \rangle$  and $| {Z}_3  \rangle$ exhibit more pronounced oscillatory evolution with higher overlaps with the initial state compared to other $| {Z}_n \rangle$ states, and this feature becomes even more prominent at larger lattice sizes.

\begin{figure}[H]
    \centering
    \subfloat[$|Z_2\rangle$]{ \label{fig:fiz2long}
    \includegraphics[width=0.9\textwidth]{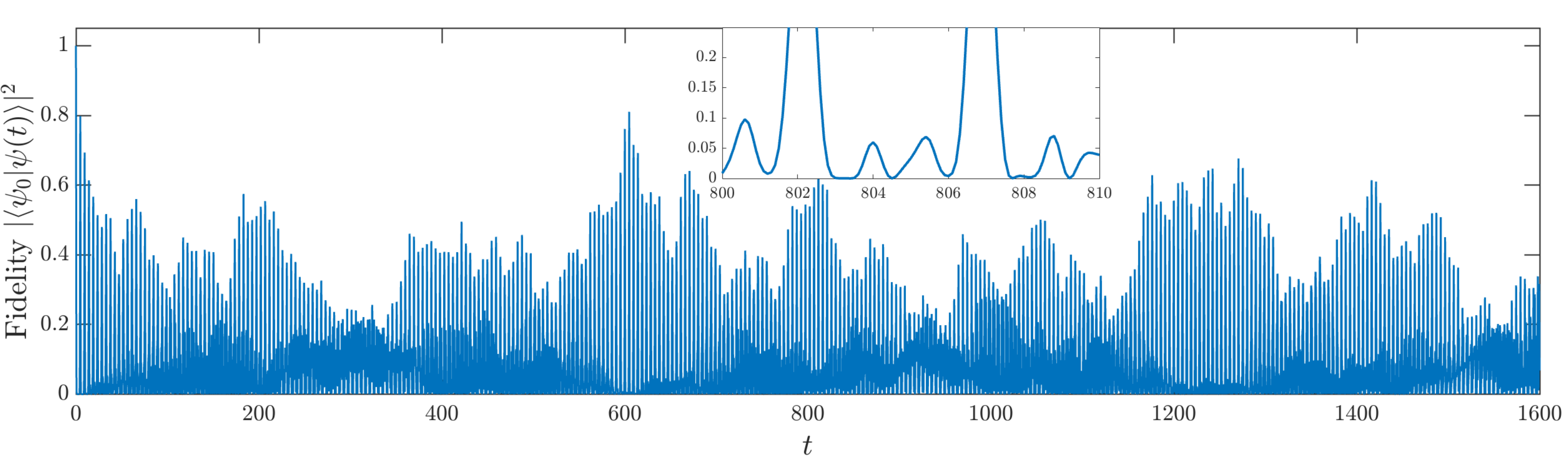}}\\
    \subfloat[$|Z_4\rangle$]{ \label{fig:fiz4long}
    \includegraphics[width=0.9\textwidth]{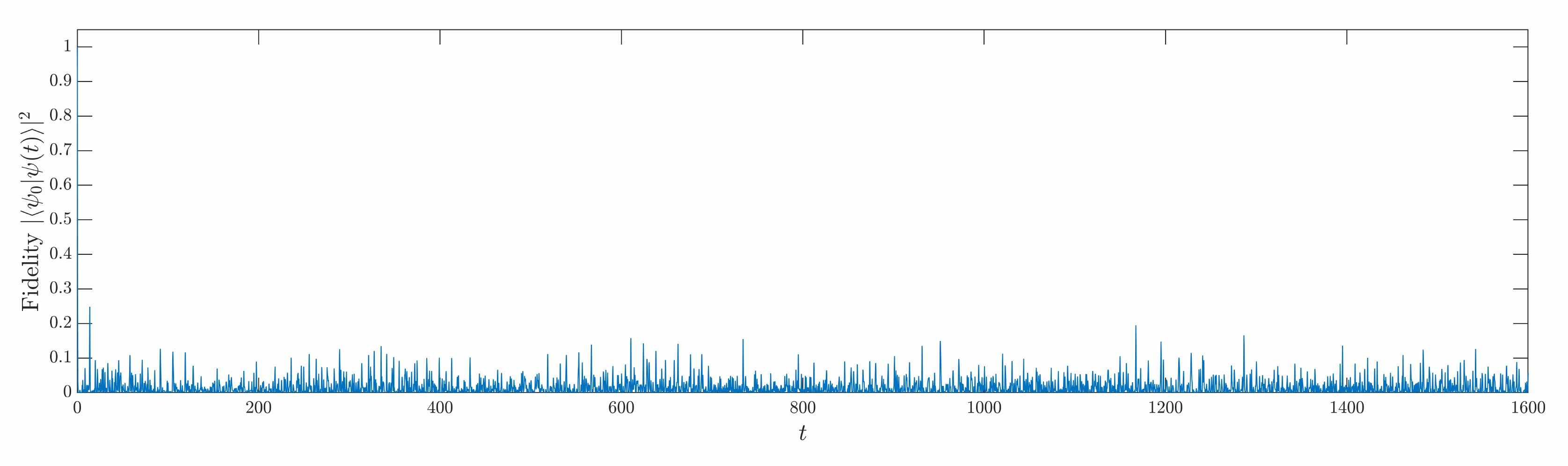}}
    \caption{Fidelity of the long-time evolution of the two initial states \ref{fig:fiz2long}) $|Z_2 \rangle$ and \ref{fig:fiz4long}) $|Z_4 \rangle$. Both are computed for the lattice size $L = 16$. Note that there is only one plot in (\ref{fig:fiz2long}) and a zoom-in view is given -- the darker parts come from the small peaks that rise up near the main peaks.}
    \label{longf}
\end{figure}

For completeness, in \autoref{longf} we also show the long-time evolution of fidelity for the initial states $|Z_2\rangle$ and $|Z_4\rangle$. It can be clearly seen that, the fidelity for $|Z_2 \rangle$ still shows a strong revivals, while the one for $|Z_4 \rangle$ rapidly losses the overlap with its initial form.

\section{Investigations of the PXP Model Conducted by Lanczos Algorithm} \label{lanczos-alg}
\subsection{Clues} \label{lanczos-numeric}
Given that our setup involves an abrupt quantum quench, as shown in \eqref{quench-setup}, where the system transitions from a free theory with a product ground state (e.g. $|Z_k\rangle$, with $k \in \mathbb{Z}^+$ or $|O\rangle$) to the PXP model, this scenario can be interpreted within the spread complexity framework \cite{Balasubramanian:2022tpr}. Specifically, if we identify the physical time $t$ with the circuit time $s$ in \eqref{evolve}, we can systematically analyze the time evolution of the system using the Krylov basis approach. This allows us to explore the dynamical properties of the system within the Krylov subspace and assess how thermalization is constrained within this restricted framework. 

Following the logic of the Lanczos algorithm, we first consider the two sets of the Lanczos coefficients $a_n$s and $b_n$s. Fortunately for the PXP Hamiltonian acting on a product state as the initial state, only $b_n$s survive. In this setup, this happens because any state generated from a product state by the PXP Hamiltonian (say $H_{\text{PXP}}^m|\psi_0 \rangle$) consists of product states that can only be transformed into each other by executing the PXP Hamiltonian an even number of times. As a result, acting the PXP Hamiltonian on $H_{\text{PXP}}^m|\psi_0 \rangle$ once does not keep any of its contained parts invariant. This observation contradicts \eqref{recurrel} unless $a_n$s are all zero.

We computed the Lanczos coefficients $b_n$s for the initial states $| {Z}_2 \rangle$, $| {Z}_3 \rangle$ and $| {Z}_4 \rangle$. It should be pointed out that, to reduce numerical errors, the algorithm that we used was not the original Lanczos algorithm but the {\it Full Orthogonalization} (FO) \cite{Parlett:1998, Rabinovici:2020ryf}. A comparison between these two algorithms and a third algorithm, {\it Partial Re-Orthogonalization} (PRO) \cite{Simon:1984, Rabinovici:2020ryf}, is given in \autoref{lanalgrev}.

\begin{figure}[b!]
    \centering
    \subfloat[\centering{Lanczos coefficients $b_n$ for $\ket{{Z}_2}$ state with enlarged arch structure and $L = 16$}\label{fig:bnZ2L16}]{
        \includegraphics[width=0.48\textwidth]{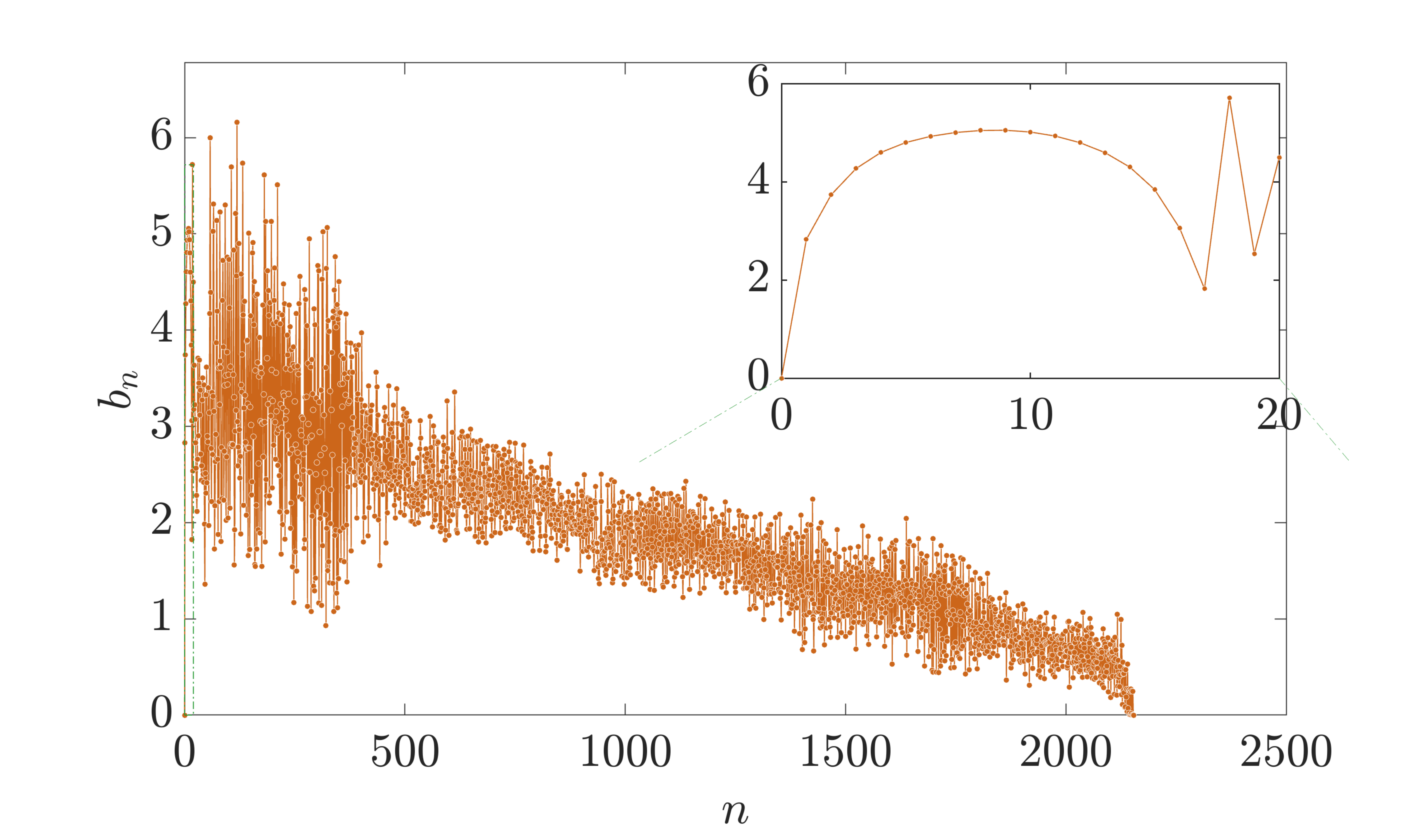}
    }
    \hfill
    \subfloat[\centering{The arch structure of $b_n$ for $\ket{{Z}_2}$ with  $L=12,\ 14, \ 16,\ 18,\ 20$}\label{fig:archZ2}]{
        \includegraphics[width=0.48\textwidth]{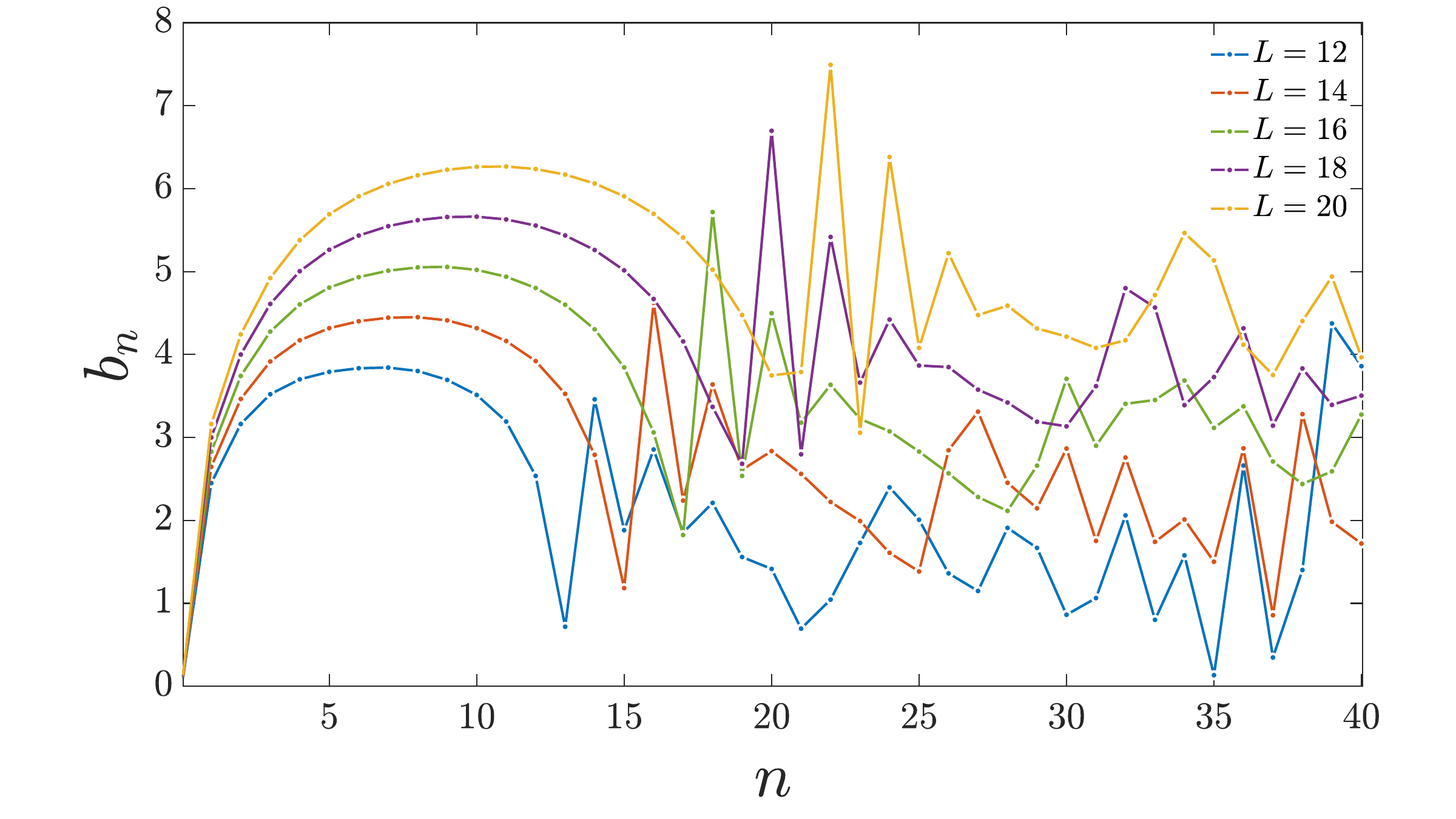}
    }
    
    \subfloat[\centering{The arch structure of $b_n$ for $\ket{{Z}_3}$ with $L=12, \ 18,\ 24$}\label{fig:archZ3}]{
        \includegraphics[width=0.48\textwidth]{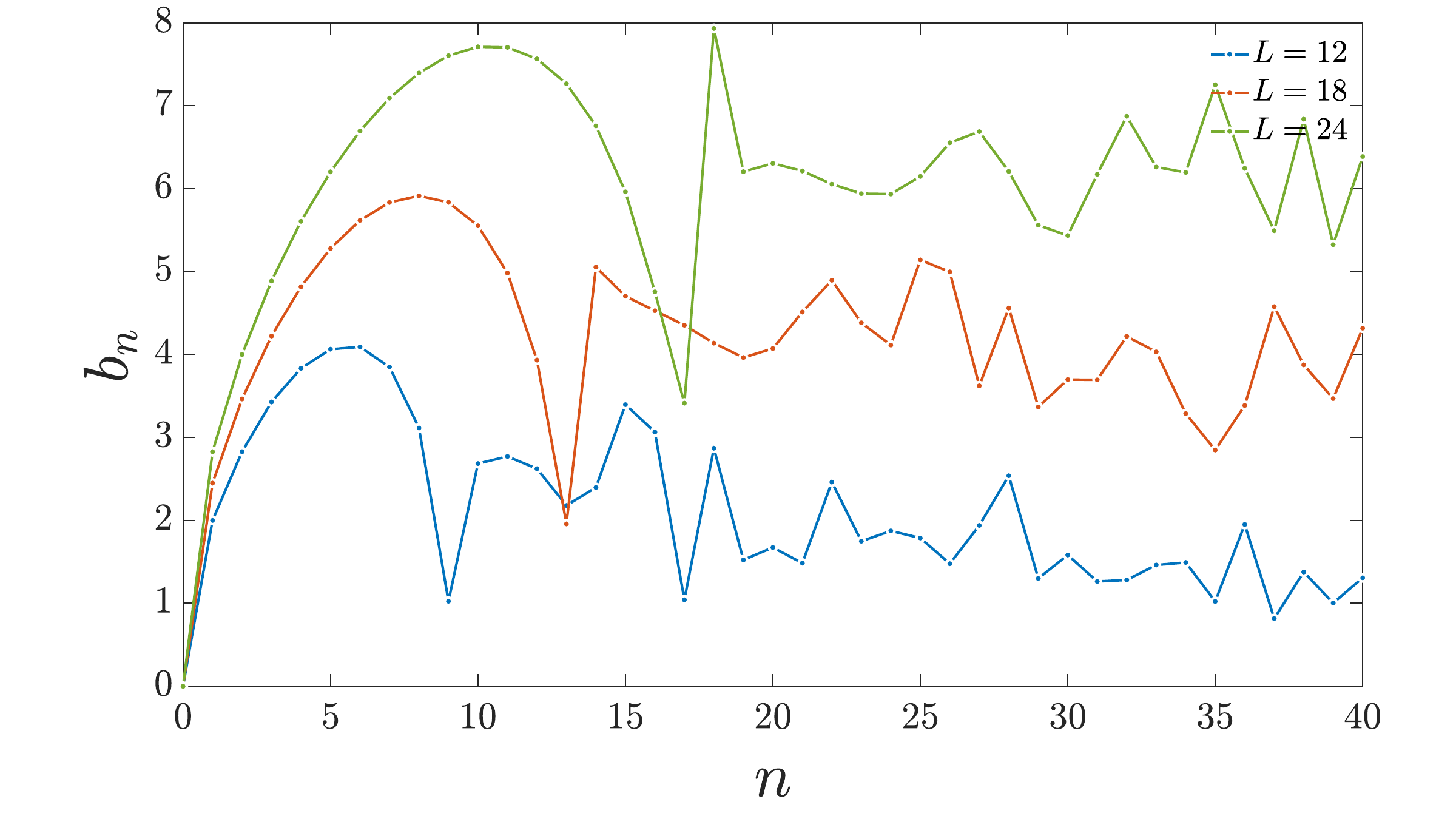}
    }    
	\hfil
    \subfloat[\centering{The arch structure of $b_n$ for $\ket{{Z}_4}$ with $L=12, \ 16,\ 24$}\label{fig:archZ4}]{
        \includegraphics[width=0.48\textwidth]{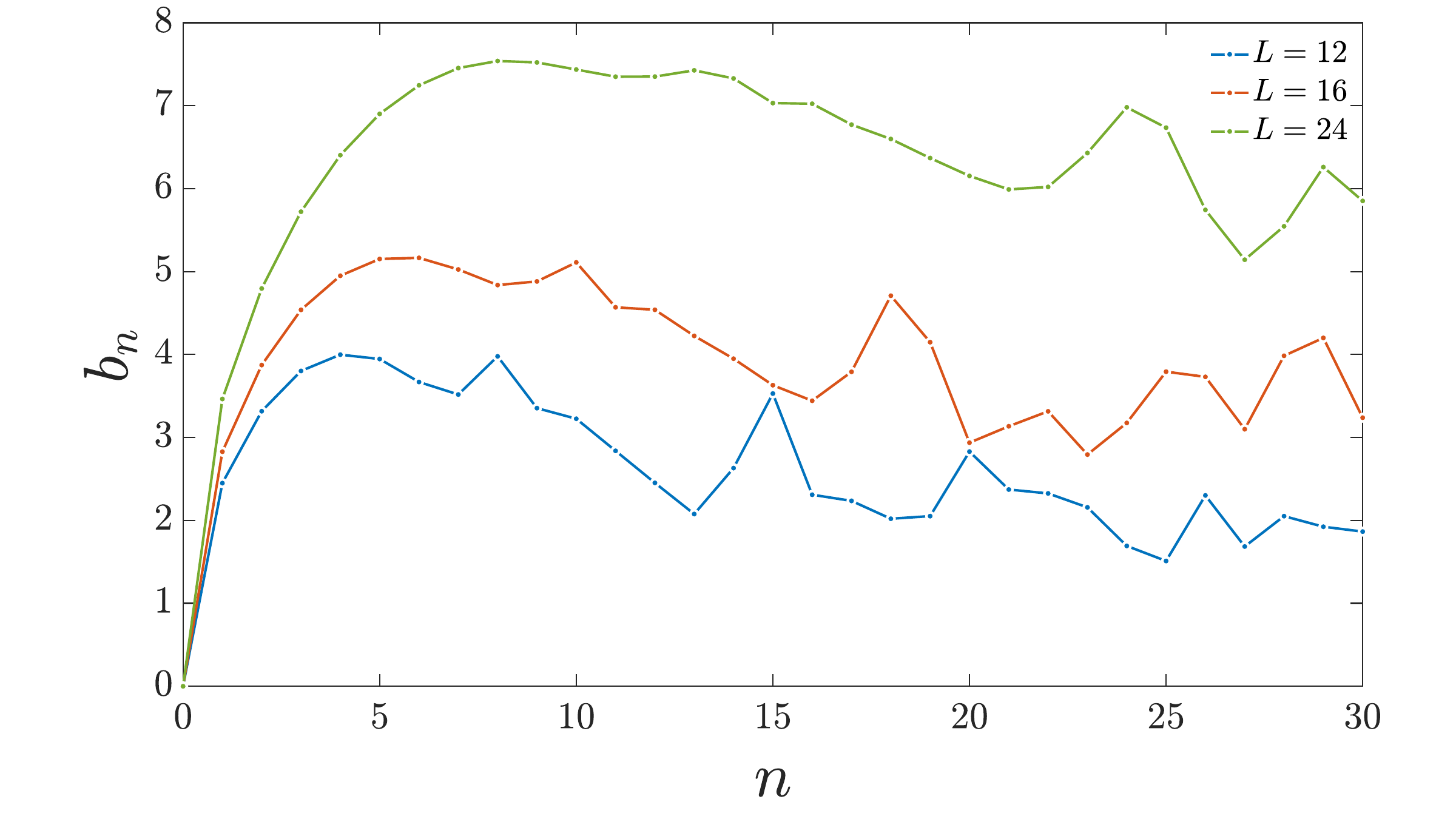}
    }
    \caption{Lanczos coefficients $b_n$ generated by applying the revised Lanczos algorithm, Full Orthogonalization (FO) \cite{Parlett:1998, Rabinovici:2020ryf}, to the PXP Hamiltonian in \eqref{pxp-hamiltonian-pbc} based on initial states \ref{fig:bnZ2L16}, \ref{fig:archZ2}) $|Z_2 \rangle$, \ref{fig:archZ3}) $|Z_3\rangle$, and \ref{fig:archZ4}) $|Z_4\rangle$. Various lattice sizes $L$ are considered.}
    \label{fig:bn}
\end{figure}

It can be clearly seen from \autoref{fig:bn} that the behavior of $b_n$ as functions of $n$, is divided into two regimes, the arch-like growth and decrease (that we simply refer to as the arch in the rest of this paper) followed by the peak and oscillations (that, using architecture analogy, we refer to as the buttress), by some $n$ of the order of the lattice size $L$.

We will first try to elucidate the origin of the arch and its meaning. However, before a more careful analysis of the PXP Hamiltonian \eqref{pxp-hamiltonian-pbc}, we first consider an analytic, intuitive example based on the Lie algebra symmetry \cite{Balasubramanian:2022tpr} that we discussed in \autoref{spreadcomplex}.
Namely, consider the $\mathfrak{su}(2)$ Lie algebra (or more exactly, the standard basis of $\mathfrak{s}l_2(\mathbb{C})$), $J_{\pm}$ and $J_z$ with commutation relations 
\begin{equation}
\left[J_z, J_{ \pm}\right]= \pm J_{ \pm}~, \quad\left[J_{+}, J_{-}\right]=2 J_z~,
\end{equation}
and the Hamiltonian of the form
\begin{equation}
H=\alpha\left(J_{+}+J_{-}\right)+\gamma J_z+\delta \mathbbm{1}~.
\label{eq:su2H}
\end{equation}
If we now select our initial state $|\psi_0\rangle$ as the lowest weight state $|j,-j\rangle$ with eigenvalue $-j$ of the $J_z$ operator (or $j$ of the Casimir), the $m^{\text{th}}$ Krylov basis vector coincides with the natural basis of the algebra obtained by acting $m$ times with the rasing operator $J_+$ on $|j,-j\rangle$ (and normalizing), i.e., it is exactly the eigenstate of $J_z$ with eigenvalue $-j+m$ and the Krylov basis dimension is $2j+1$: $|K_m\rangle = |j,-j+m\rangle,~m=0,1,2,\cdots,2j$. Moreover, Lanczos coefficients can be easily extracted from the algebra by using \eqref{recurrel} and \eqref{lanczoscoe}
\begin{equation}
a_n=\gamma(-j+n)+\delta, \qquad b_n=\alpha \sqrt{n(2 j-n+1)} .
\end{equation}
Notice that Lanczos coefficients $b_n$ above show perfect arched structure. This connection to the PXP model has also been the main observation of \cite{Bhattacharjee:2022qjw,Nandy:2023brt}.

Observe that \eqref{eq:su2H} can also be written as $H=2 \alpha J_x+\gamma J_0+\delta \mathbbm{1}$ which is understood as an analogous 
spin pointing along a direction in the $xz$-plane. By rotating this spin around the $y$-axis by an angle $\theta=\arctan{\frac{2\alpha}{\gamma}}$, the Hamiltonian becomes
\begin{equation}
H=E_0 J_z^{\prime}+\delta \mathbbm{1}~,\qquad E_0=\sqrt{4\alpha^2+\gamma^2}.
\end{equation}
This Hamiltonian is diagonal in the new rotated  basis, and its spectrum $\{E_m=(-j+m)E_0+\delta,~ m=0,1,2, \cdots, 2j \}$ has equal energy spacing $E_0$. Furthermore, we can construct new ladder operators
\begin{equation}
\begin{aligned}
& L_{+}=-\frac{2 \alpha}{E_0} J_z+\frac{1}{2}\left(\frac{\gamma+E_0}{E_0}\right) J_{+}+\frac{1}{2}\left(\frac{\gamma-E_0}{E_0}\right) J_{-}~, \\
& L_{-}=-\frac{2 \alpha}{E_0} J_z+\frac{1}{2}\left(\frac{\gamma-E_0}{E_0}\right) J_{+}+\frac{1}{2}\left(\frac{\gamma+E_0}{E_0}\right) J_{-}~,
\end{aligned}
\end{equation}
and verify their commutation relations with the Hamiltonian \eqref{eq:su2H}
\begin{equation}
{\left[H, L_{+}\right]=E_0 L_{+}} ~,\qquad {\left[H, L_{-}\right]=-E_0 L_{-}}~.
\end{equation}
Comparing with \eqref{eq:SGA}, we find that the ladder operator $L_{+}$ plays the role of the local operator $\hat{Q}^{\dagger}$ in the SGA scenario, implying the existence of a QMBS tower $\{E+mE_0, m =0,1,2,\cdots \}$, where we already know $E= -j E_0 +\delta$ according to the spectrum of the Hamiltonian. 

In conclusion, the Hamiltonian \eqref{eq:su2H} has an equally-gapped spectrum and, the fidelity of any initial state will show revivals. At the same time, we observe a perfect arched structure for the growth of the Lanczos coefficients $b_n$ versus $n$. 
A lesson we can learn from this toy example is that the appearance of the arched structure for Lanczos coefficients may reveal the presence of the QMBS tower and the scarring behavior. However, we still need more supporting evidence for this. Indeed, this analytic example is too simple to explain the buttress and oscillations in the PXP model. Hence we now move directly to the PXP Hamiltonian.
\subsection{Inferences}\label{lanczos-analytic}
Next, we try to understand the behavior of the Lanczos coefficients in the PXP model shown in \autoref{fig:bn}. For that we go back to the PXP Hamiltonian of $L= 2\ell, \ell \in \mathbb{N}$ sites in \eqref{pxp-hamiltonian-pbc}, which consists of the cubic terms $P_{i-1} X_i P_{i+1}$ that are usually not easy to study. We note that the situation can be improved by combining two consecutive qubits into one quqit. In particular, we introduce
\begin{equation}
   | \underline{2\iota_1 + \iota_2} \rangle \equiv | \iota_1 \iota_2 \rangle, \qquad \iota = 0,1,
\end{equation}
where in the RHS there is a qubit spanned by the basis $\{ |\underline{0}\rangle, |\underline{1}\rangle, |\underline{2}\rangle, |\underline{3}\rangle \}$. Here we have underlined the quqit states to distinguish them from the original qubit states. But actually the initial states we are interested\footnote{We are not interested in any initial state with a $|11\rangle$ occupation, since it results in the breakdown of the periodic boundary condition of the PXP Hamiltonian. In particular, an initial state with occupation $|1\rangle_i|1\rangle_{i+1}$ is always annihilated by the terms $P_{i-1}X_{i}P_{i+1} +P_iX_{i+1}P_{i+2}$ of the PXP Hamiltonian. As a result, we effectively have a PXP Hamiltonian with open boundary conditions where these two terms are removed.} in and the states generated from them by the PXP Hamiltonian should not contain any $|\underline{3}\rangle = |11\rangle$ -- as mentioned in the previous section, simultaneous occupation of $|1\rangle$ states on two adjacent sites is not allowed -- hence in practice we only need $\ell$ qutrits to represent a $2\ell$-qubit lattice subject to the PXP Hamiltonian\footnote{Even though we have developed this independently, we would like to note that it is somewhat similar to the idea of embedding the constrained Hilbert space of the PXP model into a subset of the full Hilbert space of the AKLT model. Readers can refer to \cite{Choi2019wqq} for the latter.}. 

As a result, each term in the PXP Hamiltonian acting on three consecutive qubits, $P_{i-1} X_i P_{i+1}$, now operates on two consecutive qutrits in the following way: 
\begin{itemize}
\item for odd $i$, we have the qubit-qutrit correspondence $|\iota_{i-2} ~\iota_{i-1} \iota_{i} \iota_{i+1} \rangle = |\underline{\eta_L ~\eta} \rangle $ :
\begin{eqnarray}
    |\iota_{i-2} ~\iota_{i-1} \iota_{i} \iota_{i+1} \rangle & \mapsto &
    \begin{cases}
        |\iota_{i-2} ~\iota_{i-1}~ (1-\iota_{i})~ \iota_{i+1} \rangle, & \iota_{i-1} =\iota_{i+1} =0 \\
        0, & \text{others}
    \end{cases}  \nonumber \\
  \Longrightarrow   |\underline{\eta_L ~ \eta} \rangle &\mapsto& 
\begin{cases}
|\underline{\eta_L } \rangle \otimes
\begin{Bmatrix}
|\underline{2-\eta)} \rangle, & \eta =0,2 \\
0, & \eta =1
\end{Bmatrix}, &\eta_L=0,2 \\
0 , & \eta_L=1
\end{cases} \quad  , \label{lmap} 
\end{eqnarray}
\item for even $i$, we have the qubit-qutrit correspondence $ |\iota_{i-1} \iota_{i} \iota_{i+1}~ \iota_{i+2} \rangle = |\underline{\eta ~\eta_R} \rangle$ :
\begin{eqnarray}
 |\iota_{i-1} \iota_{i} \iota_{i+1}~\iota_{i+2}  \rangle & \mapsto &
    \begin{cases}
        |\iota_{i-1} (1-\iota_{i}) \iota_{i+1}~\iota_{i+2}  \rangle, & \iota_{i-1} =\iota_{i+1} =0 \\
        0, & \text{others}
    \end{cases}  \nonumber \\
  \Longrightarrow 
|\underline{\eta~ \eta_R} \rangle &\mapsto &
\begin{cases}
\begin{Bmatrix}
|\underline{1-\eta)} \rangle, & \eta =0,1 \\
0, & \eta =2
\end{Bmatrix}
\otimes |\underline{\eta_R } \rangle , &\eta_R=0,1 \\
0 , & \eta_R=2
\end{cases} \quad . \label{rmap}
\end{eqnarray}
\end{itemize}
Similarly to the spin-$1/2$ operators $P$ and $X$ expressed by Pauli matrices $\frac{\mathbbm{1}_2+\sigma_3}{2}$ and $\sigma_1$, respectively, we can introduce the matrix representations of the operations in \eqref{lmap} to be $\left( \lambda_1^2 \right)_L \lambda_1$ and the operations in \eqref{rmap} to be $\lambda_6 \left( \lambda_6^2 \right)_R$, where 
\begin{equation}
    \lambda_1 = 
    \begin{pmatrix}
        0 & 1 & 0 \\
        1 & 0 & 0 \\
        0 & 0 & 0 \\
    \end{pmatrix}\,, \quad \text{and} \quad
     \lambda_6 = 
    \begin{pmatrix}
        0 & 0 & 0 \\
        0 & 0 & 1 \\
        0 & 1 & 0 \\
    \end{pmatrix}\,,
\end{equation}
are the two of the eight Gell-Mann matrices $\{\lambda_a, a =1,2,\cdots, 8 \}$, and we have let the vector space be
\begin{equation}
a |\underline{2}\rangle+ b|\underline{0}\rangle +c |\underline{1}\rangle \fallingdotseq 
\begin{pmatrix}
a \\
b \\
c \\
\end{pmatrix}~.
\label{vectorsp}
\end{equation}
In conclusion, by utilizing the equivalence between two consecutive qubits and one qutrit  in the initial states of interest, the PXP Hamiltonian can be written as  
\begin{equation}
H_{\text{PXP}} = \sum_{i}^{\ell} \left[ (\lambda_1^2)_{(i-1) \bmod \ell} (\lambda_1)_{i} + (\lambda_6)_{i} (\lambda_6^2)_{(i+1) \bmod \ell}  \right]~.
\label{pxp-hamiltonian-GM}
\end{equation}
It is an operator constructed from Gell-Mann matrices,
with periodic boundary condition $\ell \pm 1 \equiv (\ell \pm 1) \mod{\ell}$.

Notice that the Gell-Mann matrices span the Lie algebra of the $\operatorname{SU}(3)$ group in the defining representation, but we can equivalently express the PXP Hamiltonian in the standard basis of the $\operatorname{SU}(3)$ group (actually the Cartan-Weyl basis of $\mathfrak{s}l_3(\mathbb{C})$ algebra in the fundamental representation $D(1,0)$), the necessary background of which is summarized in Appendix \ref{sl3c-rootspace}. By doing this, we arrive at the ultimate form of the PXP Hamiltonian that we will work with in this paper, namely
\begin{eqnarray}
H_{\text{PXP}} &=& \sum_{i=1}^{\ell} \frac{1}{3}\left[ \left( 2\mathbbm{1}_3 + H_{12} + 2H_{23}  \right)_{(i-1) \bmod \ell} \left( E_{\alpha}+ E_{-\alpha} \right)_{i} \right.\nonumber \\
&+&  \left.\left(E_{\beta}+ E_{-\beta} \right)_{i} (2\mathbbm{1}_3 - 2H_{12} - H_{23} )_{(i+1) \bmod \ell}  \right]~,
\label{pxp-hamiltonian-sl3c}
\end{eqnarray}
where we denoted the $\mathfrak{s}l_3 (\mathbb{C})$ generators as
\begin{eqnarray}
    H_{12} = \begin{pmatrix}
        1 & & \\
        & -1 & \\
         & & 0 \\
    \end{pmatrix}~, \qquad &\qquad&
    H_{23} = \begin{pmatrix}
        0 & & \\
        & 1 & \\
         & & -1 \\
    \end{pmatrix}~, 
    \nonumber \\
    E_{\alpha} = \begin{pmatrix}
        0 & 1 & \\
        & 0 & \\
         & & 0 \\
    \end{pmatrix} = E_{-\alpha}^{\dagger}~, &\qquad& 
    E_{\beta} = \begin{pmatrix}
        0 &  & \\
        & 0 & 1 \\
         & & 0 \\
    \end{pmatrix} = E_{-\beta}^{\dagger}~,
\end{eqnarray}
and, in order to close the algebra, we also need the remaining two generators
\begin{equation}
    E_{\gamma} = \begin{pmatrix}
        0 &  & 1 \\
        & 0 &  \\
         & & 0 \\
    \end{pmatrix} = E_{-\gamma}^{\dagger}~.
\end{equation}
This way, the operators forming the PXP Hamiltonian \eqref{pxp-hamiltonian-sl3c} obey the following commutation relations of $\mathfrak{s}l_3 (\mathbb{C})$
\begin{equation}
\begin{split}
& [E_{\alpha}, E_{-\alpha}] = H_{12}~,  \quad  [E_{\beta}, E_{-\beta}] = H_{23}~, \quad   [E_{\gamma}, E_{-\gamma}] = H_{12}+H_{23} \equiv H_{13} ~, \\
& [E_{\alpha},E_{\beta}] = E_{\gamma}~,  \qquad  [E_{\beta},E_{-\gamma}] = E_{-\alpha} ~, \quad  [E_{\alpha},E_{-\gamma}] = E_{-\beta} ~. \\
\end{split}
\label{sl3c-com}
\end{equation} 

Most importantly, Hamiltonian \eqref{pxp-hamiltonian-sl3c} can be reorganized into two, non-commuting parts: the linear part
\begin{equation}
    H_{\text{PXP,lin}} =  \sum_{i =1}^{\ell} \left[  \left( E_{\alpha}+ E_{-\alpha} \right)_{i} +  \left( E_{\beta}+ E_{-\beta} \right)_{i}   \right]~,
    \label{pxp-hami-sl3c-lin}
\end{equation}
and the residual part
\begin{eqnarray}
    H_{\text{PXP,res}} &=& \sum_{i =1}^{\ell} \frac{1}{3}\left[ \left( H_{12} + 2H_{23} - \mathbbm{1}_3 \right)_{(i-1) \bmod \ell} \left( E_{\alpha}+ E_{-\alpha} \right)_{i} \right. \nonumber\\
    &-& \left.   \left( E_{\beta}+ E_{-\beta} \right)_{i} (  2H_{12} + H_{23} +\mathbbm{1}_3 )_{(i+1) \bmod \ell}  \right]~.
    \label{pxp-hami-sl3c-quad}
\end{eqnarray}
It can be seen that the linear part describes a free theory (no interactions between different sites) while the residual part contains the nearest-neighbour interactions.
One of the crucial differences between these two parts is that the latter does NOT preserve the irreducible representations, where ``preserve" can be understood in the following sense. For $\mathfrak{s}l_3 (\mathbb{C})$ in its reducible representation $D(1,0)^{\otimes \ell}$, that corresponds to $\ell$-qutrit lattice, the Cartan-Weyl basis 
\begin{equation}
    E_{\mu} \equiv \sum_{i=1}^{\ell} \left( E_{\mu} \right)_i~, \quad \mu=\pm \alpha, \pm \beta, \pm \gamma~; 
    \qquad
    H_{I} \equiv \sum_{i=1}^{\ell} \left( H_{I} \right)_i~, \quad I=12, 23, 13~,
    \label{sl3c-redrep}
\end{equation}
or equivalently the $\mathfrak{su}(3)$ basis
\begin{equation}
    \Lambda_a \equiv \sum_{i=1}^{\ell} \left( \lambda_a \right)_i, \quad a=1,2,\cdots, 8~,
    \label{su3-redrep}
\end{equation}
contains two Casimir operators that commute with $\Lambda_a$s
\begin{equation}
   \hat {C}_{\text{quad}}^{D(1,0)^{\otimes \ell}}=\sum _{a}{\Lambda}_{a}{\Lambda}_{a}\,,\qquad \qquad \hat {C}_{\text{cub}}^{D(1,0)^{\otimes \ell}}=\sum_{abc}d_{abc}{\Lambda}_{a}{\Lambda}_{b}{\Lambda}_{c}~,
   \label{casimir-redrep}
\end{equation}
where $d_{abc}$s are the symmetric coefficients of $\mathfrak{su}(3)$ given in \eqref{symm-coe}. When $D(1,0)^{\otimes \ell}$ is decomposed into irreducible representations, the matrices of the Cartan-Weyl basis and, therefore, of the Casimir operators break into diagonal blocks of irreducible representations simultaneously. This implies that 
\begin{equation}
    \left[H_{\text{PXP},\text{lin}} ~, ~\hat{C}_{\text{quad}}^{D(p,q)} \right] = \left[H_{\text{PXP},\text{lin}} ~,~ \hat{C}_{\text{cub}}^{D(p,q)} \right] = 0~,
\label{casimir-pxplin}
\end{equation}
is true for any irreducible representation of $\mathfrak{s}l_3 (\mathbb{C})$ denoted as $D(p,q)$. However, the same commutation relations do NOT hold for $H_{\text{PXP,res}}$. Derivation of this statement is provided in \autoref{casimir-pxpquad}. In the following, we will argue that such a significant difference between the linear \eqref{pxp-hami-sl3c-lin} and the residual \eqref{pxp-hami-sl3c-quad} part of the PXP Hamiltonian has strong implications on quantum dynamics and scars. 

First, starting from the initial states $|Z_k \rangle,~ k-1 \in \mathbb{Z}^+$ or $|O \rangle$ (where $|Z_k \rangle$ is now in the form of $\ell$ qutrits, and we denote $|Z_{\infty}\rangle \equiv |O \rangle$), and following the Lanczos algorithm, 
we can express the Krylov subspace generated by the PXP Hamiltonian \eqref{pxp-hamiltonian-sl3c} as
\begin{eqnarray}
    \mathscr{Kr}_{\text{PXP}} &=& \operatorname{span} \{ |Z_k \rangle, H_{\text{PXP}} |Z_k \rangle, H_{\text{PXP}}^2 |Z_k \rangle, \cdots \} \nonumber  \\
    &= & \operatorname{span} \left\{ |Z_k \rangle, ~(H_{\text{PXP,lin}} + H_{\text{PXP,res}} )|Z_k \rangle,~ (H_{\text{PXP,lin}} + H_{\text{PXP,res}})^2 |Z_k \rangle, \cdots \right\}\,.\nonumber\\
\end{eqnarray}
Then, because $H_{\text{PXP,lin}}$ preserves the irreducible representations, its repeated action on $|Z_k \rangle$ will not move the resulting states out of the particular representation that $|Z_k \rangle$ belongs to. 
Thus, the part of Krylov subspace generated only by the action of $H_{\text{PXP,lin}}$
\begin{equation}
\mathscr{Kr}_{\text{PXP,lin}}= \operatorname{span} \{ |Z_k \rangle, H_{\text{PXP,lin}} |Z_k \rangle , \cdots \}\,,
\end{equation}
should be a subspace of an irreducible representation that $|Z_k \rangle$ belongs to, i.e., $D(p,q)$. This leads to the upper bound on the dimension of $\mathscr{Kr}_{\text{PXP,lin}}$ to be (see Appendix \ref{sl3c-rootspace}) 
\begin{equation}
    \operatorname{dim} \left( \mathscr{Kr}_{\text{PXP,lin}} \right) \le \operatorname{dim} \left( D(p,q) \right) = \frac{1}{2}(p + 1)(q + 1)(p + q + 2), \quad p,q \ge 0, ~ p+2q \le \ell~.
\end{equation}
We will return to this and provide a more accurate estimation of the dimension of the Krylov subspace generated by the linear PXP Hamiltonian in section \ref{lanczos-arch}. 

In contrast, we are unable to derive a similar estimate for $H_{\text{PXP,res}}$. However, we know that the PXP Hamiltonian in \eqref{pxp-hamiltonian-pbc} does NOT generate any states, living on the $2\ell$-qubit lattice, with consecutive $|1\rangle$s. Hence we will estimate the upper bound on the dimension of $\mathscr{Kr}_{\text{PXP,res}}= \operatorname{span} \{ |Z_k \rangle, H_{\text{PXP,res}} |Z_k \rangle , \cdots \}$, or, more precisely, of $\mathscr{Kr}_{\text{PXP}}$, as the number of the $2\ell$-qubit states with no consecutive $|1\rangle$s, which is given by the Lucas number \cite{hoggatt1969fibonacci},
\begin{equation}
    \mathcal{L}_{2\ell}= \left({1+{\sqrt {5}} \over 2}\right)^{2 \ell}+\left({1-{\sqrt {5}} \over 2}\right)^{2\ell}~.
\end{equation}
We leave the details of this derivation for \autoref{lucasN}, but we would like point out that the Lucas numbers, increase exponentially with the lattice size of fermions $2\ell$. This tells us that
\begin{equation}
   \operatorname{dim} \left( \mathscr{Kr}_{\text{PXP,res}} \right) < \operatorname{dim} \left( \mathscr{Kr}_{\text{PXP}} \right) \le \mathcal{L}_{2\ell} \sim \mathcal{O} (e^{2 \log \tfrac{1+\sqrt{5}}{2} \cdot \ell})~.
   \label{krdim}
\end{equation}
Now, recall that in section \ref{eth2hilbert} we mentioned that, typically, the size of the QMBS subspace grows exponentially slower than that of the entire Hilbert space, i.e., the former either grows polynomially with the lattice size or exponentially, but in some scale smaller than the lattice size. Hence, it is natural to expect that the Krylov subspace generated by linear PXP Hamiltonian $H_{\text{PXP,lin}}$ is a QMBS subspace, while the one generated by residual Hamiltonian $H_{\text{PXP,res}}$ is more likely responsible for the thermalizing subspace. This will be our working hypothesis in the following.

The first half of this hypothesis can be verified by constructing the operator $Q^{\dagger}$, whose commutation relation with $H_{\text{PXP,lin}}$ satisfies SGA \eqref{eq:SGA}. Indeed, notice that in \eqref{pxp-hami-sl3c-lin}
\begin{equation}
    \left( E_{\alpha}+ E_{-\alpha} \right)_{i} +  \left( E_{\beta}+ E_{-\beta} \right)_{i} = 
    \begin{pmatrix}
         &1 &  \\
        1 &  & 1 \\
         & 1 &  \\
    \end{pmatrix}_i = \sqrt{2} (J_x)_i~,
    \label{su2-basis-x}
\end{equation}
where $(J_x)_i$, together with 
\begin{equation}
    (J_y)_i = \frac{1}{\sqrt{2}}
    \begin{pmatrix}
         & -i &  \\
        i &  & -i \\
         & i &  \\
    \end{pmatrix}~, \qquad
    (J_z)_i =
    \begin{pmatrix}
        1 & & \\
        & 0 & \\
        & & -1 \\
    \end{pmatrix}~,
    \label{su2-basis-yz}
\end{equation}
are the three, spin-1 matrices with commutation relations
\begin{equation}
    [J_a, J_b]_i = i \epsilon_{abc} (J_c)_i~, \qquad a,b,c \in \{ x,y,z \}~.
    \label{su2-com}
\end{equation}
Therefore, we can define
\begin{equation}
    Q^{\dagger} = \sum_{i=1}^{\ell} \left( J_z - i J_y \right)_i = \sum_{i=1}^{\ell} \left[ H_{13} + \left( E_{-\alpha} - E_{\alpha} \right) + \left( E_{-\beta} - E_{\beta} \right) \right]_i\,,
\end{equation}
such that
\begin{equation}
    \left[ H_{\text{PXP,lin}}, Q^{\dagger} \right] = \sqrt{2} Q^{\dagger}~.
    \label{SGA-PXP}
\end{equation}
Thus, following the spirit of SGA from section \ref{krtvssga}, we can construct a QMBS tower, starting from some eigenstate of $H_{\text{PXP,lin}}$, and argue that the evolution of quantum fidelity $|S(t)|^2$ will undergo revivals with period $T= \sqrt{2}\pi $. Notice that this is also a special case of the toy model considered in section \ref{lanczos-numeric}, with Hamiltonian given by \eqref{eq:su2H}, $\alpha =\frac{1}{\sqrt{2}},~\gamma=\delta=0$, and the Casimir eigenvalue $j=1$.

To summarize, in the above, we have separated out the part of the PXP Hamiltonian that generates the QMBS subspace, namely $H_{\text{PXP,lin}}$ in \eqref{pxp-hami-sl3c-lin}, as well as the residual PXP Hamiltonian, $H_{\text{PXP,res}}$ in \eqref{pxp-hami-sl3c-quad}. In the remaining parts of this section we will study how their properties and interrelations are reflected by the patterns of the growth of the Lanczos coefficients.  
In particular, in section \ref{lanczos-arch} we investigate the linear PXP Hamiltonian, where, by exploiting the idea in \cite{Caputa:2021sib}, we are able to calculate the exact Lanczos coefficients generated from the initial states $|Z_k \rangle$ without knowing the concrete Krylov basis.
In section \ref{lanczos-buttress} we take into account the influence of the residual PXP Hamiltonian, which leads to the so-called ``$|\underline{12}\rangle$ issue". We explain how this issue violates Casimir invariance and therefore introduces other irreducible representations of $D(1,0)^{\otimes \ell}$ besides $D(\ell,0)$ into the Krylov basis generated by the full PXP Hamiltonian.
\subsubsection{The Arch} \label{lanczos-arch}
Let us start by filling out the details of the qubit-to-qutrit map that we discussed above. Indeed, we mentioned that the initial state $| Z_k \rangle, ~ k -1 \in \mathbb{Z}^+$ can be treated as a $\ell$ -qutrit state, but we have not yet provided its concrete form. It is given as follows
\begin{eqnarray}
    |Z_k \rangle &=& |1 \underbrace{00\cdots 0}_{k-1}  \rangle^{\otimes 2\ell/k} = (|\underline{2} \rangle \otimes | \underbrace{\underline{00\cdots 0}}_{\left\lfloor \tfrac{k}{2} \right\rfloor-1} \rangle \otimes \left| \underline{2- (k \bmod{2})} \right\rangle \otimes | \underbrace{\underline{00\cdots 0}}_{\left\lceil \tfrac{k}{2} \right\rceil -1 } \rangle )^{\otimes \ell/k} \nonumber \\
    &=&
    \begin{cases}
        | \underline{2} \underbrace{\underline{00\cdots0}}_{m-1}\rangle^{\otimes \ell/m}, & \quad k=2m ~, \\
        | \underline{2} \underbrace{\underline{00\cdots0}}_{m-1} \underline{1} \underbrace{\underline{00\cdots 0 }}_{m}\rangle^{\otimes \ell/(2m+1)} & \quad k=2m + 1 ~, \\
    \end{cases}  \label{zk-qutrit}
\end{eqnarray}
where $m \in \mathbb{Z}^+$. 

Next, we only consider the linear term $H_{\text{PXP,lin}}$ acting on $|Z_k \rangle$ and concentrate on the resulting Lanczos coefficients rather than the specific Krylov basis. It is clear that $H_{\text{PXP,lin}}$ in \eqref{pxp-hami-sl3c-lin} describes a free theory -- no interacting terms between sites -- and therefore the order of the $\ell$ qutrits does not matter. This is a good news, because, for extracting  the Lanczos coefficients, there is no difference between $H_{\text{PXP,lin}}$ acting on $|Z_k \rangle$ or acting on
\begin{equation}
    |\underline{2} \rangle^{\otimes \ell/k} \otimes \left| \underline{2- (k \bmod{2})} \right\rangle^{\otimes \ell/k} \otimes |\underline{0} \rangle^{\otimes \left(\ell \cdot \tfrac{k-2}{k} \right) }~.
    \nonumber
\end{equation}

Furthermore, if we apply a transformation $|\underline{1} \rangle \longleftrightarrow |\underline{2} \rangle$ to \eqref{vectorsp}, which permutes the $1^{\text{st}}$ row and the $3^{\text{rd}}$ row, as well as the $1^{\text{st}}$ column and the $3^{\text{rd}}$ column of all Gell-Mann matrices, $H_{\text{PXP,lin}}$ is invariant. This implies that the Lanczos coefficients generated by $H_{\text{PXP,lin}}$ acting on $|\underline{1} \rangle$ are the same as the ones generated by $H_{\text{PXP,lin}}$ acting on $|\underline{2} \rangle$. 

\begin{figure}[H]
    \centering
    \subfloat[repeated action of $H_{\text{PXP,lin}}$ on $|\underline{22222}\rangle$]{ \label{fig:rootlattice-hami-lin-Z2}
        \includegraphics[width=0.45\linewidth]{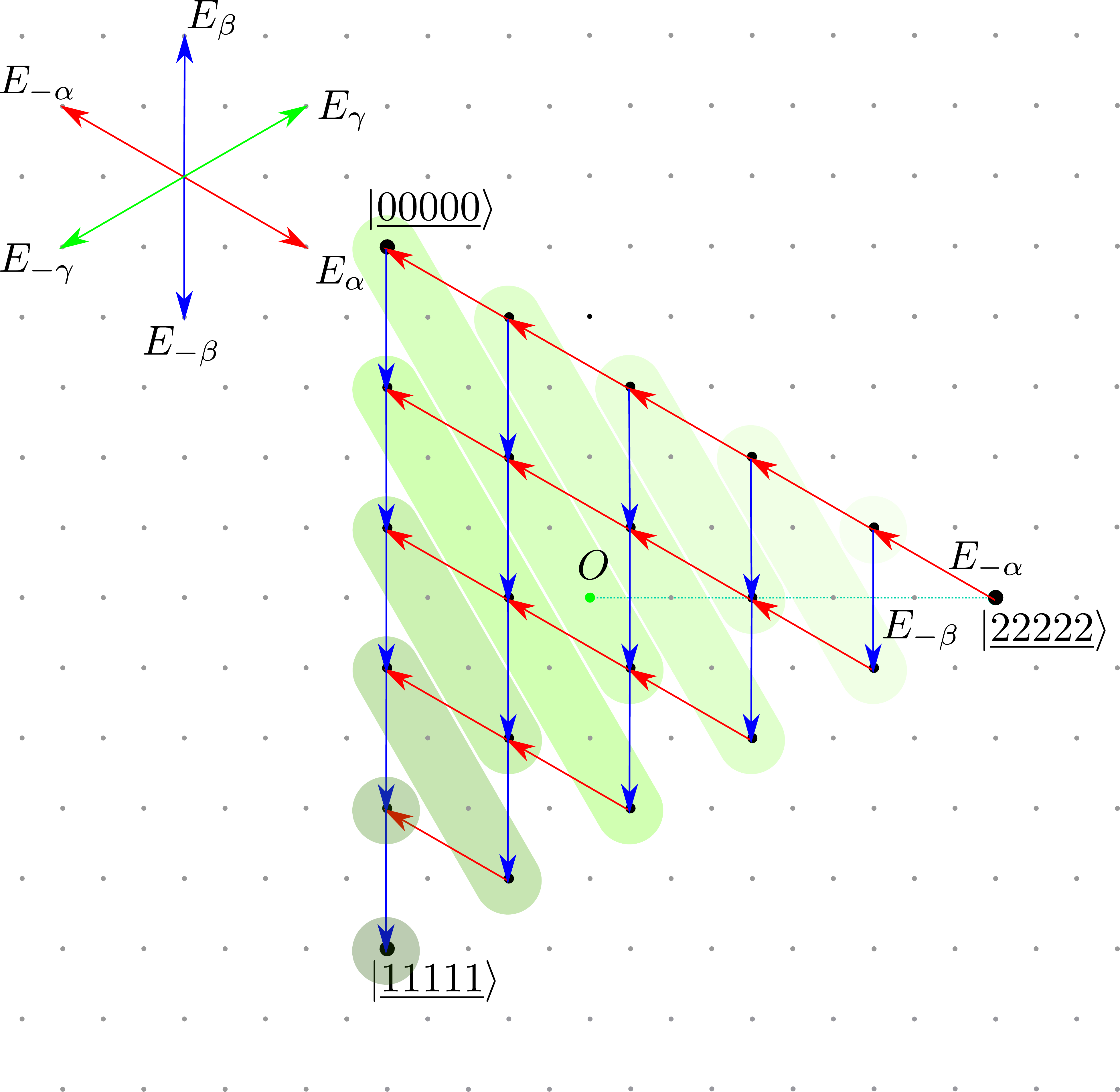}}
    \subfloat[repeated action of $H_{\text{PXP,lin}}$ on $|\underline{00000}\rangle$]{ \label{fig:rootlattice-hami-lin-O}
        \includegraphics[width=0.45\linewidth]{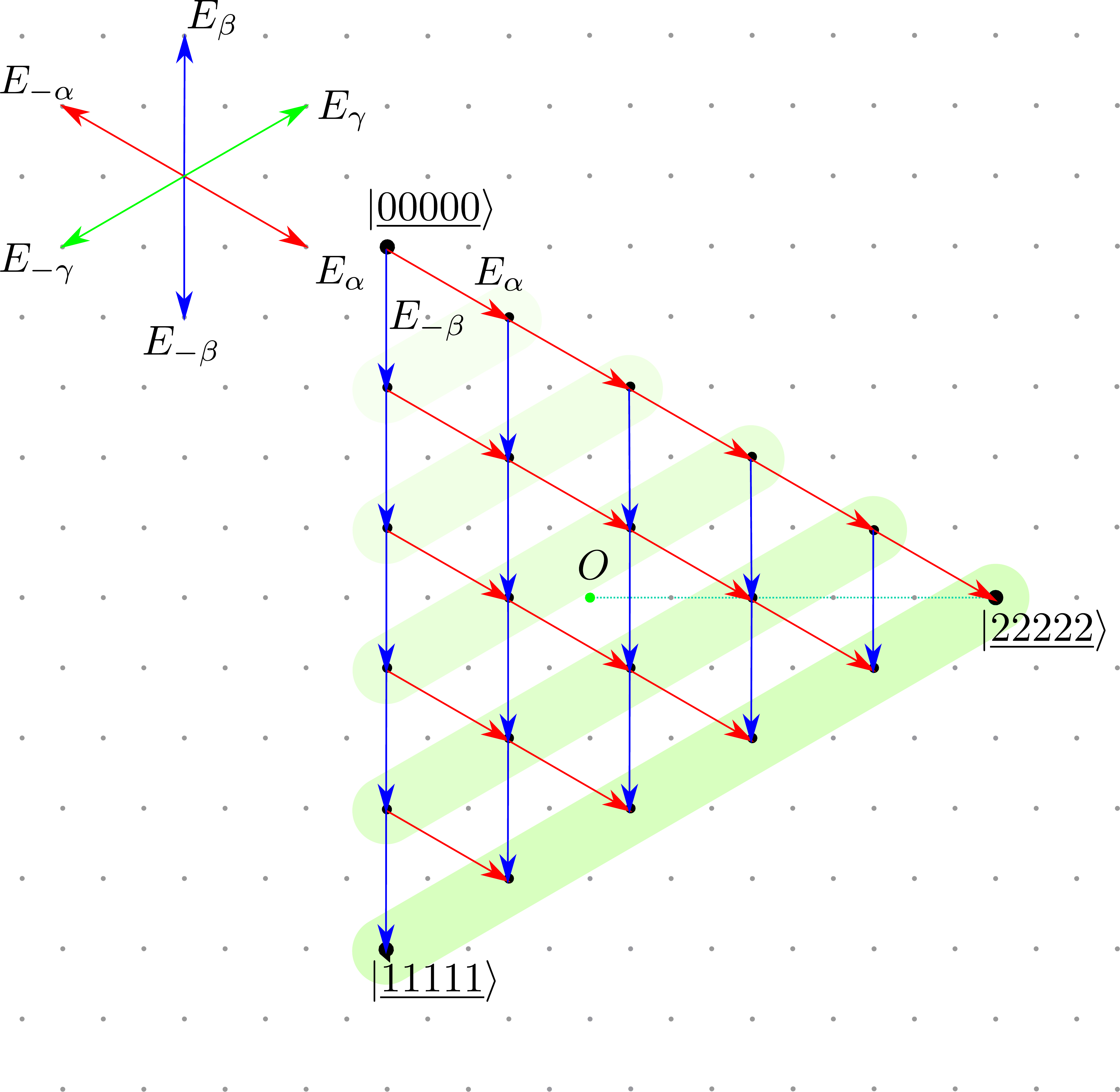}}
    \caption{A schematic diagram of how Cartan-Weyl basis in $H_{\text{PXP,lin}}$, $E_{\pm \alpha}$ and $E_{\pm \beta}$, acts as a raising or lowering ladder operators in the case where the initial states are \ref{fig:rootlattice-hami-lin-Z2}) $|\underline{2}\rangle^{\otimes n}$; \ref{fig:rootlattice-hami-lin-O}) $|\underline{0}\rangle^{\otimes n}$. Here we take $n=5$ and mark the effects of the raising ladder operators. Components that make up the same Krylov basis are circled in the same colour, where a darker colour represents the Krylov basis with a larger iteration number. }
    \label{fig:rootlattice-hami-lin}
\end{figure}

Based on these two observations, we consider an initial state 
\begin{equation}
    |Z_k' \rangle \equiv |\underline{2} \rangle^{\otimes \left(\ell \cdot \tfrac{2}{k} \right)}  \otimes |\underline{0} \rangle^{\otimes \left(\ell \cdot \tfrac{k-2}{k} \right) }~,
\end{equation}
instead of $|Z_k \rangle$ in \eqref{zk-qutrit}.
The advantage of $|Z_k' \rangle$ is that it is simply a product of two product states that can be analysed separately. Namely $|\underline{2} \rangle^{\otimes \left(\ell \cdot \tfrac{2}{k} \right)} $ and $ |\underline{0} \rangle^{\otimes \left(\ell \cdot \tfrac{k-2}{k} \right) }$. Both are totally symmetric, and belong to the $\mathfrak{s}l_3(\mathbb{C})$ irreducible representations $D \left( \ell \cdot \tfrac{2}{k},0 \right)$ and $D \left( \ell \cdot \tfrac{k-2}{k},0 \right)$ respectively (see to Appendix \ref{sl3c-young} for details). Then, since the highest weight vector of the irreducible representation $D(n,0), ~ n \in \mathbb{Z}^+$ is $n \omega_1$, the root lattice of $D(n,0)$ can be determined  using the strategy in Appendix \ref{sl3c-rootspace}, and becomes an equilateral triangle with the other two corners at $n \omega_2$ and $ n \omega_3$. Recall that, according to the conventions in \eqref{vectorsp}, $n\omega_1$, $n \omega_2$, and $n \omega_3$ correspond to the $n$-qutrit states $|\underline{2} \rangle^{\otimes n} $, $|\underline{0} \rangle^{\otimes n} $, and $|\underline{1} \rangle^{\otimes n} $, respectively. In conclusion, we can sketch the root lattice of the irreducible representation $D(n,0)$ and label the product states $|\underline{2} \rangle^{\otimes n} $ and $|\underline{0} \rangle^{\otimes n} $ that we are going to study, as \autoref{fig:rootlattice-hami-lin} shows.

Let us explain what we are going to do next. We will start from two initial product states that live in the corner of the $n$-qutrit lattice, $|\underline{2}\rangle^{\otimes n}$ and $|\underline{0}\rangle^{\otimes n}$, respectively. Utilizing the method in \cite{Caputa:2021sib}, which has been briefly reviewed in \autoref{spreadcomplex}, we reorganize the Cartan-Weyl basis of $\mathfrak{s}l_3 (\mathbb{C})$ in $H_{\text{PXP,lin}}$ into raising and lowering ladder operators
\begin{equation}
    H_{\text{PXP,lin}} = H_{\text{PXP,lin}}^+ + H_{\text{PXP,lin}}^- \,, 
    \label{pxp-ladderform}
\end{equation}
such that
\begin{equation}
    H_{\text{PXP,lin}}^+ | K_n \rangle = b_{n+1}  | K_{n+1} \rangle, \qquad
    H_{\text{PXP,lin}}^- | K_n \rangle = b_{n}  | K_{n-1} \rangle,
    \label{pxp-ladderop}
\end{equation}
where $|K_0 \rangle \equiv |\underline{2}\rangle^{\otimes n}~\text{or}~|\underline{0}\rangle^{\otimes n}$. As a result, the Lanczos coefficients can be read from the algebra.  To be more concrete, from the form of the linear PXP Hamiltonian in \eqref{pxp-hami-sl3c-lin}, the four candidates to construct $H_{\text{PXP,lin}}^{\pm}$ are the Cartan-Weyl basis $E_{\pm \alpha},~ E_{\pm \beta}$. The visualization of the actions of the Cartan-Weyl basis in the root lattice is shown by the arrows in \autoref{fig:rootlattice-hami-lin} -- the positions pointed by the arrows represent the states that result from their actions on the states at the starting positions of the arrows. Now, starting from one of the three corners of the root lattice $D(n,0)$, two of the Cartan-Weyl bases $E_{\pm \alpha},~ E_{\pm \beta}$ point to sites outside the root lattice, which means that these two operators annihilate the state corresponding to the corner and therefore play the role of the lowering ladder operator. As for the other two Cartan-Weyl bases, playing the role of raising ladder operator, their repeated actions generate the components of the Krylov basis. These components are the sites of root lattice $D(n,0)$. In \autoref{fig:rootlattice-hami-lin} we also explain how these components are generated by the Cartan-Weyl bases. Components that form the same Krylov basis are circled out in the same color. In the following we find the role that each Cartan-Weyl basis plays for initial states $|\underline{2}\rangle^{\otimes n}$ and $|\underline{0}\rangle^{\otimes n}$, respectively, and read the Lanczos coefficients.

Let us start with the initial state $|\underline{2}\rangle^{\otimes n}$. Since it corresponds to the highest weight vector $n\omega_1$ in the root lattice of $D(n,0)$, it must be annihilated by $E_{\alpha}$ and $E_{\beta}$. Therefore, the raising and lowering ladder operators in this case are (Figure \ref{fig:rootlattice-hami-lin-Z2})
\begin{eqnarray}
    H_{\text{PXP,lin}}^+ &=& \sum_{i=1}^n \left( E_{-\alpha} + E_{-\beta} \right)_i = \frac{1}{\sqrt{2}} \sum_{i=1}^n (J_-)_i ~, \nonumber \\
    H_{\text{PXP,lin}}^- &=& \sum_{i=1}^n \left( E_{\alpha} + E_{\beta} \right)_i  = \frac{1}{\sqrt{2}} \sum_{i=1}^n (J_+)_i ~.
    \label{ladderop-Z2}
\end{eqnarray}
where $(J_{\pm })_i \equiv (J_x)_i \pm i (J_y)_i$ with $(J_x)_i$ and $(J_y)_i$ the spin-1 matrices in \eqref{su2-basis-x} and \eqref{su2-basis-yz}. From \eqref{su2-com} we can derive their $\mathfrak{s}l_2 (\mathbb{C})$ (or more familiarly $\mathfrak{su}(2)$) commutation relations. 
\begin{equation}
   [ J_z, J_{\pm} ] = \pm J_{\pm}, \qquad [J_+ , J_-] = 2J_z~,e
\end{equation}
where we defined $J_a \equiv \sum_{i=1}^n (J_a)_i$ for $a \in \{x,y,z\}$. This means that we can utilize standard techniques from angular momentum in quantum mechanics. The angular momentum coupling of $n$ spin-1 particles gives the Casimir eigenvalue of the totally symmetric irreducible representation, which is $n$. Thus, the Krylov basis generated on $|\underline{2}\rangle^{\otimes n}$ consists of the eigenstates of $J_z$ with eigenvalues $-n,-n+1,\cdots, n-1,n$, i.e.
\begin{equation}
    |K_m \rangle = \left| n, n-m \right\rangle~, \qquad m = 0, 1, 2, \cdots, 2n\,.
\end{equation}
As for the Lanczos coefficients, they can be read from the action of the generators
\begin{equation}
\begin{aligned}
J_{+}|n, \mu\rangle &={\sqrt {(n-\mu)(n+\mu+1)}}|n,\mu+1\rangle ~ ,\\
J_{-}|n,\mu \rangle &= {\sqrt {(n+\mu)(n-\mu+1)}}|n,\mu-1\rangle ~,
\end{aligned}
\label{su2-ladderop}
\end{equation}
for $\mu = -n,-n+1,\cdots, n-1,n$. Comparing \eqref{su2-ladderop} with \eqref{pxp-ladderop}, we finally obtain
\begin{equation}
    b_m = \sqrt{\frac{1}{2} m (2n-m+1) }~, \qquad m =  1, 2, \cdots, 2n.
    \label{lanczos-coe-Z2}
\end{equation}
This is an arch of width $2n$ and the height $\left( n+1/2\right)/\sqrt{2}$.

Similarly, we can analyze the initial state $|\underline{0}\rangle^{\otimes n}$
which is annihilated by $E_{-\alpha}$ and $E_{\beta}$. Therefore, the raising and lowering operators in this case are (Figure \ref{fig:rootlattice-hami-lin-O})
\begin{eqnarray}
    H_{\text{PXP,lin}}^+ &=& \sum_{i=1}^n \left( E_{\alpha} + E_{-\beta} \right)_i = E_{\alpha} + E_{-\beta}~, \nonumber \\
    H_{\text{PXP,lin}}^- &=& \sum_{i=1}^n \left( E_{-\alpha} + E_{\beta} \right)_i = E_{-\alpha} + E_{\beta} ~,
\end{eqnarray}
where we have used the definitions in \eqref{sl3c-redrep}. Now, we calculate
\begin{equation}
\begin{split}
& \left(H_{\text{PXP,lin}}^+ \right)^m | n \omega_2 \rangle = ({E}_{\alpha} + {E}_{-\beta})^m | n \omega_2 \rangle
= \sum_{l=0}^m \binom{m}{l}{E}_{\alpha}^{m-l} {E}_{-\beta}^l  | n \omega_2 \rangle ~ , \qquad m \le n \\
=& \sum_{l=0}^m \binom{m}{l} \left[ \binom{n}{m-l} \binom{n-m+l}{l} \right]^{1/2} l! (m-l)! | n\omega_2 + (m-l) \omega_{12} + l \omega_{32} \rangle\,, \\
=&  \sum_{\tilde{l} = n-m}^n m! \left[ \binom{n}{\tilde{l}} \binom{\tilde{l}}{n-m}  \right]^{1/2} | n\omega_2 + (m-n+\tilde{l}) \omega_{12} + (n-\tilde{l}) \omega_{32} \rangle~,
\end{split}
\label{ladderop-O}
\end{equation}
where we have used the sites in the root lattice of the irreducible representation $D(n,0)$ to label the (normalized) states generated by $E_{\alpha}$ and $E_{\beta}$. This state is not yet normalized so we compute the norm square of $\left(H_{\text{PXP,lin}}^+ \right)^m | n \omega_2 \rangle$ that becomes
\begin{equation}
\begin{split}
\langle n \omega_2 | \left(H_{\text{PXP,lin}}^- \right)^m \left(H_{\text{PXP,lin}}^+ \right)^m | n \omega_2 \rangle 
= 2^m \frac{m!n!}{(n-m)!} ~.
\end{split}
\end{equation}
Notice that from \eqref{pxp-ladderop} we can see
\begin{equation}
\begin{split}
\left(H_{\text{PXP,lin}}^+ \right)^m | K_0 \rangle =&\left(H_{\text{PXP,lin}}^+ \right)^{m-1} b_1| K_{1} \rangle = \left(H_{\text{PXP,lin}}^+ \right)^{m-2} b_1 b_2 | K_{2} \rangle \\
=& \cdots = \prod_{l=1}^m b_l | K_{m} \rangle~,
\end{split}
\end{equation}
and the Lanczos coefficients are therefore
\begin{equation}
{b}_m = \left[ { \langle n \omega_2 | \left(H_{\text{PXP,lin}}^- \right)^m \left(H_{\text{PXP,lin}}^+ \right)^m | n \omega_2 \rangle \over \langle n \omega_2 | \left(H_{\text{PXP,lin}}^- \right)^{m-1} \left(H_{\text{PXP,lin}}^+ \right)^{m-1} | n \omega_2 \rangle } \right]^{1/2} 
=\sqrt{2 m (n-m+1)} ~, 
\label{lanczos-coe-O}
\end{equation}
with $m=1,2,\cdots,n$.
If we plot them, we can still see the arch with height $(n+1)/\sqrt{2}$, but in this case the width is $n$ rather than $2n$.  

With the ladder operators and the resulting Lanczos coefficients of the two cases in mind, we now go back to $|Z_k'\rangle$. Since we are not interested in the specific Krylov basis $|K_m \rangle$ generated from $|Z_k'\rangle$, which is not even correct for the true initial state $|Z_k \rangle$ in \eqref{zk-qutrit}, below we will use the notation $|K_m^{(2, n)} \rangle$ for the Krylov basis obtained from $|\underline{2} \rangle^{\otimes n}$ and $|K_m^{(0, n)} \rangle$ for the ones from $|\underline{0} \rangle^{\otimes n}$. Now, the linear PXP Hamiltonian can be rewritten as the sum of the raising and lowering ladder operators
\begin{eqnarray}
    H_{\text{PXP,lin}}^+ &=& \sum_{i=1}^{\ell \cdot \tfrac{2}{k}} \left( E_{-\alpha} +E_{-\beta} \right)_i  +  \sum_{j=1}^{\ell \cdot \tfrac{k-2}{k}} \left( E_{\alpha} +E_{-\beta} \right)_{\ell \cdot \tfrac{2}{k} +j } \equiv H_{\text{PXP,lin}}^{(2,\ell \cdot \tfrac{2}{k}, +)} + H_{\text{PXP,lin}}^{(0,\ell \cdot \tfrac{k-2}{k},+)} ~, \nonumber \\
     H_{\text{PXP,lin}}^- &=& \sum_{i=1}^{\ell \cdot \tfrac{2}{k}} \left( ~ E_{\alpha} ~+~E_{\beta}~ \right)_i  +  \sum_{j=1}^{\ell \cdot \tfrac{k-2}{k}} \left( E_{-\alpha} +E_{\beta} \right)_{\ell \cdot \tfrac{2}{k} +j } \equiv H_{\text{PXP,lin}}^{(2,\ell \cdot \tfrac{2}{k}, -)} + H_{\text{PXP,lin}}^{(0,\ell \cdot \tfrac{k-2}{k},-)} ~.\nonumber\\
\end{eqnarray}
Repeated action of $H_{\text{PXP,lin}}^+$ leads to
\begin{equation}
    \begin{split}
        \left( H_{\text{PXP,lin}}^+ \right)^m |K_0 \rangle = & \sum_{l=0}^m \binom{m}{l} \left[ H_{\text{PXP,lin}}^{(2,\ell \cdot \tfrac{2}{k}, +)}  \right]^l \left|K_0^{(2,\ell \cdot \tfrac{2}{k})} \right\rangle \otimes  \left[ H_{\text{PXP,lin}}^{(0,\ell \cdot \tfrac{k-2}{k},+)}\right]^{m-l} \left|K_0^{(0,\ell \cdot \tfrac{k-2}{k})} \right\rangle \\
        =& \sum_{l=0}^m \binom{m}{l} \prod_{i=1}^{l} b_{i}^{(2,\ell \cdot \tfrac{2}{k})} \prod_{j=1}^{m-l} b_{j}^{(0,\ell \cdot \tfrac{k-2}{k})} \left|K_l^{(2,\ell \cdot \tfrac{2}{k})} \right\rangle \otimes  \left|K_{m-l}^{(0,\ell \cdot \tfrac{k-2}{k})} \right\rangle~. \\
    \end{split}
\end{equation}
Thus, by calculating the inner product
\begin{equation}
    \begin{split}
       & \langle K_0 | \left( H_{\text{PXP,lin}}^- \right)^m  \left( H_{\text{PXP,lin}}^+ \right)^m |K_0 \rangle 
        = \sum_{l=0}^m \left[\binom{m}{l} \right]^2  \prod_{i=1}^{l}  \left[ b_{i}^{(2,\ell \cdot \tfrac{2}{k})}  \right]^2  \prod_{j=1}^{m-l} \left[ b_{j}^{(0,\ell \cdot \tfrac{k-2}{k})} \right]^2 \\
        =& \sum_{l=0}^m \left[\binom{m}{l} \right]^2  \prod_{i=1}^{l} \frac{1}{2} i \left( 2 \cdot \ell \cdot \tfrac{2}{k} -i +1 \right) \cdot  \prod_{j=1}^{m-l} 2 j \left( \ell \cdot \tfrac{k-2}{k} - j +1\right) \\
        =& \sum_{l=0}^m 2^{m-2l} (m!)^2 \frac{\left( \ell \cdot \frac{4}{k} \right)!}{l! \left( \ell \cdot \frac{4}{k} -l \right)!} \cdot \frac{\left( \ell \cdot \frac{k-2}{k} \right)!}{(m-l)!\left( \ell \cdot \frac{k-2}{k} -m +l \right)!} \\
       =& 2^m (m!)^2 \sum_{l=0}^m 2^{-2l} \binom{\ell \cdot \frac{4}{k}}{l} \binom{\ell \cdot \frac{k-2}{k}}{m-l}\,,
    \end{split}
    \label{Zk-innerprod}
\end{equation}
we determine the Lanczos coefficients generated from $|Z_k' \rangle$, and therefore from $|Z_k \rangle$, as
\begin{eqnarray}
b_m &=&  \left[ \frac{ \langle K_0 | \left(H_{\text{PXP,lin}}^- \right)^m \left(H_{\text{PXP,lin}}^+ \right)^m | K_0 \rangle }{ \langle K_0 | \left(H_{\text{PXP,lin}}^- \right)^{m-1} \left(H_{\text{PXP,lin}}^+ \right)^{m-1} | K_0 \rangle } \right]^{1/2}\,,\nonumber \\
&=& \sqrt{2} m \left[ \frac{\sum_{l=0}^m   \tfrac{2^{-2l}}{l! \left( \ell \cdot \frac{4}{k} -l \right)!(m-l)!\left( \ell \cdot \frac{k-2}{k} -m +l \right)!}}{\sum_{l=0}^{m-1}  \tfrac{2^{-2l}}{l! \left( \ell \cdot \frac{4}{k} -l \right)!(m-l-1)!\left( \ell \cdot \frac{k-2}{k} -m +l +1\right)!}} \right]^{1/2}\,.
    \label{lanczos-coe-Zk}
\end{eqnarray}

Some remarks about this result are in order. 
The sequence of Lanczos coefficients $\{b_n, n\in \mathbb{Z}^+\}$ is not infinite -- there exists a $n_{\text{max}}$, above which all $b_n$ vanish, and below which the plot of $b_n$ looks like an asymmetric arch since in general
\begin{equation}
    b_m \ne b_{n_{\text{max}}+1-m}, ~ m = 1, 2, \cdots, n_{\text{max}}~.
\end{equation}
This is reasonable, since the growth of the Lanczos coefficients for both $|\underline{2}\rangle^{\otimes \ell \cdot \frac{2}{k}}$ and $|\underline{0}\rangle^{\otimes \ell \cdot \frac{k-2}{k}}$, which construct the Lanczos coefficients of $|Z_k\rangle$, have this property, even though their arches are both symmetric. The threshold $n_{\text{max}}$ can be found from the second last equation of \eqref{Zk-innerprod}, where to make the factorial valid, $l$ should satisfy
\begin{equation}
    \left.
    \begin{matrix}
        \ell \cdot \frac{4}{k} - l \ge 0 \\
        \ell \cdot \frac{k-2}{k} - m + l \ge 0 \\
    \end{matrix}
    \right\} 
    ~{\Longrightarrow}~  m - \frac{k-2}{k}  \le l \le \frac{4}{k} ~.
\end{equation}
This way, $l$ taking values from a non-empty  set requires $m \le \ell \cdot \frac{k+2}{k}$, which translates to
\begin{equation}
    n_{\text{max}} = \ell \cdot \frac{k+2}{k}~, \qquad \forall~ |Z_k \rangle, ~ k-1 \in \mathbb{Z}^+~.
\end{equation}

\begin{table}[h!]
\begin{center}
\begin{tabular}{|>{$k=}l<{$}|>{$n_{\text{max}}=}c<{\ell $}|{c}|}
\hline
\multicolumn{1}{|l|}{\multirow{2}{*}{$|Z_k \rangle$ }} & \multicolumn{1}{l|}{ \multirow{2}{*}{${b}_n$}} & \multirow{2}{*}{scar or not } \\ [10pt]
\hline
\multicolumn{1}{|l|}{} & \multicolumn{1}{l|}{ } &  \\ [-10pt]
\infty ~ (|O \rangle) & 1 & N\\ [5pt]
2 & 2 & S \\  [5pt]
3 & \frac{5}{3} & S\\ [5pt]
4 & \frac{3}{2} & N\\ [5pt]
5 & \frac{7}{5} & N \\ [5pt]
6 & \frac{4}{3} & N \\ [5pt]
7 & \frac{9}{7} & N\\ [5pt]
8 & \frac{5}{4} & N\\ [5pt]
\hline
\end{tabular}
\caption{The ranges of nonzero Lanczos coefficients generated from product states $|Z_k \rangle$ by the linear PXP hamiltonian $H_{\text{PXP,lin}}$ in \eqref{pxp-hami-sl3c-lin}. }
\label{tab:lanczos-max-scar}
\end{center}
\end{table}

\autoref{tab:lanczos-max-scar} enumerates values of $n_{\text{max}}$ for the initial product states $|Z_k \rangle$. The information on whether these initial states are regarded as a quantum many-body scars is given. One may note that as the arch becomes narrower, that is, as $n_{\text{max}}$ becomes smaller, the corresponding initial state $|Z_k \rangle$ appears more ``thermal". Based on this, we give our observation here -- the pattern of the growth of the Lanczos coefficient with the arch width smaller than some number in the interval $( 3/2 \ell, 5/3 \ell)$ is no longer able to support any quantum revivals. 

\subsubsection{The Buttress} \label{lanczos-buttress}
Now its time to move to the residual part of the PXP Hamiltonian in \eqref{pxp-hami-sl3c-quad}. Unlike the linear part in \eqref{pxp-ladderform}, it cannot be easily written as the sum of raising and lowering ladder operators. Nevertheless, we can still understand better its function by exploring its algebraic structure. To begin with, we rewrite \eqref{pxp-hami-sl3c-quad} in the matrix form
\begin{eqnarray}
    H_{\text{PXP,res}}&=& -\sum_{i=1}^{\ell} \left\{
    \begin{pmatrix}
        0 & & \\
         & 0 & \\
         & & 1 \\
    \end{pmatrix}_{i} 
    \begin{pmatrix}
        0 & 1 & \\
        1 & 0 & \\
         & & 0 \\
    \end{pmatrix}_{(i+1) \bmod \ell} + \begin{pmatrix}
        0 &  & \\
         & 0 & 1 \\
         & 1 & 0 \\
    \end{pmatrix}_{i} 
    \begin{pmatrix}
        1 & & \\
         & 0 & \\
         & & 0 \\
    \end{pmatrix}_{(i+1) \bmod \ell} 
    \right\} \nonumber\\
    &\equiv& \sum_{i=1}^{\ell} \left(H_{\text{PXP,res}}\right)_i~, 
\end{eqnarray}
where the RHS of \eqref{pxp-hami-sl3c-quad} has been reorganized such that the two terms related to interaction between the $i^{\text{th}}$ and the $(i+1)^{\text{th}}$ qutrits are put together and denoted as $\left(H_{\text{PXP,res}}\right)_i$. From this form, it is not difficult to find that both terms above are nonzero if and only if $\left(H_{\text{PXP,res}}\right)_i$ acts on $|\underline{1}\rangle_i \otimes |\underline{2}\rangle_{i+1}$\footnote{Here the subscripts should be understood in the modulo sense, i.e. 
\begin{equation}
    |\underline{1}\rangle_i \otimes |\underline{2}\rangle_{i+1} \equiv |\underline{1}\rangle_{i \bmod{\ell}} \otimes |\underline{2}\rangle_{(i+1) \bmod{\ell}}~,
    \nonumber
\end{equation}
since we consider periodic boundary conditions. Below we usually simplify the notation into $|\underline{12}\rangle$ but the meaning is the same.
}. However, due to the overall minus sign, the action of $H_{\text{PXP,res}}$ on $|\underline{1}\underline{2}\rangle$ will cancel the action of $H_{\text{PXP,lin}}$ 
\begin{equation}
    H_{\text{PXP}} | \cdots \underline{12} \cdots \rangle = H_{\text{PXP,lin}} | \cdots \underline{12} \cdots \rangle + H_{\text{PXP,res}} | \cdots \underline{12} \cdots \rangle \equiv 0~,
\end{equation}
which is actually clear when we consider the state in terms of qubits, since $| \underline{12}\rangle = |0110\rangle$. An inference drawn from this is that $D(1,0)^{\otimes \ell}$, the space obtained by the tensor product of $\ell$ qutrit space, is not the Krylov space that can be generated by the entire PXP Hamiltonian in our setup -- This is why we used the Lucas numbers to estimate the dimension (see \autoref{lucasN}), rather than $3^{\ell}$ in the discussion above. 

Since such a ``$|\underline{12}\rangle$ issue" is a property of the full PXP Hamiltonian, from now on we will not consider the residual PXP Hamiltonian $H_{\text{PXP,res}}$ alone. Instead we investigate how this $|\underline{12}\rangle$ issue corrects our conclusions in section \ref{lanczos-arch}, where only the irreducible representation $D(\ell, 0)$ is involved. Before we begin, let us summarise our result here in a sentence -- the readers may have been aware that from the analysis of Casimir operators -- that in addition to $D(\ell,0)$, it is necessary to introduce the states of other irreducible representations of $D(1,0)^{\otimes \ell}$  into the Krylov basis generated by the full PXP Hamiltonian. 

To explain this claim, next we elaborate on the $|\underline{12}\rangle$ issue in the irreducible representations of $\mathfrak{s}l_3(\mathbb{C})$ decomposed from $D(1,0)^{\otimes \ell}$. Given that all discussions related to the linear PXP Hamiltonian happen to the irreducible representation $D(\ell,0)$ due to the structure of the initial states $|Z_k \rangle$, we take this irreducible representation as an example. 

In the root lattice of $D(\ell,0)$ (see \autoref{fig:rootlattice-hami-lin}), on the border with product states $|\underline{2}\rangle^{\otimes \ell}$ and $|\underline{1}\rangle^{\otimes \ell}$ at both ends, all fully symmetric states consist of qutrit states $|\underline{2}\rangle$ and $|\underline{1}\rangle$ only. Thus, the product states combining these symmetric states inevitably contain at least one 2-qutrit state $|\underline{1}\rangle_i \otimes |\underline{2}\rangle_{i+1}$ and should be completely removed from the Krylov basis. Same argument is made in other irreducible representations, for states that live on the line connecting $|\underline{2}\rangle^{\otimes \ell}$ and $|\underline{1}\rangle^{\otimes \ell}$. The aftermath of this removal is that the unnormalized Krylov basis, which were generated by $H_{\text{PXP,lin}}$ in section \ref{lanczos-arch} and therefore supposed to contain these normalized symmetric states, now consists only of the (normalized) states from other root lattice sites. The reduction in the number of normalized states results in reduction of Lanczos coefficients. Roughly speaking, due to the appearance of $H_{\text{PXP,res}}$, the arch constructed of the Lanczos coefficients collapses.

Apart from those non-$|\underline{0} \rangle$ states, another class of symmetric states, consisting of all the three qutrit levels $|\underline{2}\rangle$, $|\underline{0}\rangle$, and $|\underline{1}\rangle$ and therefore living in the interior of the root lattice, faces a more complicated situation. Given that $|\underline{12}\rangle$ issue is solved by inserting a $|\underline{0} \rangle$ in between, every totally symmetric state in this class contains at least one product state without $|\underline{12}\rangle$ structure, such as $|\underline{11\cdots 102 \cdots 22 } \rangle$. This means that the product states consisting of these symmetric states should be partially removed from the Krylov basis. However, such partial removal destroys the symmetry of the states in $D(\ell,0)$ and thus cannot be done through only one irreducible representation, namely $D(\ell,0)$. Recalling our conclusion from the discussion of the $\mathfrak{s}l_3(\mathbb{C})$ Casimir operators, namely that the residual PXP Hamiltonian does not preserve irreducible representations, we are inspired to complete the partial removal by introducing to the Krylov basis the states from other irreducible representations. In other words, $H_{\text{PXP,res}}$ activates the states in other irreducible representations. These activated states bring more degrees of freedom to the Hilbert subspace spanned by the Krylov basis and are therefore responsible for the increase of the value of Lanczos coefficients after the arch (the formation of the buttresses).

What happens to these activated states mirrors the scenario with states in $D(\ell,0)$. Under the repeated action of the linear PXP Hamiltonian $H_{\text{PXP,lin}}$, each of them generates a Krylov basis that spans a subspace of the irreducible representation where the state is located. However, the $|\underline{12} \rangle$ issue still persists and will be solved at the expense of activating more states from other irreducible representations. We may describe this process schematically as
\begin{eqnarray}
    H_{\text{PXP}} |K_m^{D(p,q)} \rangle &=& H_{\text{PXP,lin}} |K_m^{D(p,q)} \rangle + H_{\text{PXP,res}} |K_m^{D(p,q)} \rangle \nonumber \\
    &=& \left( {\tilde{b}}_{m+1}|K_{m+1}^{D(p,q)} \rangle + {\tilde{b}}_{m}|K_{m-1}^{D(p,q)} \rangle \right) + \sum_{p',q'} \beta(p',q') |K_0^{D(p',q')} \rangle~,
\end{eqnarray}
where $|K_m^{D(p,q)} \rangle$ denotes the $m^{\text{th}}$ Krylov basis that spans a subspace of the irreducible representation $D(p,q)$ and $\beta(p',q')$ is a coefficient that could be zero if the corresponding irreducible representation $D(p',q')$ is not activated. Note that we have used $\tilde{b}_m$ instead of $b_m$ as the Lanczos coefficients may be reduced due to the absence of the states forbidden by the full PXP Hamiltonian.

We admit that this discussion was somewhat abstract, so to gain better understanding of, and support for the arguments above, we now examine the collapse of arches and the formation of buttresses for small-sized lattices. In particular, we select the initial state to be $|Z_2 \rangle = |01  \rangle^{\otimes \ell} = |\underline{1 } \rangle^{\otimes \ell}$ and consider the Krylov bases generated for the lattices $L=2\ell=6$ and $L=2\ell=8$. The formulas below are a bit lengthy and cumbersome to goth though in the first reading\footnote{It also uses very technical mathematical tools and notations, such as Young tableaux and Young symmetrizers. We recommend that readers refer to Appendix \ref{sl3c-young}, \autoref{visual} and \autoref{kryb-lc} while reading, which might be helpful.}, so readers interested in the bigger picture can simply proceed to the next section. Nevertheless, given the previous qualitative and approximate works on this problem, we believe that working out the details of the examples below can be instructive. For this reason we present the detailed expressions in the following. 

 \subsubsection*{Lattice size $L = 2 \ell =6$}

\begin{figure}[h!]
\centering
\includegraphics[width=0.8\textwidth]{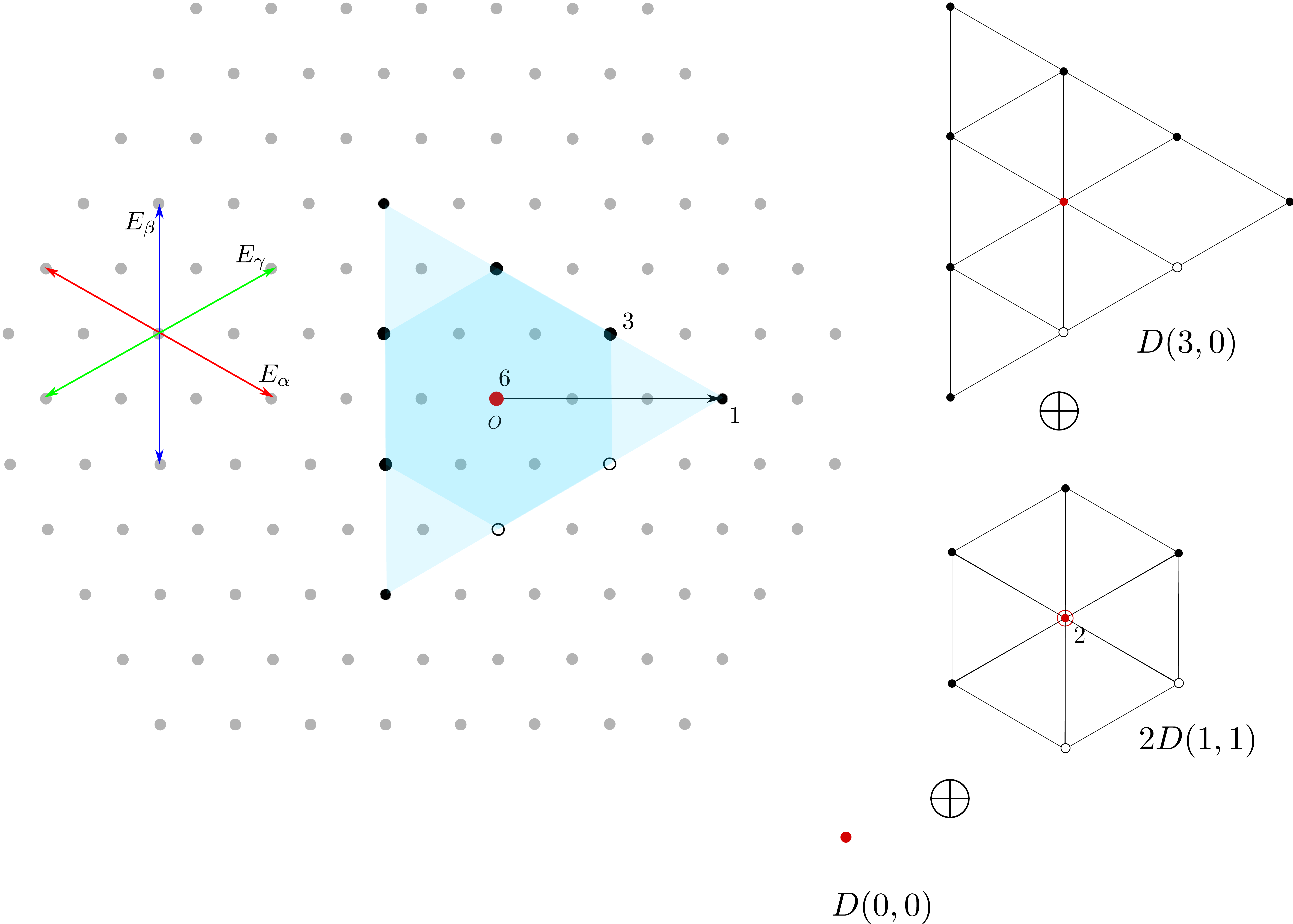} 
\caption{Decomposition of the reducible representation $D(1,0)^{\otimes 3}$ into irreducible representations $D(3,0) \oplus 2 D(1,1) \oplus D(0,0) $, corresponding to the Young tableaux ${\tiny \ydiagram{3}} ~\oplus 2 ~{\tiny \ydiagram{2,1}} ~ \oplus ~ {\tiny \ydiagram{1,1,1}} $~. The hollow circles and the red circles represent the states that cannot be or can only be partially generated by the full PXP hamiltonian (from the initial states we are interested), respectively.}
\label{rs-ydiag-n3}
\end{figure}

\noindent As \autoref{rs-ydiag-n3} shows, in the root lattice of the irreducible representation $D(3,0)$, there are two sites corresponding to the two, totally-symmetric states that cannot be generated by the full PXP Hamiltonian $H_{\text{PXP}}$. One consisting of the permutations of $|\underline{221} \rangle$, and the other consisting of the permutations of $|\underline{211} \rangle$. These states should appear in the Krylov basis $|K_2\rangle$ and $|K_4\rangle$, respectively, if the residual PXP Hamiltonian $H_{\text{PXP,res}}$ is absent. Now, without these two states, the Lanczos coefficients $b_2$ and $b_4$ decrease from $\sqrt{5}$ to $2$, since
\begin{equation}
    \sum_{i=1}^{\ell} E_{-\beta} |K_1\rangle = \sum_{i=1}^{\ell} E_{-\beta} |\underline{220}\rangle_{\tiny 
    \begin{ytableau}
        1 & 2 & 3 \\
    \end{ytableau}
    } = 2 \cdot |\underline{200}\rangle_{\tiny 
    \begin{ytableau}
        1 & 2 & 3 \\
    \end{ytableau}
    } = 2 |K_2 \rangle~,
\end{equation}
where we have introduced the standard Young tableaux as the subscripts to label the symmetrized states in the corresponding irreducible representations
\begin{eqnarray}
     |\underline{220} \rangle_{\tiny \begin{ytableau}
        1 & 2 & 3\\
        \end{ytableau} } &=&
        \frac{1}{\sqrt{3}} \left( |\underline{220}\rangle+|\underline{202}\rangle+|\underline{022}\rangle \right)  ~, \nonumber \\
    |\underline{200} \rangle_{\tiny \begin{ytableau}
        1 & 2 & 3\\
        \end{ytableau} } &=&
        \frac{1}{\sqrt{3}} \left( |\underline{200}\rangle+|\underline{020}\rangle+|\underline{002}\rangle \right)  ~.
\end{eqnarray}
The readers is referred to Appendix \ref{kryb-lc-3boxes} (and, if necessary, Appendix \ref{visual-3boxes}) for more details of this notation.

Moreover, there exists one site for the state that can only be partially generated, consisting of all permutations of $|\underline{201}\rangle$. In particular,
\begin{equation}
    |\underline{201} \rangle_{\tiny 
        \begin{ytableau}
        1 & 2 & 3\\
        \end{ytableau} } = \frac{1}{\sqrt{6}} \left( |\underline{201}\rangle+|\underline{021}\rangle+|\underline{102}\rangle +|\underline{210}\rangle+|\underline{012}\rangle+|\underline{120}\rangle \right) ~, \nonumber \\
\end{equation}
where the first three product states have the $|\underline{12}\rangle$ issue. These three terms can be eliminated by subtracting from $|K_3\rangle$ the singlet in the irreducible representation $D(0,0)$
\begin{equation}
        |\underline{201} \rangle_{\tiny 
        \begin{ytableau}
        1 \\
        2 \\ 
        3\\
        \end{ytableau} } = \frac{1}{\sqrt{6}} \left( |\underline{201}\rangle + |\underline{012}\rangle + |\underline{120}\rangle -|\underline{021}\rangle - |\underline{102}\rangle - |\underline{210}\rangle \right) ~, \nonumber \\
\end{equation}
which leads to the Krylov basis
\begin{equation}
\begin{split}
|K_3\rangle =& \frac{1}{2\sqrt{2}} |\underline{201}\rangle_{\tiny \begin{ytableau}
1 & 2 & 3\\
\end{ytableau}} - \frac{1}{2\sqrt{2}} |\underline{201}\rangle_{\tiny \begin{ytableau}
1 \\ 2 \\ 3\\
\end{ytableau}}
 + \frac{\sqrt{3}}{2}|\underline{000}\rangle_{\tiny \begin{ytableau}
1 & 2 & 3\\
\end{ytableau}} \\
 =& \frac{1}{2\sqrt{3}}\left( |\underline{210}\rangle+|\underline{102}\rangle+|\underline{021}\rangle \right)+\frac{\sqrt{3}}{2}|\underline{000}\rangle 
  ~ .
\end{split}
\end{equation}

In addition to the reduction of the Lanczos coefficient $b_3$ from $\sqrt{6}$ to $2$, the other aftermath of the activation of $|\underline{201} \rangle_{\tiny \begin{ytableau} 1 \\ 2 \\ 3\\ \end{ytableau} }$ is that the linear PXP Hamiltonian $H_{\text{PXP,lin}}$ will also act on it. Fortunately, the irreducible representation $D(0,0)$ is a singlet sector and therefore nothing will be generated from $|\underline{201} \rangle_{\tiny \begin{ytableau} 1 \\ 2 \\ 3\\ \end{ytableau} }$, and there is no contribution to the next Krylov basis vector $|K_4\rangle$.

For completeness, we enumerate all Krylov basis generated from $|Z_2 \rangle$ by the full PXP Hamiltonian, as well as the corresponding Lanczos coefficients, in Appendix \ref{kryb-lc-3boxes}. Anyway, in this case we only observe a collapse of the arch. 
    
\subsubsection*{Lattice size $L= 2 \ell =8$}

\begin{figure}[h!]
\centering
\includegraphics[width=0.8\textwidth]{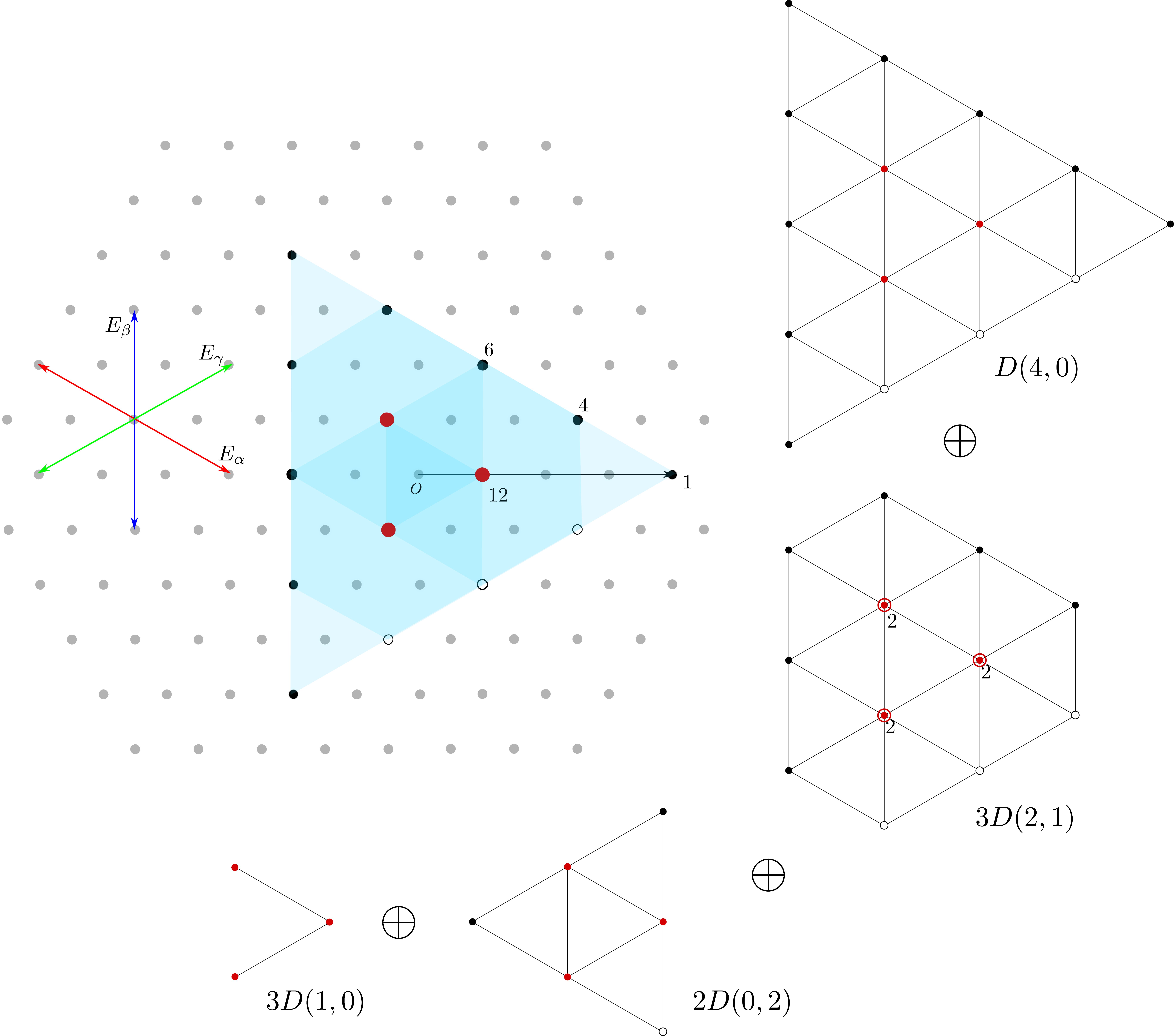}
\caption{Decomposition of the reducible representation $D(1,0)^{\otimes 4}$ into irreducible representations $D(4,0) \oplus 3 D(2,1) \oplus 2 D(0,2) \oplus 3 D(1,0)$, corresponding to Young tableaux $ {\tiny \ydiagram{4} }~ \oplus 3~{\tiny \ydiagram{3,1}} ~ \oplus 2~ {\tiny \ydiagram{2,2}}  ~ \oplus 3~ {\tiny \ydiagram{2,1,1}} $. The hollow circles and the red circles represent the states that cannot be or can only be partially generated by the full PXP hamiltonian (from the initial states we are interested), respectively.}
\label{rs-ydiag-n4}
\end{figure}

Using a similar arguments to the case of $L=2\ell =3$, we can easily find that there are three totally symmetric states that should be removed from the Krylov bases $|K_2\rangle$, $|K_4\rangle$, and $|K_6\rangle$, respectively, as \autoref{rs-ydiag-n4} shows, which results in a decrease in the corresponding Lanczos coefficients. However, from $|K_3\rangle$ that includes some product states consisting of one $|\underline{2}\rangle$, one $|\underline{0}\rangle$, and two $|\underline{1}\rangle$s, the situation is more complicated. Again we begin with the totally symmetric state in the irreducible representation $D(4,0)$
\begin{eqnarray}
    |\underline{2201}\rangle_{\tiny
\begin{ytableau}
1& 2& 3 & 4 \\
\end{ytableau}
} &=& \frac{1}{2\sqrt{3}} \left( |\underline{2201} \rangle + |\underline{2012} \rangle +|\underline{0122} \rangle +|\underline{1220} \rangle  \right. \nonumber \\
&+& |\underline{2021} \rangle +|\underline{0212}\rangle +|\underline{2120} \rangle +|\underline{1202}\rangle    \nonumber \\
&+& \left. |\underline{2210} \rangle +|\underline{2102} \rangle+|\underline{1022} \rangle  +|\underline{0221} \rangle
\right) ~,
\label{totsymm2011}
\end{eqnarray}
where only the last four terms are permitted by $H_{\text{PXP}}$. To eliminate the other eight terms, it is necessary to introduce to the Krylov basis the corresponding states in the irreducible representations $D(0,2)$ (or more exactly ~{\tiny $\begin{ytableau}
    1 & 3 \\
    2 & 4 \\
\end{ytableau}$}~) and $D(1,0)$s. This is because
\begin{eqnarray}
   |\underline{2021}\rangle_{\tiny
\begin{ytableau}
1& 3 \\
2 & 4 \\
\end{ytableau}
} &=& \frac{1}{2\sqrt{6}} \left( 2|\underline{2021} \rangle+2|\underline{2120} \rangle+ 2|\underline{0212} \rangle + 2|\underline{1202}\rangle -|\underline{0221} \rangle - |\underline{2201} \rangle \right. \nonumber \\
 &-&  \left. |\underline{0122} \rangle - |\underline{2102} \rangle - |\underline{2012} \rangle - |\underline{1022}\rangle - |\underline{2210} \rangle -|\underline{1220} \rangle \right)\,, \nonumber \\
|\underline{2012}\rangle_{\tiny
\begin{ytableau}
1& 4  \\
2 & \none \\
3 & \none \\
\end{ytableau}
} &=& \frac{1}{4} \left( |\underline{1202} \rangle+|\underline{0122} \rangle  +|\underline{2201} \rangle +|\underline{2120} \rangle+ 2|\underline{2012}\rangle 
\right. \nonumber \\
&-& \left. 
 |\underline{0212} \rangle -|\underline{1022} \rangle  - |\underline{2210} \rangle -|\underline{2021} \rangle -2 |\underline{2102} \rangle \right)\,, \nonumber \\
|\underline{2021}\rangle_{\tiny
\begin{ytableau}
1& 3  \\
2 & \none \\
4 & \none \\
\end{ytableau}
} &=& \frac{1}{4} \left( |\underline{1220} \rangle+|\underline{0122} \rangle  +|\underline{2210} \rangle +|\underline{2102} \rangle+ 2|\underline{2021}\rangle 
\right. \nonumber \\
&-& \left. 
 |\underline{0221} \rangle -|\underline{1022} \rangle  - |\underline{2201} \rangle -|\underline{2012} \rangle -2 |\underline{2120} \rangle \right)\,, \nonumber \\
|\underline{2201}\rangle_{\tiny
\begin{ytableau}
1& 2  \\
3 & \none \\
4 & \none \\
\end{ytableau}
} &=& \frac{1}{4} \left( |\underline{1220} \rangle+|\underline{0212} \rangle  +|\underline{2120} \rangle +|\underline{2012} \rangle+ 2|\underline{2201}\rangle 
\right. \nonumber \\
&-& \left. 
- |\underline{0221} \rangle -|\underline{1202} \rangle  - |\underline{2021} \rangle -|\underline{2102} \rangle -2 |\underline{2210} \rangle \right)\,, \nonumber \\
\end{eqnarray}
the second four terms of \eqref{totsymm2011} are cancelled by $|\underline{2021}\rangle_{\tiny
\begin{ytableau}
1& 3 \\
2 & 4 \\
\end{ytableau}
}$, while the first four by the sum of all the three irreducible representations $D(1,0)$, 
\begin{equation}
|\underline{2012}\rangle_{\tiny
\begin{ytableau}
1& 4  \\
2 & \none \\
3 & \none \\
\end{ytableau}}
+
|\underline{2021}\rangle_{\tiny
\begin{ytableau}
1& 3  \\
2 & \none \\
4 & \none \\
\end{ytableau}
}+
|\underline{2201}\rangle_{\tiny
\begin{ytableau}
1& 2  \\
3 & \none \\
4 & \none \\
\end{ytableau}
} ~.\nonumber
\end{equation}
Thus, the Krylov basis vector $|K_3 \rangle$ has the form
\begin{eqnarray}
    |K_3 \rangle 
&=&\frac{1}{2\sqrt{10}} \left[  \frac{2}{\sqrt{3}}\left( |\underline{2201}\rangle_{\tiny
\begin{ytableau}
1& 2& 3 & 4 \\
\end{ytableau}
} -\frac{1}{\sqrt{2}} 
|\underline{2021}\rangle_{\tiny
\begin{ytableau}
1& 3 \\
2 & 4 \\
\end{ytableau}
}  \right) \right. \nonumber \\
&-& \left.
 \left( |\underline{2012}\rangle_{\tiny
\begin{ytableau}
1& 4  \\
2 & \none \\
3 & \none \\
\end{ytableau}}
+
|\underline{2021}\rangle_{\tiny
\begin{ytableau}
1& 3  \\
2 & \none \\
4 & \none \\
\end{ytableau}
}+
|\underline{2201}\rangle_{\tiny
\begin{ytableau}
1& 2  \\
3 & \none \\
4 & \none \\
\end{ytableau}
}  \right)
\right] \nonumber \\
&+& \frac{6}{\sqrt{10}}  |\underline{2000}\rangle_{\tiny
\begin{ytableau}
1& 2& 3 & 4 \\
\end{ytableau}
}\nonumber \\
& =& \frac{1}{2\sqrt{10}} \left( |\underline{0221} \rangle+|\underline{2210} \rangle+|\underline{2102} \rangle+|\underline{1022} \rangle \right) \nonumber \\
&+& \frac{3}{\sqrt{10}} \left(|\underline{2000} \rangle+|\underline{0200} \rangle+|\underline{0020} \rangle+|\underline{0002} \rangle \right)~. \nonumber \\
\end{eqnarray}

\begin{figure}[H]
    \centering
    \includegraphics[width=0.9\linewidth]{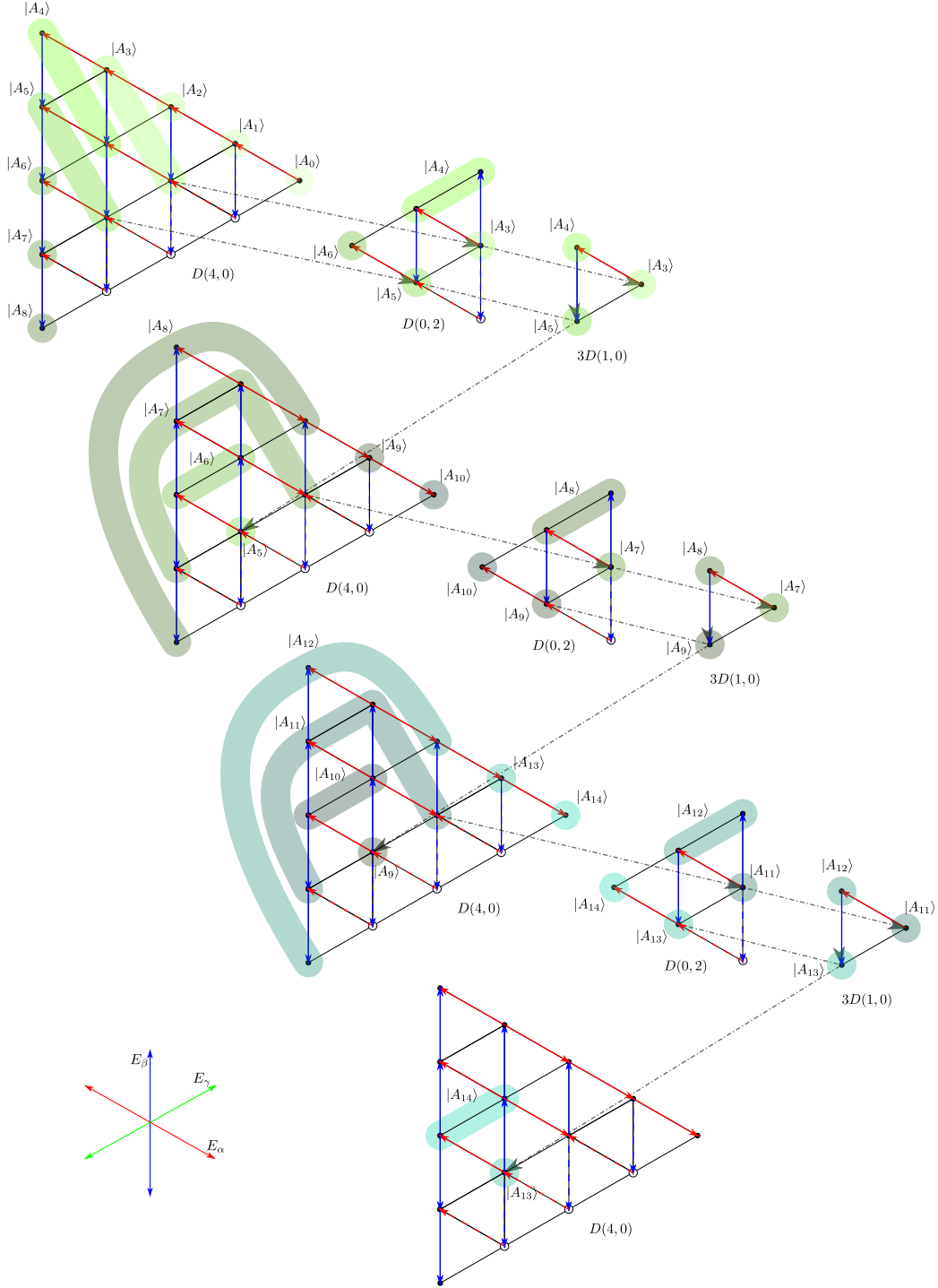}
    \caption{The ``forward" states $|A_m \rangle$ generated by the repeated actions of the linear PXP Hamiltonian $H_{\text{PXP,lin}}$ in \eqref{pxp-hami-sl3c-lin} to $|Z_2 \rangle$, after taking the two effects of the residual terms \eqref{pxp-hami-sl3c-quad} into account. The hollow circles represent the states that cannot be generated by the full PXP Hamiltonian, with the associated $\mathfrak{s}l_3(\mathbb{C})$ generators labelled by dashed arrows. Note that no Gram–Schmidt orthogonalization has be applied here. One can see that $|A_{m+4}\rangle= |A_{m} \rangle$ for all $m \ge 9$. }
    \label{fig:rootlattice-hami-Z2}
\end{figure}

Similarly to the previous case, in addition to a decrease in the Lanczos coefficient $b_3$, the activation of states in the irreducible representations $D(0,2)$ (~{\tiny $\begin{ytableau}
    1 & 3 \\
    2 & 4
\end{ytableau}$}~) and $D(1,0)$s, leads to the contributions to the next Krylov basis from action of $H_{\text{PXP,lin}}$ on these states. Since $|\underline{2012}\rangle_{\tiny
\begin{ytableau}
1& 4  \\
2 & \none \\
3 & \none \\
\end{ytableau}}~,
|\underline{2021}\rangle_{\tiny
\begin{ytableau}
1& 3  \\
2 & \none \\
4 & \none \\
\end{ytableau}
}~,
|\underline{2201}\rangle_{\tiny
\begin{ytableau}
1& 2  \\
3 & \none \\
4 & \none \\
\end{ytableau}}$ correspond to the highest weight vectors in the root lattices of the three irreducible representations $D(1,0)$s (~${\tiny
\begin{ytableau}
1& 4  \\
2 & \none \\
3 & \none \\
\end{ytableau}}~,~{\tiny
\begin{ytableau}
1& 3  \\
2 & \none \\
4 & \none \\
\end{ytableau}
}~,~{\tiny
\begin{ytableau}
1& 2  \\
3 & \none \\
4 & \none \\
\end{ytableau}}$~), respectively, the raising and lowering ladder operators in this case are exactly the same as those for $|\underline{2}\rangle^{\otimes \ell}$ (see \eqref{ladderop-Z2}). 
However, $|\underline{2021}\rangle_{\tiny
\begin{ytableau}
1& 3 \\
2 & 4 \\
\end{ytableau}
}$ is not the highest weight vector of $D(0,2)$ (~{\tiny $\begin{ytableau}
    1 & 3 \\
    2 & 4
\end{ytableau}$}~) -- it lies on the border of the root lattice. Such location allows the linear PXP Hamiltonian $H_{\text{PXP,lin}}$ to generate states in three directions, which makes it unlikely to rewrite \eqref{pxp-hami-sl3c-lin} into a sum of raising and lowering ladder operators conjugate transposed to each other. For example, for $|\underline{2021}\rangle_{\tiny
\begin{ytableau}
1& 3 \\
2 & 4 \\
\end{ytableau}
}$, the components of the next state can be generated by $E_{\alpha}$ or $E_{\pm \beta}$, although some of these components are forbidden by the full PXP Hamiltonian. As a result, after eliminating the banned states, the repeated action of $H_{\text{PXP,lin}}$ on $|\underline{2021}\rangle_{\tiny
\begin{ytableau}
1& 3 \\
2 & 4 \\
\end{ytableau}
}$ first produces two mixed symmetric states, $|\underline{2020}\rangle_{\tiny
\begin{ytableau}
1& 3 \\
2 & 4 \\
\end{ytableau}
}$ and $|\underline{2001}\rangle_{\tiny
\begin{ytableau}
1& 3 \\
2 & 4 \\
\end{ytableau}
}$, and then $|\underline{2011}\rangle_{\tiny
\begin{ytableau}
1& 3 \\
2 & 4 \\
\end{ytableau}
}$ (and eventually $|\underline{0101}\rangle_{\tiny
\begin{ytableau}
1& 3 \\
2 & 4 \\
\end{ytableau}
}$, see \autoref{fig:rootlattice-hami-Z2}).

Notice that only the forward states are listed here, where a ``forward" state, opposite to a ``backward" state, is a (permitted) state that has not yet been produced during the previous actions of the linear PXP Hamiltonian. However, it is not difficult to determine the backward states from the forward ones. Thus, after considering the states generated by the repeated actions of $H_{\text{PXP,lin}}$ on $D(1,0)$s, we obtain the Krylov basis
\begin{eqnarray}
|K_4 \rangle &=& 
 0.241943\left[ \frac{2}{\sqrt{3}}\left( 2|\underline{2001}\rangle_{\tiny
\begin{ytableau}
1& 2& 3 & 4 \\
\end{ytableau}
} -\frac{1}{\sqrt{2}} 
|\underline{2001}\rangle_{\tiny
\begin{ytableau}
1& 3 \\
2 & 4 \\
\end{ytableau}
}  \right) \right. \nonumber\\
&-& \left.   \left( |\underline{2010}\rangle_{\tiny
\begin{ytableau}
1& 4  \\
2 & \none \\
3 & \none \\
\end{ytableau}}
+
|\underline{2001}\rangle_{\tiny
\begin{ytableau}
1& 3  \\
2 & \none \\
4 & \none \\
\end{ytableau}
}+
|\underline{2001}\rangle_{\tiny
\begin{ytableau}
1& 2  \\
3 & \none \\
4 & \none \\
\end{ytableau}
}  \right) \right] \nonumber \\
&-& 0.0201619 \cdot 2\sqrt{3} |\underline{2020}\rangle_{\tiny
\begin{ytableau}
1& 3 \\
2 & 4 \\
\end{ytableau}
}  +  0.72583|\underline{0000}\rangle_{\tiny
\begin{ytableau}
1& 2& 3 & 4 \\
\end{ytableau}
} ~,
\end{eqnarray}
and
\begin{eqnarray}
 |K_5\rangle 
 &=&  0.0347245
\left[ \frac{2}{\sqrt{3}}\left( |\underline{2201}\rangle_{\tiny
\begin{ytableau}
1& 2& 3 & 4 \\
\end{ytableau}
} -\frac{1}{\sqrt{2}} 
|\underline{2021}\rangle_{\tiny
\begin{ytableau}
1& 3 \\
2 & 4 \\
\end{ytableau}
}  \right) \right. \nonumber \\
&-& \left. \left( |\underline{2012}\rangle_{\tiny
\begin{ytableau}
1& 4  \\
2 & \none \\
3 & \none \\
\end{ytableau}}
+
|\underline{2021}\rangle_{\tiny
\begin{ytableau}
1& 3  \\
2 & \none \\
4 & \none \\
\end{ytableau}
}+
|\underline{2201}\rangle_{\tiny
\begin{ytableau}
1& 2  \\
3 & \none \\
4 & \none \\
\end{ytableau}
}  \right) \right] 
\nonumber \\
&+&0.185197
\left[ \frac{2}{\sqrt{3}}\left( |\underline{2011}\rangle_{\tiny
\begin{ytableau}
1& 2& 3 & 4 \\
\end{ytableau}
} -\frac{1}{\sqrt{2}} 
|\underline{2101}\rangle_{\tiny
\begin{ytableau}
1& 3 \\
2 & 4 \\
\end{ytableau}
}  \right) \right. \nonumber \\
&-& \left.\left( |\underline{2011}\rangle_{\tiny
\begin{ytableau}
1& 4  \\
2 & \none \\
3 & \none \\
\end{ytableau}}
+
|\underline{2011}\rangle_{\tiny
\begin{ytableau}
1& 3  \\
2 & \none \\
4 & \none \\
\end{ytableau}
}+
|\underline{2101}\rangle_{\tiny
\begin{ytableau}
1& 2  \\
3 & \none \\
4 & \none \\
\end{ytableau}
}  \right) \right] \nonumber \\
 &-& 0.0115748 \cdot 2  |\underline{2000}\rangle_{\tiny
\begin{ytableau}
1& 2& 3 & 4 \\
\end{ytableau}
}
 +0.462993 \cdot 2  |\underline{0001}\rangle_{\tiny
\begin{ytableau}
1& 2& 3 & 4 \\
\end{ytableau}
}\,.
\end{eqnarray}

The crucial point about $|K_5\rangle$ is that it cannot be fully generated from $|K_4 \rangle$ by action of the linear PXP Hamiltonian $H_{\text{PXP,lin}}$ only, since 
\begin{equation}
\begin{split}
& \sum_{i=1}^{\ell} E_{-\beta}  \left[   \frac{2}{\sqrt{3}}\left( 2|\underline{2001}\rangle_{\tiny
\begin{ytableau}
1& 2& 3 & 4 \\
\end{ytableau}
} -\frac{1}{\sqrt{2}} 
|\underline{2001}\rangle_{\tiny
\begin{ytableau}
1& 3 \\
2 & 4 \\
\end{ytableau}
}  \right)  \right. \\ & \qquad \qquad \qquad  \left. - \left( |\underline{2010}\rangle_{\tiny
\begin{ytableau}
1& 4  \\
2 & \none \\
3 & \none \\
\end{ytableau}}
+
|\underline{2001}\rangle_{\tiny
\begin{ytableau}
1& 3  \\
2 & \none \\
4 & \none \\
\end{ytableau}
}+
|\underline{2001}\rangle_{\tiny
\begin{ytableau}
1& 2  \\
3 & \none \\
4 & \none \\
\end{ytableau}
}  \right)
\right] \\
=& \sum_{i=1}^{\ell} E_{-\beta} \left(|\underline{2100} \rangle+|\underline{2010} \rangle+|\underline{1020} \rangle+|\underline{1002} \rangle +|\underline{0210} \rangle+|\underline{0201} \rangle+|\underline{0102} \rangle+|\underline{0021} \rangle \right) \\
=& \left(|\underline{2110} \rangle+|\underline{1021} \rangle+|\underline{1102} \rangle+|\underline{0211} \rangle  \right)  + 2\sqrt{3} |\underline{2011}\rangle_{\tiny
\begin{ytableau}
1& 2& 3 & 4 \\
\end{ytableau}}\,,
\end{split}
\end{equation}
which implies that it is necessary to activate another copy of the root lattice of $D(4,0)$. Now, similar argument to $|\underline{2021}\rangle_{\tiny
\begin{ytableau}
1& 3 \\
2 & 4 \\
\end{ytableau}
}$, the fact that $|\underline{2011}\rangle_{\tiny
\begin{ytableau}
1& 2& 3 & 4 \\
\end{ytableau}}$ lies in the interior of the root lattice of $D(4,0)$ means that the components of the next forward state can be generated by all the four Cartan-Weyl basis $E_{\pm \alpha}$, $E_{\pm \beta}$. As a result, after eliminating the banned states, the repeated action of $H_{\text{PXP,lin}}$ on $|\underline{2011}\rangle_{\tiny
\begin{ytableau}
1& 2& 3 & 4 \\
\end{ytableau}}$ first produces (see \autoref{fig:rootlattice-hami-Z2}) two totally symmetric (forward) states, $|\underline{0011}\rangle_{\tiny
\begin{ytableau}
1& 2& 3 & 4 \\
\end{ytableau}}$ and $|\underline{2001}\rangle_{\tiny
\begin{ytableau}
1& 2& 3 & 4 \\
\end{ytableau}}$, then four states, $|\underline{0111}\rangle_{\tiny
\begin{ytableau}
1& 2& 3 & 4 \\
\end{ytableau}}$, $|\underline{0001}\rangle_{\tiny
\begin{ytableau}
1& 2& 3 & 4 \\
\end{ytableau}}$, $|\underline{0002}\rangle_{\tiny
\begin{ytableau}
1& 2& 3 & 4 \\
\end{ytableau}}$, and $|\underline{2201}\rangle_{\tiny
\begin{ytableau}
1& 2& 3 & 4 \\
\end{ytableau}}$, and then three states,
$|\underline{1111}\rangle_{\tiny
\begin{ytableau}
1& 2& 3 & 4 \\
\end{ytableau}}$, $|\underline{0000}\rangle_{\tiny
\begin{ytableau}
1& 2& 3 & 4 \\
\end{ytableau}}$, and $|\underline{2200}\rangle_{\tiny
\begin{ytableau}
1& 2& 3 & 4 \\
\end{ytableau}}$, 
contributing to the Krylov basis vectors 
\begin{eqnarray}
|K_6\rangle 
&=& 
 0.0151446\left[ \frac{2}{\sqrt{3}}\left( 2|\underline{2001}\rangle_{\tiny
\begin{ytableau}
1& 2& 3 & 4 \\
\end{ytableau}
} -\frac{1}{\sqrt{2}} 
|\underline{2001}\rangle_{\tiny
\begin{ytableau}
1& 3 \\
2 & 4 \\
\end{ytableau}
}  \right) \right. \nonumber \\
&-&\left. \left( |\underline{2010}\rangle_{\tiny
\begin{ytableau}
1& 4  \\
2 & \none \\
3 & \none \\
\end{ytableau}}
+
|\underline{2001}\rangle_{\tiny
\begin{ytableau}
1& 3  \\
2 & \none \\
4 & \none \\
\end{ytableau}
}+
|\underline{2001}\rangle_{\tiny
\begin{ytableau}
1& 2  \\
3 & \none \\
4 & \none \\
\end{ytableau}
}  \right)\right]  \nonumber \\
&+& 0.0158865 \cdot 2 \sqrt{3} |\underline{2020}\rangle_{\tiny
\begin{ytableau}
1& 3 \\
2 & 4 \\
\end{ytableau}
} - 0.023859\cdot 2 \sqrt{3} |\underline{0101}\rangle_{\tiny
\begin{ytableau}
1& 3 \\
2 & 4 \\
\end{ytableau}
}  \nonumber \\
&-& 0.03509|\underline{0000}\rangle_{\tiny
\begin{ytableau}
1& 2& 3 & 4 \\
\end{ytableau}
}
 +0.405601 \cdot \sqrt{6}  |\underline{0011}\rangle_{\tiny
\begin{ytableau}
1& 2& 3 & 4 \\
\end{ytableau}
}   ~,
\end{eqnarray}
\begin{eqnarray}
|K_7\rangle 
&=& -0.0309286 
\left[
\frac{2}{\sqrt{3}}\left( |\underline{2201}\rangle_{\tiny
\begin{ytableau}
1& 2& 3 & 4 \\
\end{ytableau}
} -\frac{1}{\sqrt{2}} 
|\underline{2021}\rangle_{\tiny
\begin{ytableau}
1& 3 \\
2 & 4 \\
\end{ytableau}
}  \right) \right. \nonumber \\
&-&\left. \left( |\underline{2012}\rangle_{\tiny
\begin{ytableau}
1& 4  \\
2 & \none \\
3 & \none \\
\end{ytableau}}
+
|\underline{2021}\rangle_{\tiny
\begin{ytableau}
1& 3  \\
2 & \none \\
4 & \none \\
\end{ytableau}
}+
|\underline{2201}\rangle_{\tiny
\begin{ytableau}
1& 2  \\
3 & \none \\
4 & \none \\
\end{ytableau}
}  \right)
 \right]
\nonumber \\
&-&0.00796488 
\left[
\frac{2}{\sqrt{3}}\left( |\underline{2011}\rangle_{\tiny
\begin{ytableau}
1& 2& 3 & 4 \\
\end{ytableau}
} -\frac{1}{\sqrt{2}} 
|\underline{2101}\rangle_{\tiny
\begin{ytableau}
1& 3 \\
2 & 4 \\
\end{ytableau}
}  \right) \right. \nonumber \\
&-&\left. \left( |\underline{2011}\rangle_{\tiny
\begin{ytableau}
1& 4  \\
2 & \none \\
3 & \none \\
\end{ytableau}}
+
|\underline{2011}\rangle_{\tiny
\begin{ytableau}
1& 3  \\
2 & \none \\
4 & \none \\
\end{ytableau}
}+
|\underline{2101}\rangle_{\tiny
\begin{ytableau}
1& 2  \\
3 & \none \\
4 & \none \\
\end{ytableau}
}  \right)
 \right]
 \nonumber \\
&+&0.00576334 \cdot 2  |\underline{0001}\rangle_{\tiny
\begin{ytableau}
1& 2& 3 & 4 \\
\end{ytableau}
}  +0.0103095 \cdot 2  |\underline{2000}\rangle_{\tiny
\begin{ytableau}
1& 2& 3 & 4 \\
\end{ytableau}
} 
 \nonumber \\
&+&0.498839 \cdot 2  |\underline{0111}\rangle_{\tiny
\begin{ytableau}
1& 2& 3 & 4 \\
\end{ytableau}
} ~, 
\end{eqnarray}
and $|K_8 \rangle$, respectively\footnote{Actually, after generating these totally symmetric states, one can continue the repeated action of $H_{\text{PXP,lin}}$ on $|\underline{2011}\rangle_{\tiny
\begin{ytableau}
1& 2& 3 & 4 \\
\end{ytableau}}$ and obtain a (forward) state $|\underline{2220}\rangle_{\tiny
\begin{ytableau}
1& 2& 3 & 4 \\
\end{ytableau}}$ and then $|\underline{2222}\rangle_{\tiny
\begin{ytableau}
1& 2& 3 & 4 \\
\end{ytableau}}$, where the banned states have already been removed. However, these two states are exactly the Krylov basis $|K_1\rangle$ and $|K_0 \rangle$, respectively and are therefore removed from $|K_9\rangle$ and $|K_{10}\rangle$, respectively, by the Gram-Schmidt process. Hence we did not mention them in the main text.}.

Note also, that in order to solve the $|\underline{12}\rangle$ issue, it is necessary to activate the contributions to $|K_7\rangle$ of $|\underline{2021}\rangle_{\tiny
\begin{ytableau}
1& 3 \\
2 & 4 \\
\end{ytableau}
}$ in another copy of the root lattice of $D(0,2)$ (~{\tiny
\begin{ytableau}
1& 3 \\
2 & 4 \\
\end{ytableau}
}~) as well as $\left( |\underline{2012}\rangle_{\tiny
\begin{ytableau}
1& 4  \\
2 & \none \\
3 & \none \\
\end{ytableau}}
+
|\underline{2021}\rangle_{\tiny
\begin{ytableau}
1& 3  \\
2 & \none \\
4 & \none \\
\end{ytableau}
}+
|\underline{2201}\rangle_{\tiny
\begin{ytableau}
1& 2  \\
3 & \none \\
4 & \none \\
\end{ytableau}
}  \right)$ in other copies of root lattices of the three irreducible representations of $D(1,0)$, just as it happened for $|K_3 \rangle$.

The occurrence of a situation similar to $|K_3\rangle$ implies that what happens to the next four Krylov bases, starting from $|K_8 \rangle$, is a repetition of what happened to the Krylov bases $|K_4 \rangle$, $|K_5 \rangle$, $|K_6 \rangle$, and $|K_7 \rangle$ discussed above. However, this does not mean that $|K_8 \rangle$ vanishes after the Gram-Schmidt process removes the components of $|K_4 \rangle$ or other Krylov bases, because in the root lattice of $D(4,0)$ where $|K_0 \rangle = |Z_2 \rangle$ is located, the states contributing to the Krylov basis are still generated by the linear PXP Hamiltonian in the way shown in Figure \ref{fig:rootlattice-hami-lin-Z2}. This generation ceases at $|K_8 \rangle$. Thus, starting from $|K_9 \rangle$, there is no longer contribution from the $D(4,0)$ where $|K_0 \rangle $ is located. This implies the vanishing of $|K_{13}\rangle$, because before Gram-Schmidt orthogonalization the states generated from $|K_{12}\rangle$ are exactly the same as those generated from $|K_8 \rangle$ was obtained (see \autoref{fig:rootlattice-hami-Z2}).

\autoref{fig:rootlattice-hami-Z2} is a sketch of generation of states by action of the linear PXP Hamiltonian on $|Z_2\rangle$, after considering the ``$|\underline{12}\rangle$ issue" brought by the residual PXP Hamiltonian. It should be noted that to avoid clutter, we only label forward states and therefore it is necessary to restore the backward states and then perform Gram-Schmidt orthogonalization. However, from \autoref{fig:rootlattice-hami-Z2} we can still find that: 1) the contribution from the original root lattice, where the initial state $|Z_2\rangle$ is located, lasts from $|K_0 \rangle$ to $|K_8\rangle$, corresponding to the collapse of the arch of Lanczos coefficients located between $b_1$ and $b_8$; 2) there exists a repeating pattern of activating the copies of the root lattices of irreducible representations $D(4,0)$, $D(0,2)$ and $D(1,0)$s with period 4, leading to the (forward) states $|A_{m+4}\rangle = |A_{m} \rangle$ for all $m >8$, hence the buttress terminates at $b_{12}$. 

For completeness, all Krylov basis generated from $|Z_2 \rangle$ by the full PXP Hamiltonian, as well as the corresponding Lanczos coefficients, are given in Appendix \ref{kryb-lc-4boxes}. In conclusion, we observe both the collapse of the arch and the formation of a buttress in this non-trivial example. 

We would like to end this section with one last comment. Until now, we have only considered the $|\underline{12} \rangle$ issue posed by the residual PXP Hamiltonian in \eqref{pxp-hami-sl3c-quad}. However, it should be pointed that, although $H_{\text{PXP,res}}$ only eliminates the product states with a $|\underline{1}\rangle_i \otimes |\underline{2}\rangle_{i+1} $, it may also reassign the coefficients of the product states with other structures -- readers may have realized that although we successfully explain the appearance of all the (mixed or totally) symmetric states in the case of $L=2\ell =8$, the ratio of the coefficients of $|\underline{2001}\rangle_{\tiny
\begin{ytableau}
1& 3 \\
2 & 4 \\
\end{ytableau}
}$ and $|\underline{2020}\rangle_{\tiny
\begin{ytableau}
1& 3 \\
2 & 4 \\
\end{ytableau}
}$ in $|K_4 \rangle$, $2\sqrt{2}$, does not match our prediction, which, according to our argument, should be $1/\sqrt{2}$. We leave this puzzle as an open question for further research.

\subsection{Evidences} \label{lanczos-numberic-random}
\begin{figure}[H]
    \centering
    \subfloat[FFT of fidelity of $|{Z}_2\rangle$]{
        \includegraphics[width=0.48\textwidth,height=0.25\textheight, keepaspectratio]{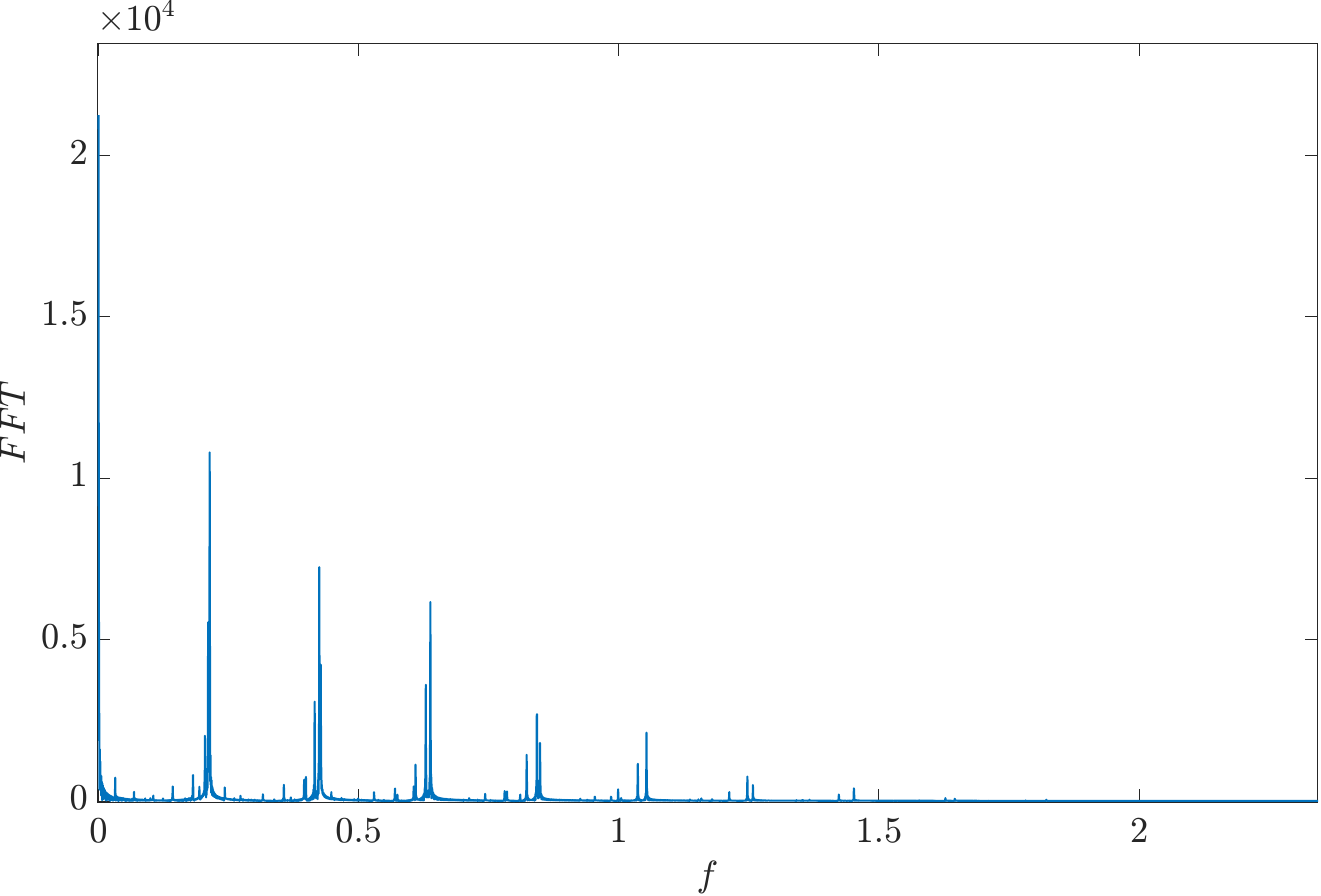}
        \label{fig:fftz2}
    }
    \hfill
    \subfloat[FFT of fidelity of $|{Z}_4\rangle$]{
        \includegraphics[width=0.48\textwidth,height=0.25\textheight, keepaspectratio]{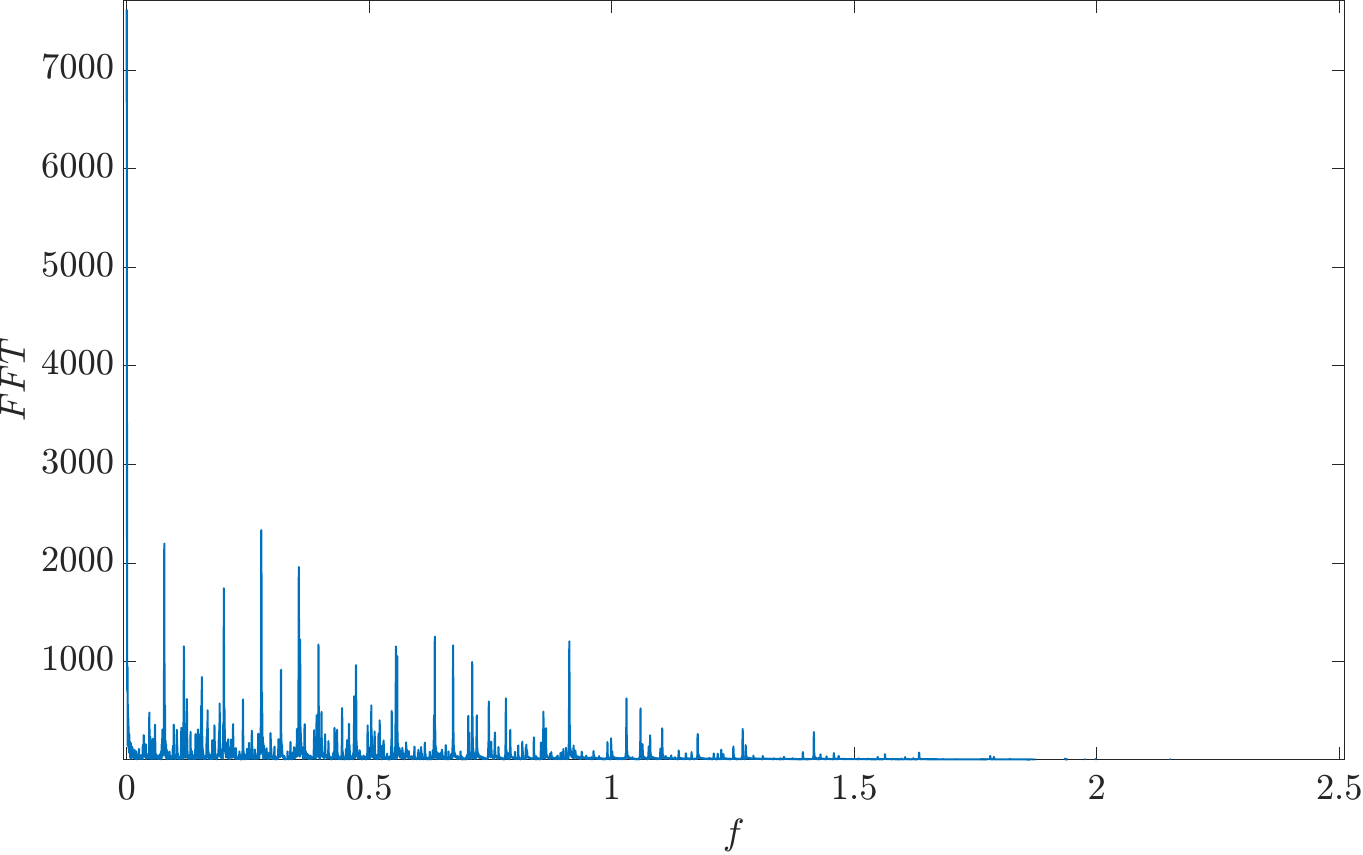}
        \label{fig:fftz4}
    }
	\caption{The FFT of the fidelity for time evolution subject to the full PXP Hamiltonian of: $|{Z}_2\rangle$ in \ref{fig:fftz2}) and  $|{Z}_4\rangle$ in \ref{fig:fftz4}). Plots for lattice size $L=12$.}
	\label{fig:fft}
\end{figure}

We close this section with serval numerical studies that support our algebraic discussions and evidences above.
To begin with, we consider the {\it Fast Fourier Transform} (FFT) \cite{vanloan} of the fidelity of the time evolution of the initial states subject to the full PXP Hamiltonian \eqref{pxp-hamiltonian-pbc}. Since, in general, the initial state $|\Psi_0 \rangle$ can be expanded in the basis of the eigenstates of the Hamiltonian as
\begin{equation}
    |\Psi_0\rangle = \sum_i c_i|E_i\rangle~,
    \label{eigenexpand}
\end{equation}
the FFT of fidelity \eqref{fidelitySGA} can be written as 
\begin{equation}
    \mathcal{F}[|S(t)|^2](f) = \sum_{m,n} |c_n c_m |^2 \mathcal{F}\left( e^{i (E_m -E_n) t} \right) = \sum_{m,n} |c_n c_m |^2 \delta \left( E_m-E_n - 2\pi f \right)~,
\end{equation}
where $f$ represents the frequency\footnote{Please note that $f$ is not the angular frequency $\omega = 2\pi f$.}.
For numerical computations, we investigate the evolution of the system in the time interval $[0,T]$ and discretize the time with a step $T/N$. As a result, the Fourier transform of fidelity becomes a discrete Fourier transform and the Dirac delta function is replaced by a Kronecker delta function with a factor $N$. Consequently, the FFT should enumerate the differences between any two energy levels whose corresponding eigenstates are contained in the initial state $|\Psi_0 \rangle$. In \autoref{fig:fft} we again compare  $|Z_2\rangle$ and $|Z_4 \rangle$ as initial states whose fidelity was presented in \autoref{longf}. It can be clearly seen that for $|Z_2\rangle$, the FFT exhibits a comb-like structure, with the distances between two adjacent teeth being approximately equal. As can be seen in Figure \ref{fig:fftz2}, this distance is $f\approx 0.21$, which corresponds to the energy difference $\omega= 2\pi f \approx 1.3$. On the contrary, a completely different pattern in FFT is obtained for $|Z_4\rangle$, and it is difficult to determine how the teeth are distributed.

 The appearance of equidistant teeth in the FFT of $|Z_2\rangle$ reveals that, in the eigen-basis of the PXP Hamiltonian, the dominant components of $|Z_2\rangle$ have are mainly supported on eigenstates with equally-spaced spectrum \cite{Turner:2018}. This is consistent with the SGA arguments and implies the revivals of fidelity, as we discussed earlier.
\begin{figure}[t!]
    \centering
    \subfloat[$|{Z}_2\rangle$]{
        \includegraphics[width=0.48\textwidth,height=0.25\textheight, keepaspectratio]{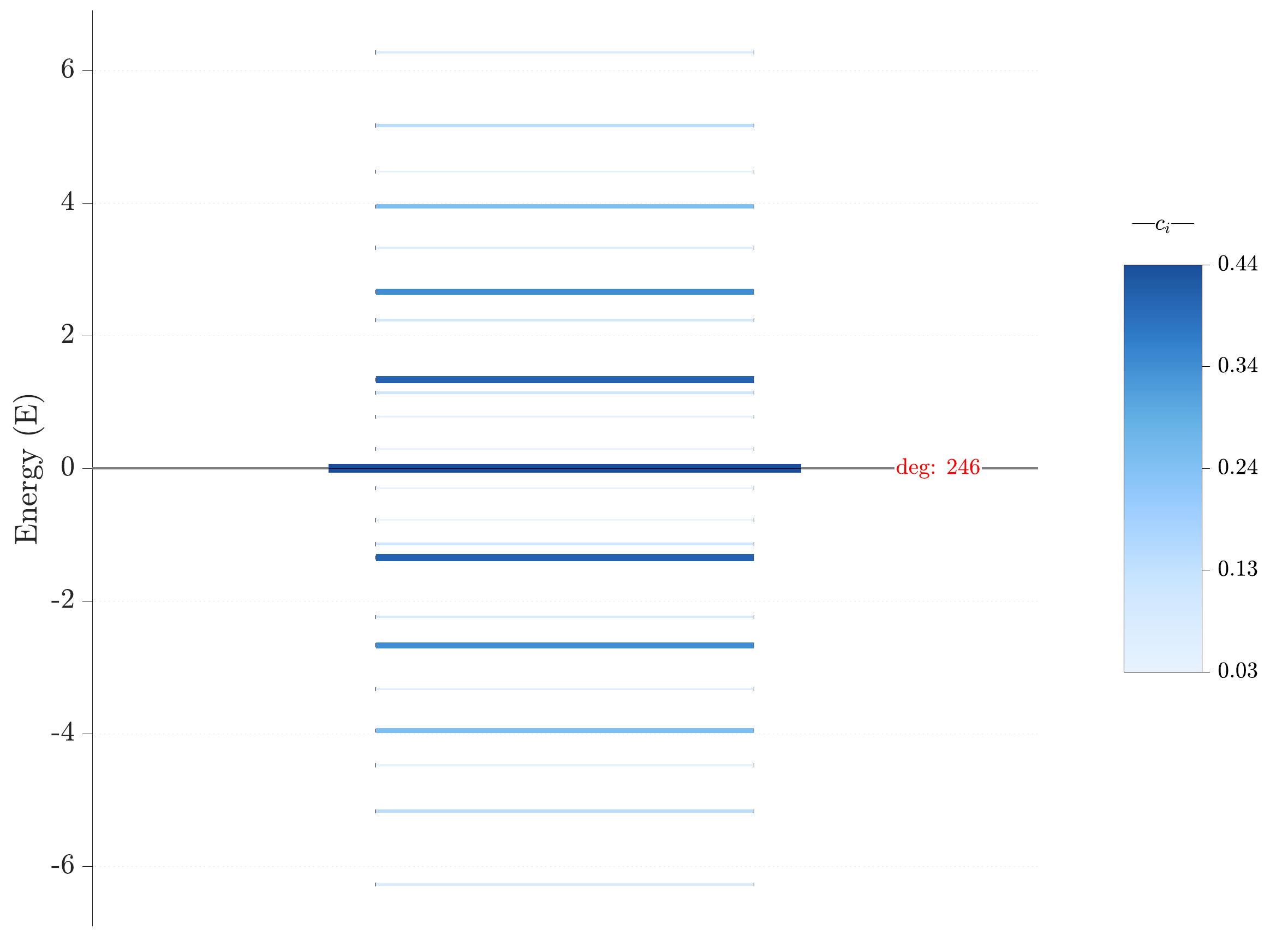}
        \label{fig:specz2}
    }
    \hfill
    \subfloat[ $|{Z}_4\rangle$]{
        \includegraphics[width=0.48\textwidth,height=0.25\textheight, keepaspectratio]{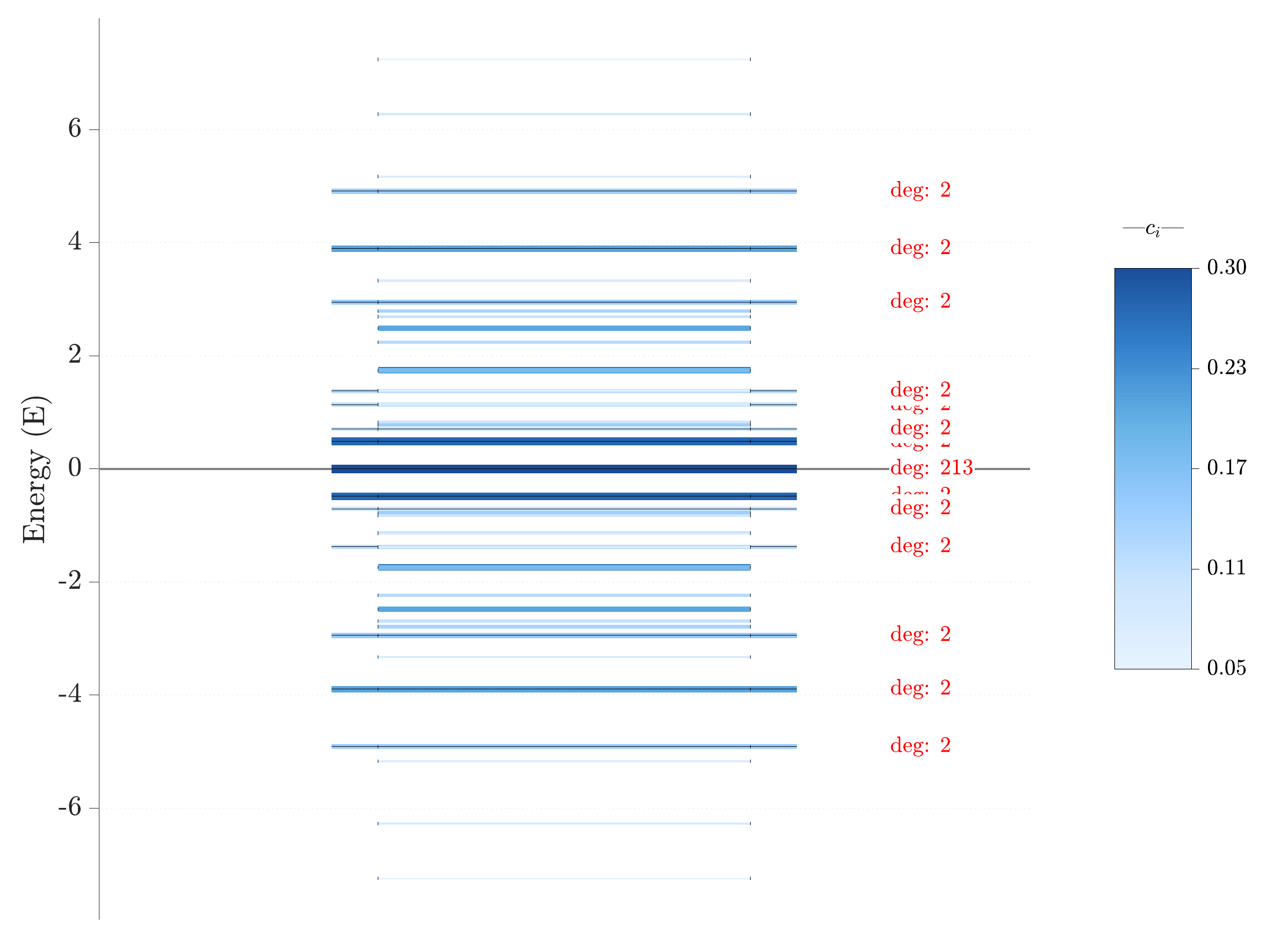}
        \label{fig:specz4}
    }
    \caption{The energy spectra $\{E_i\}$ of the eigenstates contained in the initial states $|Z_2\rangle$ (\ref{fig:specz2}) and $|Z_4\rangle$ (\ref{fig:specz2}) at system size $L = 12$. The color of the energy level indicates the component factor $c_i$ corresponding to the eigenstate $|E_i \rangle$. The $|Z_2\rangle$ state is predominantly composed of a few, nearly equally-spaced, eigenstates while $|Z_4\rangle$ exhibits a more complex and irregular distribution of dominant energy contributions.}
	\label{fig:spec}
\end{figure}

Now, to verify the existence of such an equally-spaced spectrum in $|Z_2\rangle$ (and its absence in $|Z_4\rangle$) suggested by the FFT of fidelity, we compare the energy spectra of the eigenstates contained in $|Z_2 \rangle$ and $|Z_4\rangle$ on \autoref{fig:spec}, with the components $c_i$ of these eigenstates marked with the blue brightness of the energy levels. The exact data of these two energy spectra are listed in \autoref{tab:two_tables} (see \autoref{eigenbasisZ2Z4}). Again, the energy spectrum contained in $|Z_2\rangle$ consists of a series of equally-gapped energy levels whose sum of probabilities, $\sum_i |c_i|^2$, dominate. However, we can hardly find any patterns in the energy spectrum contained in $|Z_4\rangle$. 

With these results in mind, we now consider the role played by the arched structure of the Lanczos coefficients from the perspective of the energy spectra. We still focus on the initial state $|Z_2\rangle$ and $|Z_4\rangle$ so that we can compare the energy spectra related to the arch with the full spectra in \autoref{fig:spec}.

\begin{figure}[H]
    \centering
    \subfloat[$|{Z}_2\rangle$; all $b_{n > L}$ removed]{
        \includegraphics[width=0.5\textwidth,height=0.28\textheight, keepaspectratio]{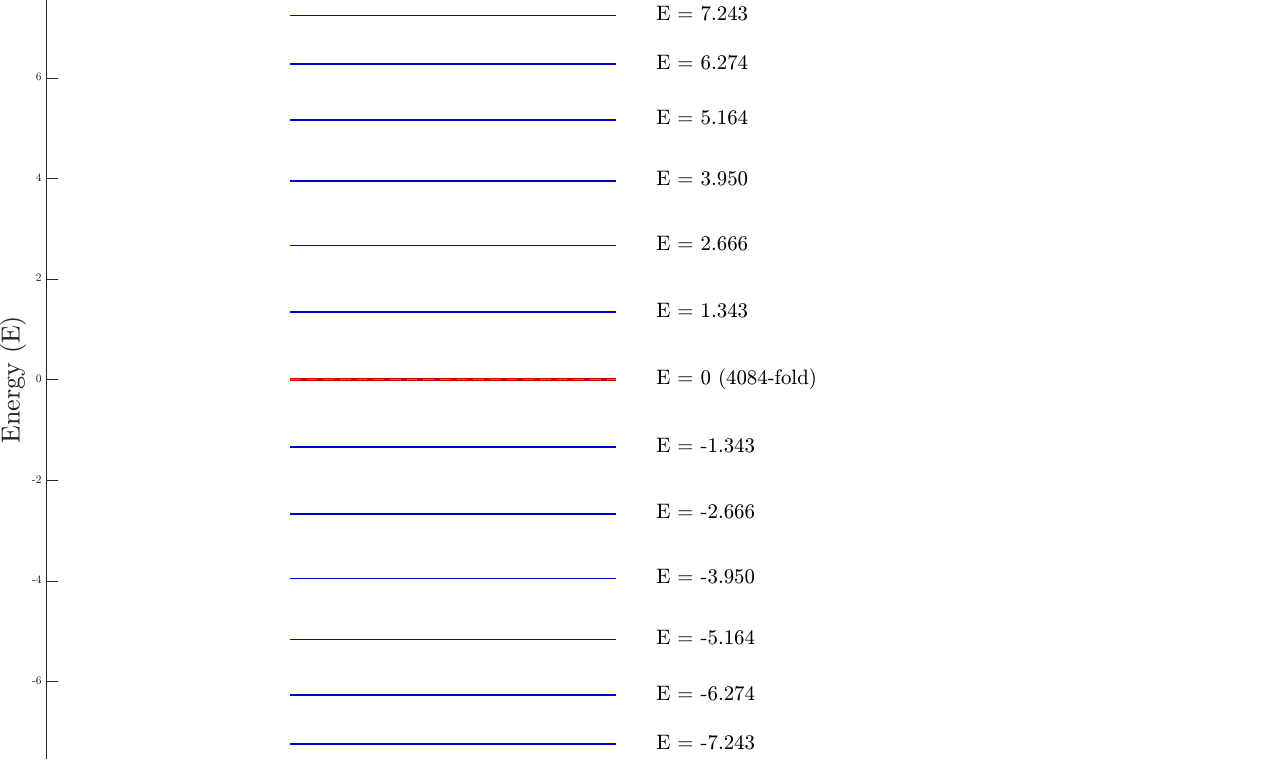}
        \label{fig:hconsz2}
    }
    \subfloat[$|{Z}_4\rangle$; all $b_{n>\tfrac{3}{4}L}$ removed]{
        \includegraphics[width=0.5\textwidth,height=0.28\textheight, keepaspectratio]{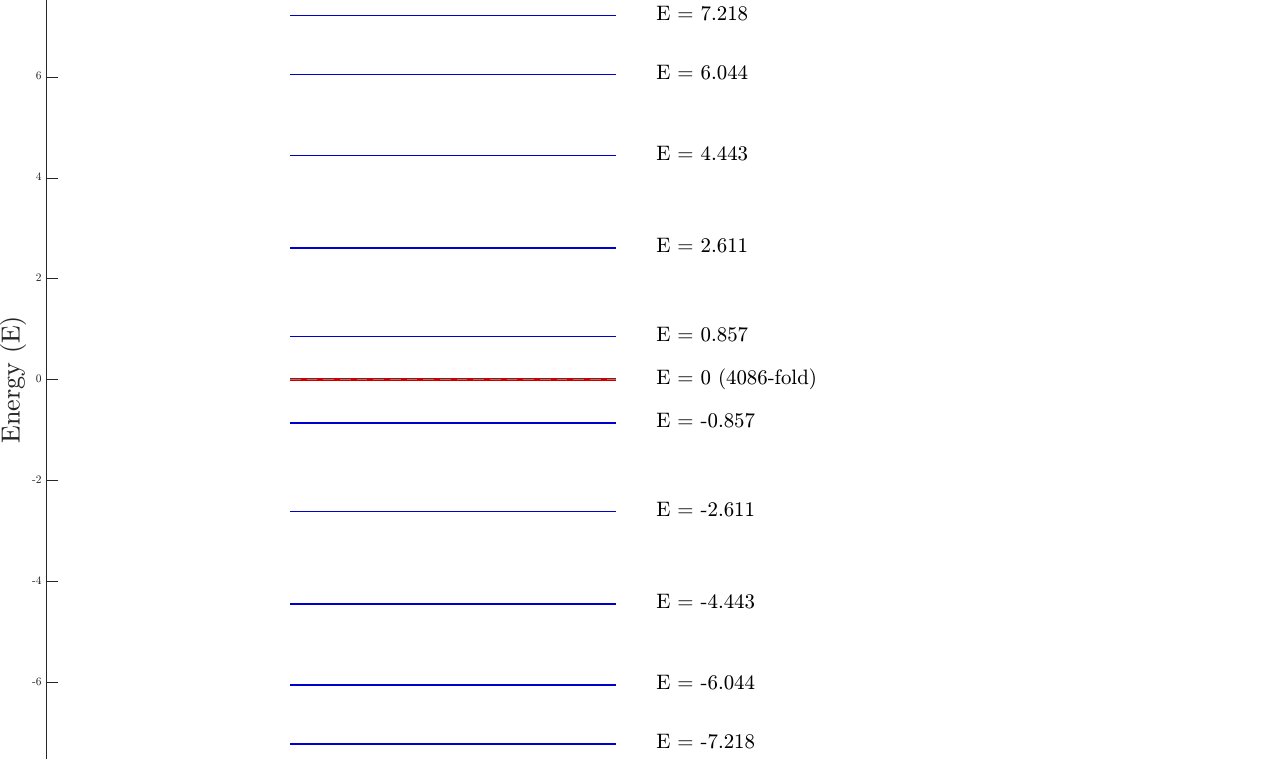}
        \label{fig:hconsz4}
    }\\
    \subfloat[$|{Z}_4\rangle$; all $b_{n>L}$ removed]{
        \includegraphics[width=0.5\textwidth,height=0.28\textheight, keepaspectratio]{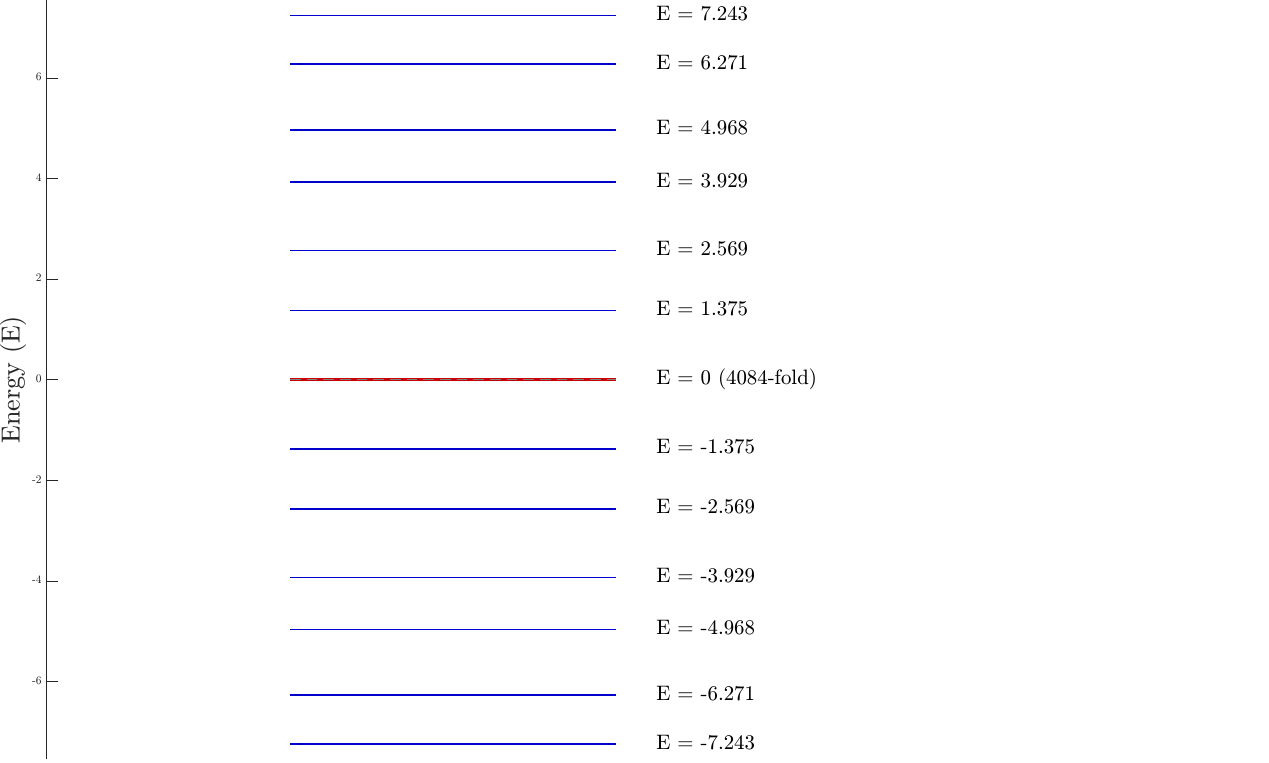}
        \label{fig:hconsz4L}
    }
    \subfloat[$E_n$ v.s. $n$]{
        \includegraphics[width=0.50\textwidth,height=0.28\textheight, keepaspectratio]{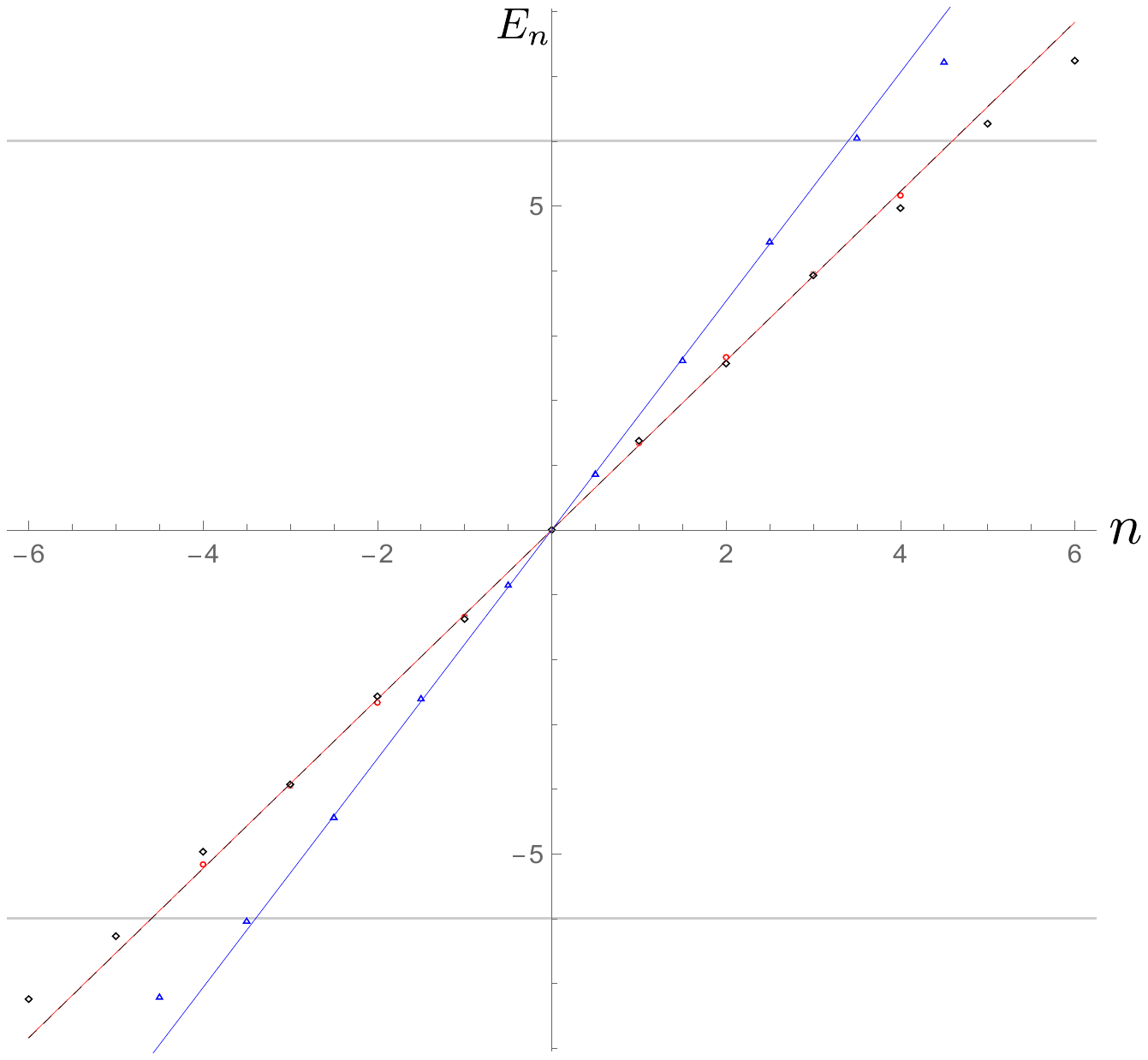}
        \label{fig:energap}
    }
	\caption{The energy spectra of the truncated PXP Hamiltonians whose tri-diagonal matrix representations have their main diagonal series vanish and their principal oﬀ-diagonal entries, namely the Lanczos coefficients $b_n$s, include only the portion associated with the entries listed in the \autoref{tab:lanczos-max-scar}. 
     }
	\label{fig:hcons}
\end{figure}

To avoid the influence of the other structure -- the buttress -- we remove all Lanczos coefficients $b_n$ for $n$ greater than $n_{\text{max}}$ given in \autoref{tab:lanczos-max-scar}. In particular, for initial state $|Z_k\rangle$ we regard all $b_n$s satisfying $n > \frac{k+2}{k}\ell = \frac{k+2}{2k}L$ as part of the buttress and remove them. Given that the two sets of the Lanczos coefficients, $a_n$s, where $a_n\equiv 0$, and $b_n$s, bring to us the tri-diagonal matrix representation of the full PXP Hamiltonian in \eqref{pxp-hamiltonian-pbc} (see \autoref{spreadcomplex}), by removing the buttress of the Lanczos coefficients, we obtain a truncated Hamiltonian whose diagonal coefficients vanish and sub-/supra-diagonal entries contain only arched structure. Now we consider the energy spectra of this truncated Hamiltonian, which results in \autoref{fig:hcons}. 
We find that for initial state $|{Z}_2\rangle$ (see Figure \ref{fig:hconsz2}) the energy gaps of the truncated Hamiltonian seem to be narrower as the absolute values of the energy $|E|$ increases while when $|E|< 6$ the energy gaps are almost equal and approximately equal to 1.31. If we ignore the zero energy levels $E=0$, we can see a similar pattern for $|{Z}_4\rangle$ (see Figure \ref{fig:hconsz4}), where the almost equal energy gaps are approximately equal to 1.77.

Without loss of completeness, we also consider the energy spectrum for $|Z_4 \rangle$ when the Lanczos coefficients of $b_{n>L}$s are removed in Figure \ref{fig:hconsz4L}. Similar to the previous two cases, when the energy $|E|> 6$ the energy gaps become narrower. In addition, we compute the standard deviations of the energy gaps of the lowest five energy levels in the three cases: $|Z_2\rangle$ without $b_{n>L}$, $|Z_4\rangle$ without $b_{n>\tfrac{3}{4}L}$, and $|Z_4\rangle$ without $b_{n>L}$, which are $0.05688, 0.09624, 0.1582$, respectively. This implies that the spectrum in the case of $|Z_4\rangle$ with $b_{n>L}$ removed is farther away from the equal-gapped energy spectrum because we retain more buttress data in this case.

In conclusion, \autoref{fig:hcons} implies a close connection between the arched structure in the growth of the Lanczos coefficients $b_n$s and the QMBS tower -- as we have argued in section \ref{lanczos-analytic} and especially in section \ref{lanczos-arch}. As for the role of the buttress, comparison of \autoref{fig:hcons} with \autoref{fig:spec} suggests that it leads to a re-distribution of the energy spectra, thus dragging the time evolution of the initial state from quantum many-body scar towards quantum thermalization.

For further discussions, we make a comment here on the narrower energy gaps of the truncated Hamiltonian for the $|Z_2 \rangle$ and the $|Z_4 \rangle$ cases: It should be noted that during the numerical computation we simply truncated the Hamiltonian such that its tri-diagonal matrix representation only contains the arched structure. This truncated Hamiltonian is different from the linear PXP Hamiltonian in \eqref{pxp-hami-sl3c-lin}, because, as revealed in section \ref{lanczos-buttress}, the residual PXP Hamiltonian in \eqref{pxp-hami-sl3c-quad} modifies the Lanczos coefficients $b_n$s from the very beginning, which causes the arch preserved by the truncated Hamiltonian to be a collapsed one. Thus, a QMBS tower, namely an equally-spaced energy spectrum, can be observed due to the contribution of the linear PXP Hamiltonian to the truncated Hamiltonian, but it is not a perfect one due to the influence of the residual PXP Hamiltonian.

Now, we can move to the width of the arch for different initial product states $|Z_k \rangle$. Analytic calculations of the linear PXP Hamiltonian tells us such a width, namely $\frac{k+2}{2k}L$, decreases with increasing $k$, which results in a narrower arch and therefore a truncated Hamiltonian with a smaller dimension. It can be seen from \autoref{fig:hcons} that for truncated Hamiltonians of smaller dimensions, the number of the energy levels that can be regarded as equally-gapped becomes smaller and harder to identify. In other words, the QMBS tower is less clear for a larger $k$ (a smaller arch width). This may be why it is difficult to see quantum revivals in the $|Z_k\rangle$ cases for $k \ge 4$. We have observed this connection between the arch width and the presence or absence of scars in section \ref{lanczos-arch}.

\begin{figure}[H]
    \centering
    \includegraphics[width=0.6\linewidth]{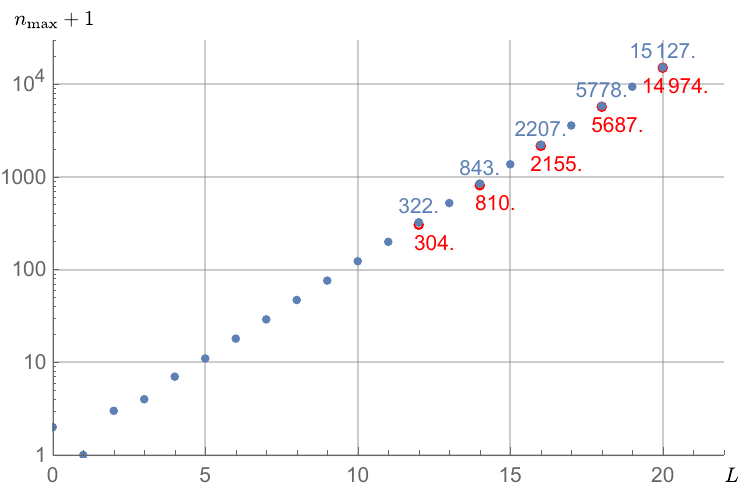}
    \caption{The first vanishing Lanczos coefficient $b_n$ for various lattice sizes (in red hollow circles, whose values are shown below), computed by the {\it full orthogonalization} (FO) algorithm. For reference, the Lucas numbers are represented by blue dots and their values are shown above. Here the initial state is chosen to be $|Z_2\rangle$.}
    \label{fig:lucas-nmax}
\end{figure}

We end this section by examining the estimates in \eqref{krdim} from section \ref{lanczos-analytic}, which provides an upper bound of the dimension of the Krylov subspace generated by acting the full PXP Hamiltonian in \eqref{pxp-hamiltonian-pbc} to arbitrary initial states. \autoref{fig:lucas-nmax} shows the relation between the first vanishing Lanczos coefficients $b_n$, $n_{\text{max}}+1$, and the lattice size $L$. It can be clearly seen that $n_{\text{max}}$ grows exponentially with $L$. To learn how far the upper bound given in \eqref{krdim} differs from the dimensions of Krylov subspaces generated by the PXP Hamiltonian, namely $n_{\text{max}}$, the Lucas numbers $\mathcal{L}_L$ are added. Although the difference between $n_{\text{max}}$ and the Lucas number increases, its ratio to $n_{\text{max}}$ decreases exponentially, which implies that for large lattice size $L$, we can use the Lucas number as an approximation to the dimension of the Krylov subspace generated by the full PXP Hamiltonian. Briefly speaking, we can modify \eqref{krdim} to 
\begin{equation}
    \operatorname{dim} \left( \mathscr{Kr}_{\text{PXP}} \right) \approx \mathcal{L}_{L}, \qquad L \equiv 2\ell \gg 1~.
\end{equation}

\section{The Dynamics of Spread Complexity} \label{pxpkcomplex}
Finally, we move to the dynamics of the spread complexity. In the previous section (in \ref{lanczos-analytic}) we have demonstrated that the full PXP Hamiltonian can be separated into two parts, the linear \eqref{pxp-hami-sl3c-lin} and the residual \eqref{pxp-hami-sl3c-quad}. We also argued that the linear part preserves a perfect QMBS tower (see \eqref{su2-basis-x}-\eqref{SGA-PXP}), dominating the arched structure in the growth of the Lanczos coefficients generated from $|Z_k \rangle$. Thus, to get some intuition for the time evolution of the spread complexity of a PXP scar, we first look at a simple model equipped with SGA in \eqref{eq:su2H} (see section \ref{lanczos-numeric}).
Starting from the initial state $|j,-j\rangle$, and identifying the circuit time $s$ as the physical time $t$, the evolution of spread complexity under such ``symmetry" Hamiltonian is given by \cite{Balasubramanian:2022tpr}:
\begin{equation}
\mathscr{C}_K(t)=\frac{2 j}{1+\frac{\gamma^2}{4 \alpha^2}} \sin ^2\left(\alpha t \sqrt{1+\frac{\gamma^2}{4 \alpha^2}}\right)~.
\label{ct}
\end{equation}

This spread complexity oscillates with time, with period $T= \frac{2\pi}{\sqrt{4\alpha^2 +\gamma^2}}$. 
We can contrast this with another typical spread complexity evolution observed for the {\it thermofield double} (TFD) initial state in models such as ${\text{SYK}}_4$ and ${\text{SYK}}_2$, where the spread complexity generally follows a linearly increasing ramp that culminates in a peak (absent in the case of ${\text{SYK}}_2$), followed by a gradual decline into a plateau \cite{Balasubramanian:2022tpr,Baggioli:2024wbz}. Hence, the long-lived oscillations in complexity that we see here could be interpreted as one of distinguishable features of scar states, highlighting their non-thermalizing dynamics.  This observation is consistent with the previous two studies in \cite{Bhattacharjee:2022qjw} and \cite{Nandy:2023brt}.  

Furthermore, we will also evaluate the {\it Krylov entropy}, or K-entropy for short, \cite{Barbon:2019wsy}\footnote{that is the Shannon entropy of the probability distribution $p_n(t)=|\varphi_n(t)|^2$.} for states as
\begin{equation}
	S_K(t)=-\sum_n|\varphi_n(t)|^2\ln|\varphi_n(t)|^2~,
\end{equation}
where $\varphi_n (t) = \langle K_n |\Psi (t) \rangle $ is the component of target state in the Krylov basis $|K_n\rangle$.
The K-entropy measures the information content or the effective dimension of the Hilbert space as function of time, and is often qualitatively related to the spread complexity\footnote{Roughly, in random matrix theories as well as chaotic models, spread complexity evolves similarly to the exponent of the K-entropy \cite{Balasubramanian:2022tpr,Bhattacharya:2023yec}}. 

\begin{figure}[h!]
    \centering
    \subfloat[$n=0,8$]{
        \includegraphics[width=0.48\textwidth]{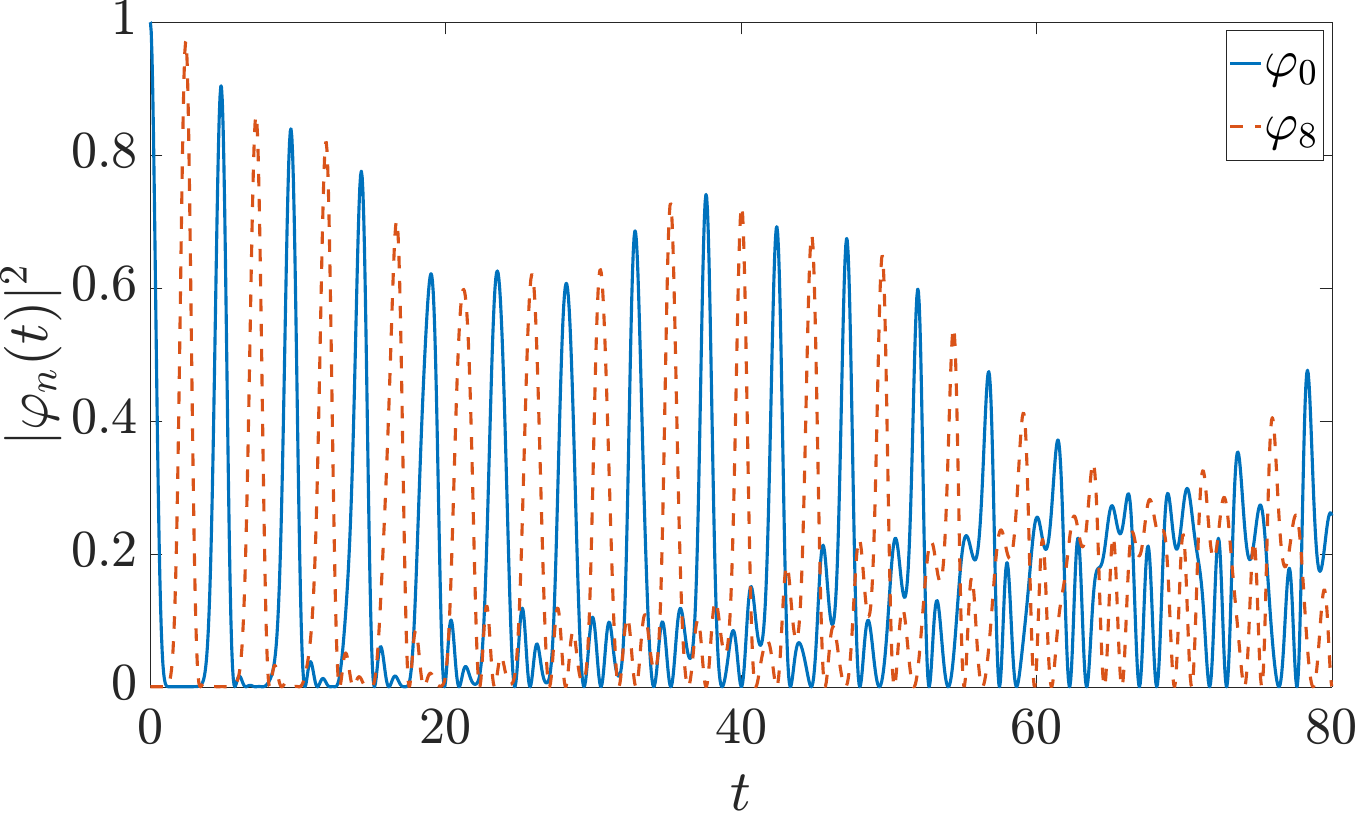}
    }
    \hfill
    \subfloat[$n=1,2,6,7$]{
        \includegraphics[width=0.48\textwidth]{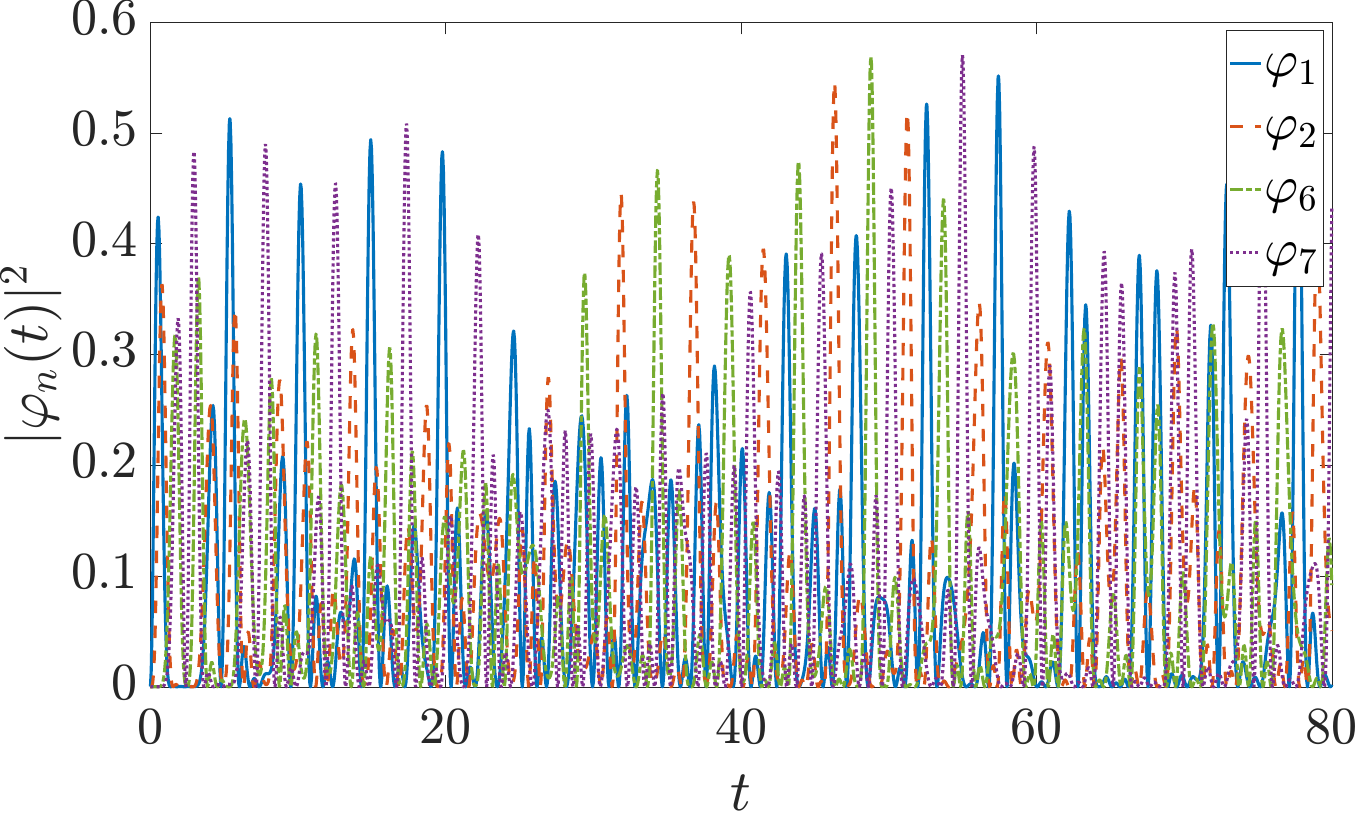}
    }

    \subfloat[$n=3,4,5$]{
        \includegraphics[width=0.48\textwidth]{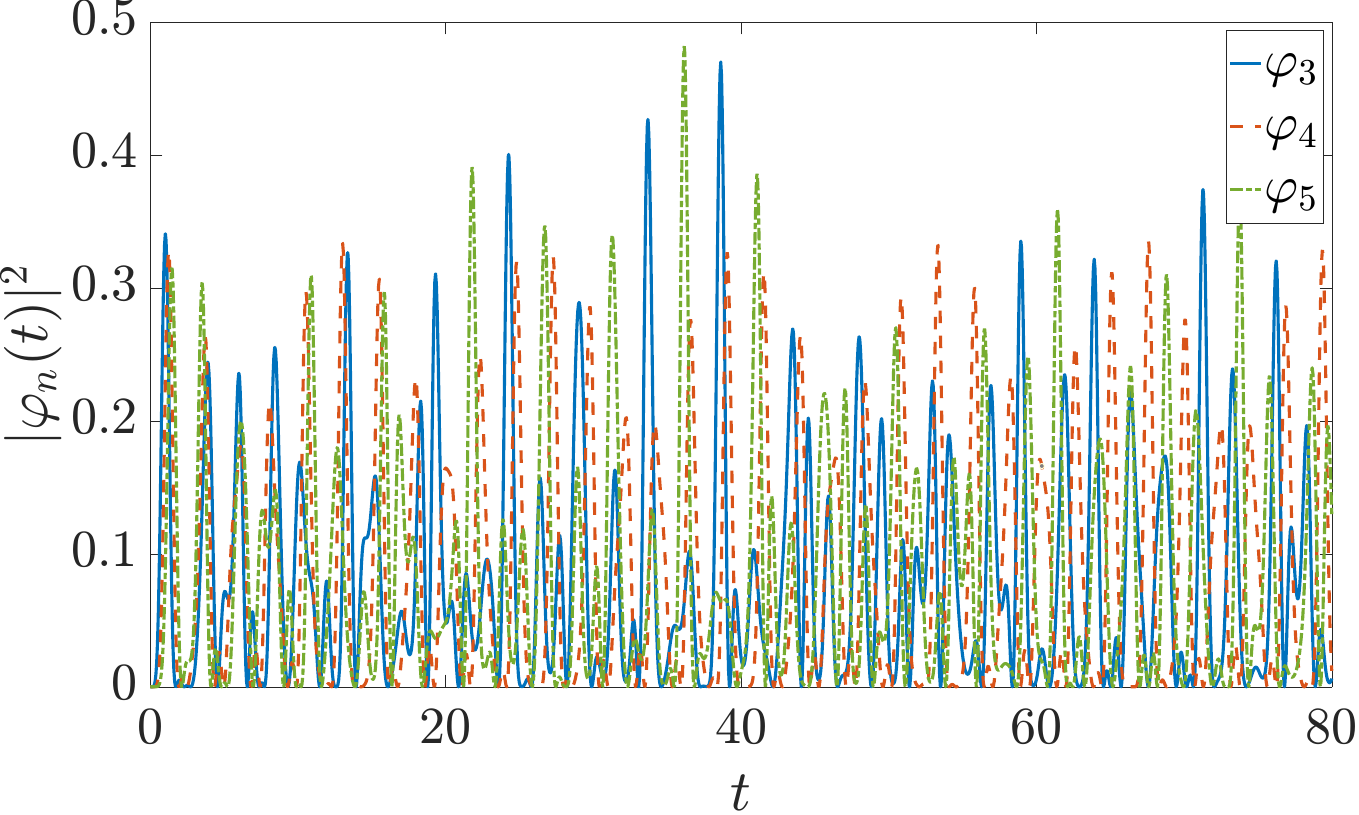}
    }    
	\hfill
    \subfloat[$n=9,10,11,12$]{
        \includegraphics[width=0.48\textwidth]{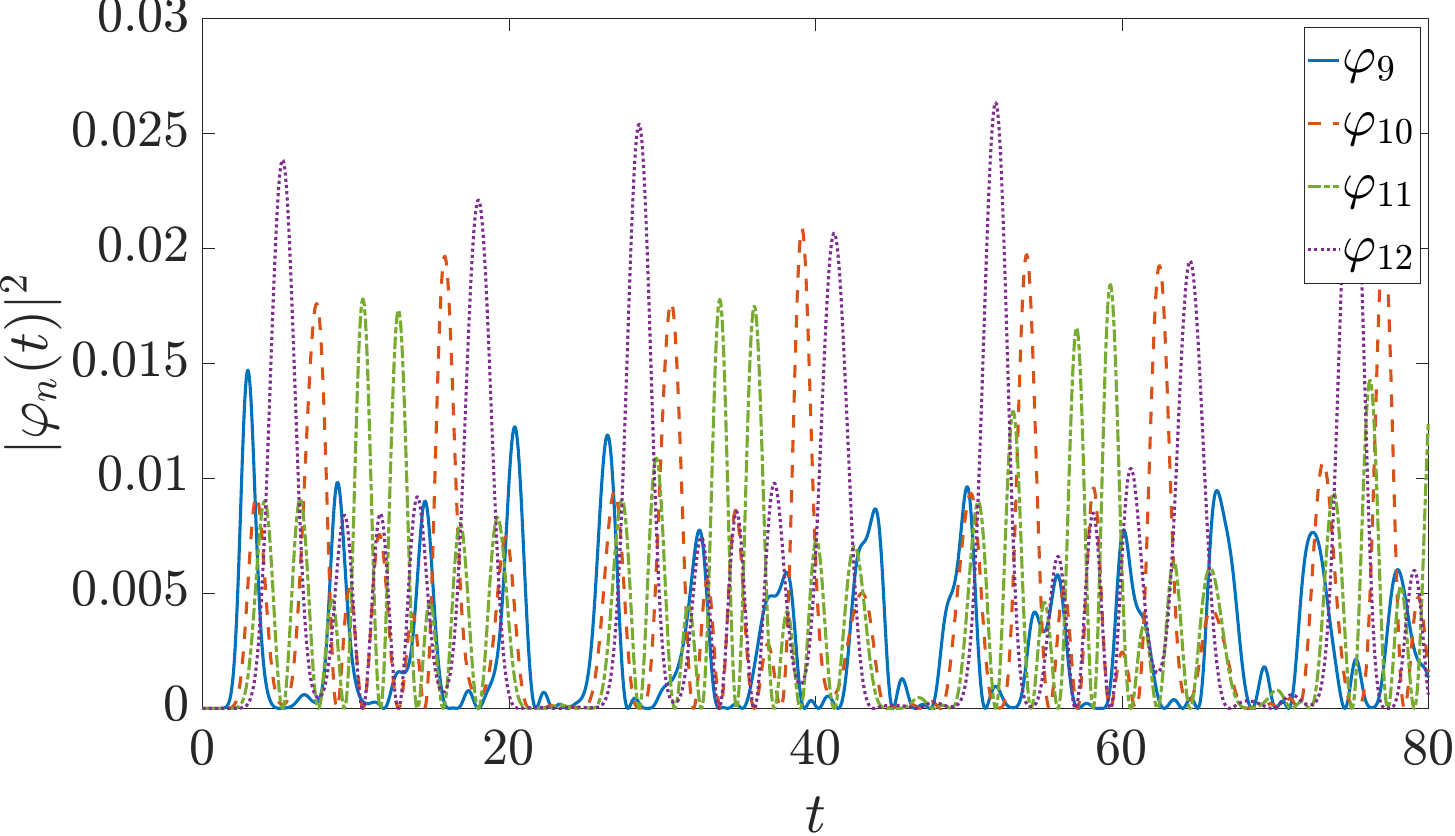}
    }
    \caption{Time evolution of probabilities of $p_n(t)=|\varphi_n (t)|^2$s, for the full PXP Hamiltonian evolution of the initial state $|Z_2\rangle$ and $L=8$. }
    \label{fig:varphi}
\end{figure}

Obviously, the components of target state in the Krylov basis, $\varphi_n(t)$ (or $p_n(t)=|\varphi_n(t)|^2$), play a key role in the calculation of spread complexity as well as spread entropy -- they actually build a bridge between the Lanczos coefficients and these two quantities. Hence we analyze the evolution of $\varphi_n(t)$s first. 
In \autoref{fig:varphi} we plot  $|\varphi_n(t)|^2$ for the scar state $|Z_2\rangle$ on a lattice of size $L=8$, governed by the full PXP Hamiltonian. In this case there are only twelve nonzero Lanczos coefficients, as shown in Appendix \ref{kryb-lc-4boxes}, and we only need to visualize thirteen probabilities $|\varphi_n(t)|^2$s. Clearly, all probabilities show oscillations. Furthermore, it can be found that each probability consists of simple oscillations. This is because
\begin{equation}
    \varphi_n(t) 
    = \sum_m \langle K_n |E_m \rangle \langle E_m | e^{-iHt} |\Psi_0 \rangle = \sum_m \langle K_n |E_m \rangle \langle E_m  |\Psi_0 \rangle \cdot e^{-iE_mt}~,
\end{equation}
where the sum is over all the eigenstates spanning the initial state $|\Psi_0 \rangle$. As a result,
\begin{equation}
    |\varphi_n(t)|^2 = \sum_{l,m} \langle \Psi_0|E_l\rangle \langle E_l | K_n \rangle\langle K_n |E_m \rangle \langle E_m  |\Psi_0 \rangle e^{i(E_l-E_m)t}~.
\end{equation}
Note that we obtain fidelity when taking $n=0$, and, like fidelity, other probabilities will also show revivals for the initial scar state $|Z_2\rangle$. 

Let us elaborate on this a bit more. One point to note is that $\varphi_n(t)$s, where $n \le 8$, can be divided into five groups, $\varphi_0(t)$ and $\varphi_8(t)$, $\varphi_1(t)$ and $\varphi_7(t)$, $\varphi_2(t)$ and $\varphi_6(t)$, $\varphi_3(t)$ and $\varphi_5(t)$, and finally $\varphi_4(t)$. The plots of these probabilities in each group have almost the same envelope (the smooth curve outlining the extremes of the oscillations). This should be correlated with the symmetry Lanczos coefficients 
\begin{equation}
    b_m = b_{n_0+1-m}, \quad m=1,2,\cdots,n_0
    \label{lc-symm}
\end{equation}
for some $n_0$, such that $\varphi_m(t)$ and $\varphi_{n_0-m}(t)$, where $m=0,1,\cdots, n_0$, are identical up to a phase factor, resulting in the same envelope of $|\varphi_n(t)|^2$. Obviously the Lanczos coefficients generated by repeatedly action of the linear PXP Hamiltonian \eqref{pxp-hami-sl3c-lin} on $|Z_2 \rangle$, that is, the Lanczos coefficients in \eqref{lanczos-coe-Z2}, satisfy such this relation with $n_0=L$. Since for $|Z_2 \rangle$ the contribution from the linear term dominates the Lanczos coefficients generated by the full PXP Hamiltonian (see Appendix \ref{kryb-lc-4boxes}), we get very similar envelopes for $|\varphi_n(t)|^2$ and $|\varphi_{8-n}(t)|^2$.

\begin{figure}[h!]
    \centering
    \subfloat[$n=0,10$]{
        \includegraphics[width=0.48\textwidth]{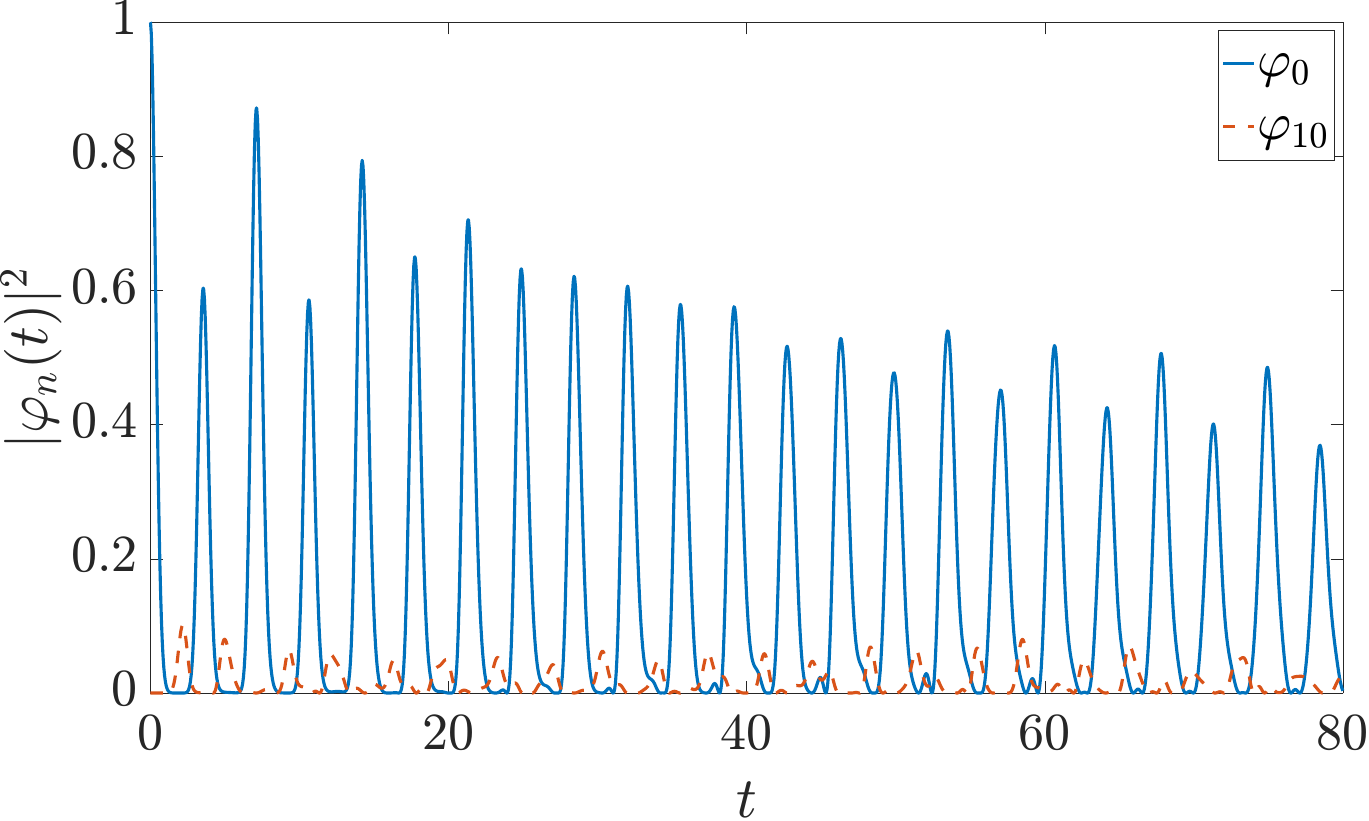}
    }
    \hfill
    \subfloat[$n=2,8$]{
        \includegraphics[width=0.48\textwidth]{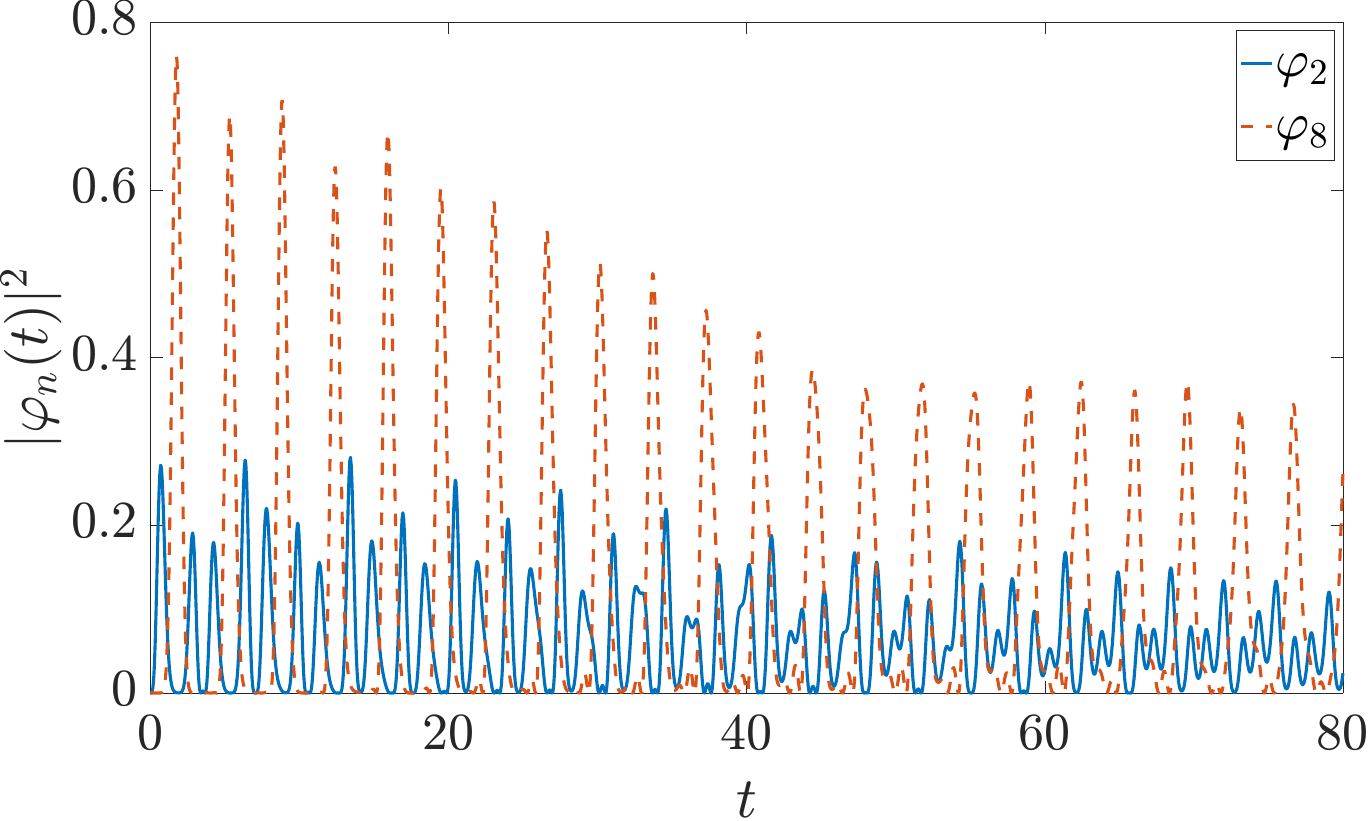}
    }
    \caption{Time evolution of several of the probabilities $p_n(t)=|\varphi_n (t)|^2 $, for the full PXP Hamiltonian evolution of the initial state $|Z_3\rangle$ and $L=12$.}
    \label{fig:varphiZ3}
\end{figure}

However, a qualitatively different pattern in the evolution of probabilities can be seen for the $|Z_3 \rangle$ initial state. \autoref{fig:varphiZ3} show time evolution of some of the probabilities in this case. According to our argument in section \ref{lanczos-arch}, the growth of the Lanczos coefficients contains an arch with width $\frac{5}{6}L$. This arch is not symmetric -- the Lanczos coefficients generated by the linear PXP Hamiltonian in \eqref{pxp-hami-sl3c-lin} do not satisfy the relation \eqref{lc-symm} with $n_0=\frac{5}{6}L$, as implied in \eqref{lanczos-coe-Zk}. This is actually what happens in general $|Z_k\rangle$, since \eqref{lanczos-coe-Zk} is valid for all $k$s. This way, allows us to conclude that whether the envelopes of $|\varphi_n(t)|^2$ and $|\varphi_{\frac{k+2}{2k}L-n}(t)|^2$ are almost identical has little to do with the scarring or thermalization of $|Z_k \rangle$.

Last but not least, we plot the spread complexity and Krylov entropy for the time evolution of three representative initial states, the two scar states $|{Z}_2\rangle$ and $|{Z}_3\rangle$, and the thermal state $|{Z}_4\rangle$, subjected to the full PXP Hamiltonian. Notably, both the complexity and the entropy of the time evolution of $|{Z}_2\rangle$ exhibit pronounced periodic oscillations. For the case of $|{Z}_3\rangle$ a similar behavior can be seen on a larger time scales, whereas for $|{Z}_4\rangle$, such oscillations are significantly suppressed, appearing as minor fluctuations around the plateau. This observation is consistent with the results in \eqref{ct}, where scar states are characterized by long-lived oscillatory dynamics, in contrast to non-scar states, which evolve towards a steady regime with only small-amplitude fluctuations at late times.
\begin{figure}[h!]
    \subfloat[Spread complexity and entropy for $|{Z}_2\rangle$]{
        \begin{minipage}{\textwidth}
            \centering
            \includegraphics[width=0.9\textwidth]{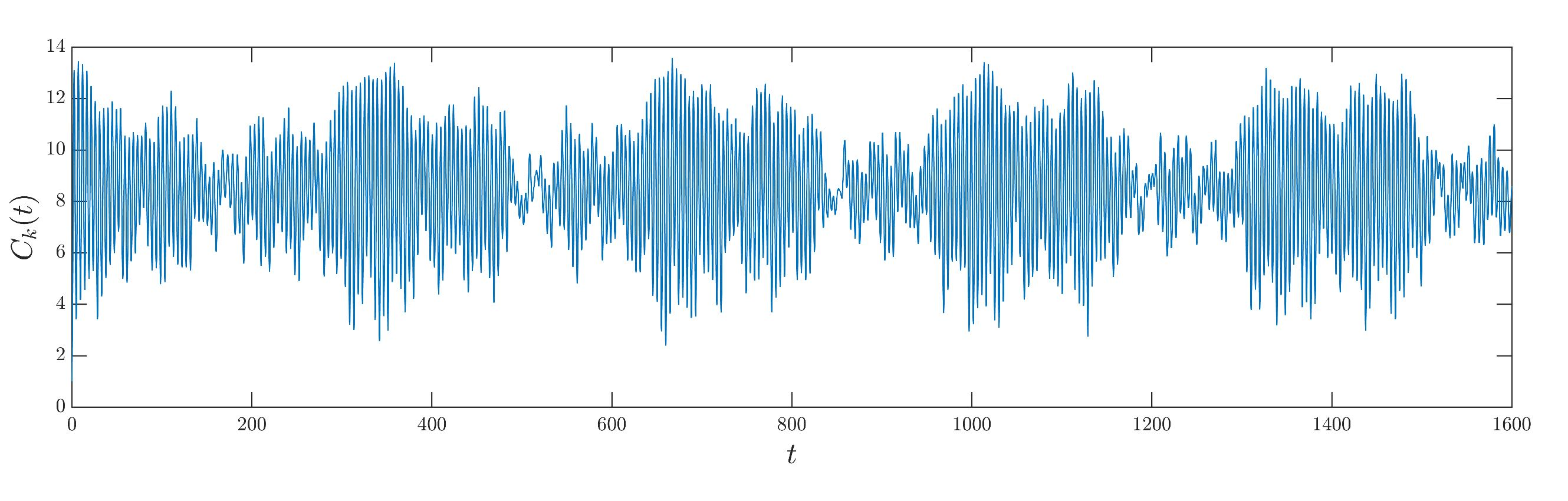}
            
            \vspace{-0.2cm} 
            \includegraphics[width=0.9\textwidth]{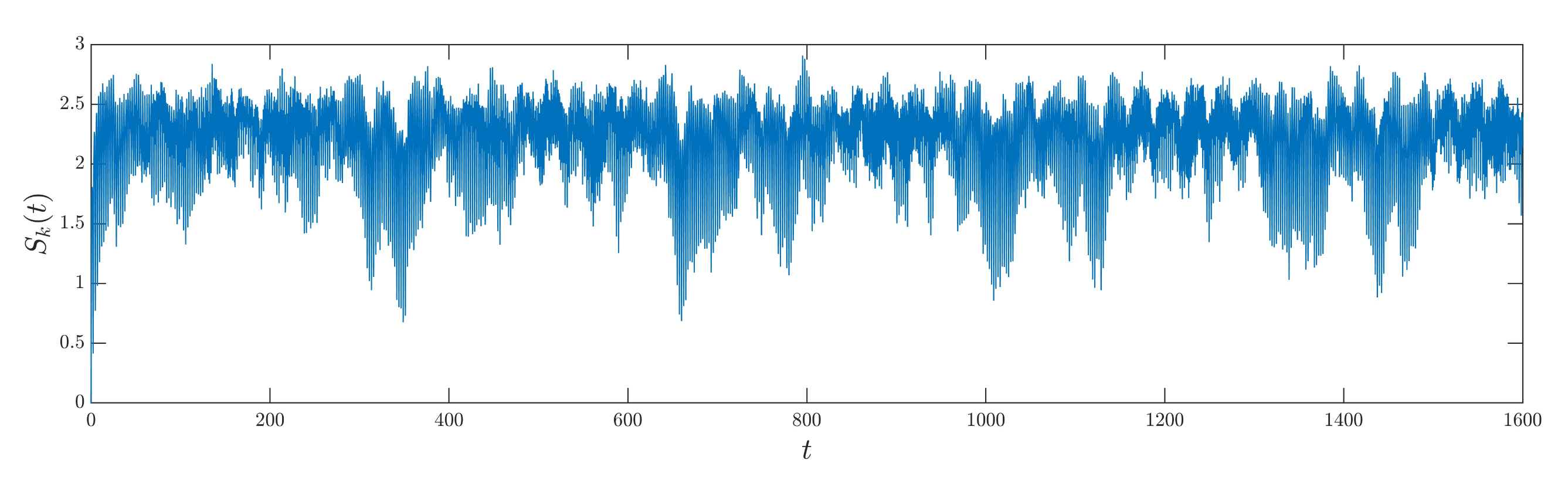}
        \end{minipage}
    }
    
    \subfloat[Spread complexity and entropy for $|{Z}_3\rangle$]{
        \begin{minipage}{\textwidth}
            \centering
            \includegraphics[width=0.9\textwidth]{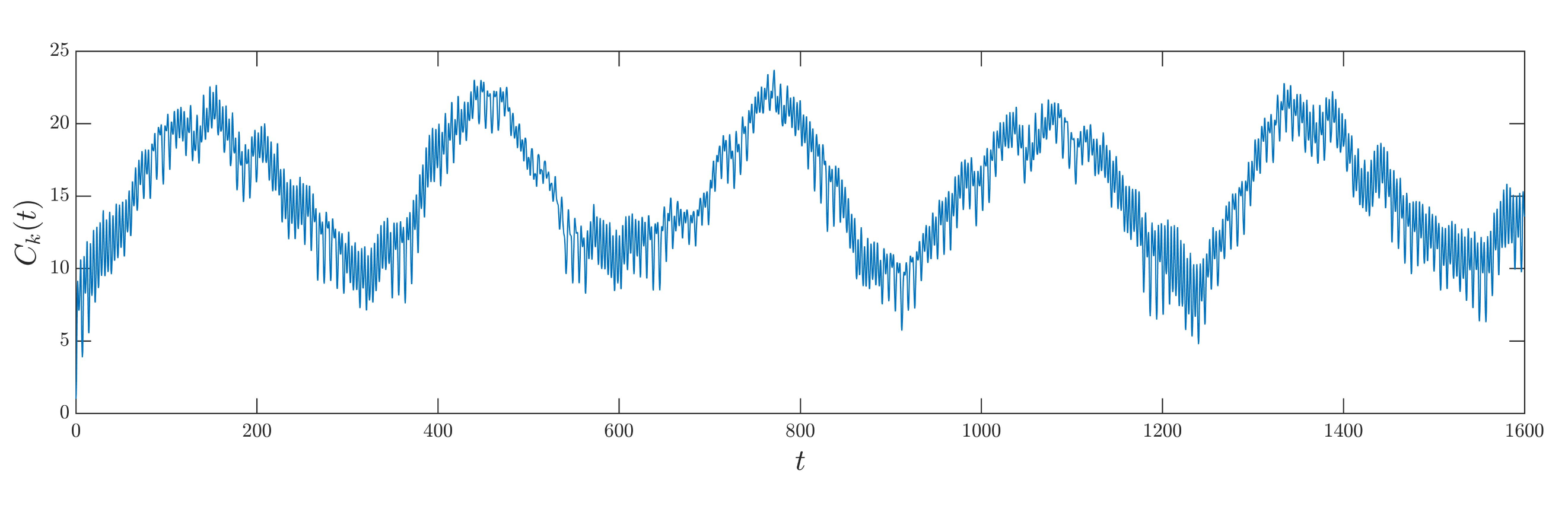}
            
            \vspace{-0.2cm}
            \includegraphics[width=0.9\textwidth]{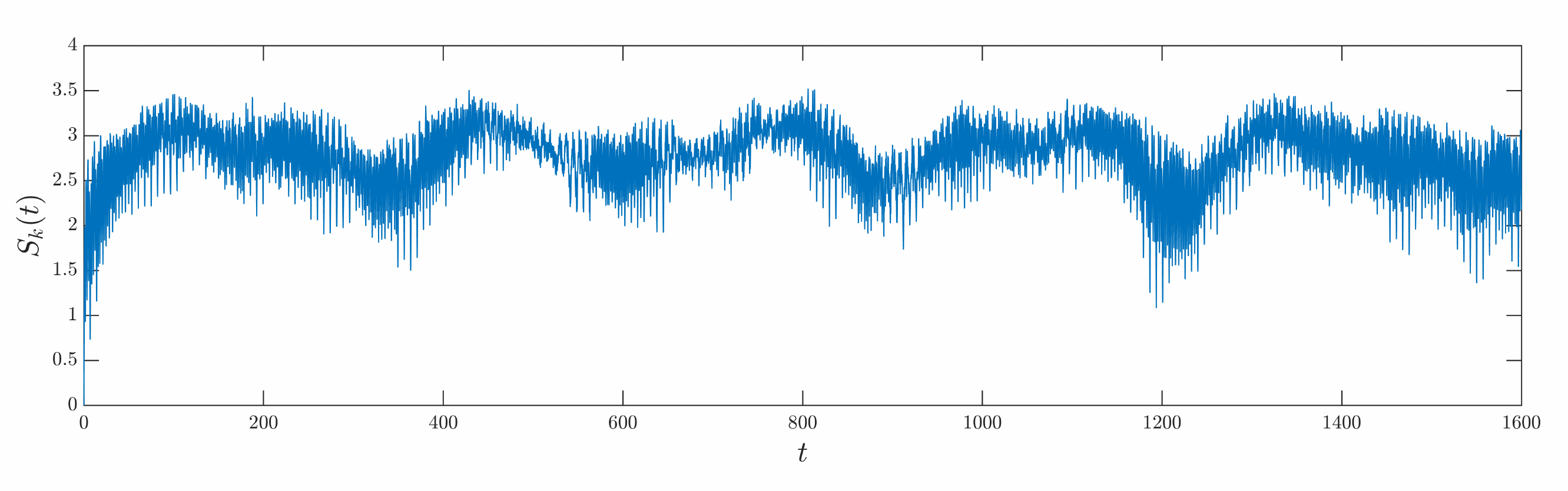}
        \end{minipage}
    }
    \caption{Spread complexity and K-entropy for the two known scar initial states $|{Z}_2\rangle$ and $|{Z}_3\rangle$ for size $L = 12$. Their complexity and entropy both show periodic oscillations, indicating non-thermal properties.} 
    \end{figure}
    \clearpage
\begin{figure}

    \subfloat[Spread complexity and entropy for $|{Z}_4\rangle$]{
        \begin{minipage}{\textwidth}
            \centering
            \includegraphics[width=0.9\textwidth]{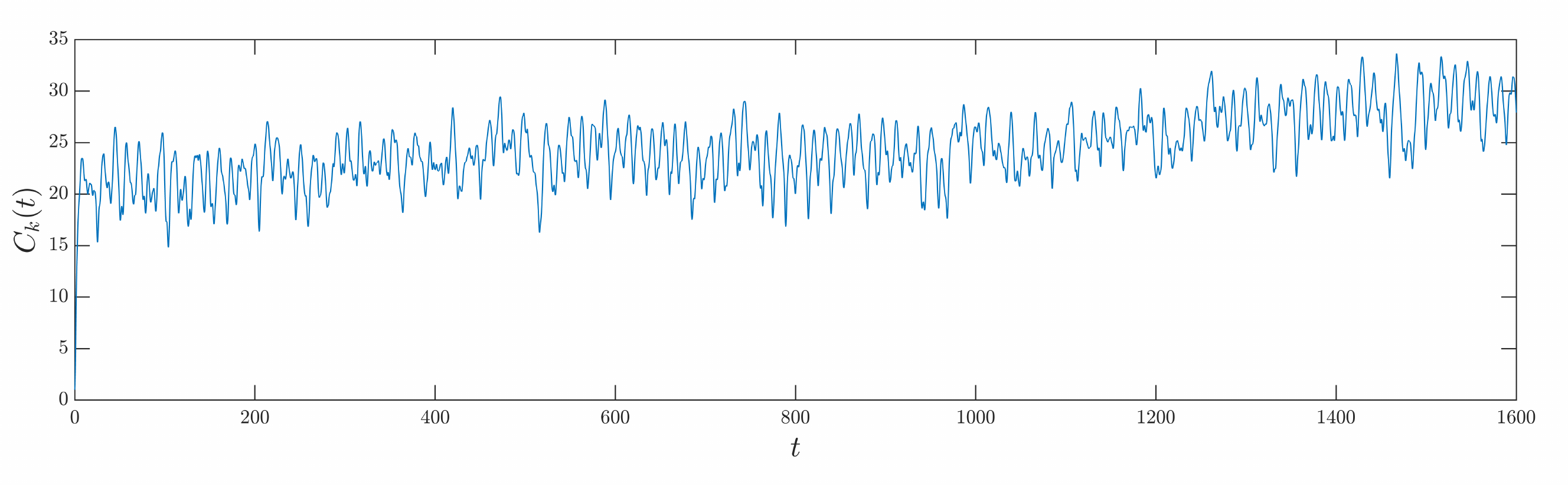}
            
            \vspace{-0.2cm}
            \includegraphics[width=0.9\textwidth]{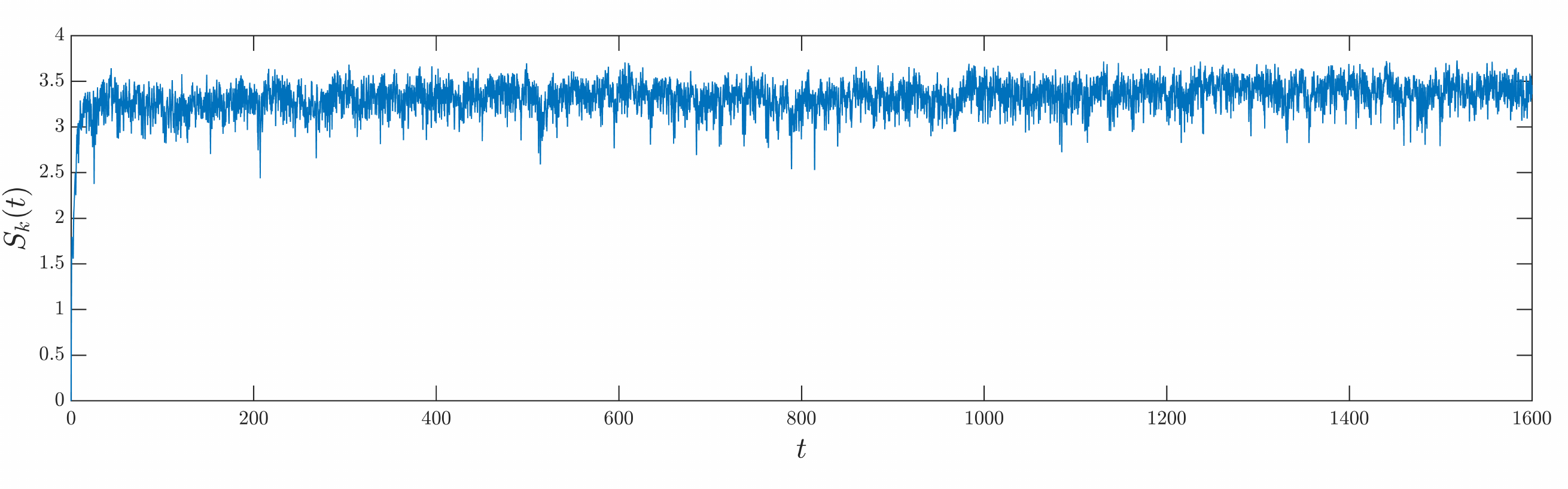}
        \end{minipage}
    }
    \caption{Spread complexity and K-entropy for non-scar initial state $|{Z}_4\rangle$ for size $L = 12$, which exhibit suppressed oscillations, indicating thermal properties.}
    
\end{figure}

\section{Conclusions} \label{conclusion}
In this work, we revisited quantum many-body scars in the PXP model and their features from the point of view of the spread complexity focusing on true symmetry structure of the model. More precisely, we considered a quench setup \eqref{quench-setup} where some product initial state, scarring or not, is taken out of equilibrium and evolves with the PXP Hamiltonian. We focused on size $L$ lattice chains with periodic boundary conditions \eqref{pxp-hamiltonian-pbc}. Identifying the circuit time $s$ with the physical time $t$, such a setup can be regarded as a quantum computing protocol \eqref{evolve} where the spread complexity $\mathscr{C}_K(t)$ can be computed. We then studied the time evolution of the spread complexity (\autoref{pxpkcomplex}), as well as the behaviors of the by-product quantities from the Krylov basis approach, such as the Lanczos coefficients $b_n$s (\autoref{lanczos-alg}).

We found that, in the PXP model, typical behavior of Lanczos coefficients $b_n$ (as functions of $n$) consists of two regimes, an arch and a buttress (section \ref{lanczos-numeric}). The arched structure can be attributed to the quantum many-body scar subspace, while the buttress is related to the thermal subspace. We also managed to estimate the upper bound on the dimension of the Krylov basis $n_{\text{max}}$ by Lucas numbers. 

Utilizing the representation theory of $\mathfrak{s}l_3(\mathbb{C})$, we separated the full PXP Hamiltonian with periodic boundary conditions in \eqref{pxp-hamiltonian-pbc} into a linear part \eqref{pxp-hami-sl3c-lin}, and a residual part \eqref{pxp-hami-sl3c-quad} (section \ref{lanczos-analytic}). We analyzed the influence of the linear part on Lanczos coefficients, estimated the width of the arch and observed that, the fact that the width of the arch is greater than some number in the interval $(3/4L,5/6L)$ can be used as a signal for the corresponding initial state to be a scar (section \ref{lanczos-arch}). Furthermore, from the residual PXP Hamiltonian, we managed to explain the collapse of the arch and the formation of the buttress (section \ref{lanczos-buttress}). We provided numerical evidence to support our findings in section \ref{lanczos-numberic-random}.

For spread complexity $\mathscr{C}_K(t)$ and Krylov entropy $S_K(t)$, we find that both quantities show periodic oscillations over time when the initial state is a scar state (e.g. $|Z_2\rangle$ and $|Z_3\rangle$), while for the initial thermalizing state (e.g. $|Z_4\rangle$) oscillations are suppressed. Thus, in addition to fidelity, spread complexity and Krylov entropy are two good probes of quantum revivals.

There are several open questions that warrant further investigation. First, as mentioned at the end of section \ref{lanczos-analytic}, our study in the residual PXP Hamiltonian is qualitative, which helps us understand the components of the Krylov basis in each irreducible representation of $D(1,0)^{\otimes \ell}$ (where $L=2\ell$), but the origin of the coefficients of these components remains unclear -- we may need more quantitative analysis of the residual PXP Hamiltonian to answer this question.

Second, it is worthwhile to have a more careful look at the components of the target state in the Krylov basis $\varphi_n(t) = \langle K_n | \Psi(t) \rangle$. We have found that, for initial state $|Z_2\rangle$, all probabilities $|\varphi_n(t)|^2$, including fidelity $|\varphi_0(t)|^2$, exhibit oscillations. We saw that these oscillations consist of simple harmonic oscillations whose frequency is the difference between the energies of the eigenstates contained in the initial state. This implies that, similarly to fidelity, higher probabilities $|\varphi_n(t)|^2$ can also distinguish between scars and thermal states. Furthermore, since $\varphi_n(t)$s are solutions to the discrete Schr\"odinger equations in \eqref{dschrodinger} parametrized by the Lanczos coefficients, it would be interesting to decode how the arch and the buttress are reflected in particular probabilities. One example, as shown in \autoref{pxpkcomplex}, is the similarity of the envelope of $|\varphi_n(t)|^2$ and the envelope of $|\varphi_{L-n}(t)|^2$ in $|Z_2\rangle$ case.

Finally, there is still a lot to understand about dynamics of spread complexity and Krylov entropy in systems that violate ETH. For example, for oscillatory time evolution, we can ask if there are periodic patterns, and what are their periods. Indeed, it might be interesting to zoom in and consider complexity for one period and compare it quantitatively with time evolution of some Hamiltonian acting on a specific initial state whose generated Lanczos coefficients form a perfect arch (see \eqref{ct}). Also analyzing scars in setups amenable to integrability techniques \cite{Sanada:2024gqs}, or in holographic models \cite{Liska:2022vrd,Caputa:2022zsr,Milekhin:2023was} will provide important tests for findings and discussions, and we leave it for the future.

\bigskip
\section*{Acknowledgments}

SL would like to thank Ling-Yan Hung, Yikun Jiang, Ce Shen, and Yixu Wang for their very useful suggestions to improve this manuscript. 
This work is supported by the ERC Consolidator grant (number: 101125449/acronym: QComplexity).  Views and opinions expressed are however those of the authors only and do not necessarily reflect those of the European Union or the European Research Council. Neither the European Union nor the granting authority can be held responsible for them. PC is supported by the NCN Sonata Bis 9 2019/34/E/ST2/00123 grant.

\appendix

\section{Lanczos Algorithm: The Origin and the Revisions } \label{lanalgrev}

\begin{figure}[H]
    \centering
    \subfloat[]{\includegraphics[width=0.8\textwidth]{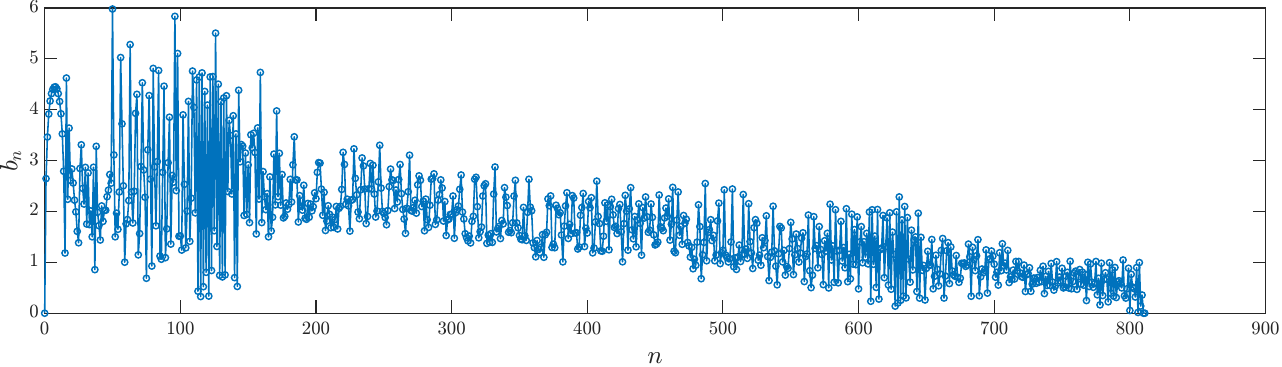} \label{fo}} \\
    \subfloat[]{\includegraphics[width=0.8\textwidth]{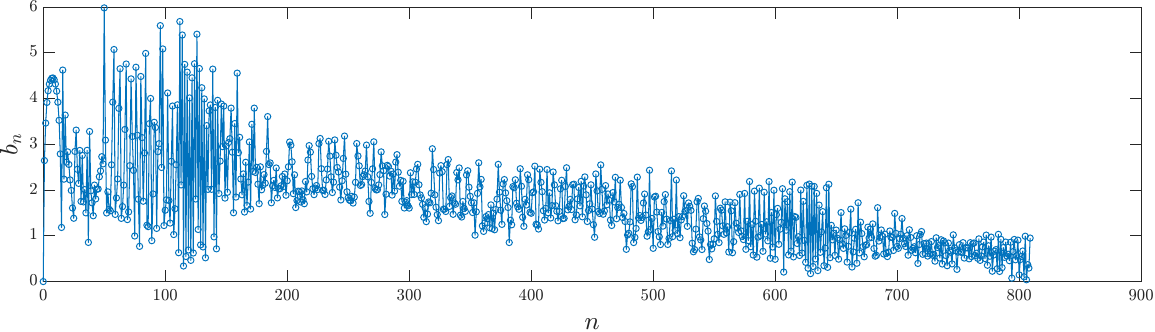}
    \label{pro}} \\
    \subfloat[]{\includegraphics[width=0.8\textwidth]{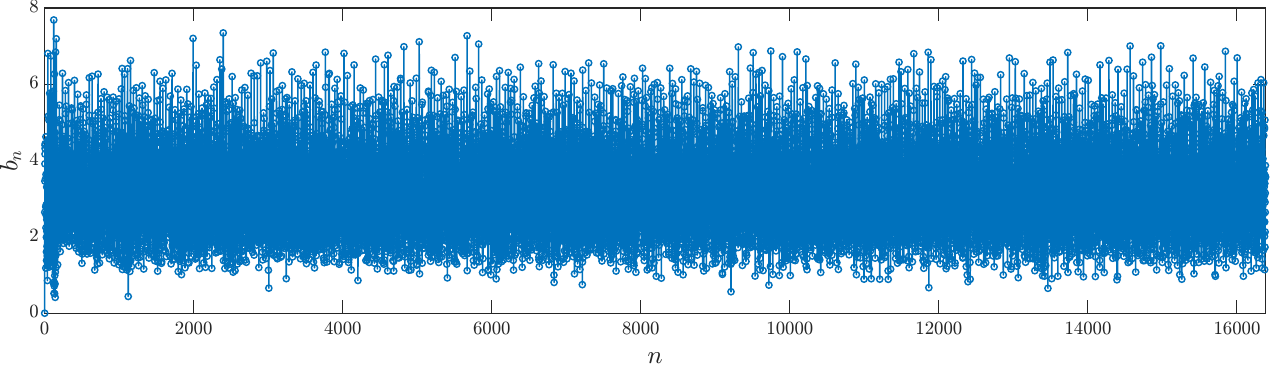}
    \label{stdlancz}} \\
    \caption{Lanczos coefficient $b_n$ for $|Z_2\rangle$ state generated by different algorithms with lattice size $L = 14$, the result by applying Full Orthogonalization Algorithm (FO) is shown in the top panel (\ref{fo}), Partial Re-Orthogonalization (PRO) is shown in the middle panel (\ref{pro}), and the result of normal lanczos algorithm is shown in the bottom panel (\ref{stdlancz}). In our computations, both the FO and PRO yielded nearly the same dimension of the Krylov subspace, with the values of $b_n$ also being nearly identical, except for slight numerical variations. However, due to severe numerical errors, the standard Lanczos algorithm fails to properly terminate the Krylov subspace construction, resulting in iterations that persist beyond the original system dimension. }
    \label{foprostd}
\end{figure}

When performing numerical computations, the standard Lanczos algorithm is typically not applied directly. This is because, after multiple iterations, numerical errors accumulate, leading to a rapid loss of orthogonality in the generated Krylov basis. To address this issue and ensure the stable generation of Lanczos coefficients and the corresponding Krylov basis, we employ the two numerical techniques \cite{Simon:1984,Parlett:1998} in our approach, a review of which can be found in \cite{Rabinovici:2020ryf}: {\it Full Orthogonalization} (FO) \cite{Parlett:1998} performs re-orthogonalization at every iteration step, offering strong numerical stability at the cost of higher computational and memory demands. In contrast, {\it Partial Re-orthogonalization} (PRO) \cite{Simon:1984} only performs re-orthogonalization under certain conditions, significantly reducing computational overhead and making it well-suited for large-scale sparse Hamiltonians. However, its numerical stability is generally lower than that of FO. In our paper, we have employed the FO algorithm to construct the Krylov subspace, in order to ensure the accuracy of the results. In this appendix, we enumerate in \autoref{foprostd} the Lanczos coefficients obtained using different computational methods, FO, PRO, and the original Lanczos algorithm, to evaluate their respective accuracy and stability.

\subsection{Full Orthogonalization (FO)}

\begin{algorithm}[H]
\caption{Full Orthogonalization Algorithm}
\begin{tabular}{|p{0.05\textwidth}|p{0.85\textwidth}|}
\hline
\multicolumn{2}{|c|}{\textbf{Algorithm: Full Orthogonalization}} \\
\hline
\textbf{Step} & \textbf{Description} \\
\hline
1 & \textbf{Initialization:} \\
  & $|K_0\rangle=\frac{1}{\sqrt{\langle\Psi_0|\Psi_0\rangle}}|\Psi_0\rangle$ \\
\hline
2 & \textbf{Initial Parameters:} \\
  & Set $a_0 = \langle K_0|H|K_0\rangle$ and $b_0=0$ \\
\hline
3 & \textbf{Main Loop:} For $n\geq1$ do: \\
  & \quad a. $|A_n\rangle=H|K_{n-1}\rangle$ \\
  & \quad b. Doing Gram-Schmidt with respect to all generated Krylov basis: \\
  & \quad\quad $|A_n'\rangle = |A_n\rangle-\sum_{i=0}^{n-1}|K_i\rangle\langle K_i|A_n\rangle$ \\
  & \quad c. Repeat Gram-Schmidt orthogonalization again (for numerical stability) \\
  & \quad d. $b_n = \sqrt{\langle A_n'|A_n'\rangle}$ \\
  & \quad e. If $b_n < \epsilon$ break (where $\epsilon \to 0$ is a threshold we set) \\
  & \quad f. Otherwise: \\
  & \quad\quad i. $|K_n\rangle=\frac{1}{b_n}|A_n'\rangle$ \\
  & \quad\quad ii. $a_n=\langle K_n|H|K_n\rangle$ \\
  & \quad\quad iii. Continue with next iteration \\
\hline
\end{tabular}
\end{algorithm}

\subsection{Partial Re-Orthogonalization (PRO)}

To perform Partial Re-Orthogonalization, we introduce a matrix $W$ with elements $W_{ij}=\langle K_i|K_j\rangle$ and a threshold $\epsilon$, for more details, please refer to \cite{Rabinovici:2020ryf}.

\begin{algorithm}[H]
\caption{Partial Re-Orthogonalization Algorithm}
\begin{tabular}{|p{0.05\textwidth}|p{0.85\textwidth}|}
\hline
\multicolumn{2}{|c|}{\textbf{Algorithm: Partial Re-Orthogonalization}} \\
\hline
\textbf{Step} & \textbf{Description} \\
\hline
1 & \textbf{Initialization:} \\
  & $|K_0\rangle=\frac{1}{\sqrt{\langle\Psi_0|\Psi_0\rangle}}|\Psi_0\rangle$ \\
  & Set $a_0 = \langle K_0|H|K_0\rangle$, $b_0=0$ and $W_{00} = 1$ \\
\hline
2 & \textbf{First Iteration:} \\
  & $|A_1\rangle = H|K_0\rangle$ \\
  & Do GS orthogonalization with respect to $|K_0\rangle$ \\
  & $b_1 = \sqrt{\langle A_1|A_1\rangle}$ \\
  & If $b_1<\sqrt{\epsilon}$ stop. Otherwise $|K_1\rangle = \frac{1}{b_1}|A_1\rangle$ and $a_1 =\langle K_1|H|K_1\rangle$ \\
  & Set $W_{01}=\epsilon$ and $W_{11}=1$ \\
\hline
3 & \textbf{Main Loop:} Loop for $n\geq2$, and for every n do: \\
  & \quad a. $|A_n\rangle=H |K_{n-1}\rangle-a_{n-1}|K_{n-1}\rangle-b_{n-1}|K_{n-2}\rangle$ \\
  & \quad b. Compute the a-priori Lanczos coefficent $b_n=\sqrt{\langle A_n|A_n\rangle}$ \\
  & \quad c. if $b_n<\sqrt{\epsilon}$ break, otherwise continue \\
  & \quad d. Orthogonalize explicitly $|A_n\rangle$ with respect to $|K_{n-1}\rangle$ \\
  & \quad e. Set $W_{nn} = 1$ and $W_{n-1,n} = \epsilon$ \\
  & \quad f. Loop for all $k=0,...,n-2$, determine $W_{kn}$ doing: \\
  & \quad\quad $\tilde{W} = b_{k+1} W_{k+1, n-1}^*+b_k W_{k-1, n-1}^*+a_kW_{k,n-1}^*-a_{n-1}W_{k,n-1}-b_{n-1} W_{k, n-2}$ \\
  & \quad\quad $W_{k n}=\frac{1}{b_n}\left[\widetilde{W}+\frac{\widetilde{W}}{|\widetilde{W}|} \cdot 2 \epsilon\|H\|\right]$ \\
  & \quad g. if for some $k\leq n-2$ $W_{kn}>\sqrt{\epsilon}$, do: \\
  & \quad\quad i. Re-orthogonalize explicitly $|A_n\rangle$ and $|A_{n-1}\rangle$ with respect to all previous Krylov basis \\
  & \quad\quad ii. From the new $|A_{n-1}\rangle$, re-compute $b_{n-1}$ and $a_{n-1}$. Break if $b_{n-1}<\sqrt{\epsilon}$, \\
  & \quad\quad\quad otherwise re-compute $|K_{n-1}\rangle$ \\
  & \quad\quad iii. From the new $|A_n\rangle$, re-compute $b_n$ and $a_{n}$. Break if $b_n<\sqrt{\varepsilon_M}$, \\
  & \quad\quad\quad otherwise compute $|K_n\rangle=\frac{1}{b_n}|A_n\rangle$ \\
  & \quad\quad iv. Set $W_{a, n-1}=\delta_{a, n-1}+\left(1-\delta_{a, n-1}\right) \varepsilon_M$, for all $a=0, \ldots, n-1$ \\
  & \quad\quad v. Set $W_{a, n}=\delta_{a, n}+\left(1-\delta_{a, n}\right) \varepsilon_M$, for all $a=0, \ldots, n$ \\
  & \quad h. Otherwise $|K_n\rangle=\frac{1}{b_n}|A_n\rangle$ \\
\hline
\end{tabular}
\end{algorithm}

\section{The Lie Algebra $\mathfrak{s}l_3 (\mathbb{C})$ and its Representations} \label{sl3c}

In this appendix we provide the mathematical background for section \ref{lanczos-analytic} -- the Lie algebra representation theory of $\operatorname{SU}(3)$ or more precisely $\mathfrak{s}l_3 (\mathbb{C})$. The appendix consists of two parts: 
In section \ref{sl3c-rootspace} we introduce the idea of $\mathfrak{s}l_3 (\mathbb{C})$ root space, from which we construct Casimir operators and determine irreducible representations \cite{Fulton2004-12,Fulton2004-13}.
In section \ref{sl3c-young} we label those irreducible representations by Young tableaux and determine corresponding Young symmetrizers, by which a construction of the irreducible representations can be provided \cite{Fulton2004-4}. 

\subsection{From Gell-Mann Matrices to Root Space of $\mathfrak{s}l_3 (\mathbb{C})$} \label{sl3c-rootspace}

We begin with the Gell-Mann matrices that appeared in \eqref{pxp-hamiltonian-GM}. Without loss of completeness, we list all the eight Gell-Mann matrices below:
\begin{eqnarray}
\lambda_{1}=  
    \begin{pmatrix}
    0&1&0\\
    1&0&0\\
    0&0&0\\
    \end{pmatrix}, \quad
& ~~\lambda_{2}=
    \begin{pmatrix}
    0&i&0\\
    i&0&0\\
    0&0&0\\
    \end{pmatrix}, \quad
&\lambda _{3}=
    \begin{pmatrix}
    1&0&0\\
    0&-1&0\\
    0&0&0\\
    \end{pmatrix}, \nonumber \\
\lambda_{4}=
    \begin{pmatrix}
    0&0&1\\
    0&0&0\\
    1&0&0\\
    \end{pmatrix}, \quad
&\lambda _{5}=
    \begin{pmatrix}
    0&0&-i\\
    0&0&0\\
    i&0&0\\
    \end{pmatrix}, &\\
\lambda_{6}=
    \begin{pmatrix}
    0&0&0\\
    0&0&1\\
    0&1&0\\
    \end{pmatrix}, \quad
& ~~\lambda_{7}=
    \begin{pmatrix}
    0&0&0\\
    0&0&-i\\
    0&i&0\\
    \end{pmatrix}, \quad
&\lambda_{8}=\frac {1}{\sqrt {3}}
    \begin{pmatrix}
    1&0&0\\
    0&1&0\\
    0&0&-2\\
    \end{pmatrix}. \nonumber
    \label{GM-com}
\end{eqnarray}
The eight Gell-Mann matrices $\{ \lambda_a, a=1,2, \cdots, 8 \}$ form basis that span the Lie algebra of the $\operatorname{SU}(3)$ group in the defining representation. They obey the commutation relation
\begin{equation}
    \left[\lambda _{a},\lambda _{b}\right]=2i\sum_{c=1}^{8}f_{abc}\lambda _{c}
\end{equation}
as well as the anti-commutation relation
\begin{equation}
\{\lambda _{a},\lambda _{b}\}={\frac {4}{3}}\delta _{ab}I_{3}+2\sum _{c=1}^{8}{d_{abc}\lambda _{c}},
\label{GM-anticom}
\end{equation}
where the structure constants, $f_{abc}$, are
\begin{equation}
    \begin{aligned}
    f_{123}&=1,\\
    f_{147}=-f_{156}=f_{246}=f_{257}=f_{345}=-f_{367}&={\frac {1}{2}},\\
    f_{458}=f_{678}&={\frac {\sqrt {3}}{2}},
    \end{aligned}
    \label{struct-const}
\end{equation}
and the symmetric coefficients, $d_{abc}$, are
\begin{equation}
    \begin{aligned}
    d_{118}=d_{228}=d_{338}=-d_{888}&={\frac {1}{\sqrt {3}}}\\
    d_{448}=d_{558}=d_{668}=d_{778}&=-{\frac {1}{2{\sqrt {3}}}}\\
    d_{344}=d_{355}=-d_{366}=-d_{377}=-d_{247}=d_{146}=d_{157}=d_{256}&={\frac {1}{2}}~.
    \end{aligned}
    \label{symm-coe}
\end{equation}
The symmetric coefficients are closely related to the cubic Casimir operator, which, together with the quadratic Casimir operator, are the two independent Casimir operators that can be constructed in the case of $\operatorname{SU}(3)$ group. In particular,
\begin{equation}
   \hat {C}_{\text{quad}}=\sum _{a}\hat {F}_{a}\hat {F}_{a}\qquad \qquad \hat {C}_{\text{cub}}=\sum_{abc}d_{abc}\hat {F}_{a}\hat {F}_{b}\hat {F}_{c}~,
   \label{casimir-def}
\end{equation}
where $\hat{F}_a$ can be any representation of $\operatorname{SU}(3)$ with structure constants $f_{ijk}$ given in \eqref{struct-const} and symmetric coefficients $d_{abc}$ given in \eqref{symm-coe}, for example $\hat{F}_a = \frac{1}{2} \lambda_a, ~ a=1,2,\cdots, 8$.

Now we let 
\begin{eqnarray}
    && H_{12} = 2 \lambda_3, \quad  H_{23} = \sqrt{3} \lambda_8 -\lambda_3, \label{cartan-subalg} \\
    && E_{12} = \lambda_1 + i \lambda_2, \quad 
     E_{23} = \lambda_6 + i \lambda_7, \quad
    E_{13} = \lambda_4 + i \lambda_5, \nonumber \\
    && E_{21} = \lambda_1 - i \lambda_2, \quad 
    E_{32} = \lambda_6 - i \lambda_7, \quad
    E_{31} = \lambda_4 - i \lambda_5, \label{root-sp}
\end{eqnarray}
where the subscripts are taken in such way that $E_{ij}$ denotes a matrix with a $1$ in the $i^{\text{th}}$ row and $j^{\text{th}}$ column,
\begin{equation}
    (E_{ij})_{mn} = \delta_{mi} \delta_{nj} ~,
\end{equation}
and $H_{ij}$ denotes a diagonal matrix with $1$ and $-1$ as the $i^{\text{th}}$ and $j^{\text{th}}$ diagonal elements, respectively, and therefore
\begin{equation}
    H_{ij} = E_{ii}- E_{jj}~.
\end{equation}
Plug \eqref{cartan-subalg} and \eqref{root-sp} into \eqref{pxp-hamiltonian-GM}, and we obtain \eqref{pxp-hamiltonian-sl3c}, notice that $\lambda_1^2$ and $\lambda_6^2$ can be rewritten by \eqref{GM-anticom} into
\begin{equation}
    \lambda_1^2 = \frac{4}{3} \mathbbm{1}_3 + \frac{2}{\sqrt{3}} \lambda_8~, \qquad 
    \lambda_6^2 = \frac{4}{3} \mathbbm{1}_3 - \lambda_3 - \frac{1}{\sqrt{3}} \lambda_8 ~.
\end{equation}

\eqref{cartan-subalg} and \eqref{root-sp} form the standard basis of the Lie algebra $\mathfrak{s}l_3 (\mathbb{C})$:
The two matrices in \eqref{cartan-subalg}, $H_{12}$ and $H_{23}$ are known as the basis of {\it Cartan subalgebra} $\mathfrak{h}$, the commutator between an arbitrary element of which and $E_{ij}$ is
\begin{equation}
    [H,E_{ij}] = ( h_i - h_j)E_{ij} ~, \qquad \forall ~ H = \operatorname{diag} (h_1, h_2, h_3) \in \mathfrak{h}~.
    \label{eigenvec}
\end{equation}
The other six basis in \eqref{root-sp}, usually also denoted as
\begin{eqnarray}
   && E_{\alpha} = E_{12}~, ~\qquad \qquad E_{\beta} = E_{23}~, \qquad \qquad E_{\gamma} = E_{13}~, \nonumber \\
   && E_{-\alpha} = E_{21} = E_{\alpha}^{\dagger} ~, \quad E_{-\beta} = E_{32} = E_{\beta}^{\dagger} ~, \quad E_{-\gamma} = E_{31}= E_{\gamma}^{\dagger}~,
\end{eqnarray}
satisfy the commutation relation
\begin{eqnarray}
 [E_{ij}, E_{kl}]
&=&
\begin{cases}
E_{ii}-E_{jj} =H_{ij}, & i=l,k=j\\
E_{il} ~\text{or}~ -E_{kj}, & k=j, i\ne l ~\text{or}~ i=l, k\ne j\\
0 & i\ne l, k\ne j
\end{cases}~,
\label{adjact}
\end{eqnarray}
the nonzero ones of which are
\begin{equation}
\begin{split}
& [E_{\alpha}, E_{-\alpha}] = H_{12}~,  \quad  [E_{\beta}, E_{-\beta}] = H_{23}~, \quad   [E_{\gamma}, E_{-\gamma}] = H_{13} ~, \\
& [E_{\alpha},E_{\beta}] = E_{\gamma}~,  \qquad  [E_{\beta},E_{-\gamma}] = E_{-\alpha} ~, \quad  [E_{\alpha},E_{-\gamma}] = E_{-\beta} ~. \\  
\end{split}
\tag{\ref{sl3c-com}}
\end{equation}

Of an element of Cartan subalgebra $\mathfrak{h}$ whose basis are given in \eqref{cartan-subalg}, we define a functional $\omega_i,~i=1,2,3$, such that 
\begin{equation}
    \omega_i (H) = h_i~, \qquad \forall~ H= \operatorname{diag} (h_1, h_2, h_3) \in \mathfrak{h}~
\end{equation}
is an inner product.
It is not difficult to see that $\{ \omega_1, \omega_2,\omega_3 \}$ span a dual (vector) space of the Cartan subalgebra $\mathfrak{h}$, denoted as $\mathfrak{h}^*$. However, $\omega_i$s are not linearly independent -- from the two basis of $\mathfrak{h}$, $H_{12}$ and $H_{23}$, we can see $h_1 + h_2 +h_3 \equiv 0$, which leads to the constraint
\begin{equation}
    \omega_1 +\omega_2 + \omega_3 \equiv 0 ~.
    \label{linear-dep}
\end{equation}
The introduction of $\omega_i$ turns \eqref{eigenvec} into
\begin{equation}
    [H,E_{ij}] = ( \omega_i - \omega_j)(H) E_{ij}~, \qquad \forall H \in \mathfrak{h}~.
\end{equation}
This implies that $E_{ij,i\ne j}$ is an eigenvector of any $H \in \mathfrak{h}$ with eigenvalue $( \omega_i - \omega_j)(H)$ and is therefore called an eigenvector of $\mathfrak{h}$ with eigenvalue $\omega_{ij} = \omega_i -\omega_j$. In particular, $E_{\alpha}$, $E_{\beta}$, and $E_{\gamma}$ have eigenvalues $\omega_{12}$, $\omega_{23}$, and $\omega_{13}$, respectively. Furthermore, for any two eigenvectors of $\mathfrak{h}$, $X$ and $Y$ with $[H,X] = \alpha (H) \cdot X, ~ [H,Y] = \alpha (H) \cdot Y$, their commutator
\begin{equation}
    [H, [X, Y ]] = [X, [H, Y ]] + [[H, X], Y ] = [X, \beta(H)Y ] + [\alpha(H)X, Y ] = (\alpha(H) + \beta(H))[X, Y ]~,
    \label{adjadd}
\end{equation}
where we have applied Jacobi identity at the first equation, is also an eigenvector of $\mathfrak{h}$ with eigenvalue $\alpha+\beta$. 

\begin{figure}[H]
    \centering
    \includegraphics[width=0.4\linewidth]{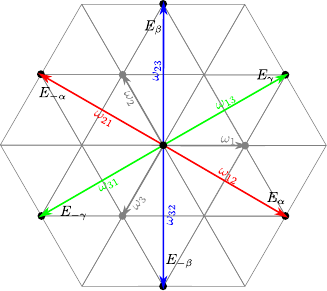}
    \caption{The hexagonal lattice spanned by $\{ \omega_1, \omega_2, \omega_3 \}$.The three generators of the root lattice $\Lambda_R$, $\pm \omega_{12}$, $\pm \omega_{23}$, and $\pm \omega_{13}$, where $\omega_{ij}= \omega_i -\omega_j$, are labeled by the corresponding $\mathfrak{s}l_3 (\mathbb{C})$ basis $E_{\pm \alpha}$, $E_{\pm \beta}$, and $E_{\pm \gamma}$.}
    \label{fig:root-lattice}
\end{figure}

Based on the facts above, we can interpret the dual space $\mathfrak{h}^*$ as a hexagonal lattice. The hexagonal lattice is generated by the three linearly dependent unit vectors $\omega_1,~\omega_2,~\omega_3$, the angle between two of which is $\frac{\pi}{3}$ according to \eqref{linear-dep}. Hence a lattice site can be labeled as $p \omega_1 -q \omega_3, ~ p,q \in \mathbb{Z}$. As for the eigenvector of $\mathfrak{h}$ with an eigenvalue $\alpha$, it is interpreted as a translation of the root lattice by vector $\alpha$. This translation is addable, as \eqref{adjadd} shows. Thus, starting from the origin of the root lattice where $p,q=0$, repeated translations ultimately generate a subset of $\mathfrak{h}^*$ called the {\it root lattice} $\Lambda_R$ of $\mathfrak{s}l_3 (\mathbbm{C})$. Figure \ref{fig:root-lattice} gives a schematic diagram as a summary of this paragraph. 

\begin{figure}[b!]
    \centering
    \subfloat[]{ \label{fig:highest-weight-vec-a}
    \includegraphics[width=0.4\linewidth]{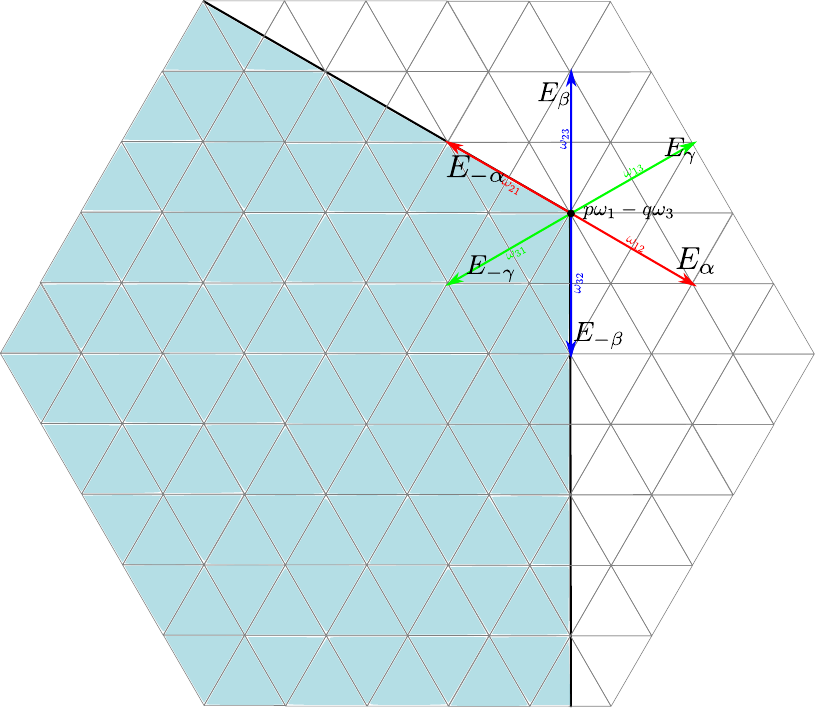}}
    \subfloat[]{ \label{fig:highest-weight-vec-b} 
    \includegraphics[width=0.4\linewidth]{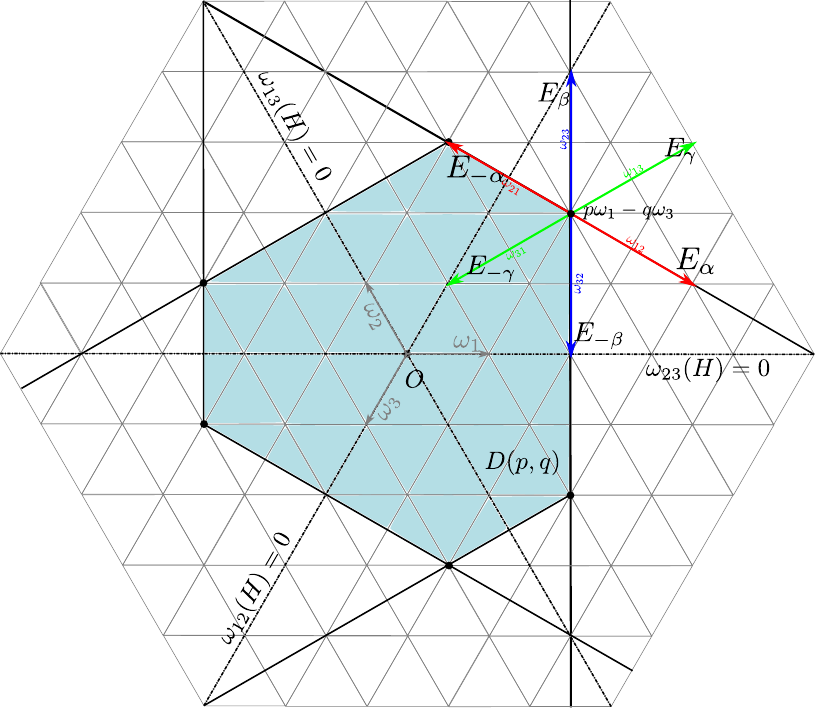}} \\
    \subfloat[]{ \label{fig:highest-weight-vec-c}
    \includegraphics[width=0.4\linewidth]{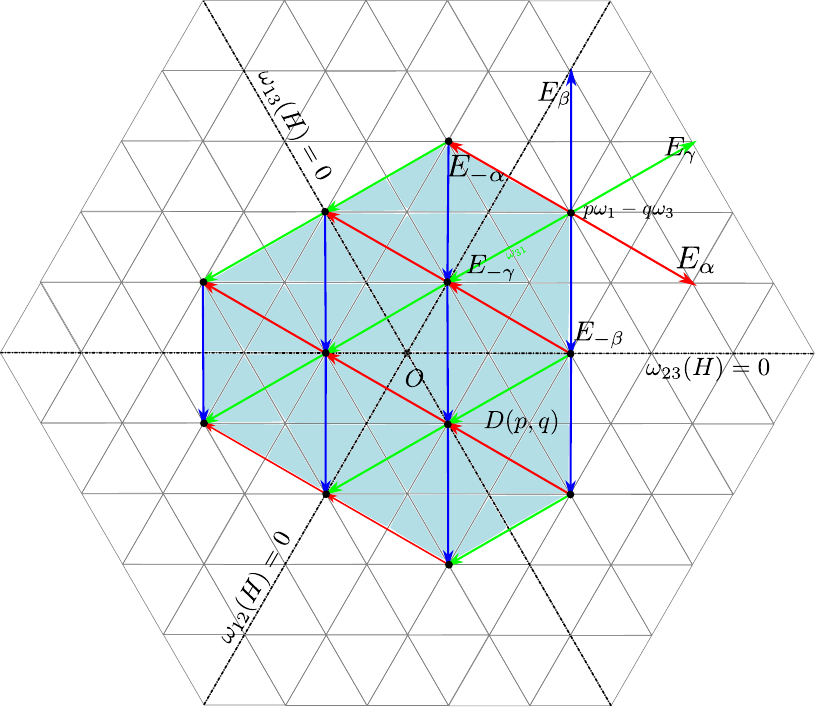}}
    \caption{The highest weight vector $p\omega_1-q\omega_3$ and the circled root lattice $\Lambda_R$, corresponding to irreducible representation $D(p,q)$ (In this specific case, $p=1,q=2$).}
    \label{fig:highest-weight-vec}
\end{figure}

However, the translation that generates the root space cannot be ceaseless -- it will stop at some lattice sites $p \omega_1 - q\omega_3$ determined by the two Casimir operators in \eqref{casimir-def}, and produces a root lattice that corresponds to an irreducible representation of $\mathfrak{s}l_3 (\mathbb{C})$ (denoted by $D(p,q)$ in the Dynkin basis), as we have learned from the representation theory of $\operatorname{SU}(2)$. To be concrete, the corresponding blocks of the quadratic and cubic Casimir operators to the irreducible representation $D(p,q)$ are
\begin{eqnarray}
     \hat{C}_{\text{quad}} &\fallingdotseq& \frac{1}{3}(p^2+q^2+3p+3q+pq) \mathbbm{1}_{\operatorname{dim} D(p,q)} ~,\nonumber \\
     \hat{C}_{\text{cub}} ~~ &\fallingdotseq& \frac{1}{18}(p-q)(3+p+2q)(3+q+2p) \mathbbm{1}_{\operatorname{dim} D(p,q)}~,
     \label{casimir-repr}
\end{eqnarray}
respectively, where $\operatorname{dim} D(p,q)$ is the dimension of the irreducible representation
\begin{equation}
    \operatorname{dim}(D(p,q)) = \frac{1}{2}(p + 1)(q + 1)(p + q + 2)~.
\end{equation}
By solving \eqref{casimir-repr}, the site $p\omega_1 -q\omega_3$ determined by the Casimir operators is the so-called highest weight vector. It is annihilated by $E_{\alpha},~E_{\beta}$ and therefore $E_{\gamma}$ as well. This fact help us circle the root lattice of the irreducible representation $D(p,q)$ as follows: First, draw lines parallel to $\omega_{12}$ and $\omega_{23}$ through the highest weight vector $p\omega_1 -q \omega_3$, respectively (see Figure \ref{fig:highest-weight-vec-a}), by doing so two borders are prepared. We then notice that $\{ H_{12}, E_{\pm \alpha} \}$ (or $\{ H_{23}, E_{\pm \beta} \}$) are the three generators of the $\mathfrak{su}(2)$ algebra (or, more precisely, the $\mathfrak{s}l_2(\mathbb{C})$ algebra). Our knowledge of $\mathfrak{s}l_2(\mathbb{C})$ algebra tells us that along the border parallel to $\omega_{12}$ (or $\omega_{23}$), the highest weight vector and the vector annihilated by $E_{-\alpha}$ (or by $E_{-\beta}$) should be symmetric about $\omega_{12}(H)=0$ (or $\omega_{23}(H)=0$). Thus, the other four borders can be determined by reflecting the two borders with respect to the three lines $\omega_{12}(H)=0$ and $\omega_{23}(H)=0$, respectively (see Figure \ref{fig:highest-weight-vec-b}). Eventually the sites on the root lattice $\Lambda_R$, whose range has been circled by the six borders, are generated by $E_{-\alpha}$, $E_{-\beta}$, and $E_{-\gamma}$ from the highest weight vector (Figure \ref{fig:highest-weight-vec-c}).

To sum up, from the eigenvalues of the two Casimir operators \eqref{casimir-def}, we can determine by \eqref{casimir-repr} the highest weight vector $p \omega_1-q\omega_3$ and circle the root lattice of the corresponding irreducible representation $D(p,q)$ (figure \ref{fig:highest-weight-vec}). However, the multiplicities of the lattice sites or, in a more physical sense, the degeneracies of the states remain unclear. We leave this issue for the next section.

\subsection{The Symmetric Group and Young Tableaux} \label{sl3c-young}

The root lattice $\Lambda_R$ constructed in the previous section does not tell us whether a lattice site is occupied by a singlet or a multiplet. Furthermore, the Hilbert space of the PXP model is not any irreducible representation of $\mathfrak{s}l_3(\mathbb{C})$ -- it is a subspace of the $\ell$-qutrit Hilbert $D(1,0)^{\otimes \ell}$. Here $D(1,0)$ denotes the only (irreducible) representation of the one-qutrit Hilbert space, the highest weight vector of which, corresponding to $| \underline{2} \rangle$, is $\omega_1$. Therefore, we also need details of decomposing $D(1,0)^{\otimes \ell}$ into irreducible representations. 

We utilize {\it Young diagrams} and {\it Young tableaux} to achieve these two goals, based on two facts: One is that the decomposition of $D(1,0)^{\otimes \ell}$ into irreducible representations is to obtain all possible Young tableaux of $\ell$ boxes. The other is that different irreducible representations correspond to different Young diagrams of $\ell$ boxes.
In particular, a Young diagram of $\ell$ boxes, where $\ell$ should be a non-negative integer, gives a partition of $\ell$ by arranging the boxes in left-aligned rows, with row lengths in non-increasing order. Then a Young tableau is a filling of a Young diagram with one positive integer in each box such that these integers are 
\begin{enumerate}
\item weakly increasing across each row from left to right, i.e. equalities are allowed;
\item strictly increasing down each column, i.e. equalities are not allowed.
\end{enumerate}
Of course one can require the increasing of integers across each row from left to right to be strict as well, which leads to a {\it standard} Young tableau, one where each number in the set $\{1, 2, \cdots, \ell\}$ appears only once.

It should be pointed out that by setting the two filling rules above, it is natural to construct the {\it Young symmetrizer} that preserves the symmetry on each row and the anti-symmetry on each column. To see that, we consider a standard Young tableau labeled by $\lambda$. For the integers of each row, we take out the elements of the permutation group $S_{\ell}$ that can be applied to them. Therefore, each row contributes a subset of $S_{\ell}$ and we can define the union of these subsets to be
\begin{equation}
    P_{\lambda} = \{\sigma \in S_{\ell} : \sigma ~\text{preserves each row} \} ~.
\end{equation}
Similarly, for columns we define
\begin{equation}
    Q_{\lambda} = \{ \sigma \in S_{\ell} : \sigma~\text{preserves each column} \} ~. 
\end{equation}
Here the meaning of ``preserves each row/column" have been explained above. Then we symmetrize each row by defining
\begin{equation}
    a_{\lambda} = \sum_{\sigma \in P_{\lambda}}e_{\sigma}~\,,
\end{equation}
and anti-symmetrize each column by defining
\begin{equation}
    b_{\lambda} = \sum_{ \sigma \in Q_{\lambda}} \operatorname{sgn}(\sigma)e_{\sigma} ~, 
\end{equation}
where $e_{\sigma}$ means the permutation operation $\sigma$, i.e.
\begin{equation}
    e_{\sigma}~: ~ i \mapsto \sigma (i), \quad \forall~ i \in \{ 1, 2, \cdots, \ell\}~,
\end{equation}
and $\operatorname{sgn}(\sigma) = (-1)^{\sigma}$ is the sign of the permutation $\sigma \in S_{\ell}$. The full Young symmetrizer associated to the standard Young tableau $\lambda$ is then given by 
\begin{equation}
    c_{\lambda} = a_{\lambda} \cdot b_{\lambda} 
    ~.
\end{equation}
One can use this strategy to symmetrize states of a given irreducible representation. Notice that for a given Young diagram, different standard Young tableaux give different Young symmetrizers and therefore symmetrize states in different ways, the copies of a specific irreducible representation into which $D(1,0)^{\otimes \ell}$ is decomposed are labeled by the standard Young tableaux.

For irreducible representations of $\mathfrak{s}l_3 (\mathbb{C})$, the positive integers used to fill a Young diagram must belong to the set $\{1,2,3\}$, the elements of which correspond to the three states of the qutrit $|2\rangle$, $|0\rangle$, and $|1\rangle$. This implies that any column with three boxes has unique filling, and furthermore, any column with more than three boxes is not allowed by $\mathfrak{s}l_3 (\mathbb{C})$. Thus, a Young diagram can be denoted by the number of the single-box columns, $p$, and that of the double-box columns, $q$, since the fillings of both are nontrivial. This leads to the notation for irreducible representations we used in the previous section, namely $D(p,q)$.
Enumerate all possible fillings (Young tableaux) of a Young diagram with $\{1,2,3\}$ and we obtain the root lattice of $D(p,q)$ that each site is described by one filling (or more, if the multiplicity of the site is nontrivial).

In conclusion, the unresolved issue left in the previous section can be solved by Young diagrams and Young tableaux. In particular, every irreducible representation of $\mathfrak{s}l_3 (\mathbb{C})$ corresponds to a Young diagram. The decomposition of $D(1,0)^{\otimes \ell}$ into irreducible representations can be expressed by enumerating all possible standard Young tableaux of all Young diagrams. As for the details of the root lattice, it is revealed by listing all the fillings of a given Young diagram with integers in $\{1,2,3\}$.

\section{The PXP Hamiltonian and the $\operatorname{SU}(3)$ Casimir Operators} \label{casimir-pxpquad}

In this appendix we calculate the commutation relations between the residual PXP Hamiltonian in \eqref{pxp-hami-sl3c-quad} and the Casimir operators defined in \eqref{casimir-redrep}. 

We begin with the reducible representation $D(1,0)^{\otimes \ell}$. Assume that it can be decomposed into irreducible representations $D(p,q)$s
\begin{equation}
    D(1,0)^{\otimes \ell} = \bigoplus_{p,q \ge 0, ~ p+2q \le \ell} \sigma(p,q) D(p,q), \quad p,q \in \mathbb{N}~.
\end{equation}
This implies there exists a unitary matrix $P$ to break the Cartan-Weyl basis \eqref{sl3c-redrep} into diagonal blocks, i.e.
\begin{eqnarray}
    E_{\mu}^{D(1,0)^{\otimes \ell}} =P \left( \bigoplus_{p,q \ge 0, ~ p+2q \le \ell} \sigma(p,q) E_{\mu}^{D(p,q)} \right) P^{\dagger}, \quad \mu = \pm \alpha, \pm \beta, \pm \gamma~; \nonumber \\
    H_{I}^{D(1,0)^{\otimes \ell}} =P \left( \bigoplus_{p,q \ge 0, ~ p+2q \le \ell} \sigma(p,q) H_I^{D(p,q)} \right) P^{\dagger}, \quad I = 12,23,13~.
\end{eqnarray}
The same thing happens with the $\operatorname{SU}(3)$ basis 
\begin{equation}
    \Lambda_{a}^{D(1,0)^{\otimes \ell}} =P \left( \bigoplus_{p,q \ge 0, ~ p+2q \le \ell} \sigma(p,q) \Lambda_{a}^{D(p,q)} \right) P^{\dagger}, \quad a = 1, 2, \cdots, 8~.
\end{equation}
Thus, for the Casimir operators in \eqref{casimir-redrep}, we have
\begin{eqnarray}
    &&\begin{split}
        &\hat {C}_{\text{quad}}^{D(1,0)^{\otimes \ell}}=\sum _{a}{\Lambda}_{a}^{D(1,0)^{\otimes \ell}} {\Lambda}_{a}^{D(1,0)^{\otimes \ell}} \\
        &=  \sum _{a} P \left( \bigoplus_{p,q \ge 0, ~ p+2q \le \ell} \sigma(p,q) \Lambda_{a}^{D(p,q)} \right) P^{\dagger} \cdot P \left( \bigoplus_{p,q \ge 0, ~ p+2q \le \ell} \sigma(p,q) \Lambda_{a}^{D(p,q)} \right) P^{\dagger} \\
        &=  P \left[ \bigoplus_{p,q \ge 0, \atop p+2q \le \ell} \sigma(p,q) \left( \sum _{a} \Lambda_{a}^{D(p,q)} \Lambda_{a}^{D(p,q)} \right) \right] P^{\dagger} 
        =  P \left[ \bigoplus_{p,q \ge 0, \atop p+2q \le \ell} \sigma(p,q) \hat {C}_{\text{quad}}^{D(p,q)} \right] P^{\dagger} ~, 
    \end{split} \nonumber \\
    &&\begin{split}
        & \hat {C}_{\text{cub}}^{D(1,0)^{\otimes \ell}}= \sum_{abc}d_{abc}{\Lambda}_{a}^{D(1,0)^{\otimes \ell}}{\Lambda}_{b}^{D(1,0)^{\otimes \ell}}{\Lambda}_{c}^{D(1,0)^{\otimes \ell}} \\
        &=  \sum_{abc} d_{abc} \cdot  P \left( \bigoplus_{p,q \ge 0, ~ p+2q \le \ell} \sigma(p,q) \Lambda_{a}^{D(p,q)} \right) P^{\dagger} \cdot P \left( \bigoplus_{p,q \ge 0, ~ p+2q \le \ell} \sigma(p,q) \Lambda_{b}^{D(p,q)} \right) P^{\dagger} \\
        & \qquad \qquad \quad \cdot P \left( \bigoplus_{p,q \ge 0, ~ p+2q \le \ell} \sigma(p,q) \Lambda_{c}^{D(p,q)} \right) P^{\dagger}
        \\
        &=  P \left[ \bigoplus_{p,q \ge 0, \atop p+2q \le \ell} \sigma(p,q) \left( \sum _{abc} d_{abc} \Lambda_{a}^{D(p,q)} \Lambda_{a}^{D(p,q)} \Lambda_{c}^{D(p,q)}  \right) \right] P^{\dagger} 
        =  P \left[ \bigoplus_{p,q \ge 0, \atop p+2q \le \ell} \sigma(p,q) \hat {C}_{\text{cub}}^{D(p,q)} \right] P^{\dagger} ~.
    \end{split} \nonumber \\
    \label{casimir-iredrep}
\end{eqnarray}

Since the commutation relations in \eqref{GM-anticom} still hold for $\Lambda_a^{D(1,0)^{\otimes \ell}}$, one can verify
\begin{eqnarray}
    ~&& \begin{split}
    \left[ \hat {C}_{\text{quad}}^{D(1,0)^{\otimes \ell}} ~, ~ {\Lambda}_{a}^{D(1,0)^{\otimes \ell}} \right] 
    =& \sum _{b}\left[ {\Lambda}_{b}^{D(1,0)^{\otimes \ell}} {\Lambda}_{b}^{D(1,0)^{\otimes \ell}} ~, ~ {\Lambda}_{a}^{D(1,0)^{\otimes \ell}}  \right] \\
    = i 2 \sum_b & \left\{ f_{bac} {\Lambda}_{c}^{D(1,0)^{\otimes \ell}}~, ~ {\Lambda}_{a}^{D(1,0)^{\otimes \ell}}\right\} \equiv 0~, 
    \end{split}
    \nonumber \\
    && \begin{split}
        \left[ \hat {C}_{\text{cub}}^{D(1,0)^{\otimes \ell}} ~, ~ {\Lambda}_{a}^{D(1,0)^{\otimes \ell}} \right] 
        =& \sum_{bce}d_{bce} \left[ {\Lambda}_{b}^{D(1,0)^{\otimes \ell}}{\Lambda}_{c}^{D(1,0)^{\otimes \ell}}{\Lambda}_{e}^{D(1,0)^{\otimes \ell}} ~, ~ {\Lambda}_{a}^{D(1,0)^{\otimes \ell}} \right] \\
        = i2 \sum_{bce}& \left( d_{gce} f_{gab} + d_{bge}f_{gac} + d_{bcg}f_{gae}\right) {\Lambda}_{b}^{D(1,0)^{\otimes \ell}}{\Lambda}_{c}^{D(1,0)^{\otimes \ell}}{\Lambda}_{e}^{D(1,0)^{\otimes \ell}} 
        \equiv 0~,
    \end{split}
    \nonumber \\
\end{eqnarray}
where we have applied the Jacobi identity
\begin{equation}
    f_{abj}d_{jcd} + f_{dbj}d_{jca} + f_{cbj} d_{jad} =0~.
\end{equation}
Now left-multiply $P^{\dagger}$ and right-multiply $P$, and we have
\begin{eqnarray}
    0 &\equiv& \left[ P^{\dagger} \hat {C}_{\text{quad / cub}}^{D(1,0)^{\otimes \ell}} P ~, ~ P^{\dagger} {\Lambda}_{a}^{D(1,0)^{\otimes \ell}} P \right] \nonumber \\
    &=& \left[ \bigoplus_{p,q \ge 0, ~ p+2q \le \ell} \sigma(p,q) \hat {C}_{\text{quad / cub}}^{D(p,q)} ~, ~ \bigoplus_{p,q \ge 0, ~ p+2q \le \ell} \sigma(p,q) \Lambda_{a}^{D(p,q)} \right] \\
    & =& \bigoplus_{p,q \ge 0, ~ p+2q \le \ell} \sigma(p,q) \left[ \hat {C}_{\text{quad / cub}}^{D(p,q)}~, ~ \Lambda_{a}^{D(p,q)} \right] ~,  \qquad  a = 1,2, \cdots, 8~.
\end{eqnarray}
Therefore, 
\begin{equation}
    \left[ \hat {C}_{\text{quad / cub}}^{D(p,q)}~, ~ \Lambda_{a}^{D(p,q)} \right] \equiv 0, \quad a = 1,2, \cdots, 8\,,
\end{equation}
and equivalently
\begin{equation}
    \left[ \hat {C}_{\text{quad / cub}}^{D(p,q)}~, ~ E_{\mu}^{D(p,q)} \right] = \left[ \hat {C}_{\text{quad / cub}}^{D(p,q)}~, ~ H_{I}^{D(p,q)} \right] \equiv 0, \quad \mu = \pm \alpha, \pm \beta, \pm \gamma~ ; I = 12,23,13~,
\end{equation}
for all irreducible representations $D(p,q)$. By Schur's lemma, $\hat {C}_{\text{quad / cub}}^{D(p,q)}$ is proportional to the identity, i.e.
\begin{equation}
    \hat {C}_{\text{quad / cub}}^{D(p,q)} = {C}_{\text{quad / cub}}^{D(p,q)} \mathbbm{1}_{\operatorname{dim} (D(p,q))}~. 
\end{equation}

Go back to the PXP Hamiltonian in \eqref{pxp-hamiltonian-sl3c}. For the linear part $H_{\text{PXP},\text{lin}}$ \eqref{pxp-hami-sl3c-lin}, the Hamiltonian is a linear combination of Cartan-Weyl basis and obviously
\begin{equation}
     \left[ \hat {C}_{\text{quad / cub}}^{D(p,q)}~, ~ H_{\text{PXP}, \text{lin}} \right] \equiv 0
\tag{\ref{casimir-pxplin}'}\,,
\end{equation}
is true for any irreducible representation of $\mathfrak{s}l_3 (\mathbb{C})$, as we have pointed out in section \ref{lanczos-analytic}. As for the residual part $H_{\text{PXP,res}}$ \eqref{pxp-hami-sl3c-quad}, since the unitary transformation $P$ does not necessarily decompose $(\lambda_a)_i$ into diagonal blocks, we shall consider whether the commutation between the Casimir operators in diagonal blocks and $P^{\dagger} (\lambda_a)_i (\lambda_b)_j P$ vanish, i.e. whether
\begin{equation}
    \begin{split}
        \left[ \hat {C}_{\text{quad / cub}}^{D(p,q)}~, ~ P^{\dagger} (\lambda_a)_i (\lambda_b)_j P \right] 
        = P^{\dagger} \left[ P \hat {C}_{\text{quad / cub}}^{D(p,q)} P^{\dagger}~,~ (\lambda_a)_i (\lambda_b)_j  \right] P = 0~, \qquad  & 
    \end{split}
    \label{casimir-pxpquad-exa}
\end{equation}
for every irreducible representation of $D(p,q)$. It is not easy to figure this commutator out directly. Hence we instead consider the proposition
\begin{equation}
\left[ \hat {C}_{\text{quad / cub}}^{D(1,0)^{\otimes \ell}} ~,~ (\lambda_a)_i (\lambda_b)_j  \right]=0~.
\label{casimir-pxpquad-propos}
\end{equation}
According to \eqref{casimir-iredrep} it is a necessary condition for \eqref{casimir-pxpquad-exa}. 

The proposition \eqref{casimir-pxpquad-propos} is false, to prove which we take the case of the quadratic Casimir operator as an example:
\begin{equation}
    \begin{split}
        &\left[ \hat {C}_{\text{quad}}^{D(1,0)^{\otimes \ell}} ~,~ (\lambda_b)_i (\lambda_c)_j  \right]
        = \sum_a \sum_{i} [\Lambda_a^{D(1,0)^{\otimes \ell}} \Lambda_a^{D(1,0)^{\otimes \ell}}~,~ (\lambda_b)_i (\lambda_c)_{i+1}]  \\
        =& \sum_a\sum_{i} \left( \left\{ \Lambda_a^{D(1,0)^{\otimes \ell}},~ [ \Lambda_a^{D(1,0)^{\otimes \ell}}~, ~(\lambda_b)_i ] \right\} (\lambda_c)_{i+1} 
        + (\lambda_b)_i \left\{ \Lambda_a^{D(1,0)^{\otimes \ell}}, ~[ \Lambda_a^{D(1,0)^{\otimes \ell}} ~,~ (\lambda_c)_{i+1}] \right\} \right) \\
        =& \sum_{ad} \sum_{i} \left( \left\{ \Lambda_a^{D(1,0)^{\otimes \ell}}~, i f_{abd} (\lambda_d)_i \} (\lambda_c)_{i+1} +  (\lambda_b)_i \{ \Lambda_a, if_{acd} (\lambda_{d})_{i+1} \right\}   \right) \\
        =&  \sum_{ad} \left( i 2 f_{abd} \sum_{i, j \atop j \ne i} (\lambda_a)_j  (\lambda_d)_i  (\lambda_c)_{i+1}  + i2 f_{acd} \sum_{i, j \atop j \ne i}  (\lambda_b)_{i-1} (\lambda_a)_j   (\lambda_d)_{i}  \right)~,
    \end{split}
\end{equation}
where in general
\begin{equation}
    \begin{split}
        & \sum_{ad} \sum_{i, j \atop j \ne i} f_{abd}  (\lambda_a)_j  (\lambda_d)_i (\lambda_c)_{i+1} \\
        =&\frac{1}{4}  \sum_{i. j \atop j \ne i} \sum_{a,d \atop a\ne d} f_{abd} \left[ (\lambda_a)_j  (\lambda_d)_i - (\lambda_d)_j  (\lambda_a)_i \right] (\lambda_c)_{i+1} \\
         &+ \frac{1}{4}  \sum_{i. j \atop j \ne i} \sum_{a,d \atop a\ne d} f_{abd} \left[ (\lambda_a)_i  (\lambda_d)_j - (\lambda_d)_i  (\lambda_a)_j \right] (\lambda_c)_{j+1} \\
        =&\frac{1}{4}  \sum_{i. j \atop j \ne i} \sum_{a,d \atop a\ne d} f_{abd} \left[ (\lambda_a)_j  (\lambda_d)_i - (\lambda_d)_j  (\lambda_a)_i \right] \left[ (\lambda_c)_{i+1}- (\lambda_c)_{j+1} \right] \ne 0 ~.\\
    \end{split}
\end{equation}
Thus, as we have pointed out in section \ref{lanczos-analytic}, the residual PXP Hamiltonian does NOT necessarily preserve the sectors of the Casimir operators.

\section{PXP Krylov subspace and Lucas Numbers} 
\label{lucasN}

In this appendix we try to count the maximal number of Krylov basis able to be generated by the entire PXP Hamiltonian in \eqref{pxp-hamiltonian-pbc}. Basically it is the number of the tensor product states of $2\ell$ qubits that do not have any consecutive $|1\rangle$, as we have discussed in section \ref{lanczos-analytic}. It will be more convenient to consider this question in the qubit description, where each $|1\rangle$ can be regarded as an insertion into the $|00\cdots 0\rangle$ state. This is a combinatorial question.

Since it is combinatorics, we translate it into the following problem: Consider beadwork of $n$ (here we generalize the case to arbitrary size of qubit lattice, rather than the $2\ell$-qubit lattice in the main body of this paper) white or black beads. The rule of stringing them onto a thin wire is that no two consecutive beads are both in black after tying the two ends of the wire. The question is, how many ways can we attach the beads onto the wire? 
\begin{figure}[H]
\centering
\includegraphics[width=0.6\textwidth]{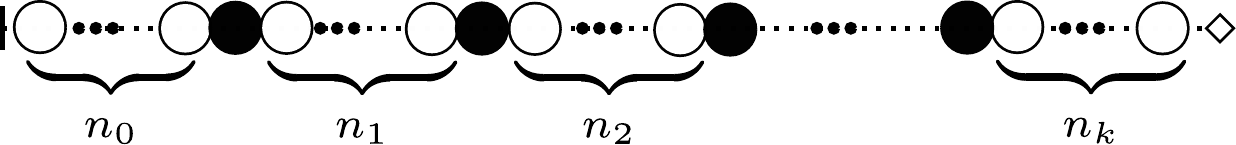}
\caption{A schematic diagram of a beadwork, the stringing rules of which are $n_0+n_k, n_1, n_2, \cdots, n_{k-1} \ge 1$.}
\end{figure}
Notice this is NOT equivalent to a question about composition or weak composition -- since at the two ends of the wire, there can be zero white beads at one end. Hence there are two situations: both end are white beads, or only one end has a black bead (left or right). For each case it is a composition. Thus, we have for $k$ black beads in $n$ beads:
\begin{equation}
\mathcal{L}(n) \equiv \sum_{k} \mathcal{L}(n,k), \qquad \mathcal{L}(n, k) = \binom{n-(k+1)}{k}  +2 \times \binom{n-1-k}{k-1}\,.
\end{equation}
Now, utilize the relation to $n^{\text{th}}$ Fibonacci number $F(n)$, which is
\begin{equation}
\sum _{k=0}^{\lfloor n/2\rfloor }{\binom {n-k}{k}}=F(n+1),
\end{equation}
and we obtain
\begin{equation}
\begin{split}
\mathcal{L}(n)  =& \sum_{k=0}^{\left\lfloor \frac{n-1}{2} \right\rfloor} \binom{n-(k+1)}{k} + 2\times \sum_{k=1}^{\left\lfloor \frac{n}{2} \right\rfloor} \binom{n-1-k}{k-1} \\
=& \sum_{k=0}^{\left\lfloor \frac{n-1}{2} \right\rfloor} \binom{n-1-k}{k} + 2\times \sum_{k=0}^{\left\lfloor \frac{n-2}{2} \right\rfloor} \binom{n-2-k}{k} \\
=& F(n) + 2 F(n-1) = F(n+1) + F(n-1)~.
\end{split}
\end{equation}
$\mathcal{L}(n)$ is known as the Lucas number \cite{hoggatt1969fibonacci} defined by recursion relation
\begin{equation}
\mathcal{L}_{n} \equiv
\begin{cases}
2&{\text{if }}n=0;\\
1&{\text{if }}n=1;\\
\mathcal{L}_{n-1}+\mathcal{L}_{n-2}&{\text{if }}n>1.
\end{cases}
\end{equation}
or by closed formula
\begin{equation}
\mathcal{L}_{n}=\varphi ^{n}+(1-\varphi )^{n}=\varphi ^{n}+(-\varphi )^{-n}=\left({1+{\sqrt {5}} \over 2}\right)^{n}+\left({1-{\sqrt {5}} \over 2}\right)^{n}~,
\end{equation}
where $\varphi$ is the golden ratio. The first few Lucas numbers \cite{sloane1995encyclopedia} are: 
2, 1, 3, 4, 7, 11, 18, 29, 47, 76, 123, 199, 322, 521, 843, 1364, 2207, 3571, 5778, 9349, 15127, 24476, 39603, 64079, 103682, 167761, 271443, 439204, 710647, 1149851, 1860498, 3010349, 4870847, 7881196, 12752043, 20633239, 33385282, 54018521, 87403803, ...

\section{Small-Sized Qutrit Lattice and representations of $\mathfrak{s}l_3(\mathbb{C})$} \label{visual}

In this appendix we give details of the decomposition of the reducible representation $D(1,0)^{\otimes \ell}$ corresponding to the $\ell$-qutrit lattice, in the cases where $\ell$ is small ($\ell = 3,4$). The appendix is closely related to \autoref{sl3c}.

\subsection{$\ell=3$}
\label{visual-3boxes}

In this case, we have all the symmetric groups given by the tensor product of three Young boxes. Following the rules of constructing standard Young tableaux one find
\begin{equation}
\begin{ytableau}
1 \\
\end{ytableau} 
\otimes
\begin{ytableau}
2 \\
\end{ytableau}
\otimes
\begin{ytableau}
3 \\
\end{ytableau} =
\begin{ytableau}
1 & 2 & 3\\
\end{ytableau} 
\oplus
\begin{ytableau}
1 & 2 \\
 3 & \none \\
\end{ytableau}
 \oplus
\begin{ytableau}
1 & 3 \\
 2 & \none \\
\end{ytableau}
\oplus 
\begin{ytableau}
1 \\ 
2 \\
3\\
\end{ytableau} 
\label{ytable3}
\end{equation}
In Young tableaux language every state belonging to a particular symmetric group corresponds to a filling of the Young diagram that describes the symmetric group. Now, fill the numbers in $\{ 1,2,3 \}$, which represent the state of one qutrit site $\{ |\underline{2} \rangle, |\underline{0} \rangle, |\underline{1} \rangle \}$, respectively, into each Young diagram in the RHS of \eqref{ytable3}. It results in 10, 8, 1 filling(s) in total for Young diagrams {\tiny{\ydiagram{3}}}~, {\tiny{\ydiagram{2,1}}}~, and {\tiny{\ydiagram{1,1,1}}}~, respectively. Utilizing that, we can decompose the states into four root space that correspond to three irreducible representations, $D(3,0)$, $D(1,1)$, and $D(0,0)$, respectively , as \autoref{rs-ydiag-n3} shows. They have different eigenvalues when being acted by the two Casimir operators.

Based on the standard Young tableaux, we write down of each partition the subgroups of the symmetric group, $P_{\lambda}$ and $Q_{\lambda}$, and the corresponding group elements in the group algebra, $a_{\lambda}$ and $b_{\lambda}$:

\begin{itemize}
\item $\lambda =(3)$~: 

\begin{equation}
\begin{ytableau}
1 & 2 & 3 \\
\end{ytableau} \Longrightarrow
\begin{cases}
\begin{split}
P_{(3)} = S_3  \Rightarrow a_{(3)} = \sum_{\sigma \in S_3} e_{\sigma} = &e_{()} + e_{(12)} + e_{(13)} \\
&+ e_{(23)} +e_{(123)} +e_{(132)}~, 
\end{split}
\\ 
Q_{(3)} =\{()\} \Rightarrow b_{(3)} = e_{()} ~.\\
\end{cases}
\end{equation}

\item $\lambda =(2,1)$~:

\begin{eqnarray}
\begin{ytableau}
1 & 2  \\
3 & \none \\
\end{ytableau}
&\Longrightarrow&
\begin{cases}
P_{(2,1)} = \{ (), (12) \}  \Rightarrow a_{(2,1)} =  e_{()} + e_{(12)} ~, \\ 
Q_{(2,1)} =\{ (), (13) \} \Rightarrow b_{(2,1)} =  e_{()} - e_{(13)} ~; \\
\end{cases}
\\
\begin{ytableau}
1 & 3  \\
2 & \none \\
\end{ytableau}
&\Longrightarrow&
\begin{cases}
P_{(2,1)} = \{ (), (13) \}  \Rightarrow a_{(2,1)} =  e_{()} + e_{(13)} ~,\\ 
Q_{(2,1)} =\{ (), (12) \} \Rightarrow b_{(2,1)} =  e_{()} - e_{(12)} ~.\\
\end{cases}
\end{eqnarray}

\item $\lambda =(1,1,1)$~:

\begin{equation}
\begin{ytableau}
1 \\
2  \\
3  \\
\end{ytableau}
\Longrightarrow
\begin{cases}
P_{(1,1,1)} =\{()\} \Rightarrow a_{(1,1,1)} = e_{()} ~,\\
\begin{split}
Q_{(1,1,1)} = S_3  \Rightarrow b_{(1,1,1)} = \sum_{\sigma \in S_3} (-1)^{\sigma}e_{\sigma} =& e_{()} - e_{(12)} - e_{(13)} \\
& - e_{(23)} +e_{(123)} +e_{(132)}~. \\
\end{split}
\\ 
\end{cases}
\end{equation}

\end{itemize}

Thus, the Young symmetrizers are defined to be
\begin{itemize}
    \item $\lambda=(3)$~:
    \begin{equation}
        c_{(3)} = e_{()} + e_{(12)} + e_{(13)} + e_{(23)} +e_{(123)} +e_{(132)} ~.
    \end{equation}
    \item $\lambda=(2,1)$~:
    \begin{eqnarray}
        c_{(2,1)} & =& \left(  e_{()} + e_{(12)} \right) \cdot \left( e_{()} - e_{(13)} \right) = e_{()} + e_{(12)} - e_{(13)} - e_{(132)} ~, \\ 
        {\text{or}} \quad c_{(2,1)} & =& \left(  e_{()} + e_{(13)} \right) \cdot \left( e_{()} - e_{(12)} \right) = e_{()} + e_{(13)} - e_{(12)} - e_{(123)} ~.
    \end{eqnarray}
    \item $\lambda = (1,1,1)$~:
    \begin{equation}
        c_{(1,1,1)}  = e_{()} - e_{(12)} - e_{(13)} - e_{(23)} +e_{(123)} +e_{(132)}~.
    \end{equation}
\end{itemize}

\subsection{$\ell=4$}
\label{visual-4boxes}

In this case, we have all the symmetric groups given by the tensor product of three Young boxes. Following the rules of constructing the standard Young tableaux one find
\begin{eqnarray}
\begin{ytableau}
1 \\
\end{ytableau} 
\otimes
\begin{ytableau}
2 \\
\end{ytableau}
\otimes
\begin{ytableau}
3 \\
\end{ytableau} 
\otimes
\begin{ytableau}
4 \\
\end{ytableau} 
&=&
\begin{ytableau}
1 & 2 & 3 & 4\\
\end{ytableau} 
\oplus
\begin{ytableau}
1 & 2 & 3\\
 4 & \none & \none \\
\end{ytableau}
\oplus
\begin{ytableau}
1 & 2 & 4\\
 3 & \none & \none \\
\end{ytableau}
\oplus
\begin{ytableau}
1 & 3 & 4\\
 2 & \none & \none \\
\end{ytableau} 
\oplus 
\begin{ytableau}
1 & 2 \\
 3 & 4 \\
\end{ytableau} \nonumber \\
 && \oplus~
\begin{ytableau}
1 & 2 \\
 3 & 4 \\
\end{ytableau} 
 \oplus 
\begin{ytableau}
1 & 2 \\ 
3 & \none \\
4 & \none \\
\end{ytableau} 
\oplus 
\begin{ytableau}
1 & 3 \\ 
2 & \none \\
4 & \none \\
\end{ytableau} 
\oplus 
\begin{ytableau}
1 & 4 \\ 
2 & \none \\
3 & \none \\
\end{ytableau} 
\oplus 
\begin{ytableau}
1  \\ 
2  \\
3  \\
4 \\
\end{ytableau} 
\label{ytable4}
\end{eqnarray}
Filling the numbers in $\{ 1,2,3 \}$ into each Young diagram in the RHS of \eqref{ytable4} results in 15, 15, 6, 3, 0 filling(s) in total for Young diagrams {\tiny{\ydiagram{4}}}~, {\tiny{\ydiagram{3,1}}}~, {\tiny{\ydiagram{2,2}}}~, {\tiny{\ydiagram{2,1,1}}}~, and {\tiny{\ydiagram{1,1,1,1}}}~, respectively. Hence we can decompose the states into nine root space that correspond to four irreducible representations, $D(4,0)$, $D(2,1)$, $D(0,2)$, and $D(1,0)$, respectively, as \autoref{rs-ydiag-n4} shows. They have different eigenvalues when being acted by Casimir operators.

Based on the standard Young tableaux, we write down of each partition the subgroups of the symmetric group, $P_{\lambda}$ and $Q_{\lambda}$, and the corresponding group elements in the group algebra, $a_{\lambda}$ and $b_{\lambda}$:

\begin{itemize}
\item $\lambda =(4)$~: 

\begin{equation}
\begin{ytableau}
1 & 2 & 3 & 4\\
\end{ytableau} 
\Longrightarrow
\begin{cases}
P_{(4)} = S_4  \Rightarrow a_{(4)} = \sum_{\sigma \in S_4} e_{\sigma}~, \\ 
Q_{(4)} =\{()\} \Rightarrow b_{(4)} = e_{()}~. \\
\end{cases}
\end{equation}

\item $\lambda =(3,1)$~:

\begin{eqnarray}
\begin{ytableau}
1 & 2 & 3 \\
4 & \none & \none \\
\end{ytableau}
&\Longrightarrow&
\begin{cases}
\begin{split}
P_{(3,1)} = S_3  \Rightarrow a_{(3,1)} =& e_{()} + e_{(12)} + e_{(13)} \\
&+ e_{(23)} +e_{(123)} +e_{(132)} ~, \\ 
\end{split}
\\
Q_{(3,1)} =\{ (), (14) \} \Rightarrow b_{(3,1)} =  e_{()} - e_{(14)}~; \\
\end{cases}
\\
\begin{ytableau}
1 & 2 & 4  \\
3 & \none & \none \\
\end{ytableau}
&\Longrightarrow&
\begin{cases}
\begin{split}
P_{(3,1)} = S_3 \big|_{3\to 4}  \Rightarrow a_{(3,1)} =& e_{()} + e_{(12)} + e_{(14)} \\
& + e_{(24)} +e_{(124)} +e_{(142)}~, \\
\end{split}
\\ 
Q_{(3,1)} =\{ (), (13) \} \Rightarrow b_{(3,1)} =  e_{()} - e_{(13)}~; \\
\end{cases}
\\
\begin{ytableau}
1 & 3 & 4  \\
2 & \none & \none \\
\end{ytableau}
&\Longrightarrow&
\begin{cases}
\begin{split}
P_{(3,1)} = S_3 \big|_{3\to 4, 2\to 3 }  \Rightarrow a_{(3,1)} =& e_{()} + e_{(13)} + e_{(14)}\\ 
&+ e_{(34)} +e_{(134)} +e_{(143)} ~, \\
\end{split}
\\ 
Q_{(3,1)} =\{ (), (12) \} \Rightarrow b_{(3,1)} =  e_{()} - e_{(12)}~. \\
\end{cases} \nonumber \\
\end{eqnarray}

\item $\lambda =(2,2)$~:

\begin{eqnarray}
\begin{ytableau}
1 & 2\\
3 & 4 \\
\end{ytableau}
&\Longrightarrow&
\begin{cases}
\begin{split}
P_{(2,2)} =& \{ (), (12), (34), (12)(34) \} \\
&\Rightarrow a_{(2,2)} = e_{()} + e_{(12)} + e_{(34)} + e_{(12)(34)} ~, \\ 
\end{split}\\
\begin{split}
Q_{(2,2)} =& \{  (), (13), (24), (13)(24) \} \\
& \Rightarrow b_{(2,2)} = e_{()} - e_{(13)} - e_{(24)} + e_{(13)(24)} ~; \\
\end{split} \\
\end{cases}
\nonumber \\
\\
\begin{ytableau}
1 & 3\\
2 & 4 \\
\end{ytableau}
&\Longrightarrow&
\begin{cases}
\begin{split}
P_{(2,2)} =& \{  (), (13), (24), (13)(24) \} \\
&\Rightarrow a_{(2,2)} =e_{()} + e_{(13)} + e_{(24)} + e_{(13)(24)} ~,\\ 
\end{split} \\
\begin{split}
Q_{(2,2)} =& \{ (), (12), (34), (12)(34) \} \\
& \Rightarrow b_{(2,2)} =  e_{()} - e_{(12)} - e_{(34)} + e_{(12)(34)} ~. \\
\end{split} \\
\end{cases}
\nonumber \\
\end{eqnarray}

\item $\lambda =(2,1,1)$~:

\begin{eqnarray}
\begin{ytableau}
1 & 2  \\
3 & \none \\
4 & \none  \\
\end{ytableau}
&\Longrightarrow&
\begin{cases}
P_{(2,1,1)} =\{ (), (12) \} \Rightarrow a_{(2,1,1)} =  e_{()} + e_{(12)}~, \\
\begin{split}
Q_{(2,1,1)} = S_3 \big|_{3\to 4, 2\to 3 }  \Rightarrow b_{(2,1,1)} =& e_{()} - e_{(13)} - e_{(14)} \\
& - e_{(34)} +e_{(134)} +e_{(143)} ~;\\ 
\end{split} \\
\end{cases}
\\
\begin{ytableau}
1 & 3  \\
2 & \none \\
4 & \none  \\
\end{ytableau}
&\Longrightarrow&
\begin{cases}
P_{(2,1,1)} =\{ (), (13) \} \Rightarrow a_{(2,1,1)} =  e_{()} + e_{(13)}~, \\
\begin{split}
Q_{(2,1,1)} = S_3 \big|_{3\to 4}  \Rightarrow b_{(2,1,1)} =& e_{()} - e_{(12)} - e_{(14)} \\
& - e_{(24)} +e_{(124)} +e_{(142)}~; \\ 
\end{split} \\
\end{cases}
\\
\begin{ytableau}
1 & 4  \\
2 & \none \\
3 & \none  \\
\end{ytableau}
&\Longrightarrow&
\begin{cases}
P_{(2,1,1)}  =\{ (), (14) \} \Rightarrow a_{(2,1,1)} =  e_{()} + e_{(14)} ~,\\
\begin{split}
Q_{(2,1,1)} = S_3  \Rightarrow b_{(2,1,1)} =& e_{()} - e_{(12)} - e_{(13)} \\
& - e_{(23)} +e_{(123)} +e_{(132)}~. \\ 
\end{split} \\
\end{cases}
\end{eqnarray}
\end{itemize}
Without loss of completeness, the totally antisymmetric one is
\begin{itemize}
\item $\lambda =(1,1,1,1)$~:

\begin{equation}
\begin{ytableau}
1 \\
2  \\
3  \\
4 \\
\end{ytableau}
\Longrightarrow
\begin{cases}
P_{(1,1,1)} =\{()\} \Rightarrow a_{(1,1,1,1)} = e_{()} ~\\
Q_{(1,1,1)} = S_4  \Rightarrow b_{(1,1,1,1)} = \sum_{\sigma \in S_4} (-1)^{\sigma}e_{\sigma}~. \\ 
\end{cases}
\end{equation}

\end{itemize}

Thus, the Young symmetrizers are defined to be

\begin{itemize}
\item $\lambda = (4)$~ :
\begin{eqnarray}
c_{(4)} &=& \sum_{\sigma \in S_4} e_{\sigma} ~.
\end{eqnarray}

\item $\lambda = (3,1)$~ :
\begin{eqnarray}
c_{(3,1)} &=&  \left( e_{()} + e_{(12)} + e_{(13)} + e_{(23)} +e_{(123)} +e_{(132)} \right) \cdot \left(  e_{()} - e_{(14)} \right) \\
&=&  e_{()} + e_{(12)} + e_{(13)} + e_{(23)} +e_{(123)} +e_{(132)} \nonumber \\
&& - e_{(14)} - e_{(142)} - e_{(143)} -e_{(14)(23)} - e_{(1423)} - e_{(1432)}~; \nonumber  \\
{\text{or} } \quad &=& \left(  e_{()} + e_{(12)} + e_{(14)} + e_{(24)} +e_{(124)} +e_{(142)} \right) \cdot  \left(   e_{()} - e_{(13)} \right)  \\
&=&  e_{()} + e_{(12)} + e_{(14)} + e_{(24)} +e_{(124)} +e_{(142)} \nonumber \\
&&- e_{(13)} - e_{(132)} - e_{(134)} -e_{(13)(24)} - e_{(1324)} - e_{(1342)}  ~;\nonumber \\
{\text{or} } \quad &=& \left(  e_{()} + e_{(13)} + e_{(14)} + e_{(34)} +e_{(134)} +e_{(143)}  \right) \cdot \left( e_{()} - e_{(12)} \right) \\
&=&  e_{()} + e_{(13)} + e_{(14)} + e_{(34)} +e_{(134)} +e_{(143)} \nonumber \\
&&- e_{(12)} - e_{(123)} - e_{(124)} -e_{(12)(34)} - e_{(1234)} - e_{(1243)} ~. \nonumber \\
\end{eqnarray}

\item $\lambda = (2,2)$~ :
\begin{eqnarray}
c_{(2,2)} & =& \left( e_{()} + e_{(12)} + e_{(34)} + e_{(12)(34)}  \right) \cdot \left( e_{()} - e_{(13)} - e_{(24)} + e_{(13)(24)} \right)  \\
&=&  e_{()} + e_{(12)} + e_{(34)} + e_{(12)(34)}  - e_{(13)} - e_{(132)} - e_{(143)} - e_{(1432)}\nonumber \\ 
&& - e_{(24)} - e_{(124)} - e_{(234)} - e_{(1234)} 
+ e_{(13)(24)} + e_{(1324)} + e_{(1423)}  + e_{(14)(23)} ~;
\nonumber \\
{\text{or} } \quad  &=& \left( e_{()} + e_{(13)} + e_{(24)} + e_{(13)(24)} \right) \cdot \left(  e_{()} - e_{(12)} - e_{(34)} + e_{(12)(34)} \right)   \\ 
&=&  e_{()} + e_{(13)} + e_{(24)} + e_{(13)(24)}  - e_{(12)} - e_{(123)} - e_{(142)} - e_{(1423)} \nonumber \\ 
&& - e_{(34)} - e_{(134)} - e_{(243)} - e_{(1324)} 
+ e_{(12)(34)} + e_{(1234)} + e_{(1432)}  + e_{(14)(23)} ~.
\nonumber
\end{eqnarray}

\item $\lambda = (2,1,1)$~ :
\begin{eqnarray}
c_{(2,1,1)} &=& \left( e_{()} + e_{(12)}  \right) \cdot \left(  e_{()} - e_{(13)} - e_{(14)} - e_{(34)} +e_{(134)} +e_{(143)} \right)  \\
&=& e_{()} - e_{(13)} - e_{(14)} - e_{(34)} +e_{(134)} +e_{(143)} \nonumber \\
&&+ e_{(12)} - e_{(132)} - e_{(142)} - e_{(12)(34)} + e_{(1342)} + e_{(1432)}~;
\nonumber \\
{\text{or} } \quad &=& \left( e_{()} + e_{(13)} \right) \cdot \left( e_{()} - e_{(12)} - e_{(14)} - e_{(24)} +e_{(124)} +e_{(142)} \right)  \\
&=& e_{()} - e_{(12)} - e_{(14)} - e_{(24)} +e_{(124)} +e_{(142)} \nonumber \\
&&+ e_{(13)} - e_{(123)} - e_{(143)} - e_{(13)(24)} + e_{(1243)} + e_{(1423)}~;
\nonumber \\
{\text{or} } \quad  &=& \left(  e_{()} + e_{(14)}  \right) \cdot \left(  e_{()} - e_{(12)} - e_{(13)} - e_{(23)} +e_{(123)} +e_{(132)} \right)  \\
&=& e_{()} - e_{(12)} - e_{(13)} - e_{(23)} +e_{(123)} +e_{(132)} \nonumber \\
&&+ e_{(14)} - e_{(124)} - e_{(134)} - e_{(14)(23)} + e_{(1234)} + e_{(1324)}~.
\nonumber 
\end{eqnarray}

\item $\lambda = (1,1,1,1)$~ :
\begin{eqnarray}
c_{(1,1,1,1)} &=&  \sum_{\sigma \in S_4} (-1)^{\sigma}e_{\sigma}~.
\end{eqnarray}

\end{itemize}

\section{Krylov Basis and Lanczos Coefficients of the Small-Sized PXP Model} \label{kryb-lc}

In this appendix we list the Krylov basis and the corresponding Lanczos coefficients generated by repeated action of $H_{\text{PXP}}$ on the initial state $|Z_k \rangle$ in the cases where the lattice size $L$ is small ($L=2\ell =6,8$).

To make the structures of the Krylov bases clear, we introduce the standard Young tableau as the subscript to label the states symmetrized by the corresponding Young symmetrizer (given in \autoref{visual}). Here please note that the Young symmetrizer actually permute the sequence of the position space, i.e.
\begin{equation}
e_{\sigma}: ~ ( x_1, x_2 , \cdots , x_n ) \mapsto ( x_{\sigma(1)},  x_{\sigma(2)}, \cdots, x_{\sigma(n)} )~,
\end{equation}
and therefore the wavefunction of a product state $\langle x_1, \cdots, x_n | \phi_1, \cdots, \phi_n \rangle = \prod_{i=1}^n \phi_i (x_i )$ becomes
\begin{equation}
e_{\sigma}:~ \prod_{i=1}^n \phi_i (x_i) \mapsto \prod_{i=1}^n \phi_i \left( x_{\sigma(i)} \right) = \prod_{j=1}^n \phi_{\sigma^{-1}(j)} \left( x_{j} \right)~.
\end{equation}
This implies that when the Young symmetrizer acts on the product state, it results in
\begin{equation}
e_{\sigma}:~ | \phi_1, \phi_2, \cdots, \phi_n \rangle \mapsto | \phi_{\sigma^{-1}(1)}, \phi_{\sigma^{-1}(2)}, \cdots, \phi_{\sigma^{-1}(n)} \rangle~.
\end{equation}

\subsection{$L=2\ell =6$} \label{kryb-lc-3boxes}

We first list some of the symmetrized states in the cases where $\ell=3$.
\begin{itemize}
    \item $|\underline{222}\rangle$ (similar to $|\underline{000}\rangle$ and $|\underline{111}\rangle$) :
    \begin{equation}
         |\underline{222}\rangle_{\tiny 
        \begin{ytableau}
        1 & 2 & 3\\
        \end{ytableau} } =
        |\underline{222}\rangle  ~.
    \end{equation}
    \item $|\underline{220} \rangle$ (similar to $|\underline{001} \rangle$, $|\underline{200} \rangle$, and $|\underline{011} \rangle$) :
    \begin{eqnarray}
        |\underline{220} \rangle_{\tiny \begin{ytableau}
        1 & 2 & 3\\
        \end{ytableau} } &=&
        \frac{1}{\sqrt{3}} \left( |\underline{220}\rangle+|\underline{202}\rangle+|\underline{022}\rangle \right)  ~; \nonumber \\
        |\underline{220} \rangle_{\tiny \begin{ytableau}
        1 & 2 \\
        3 & \none \\
        \end{ytableau} } &=&
        \frac{1}{\sqrt{6}} \left( 2|\underline{220}\rangle-|\underline{022}\rangle- |\underline{202}\rangle \right)  ~, \nonumber \\
        |\underline{202} \rangle_{\tiny \begin{ytableau}
        1 & 3 \\
        2 & \none \\
        \end{ytableau} } &=& 
        \frac{1}{\sqrt{6}} \left( 2|\underline{202}\rangle - |\underline{022}\rangle- |\underline{220}\rangle \right)  ~. \nonumber \\
    \end{eqnarray}
    \item $|\underline{201} \rangle$ :
    \begin{eqnarray}
        |\underline{201} \rangle_{\tiny 
        \begin{ytableau}
        1 & 2 & 3\\
        \end{ytableau} } &=& \frac{1}{\sqrt{6}} \left( |\underline{201}\rangle+|\underline{021}\rangle+|\underline{102}\rangle +|\underline{210}\rangle+|\underline{012}\rangle+|\underline{120}\rangle \right) ~; \nonumber \\
        |\underline{201} \rangle_{\tiny 
        \begin{ytableau}
        1 & 2 \\
        3 & \none \\
        \end{ytableau} } &=& \frac{1}{2} \left( |\underline{201}\rangle+|\underline{021}\rangle- |\underline{102}\rangle -|\underline{012}\rangle \right) ~, 
        \nonumber \\
        |\underline{210} \rangle_{\tiny 
        \begin{ytableau}
        1 & 2 \\
        3 & \none \\
        \end{ytableau} } &=& \frac{1}{2} \left( |\underline{210}\rangle+|\underline{120}\rangle- |\underline{012}\rangle -|\underline{102}\rangle \right) ~, 
        \nonumber \\
        |\underline{210} \rangle_{\tiny 
        \begin{ytableau}
        1 & 3 \\
        2 & \none \\
        \end{ytableau} } &=& \frac{1}{2} \left( |\underline{210}\rangle+|\underline{012}\rangle- |\underline{120}\rangle -|\underline{021}\rangle \right) ~, 
        \nonumber \\
        |\underline{201} \rangle_{\tiny 
        \begin{ytableau}
        1 & 3 \\
        2 & \none \\
        \end{ytableau} } &=& \frac{1}{2} \left( |\underline{201}\rangle+|\underline{102}\rangle- |\underline{021}\rangle -|\underline{120}\rangle \right) ~; 
        \nonumber \\
        |\underline{201} \rangle_{\tiny 
        \begin{ytableau}
        1 \\
        2 \\ 
        3\\
        \end{ytableau} } &=& \frac{1}{\sqrt{6}} \left( |\underline{201}\rangle + |\underline{012}\rangle + |\underline{120}\rangle -|\underline{021}\rangle - |\underline{102}\rangle - |\underline{210}\rangle \right) ~. \nonumber \\
    \end{eqnarray}
\end{itemize}

The Krylov basis generated from $|Z_2 \rangle$ are 

\begin{eqnarray}
 |K_0\rangle &=& |101010\rangle 
 = |\underline{222}\rangle = |\underline{222}\rangle_{\tiny \begin{ytableau}
1 & 2 & 3\\
\end{ytableau} } ~; \\
 |K_1\rangle &=& \frac{1}{\sqrt{3}} \left( |101000\rangle+|100010\rangle+|001010\rangle \right) \nonumber \\
 &=& \frac{1}{\sqrt{3}} \left( |\underline{220}\rangle+|\underline{202}\rangle+|\underline{022}\rangle \right)  = |\underline{220} \rangle_{\tiny \begin{ytableau}
1 & 2 & 3\\
\end{ytableau} } ~, \nonumber \\ \\
 |K_2\rangle &=& \frac{1}{\sqrt{3}} \left(|100000\rangle+|001000\rangle+|000010\rangle \right) \nonumber \\
 &=& \frac{1}{\sqrt{3}}  \left( |\underline{200}\rangle+|\underline{020}\rangle+|\underline{002}\rangle \right) = |\underline{200} \rangle_{\tiny \begin{ytableau}
1 & 2 & 3\\
\end{ytableau} } ~; \nonumber \\ \\
 |K_3\rangle &=& 0.288675\left( |001001\rangle+|100100\rangle+|010010\rangle \right)+0.866025|000000\rangle \nonumber \\
 &=& \frac{1}{2\sqrt{3}}\left( |\underline{021}\rangle+|\underline{210}\rangle+|\underline{102}\rangle \right)+\frac{\sqrt{3}}{2}|\underline{000}\rangle \nonumber \\
 &=& \frac{1}{2\sqrt{2}} |\underline{201}\rangle_{\tiny \begin{ytableau}
1 & 2 & 3\\
\end{ytableau}} - \frac{1}{2\sqrt{2}} |\underline{201}\rangle_{\tiny \begin{ytableau}
1 \\ 2 \\ 3\\
\end{ytableau}}
 + \frac{\sqrt{3}}{2}|\underline{000}\rangle_{\tiny \begin{ytableau}
1 & 2 & 3\\
\end{ytableau}} ~ , \nonumber \\
 \\
 |K_4\rangle &=&  \frac{1}{\sqrt{3}} \left(|000001\rangle+|010000\rangle+|000100\rangle \right) \nonumber \\ 
 &=& \frac{1}{\sqrt{3}} \left( |\underline{001}\rangle+|\underline{100}\rangle+|\underline{010}\rangle \right) = |\underline{001} \rangle_{\tiny \begin{ytableau}
1 & 2 & 3\\
\end{ytableau} } ~ ; \nonumber \\ \\
|K_5 \rangle &=&\frac{1}{\sqrt{3}} \left(|010001\rangle+|000101\rangle+|010100\rangle\right) \nonumber \\
&=&
\frac{1}{\sqrt{3}} \left( |\underline{101}\rangle+|\underline{011}\rangle+|\underline{110}\rangle \right) =  |\underline{011} \rangle_{\tiny \begin{ytableau}
1 & 2 & 3\\
\end{ytableau} } ~,\nonumber \\ \\
 |K_6 \rangle &=& |010101 \rangle = |\underline{111} \rangle =  |\underline{111} \rangle_{\tiny \begin{ytableau}
1 & 2 & 3\\
\end{ytableau} } ~.
\end{eqnarray}
The Lanczos coefficients are
\begin{eqnarray}
    &&b_1 = \sqrt{3}= b_6 ~; ~b_2= 2 = b_3 =b_4 =b_5 ~; \nonumber \\
    && b_{n\ge 7} =0~.
\end{eqnarray}

\subsection{$L=2\ell=8$} \label{kryb-lc-4boxes}

We first list some of the symmetrized states in the cases where $\ell=4$ as follows.
\begin{itemize}
\item $|\underline{2222}\rangle$ (similar to $|\underline{0000}\rangle$ and $|\underline{1111}\rangle$) :
    \begin{eqnarray}
    |\underline{2222}\rangle_{\tiny
        \begin{ytableau}
        1& 2& 3 & 4 \\
        \end{ytableau}
    } &=& |\underline{2222}\rangle ~. 
    \end{eqnarray}
    
\item $|\underline{2220}\rangle$ (similar to $|\underline{0111}\rangle$, $|\underline{2000}\rangle$ and $|\underline{01111}\rangle$) :
    \begin{eqnarray}
    |\underline{2220}\rangle_{\tiny
        \begin{ytableau}
        1& 2& 3 & 4 \\
        \end{ytableau}
    } &=& \frac{1}{2} \left( |\underline{2220} \rangle + |\underline{2202} \rangle + |\underline{2022} \rangle+|\underline{0222} \rangle \right)~; \nonumber \\
    |\underline{2220}\rangle_{\tiny
        \begin{ytableau}
        1& 2& 3  \\
        4 & \none & \none \\
        \end{ytableau}
    } &=& \frac{1}{2\sqrt{3}} \left( 3 |\underline{2220} \rangle -|\underline{0222} \rangle - |\underline{2022} \rangle - |\underline{2202} \rangle   \right)~, \nonumber \\
    |\underline{2202}\rangle_{\tiny
        \begin{ytableau}
        1& 2& 4  \\
        3 & \none & \none \\
        \end{ytableau}
    } &=& \frac{1}{2\sqrt{3}} \left( 3 |\underline{2202} \rangle -|\underline{0222} \rangle - |\underline{2022} \rangle - |\underline{2220} \rangle   \right)~, \nonumber \\
     |\underline{2022}\rangle_{\tiny
        \begin{ytableau}
        1& 3& 4  \\
        2 & \none & \none \\
        \end{ytableau}
    } &=& \frac{1}{2\sqrt{3}} \left( 3 |\underline{2022} \rangle- |\underline{0222} \rangle - |\underline{2202} \rangle - |\underline{2220} \rangle   \right) ~. \nonumber \\
    \end{eqnarray} 
    
\item $|\underline{2200}\rangle$ (similar to $|\underline{0011}\rangle$) :
\begin{eqnarray}
 |\underline{2200}\rangle_{\tiny
\begin{ytableau}
1& 2& 3 & 4 \\
\end{ytableau}
} &=& \frac{1}{\sqrt{6}} \left( |\underline{2200} \rangle+|\underline{2020} \rangle+|\underline{2002} \rangle+|\underline{0220}\rangle +|\underline{0202} \rangle+|\underline{0022} \rangle \right) ~; \nonumber \\
 |\underline{2200}\rangle_{\tiny
\begin{ytableau}
1& 2& 3  \\
4 & \none & \none \\
\end{ytableau}
} &=&  \frac{1}{\sqrt{6}} \left( |\underline{2200} \rangle+|\underline{0220} \rangle+|\underline{2020} \rangle - |\underline{0202}\rangle -|\underline{2002} \rangle - |\underline{0022} \rangle \right) ~, \nonumber \\
|\underline{2200}\rangle_{\tiny
\begin{ytableau}
1& 2& 4  \\
3 & \none & \none \\
\end{ytableau}
} &=&  \frac{1}{\sqrt{6}} \left( |\underline{2200} \rangle+|\underline{0202} \rangle+|\underline{2002} \rangle - |\underline{0220}\rangle -|\underline{2020} \rangle - |\underline{0022} \rangle \right)  ~, \nonumber \\
|\underline{2020}\rangle_{\tiny
\begin{ytableau}
1&  3 & 4  \\
2 & \none & \none \\
\end{ytableau}
} &=&  \frac{1}{\sqrt{6}} \left( |\underline{2020} \rangle+|\underline{0022} \rangle+|\underline{2002} \rangle - |\underline{0220}\rangle -|\underline{2200} \rangle - |\underline{0202} \rangle \right) ~; \nonumber \\
|\underline{2200}\rangle_{\tiny
\begin{ytableau}
1& 2 \\
3 & 4 \\
\end{ytableau}
} &=& \frac{1}{2\sqrt{3}} \left( 2|\underline{2200} \rangle+ 2|\underline{0022} \rangle -|\underline{0220} \rangle - |\underline{2002}\rangle -|\underline{2020} \rangle-|\underline{0202} \rangle \right) ~, \nonumber \\
|\underline{2020}\rangle_{\tiny
\begin{ytableau}
1& 3 \\
2 & 4 \\
\end{ytableau}
} &=& \frac{1}{2\sqrt{3}} \left( 2|\underline{2020} \rangle+ 2|\underline{0202} \rangle -|\underline{0220} \rangle - |\underline{2002}\rangle -|\underline{2200} \rangle-|\underline{0022} \rangle \right) ~. \nonumber \\
\end{eqnarray}

\item $|\underline{2201}\rangle$ :
\begin{eqnarray}
 |\underline{2201}\rangle_{\tiny
\begin{ytableau}
1& 2& 3 & 4 \\
\end{ytableau}
} &=& \frac{1}{2\sqrt{3}} \left( |\underline{2201} \rangle+|\underline{0221} \rangle+|\underline{1202} \rangle+|\underline{2102}\rangle +|\underline{2210} \rangle+|\underline{2021} \rangle \right. \nonumber \\
&& \qquad \left. + |\underline{2012} \rangle+|\underline{0212} \rangle+|\underline{1220} \rangle+|\underline{0122}\rangle +|\underline{2120} \rangle+|\underline{1022} \rangle \right) \nonumber \\
|\underline{2201}\rangle_{\tiny
\begin{ytableau}
1& 2& 3  \\
4 & \none & \none \\
\end{ytableau}
} &=& \frac{1}{3\sqrt{2}} \left(  2|\underline{2201} \rangle +2 |\underline{0221} \rangle + 2|\underline{2021} \rangle \right. \nonumber \\
&& \qquad \qquad \left. -|\underline{1202} \rangle -|\underline{2102} \rangle -|\underline{0212} \rangle -|\underline{1022}\rangle -|\underline{0122} \rangle -|\underline{2012} \rangle   \right) ~, \nonumber \\
|\underline{2210}\rangle_{\tiny
\begin{ytableau}
1& 2& 3  \\
4 & \none & \none \\
\end{ytableau}
} &=& \frac{1}{3\sqrt{2}} \left(  2|\underline{2210} \rangle +2 |\underline{1220} \rangle + 2|\underline{2120} \rangle 
\right. \nonumber \\
&& \qquad \qquad \left.
-|\underline{0212} \rangle -|\underline{2012} \rangle -|\underline{1202} \rangle -|\underline{0122}\rangle -|\underline{1022} \rangle -|\underline{2102} \rangle   \right)~, \nonumber \\
|\underline{2210}\rangle_{\tiny
\begin{ytableau}
1& 2& 4  \\
3 & \none & \none \\
\end{ytableau}
} &=& \frac{1}{3\sqrt{2}} \left(  2|\underline{2210} \rangle +2 |\underline{0212} \rangle + 2|\underline{2012} \rangle
\right. \nonumber \\
&& \qquad \qquad \left.
-|\underline{1220} \rangle -|\underline{2120} \rangle -|\underline{0221} \rangle -|\underline{1022}\rangle -|\underline{0122} \rangle -|\underline{2021} \rangle   \right)~, \nonumber \\
|\underline{2201}\rangle_{\tiny
\begin{ytableau}
1& 2& 4  \\
3 & \none & \none \\
\end{ytableau}
} &=& \frac{1}{3\sqrt{2}} \left(  2|\underline{2201} \rangle +2 |\underline{1202} \rangle + 2|\underline{2102} \rangle
\right. \nonumber \\
&& \qquad \qquad \left.
-|\underline{0221} \rangle -|\underline{2021} \rangle -|\underline{1220} \rangle -|\underline{0122}\rangle -|\underline{1022} \rangle -|\underline{2120} \rangle   \right) ~, \nonumber \\
|\underline{2120}\rangle_{\tiny
\begin{ytableau}
1& 3& 4  \\
2 & \none & \none \\
\end{ytableau}
} &=& \frac{1}{3\sqrt{2}} \left(  2|\underline{2120} \rangle +2 |\underline{0122} \rangle + 2|\underline{2102} \rangle
\right. \nonumber \\
&& \qquad \qquad \left.
-|\underline{1220} \rangle -|\underline{2210} \rangle -|\underline{0221} \rangle -|\underline{1202}\rangle -|\underline{0212} \rangle -|\underline{2201} \rangle   \right) ~, \nonumber \\
|\underline{2021}\rangle_{\tiny
\begin{ytableau}
1& 3& 4  \\
2 & \none & \none \\
\end{ytableau}
} &=& \frac{1}{3\sqrt{2}} \left(  2|\underline{2021} \rangle +2 |\underline{1022} \rangle + 2|\underline{2012} \rangle
\right. \nonumber \\
&& \qquad \qquad \left.
-|\underline{0221} \rangle -|\underline{2201} \rangle -|\underline{1220} \rangle -|\underline{0212}\rangle -|\underline{1202} \rangle -|\underline{2210} \rangle   \right)  ~; \nonumber \\
|\underline{2201}\rangle_{\tiny
\begin{ytableau}
1& 2 \\
3 & 4 \\
\end{ytableau}
} &=&  \frac{1}{2\sqrt{6}} \left( 2|\underline{2201} \rangle+2|\underline{2210} \rangle+ 2|\underline{0122} \rangle + 2|\underline{1022}\rangle -|\underline{0221} \rangle - |\underline{2021} \rangle \right. \nonumber \\
&& \qquad \left.  - |\underline{0212} \rangle - |\underline{2012} \rangle - |\underline{2102} \rangle - |\underline{1202}\rangle - |\underline{2120} \rangle -|\underline{1220} \rangle \right) \nonumber \\
|\underline{2021}\rangle_{\tiny
\begin{ytableau}
1& 3 \\
2 & 4 \\
\end{ytableau}
} &=& \frac{1}{2\sqrt{6}} \left( 2|\underline{2021} \rangle+2|\underline{2120} \rangle+ 2|\underline{0212} \rangle + 2|\underline{1202}\rangle -|\underline{0221} \rangle - |\underline{2201} \rangle  \right. \nonumber \\
&& \qquad \left.  - |\underline{0122} \rangle - |\underline{2102} \rangle - |\underline{2012} \rangle - |\underline{1022}\rangle - |\underline{2210} \rangle -|\underline{1220} \rangle \right) \nonumber \\
|\underline{2012}\rangle_{\tiny
\begin{ytableau}
1& 4  \\
2 & \none \\
3 & \none \\
\end{ytableau}
} &=& \frac{1}{4} \left( |\underline{1202} \rangle+|\underline{0122} \rangle  +|\underline{2201} \rangle +|\underline{2120} \rangle+ 2|\underline{2012}\rangle 
\right. \nonumber \\
&& \qquad \left. 
- |\underline{0212} \rangle -|\underline{1022} \rangle  - |\underline{2210} \rangle -|\underline{2021} \rangle -2 |\underline{2102} \rangle \right) \nonumber \\
|\underline{2021}\rangle_{\tiny
\begin{ytableau}
1& 3  \\
2 & \none \\
4 & \none \\
\end{ytableau}
} &=& \frac{1}{4} \left( |\underline{1220} \rangle+|\underline{0122} \rangle  +|\underline{2210} \rangle +|\underline{2102} \rangle+ 2|\underline{2021}\rangle 
\right. \nonumber \\
&& \qquad \left. 
- |\underline{0221} \rangle -|\underline{1022} \rangle  - |\underline{2201} \rangle -|\underline{2012} \rangle -2 |\underline{2120} \rangle \right) \nonumber \\
|\underline{2201}\rangle_{\tiny
\begin{ytableau}
1& 2  \\
3 & \none \\
4 & \none \\
\end{ytableau}
} &=& \frac{1}{4} \left( |\underline{1220} \rangle+|\underline{0212} \rangle  +|\underline{2120} \rangle +|\underline{2012} \rangle+ 2|\underline{2201}\rangle 
\right. \nonumber \\
&& \qquad \left. 
- |\underline{0221} \rangle -|\underline{1202} \rangle  - |\underline{2021} \rangle -|\underline{2102} \rangle -2 |\underline{2210} \rangle \right) \nonumber \\
\end{eqnarray}

\item $|\underline{2001}\rangle$ :
\begin{eqnarray}
|\underline{2001}\rangle_{\tiny
\begin{ytableau}
1& 2& 3 & 4 \\
\end{ytableau}
} &=& \frac{1}{2\sqrt{3}} \left( |\underline{2001} \rangle+|\underline{0201} \rangle+|\underline{0021} \rangle+|\underline{1002}\rangle +|\underline{2100} \rangle+|\underline{2010} \rangle \right. \nonumber \\
&& \qquad \left. + |\underline{1200} \rangle+|\underline{0120} \rangle+|\underline{0012} \rangle+|\underline{0210}\rangle +|\underline{1020} \rangle+|\underline{0102} \rangle \right) ~; \nonumber \\
|\underline{2001}\rangle_{\tiny
\begin{ytableau}
1& 2& 3  \\
4 & \none & \none \\
\end{ytableau}
} &=& \frac{1}{\sqrt{6}} \left( |\underline{2001} \rangle+|\underline{0201} \rangle+|\underline{0021} \rangle - |\underline{1002} \rangle - |\underline{0102} \rangle - |\underline{0012} \rangle \right) ~, \nonumber \\
|\underline{2010}\rangle_{\tiny
\begin{ytableau}
1& 2& 3  \\
4 & \none & \none \\
\end{ytableau}
} &=& \frac{1}{3\sqrt{2}} \left( |\underline{2010} \rangle+|\underline{0210} \rangle+|\underline{1020} \rangle+|\underline{2100}\rangle +|\underline{1200} \rangle+|\underline{0120} \rangle 
\right. \nonumber \\
&& \qquad \qquad \left.
- 2|\underline{0012} \rangle -2 |\underline{1002} \rangle - 2|\underline{0102} \rangle \right) ~, \nonumber \\
|\underline{2010}\rangle_{\tiny
\begin{ytableau}
1& 2& 4  \\
3 & \none & \none \\
\end{ytableau}
} &=& \frac{1}{\sqrt{6}} \left( |\underline{2010} \rangle+|\underline{0210} \rangle+|\underline{0012} \rangle - |\underline{1020} \rangle - |\underline{0120} \rangle - |\underline{0021} \rangle \right) ~, \nonumber \\
|\underline{2001}\rangle_{\tiny
\begin{ytableau}
1& 2& 4  \\
3 & \none & \none \\
\end{ytableau}
} &=& \frac{1}{3\sqrt{2}} \left( |\underline{2001} \rangle+|\underline{0201} \rangle+|\underline{1002} \rangle+|\underline{2100}\rangle +|\underline{1200} \rangle+|\underline{0102} \rangle 
\right. \nonumber \\
&& \qquad \qquad \left.
- 2|\underline{0021} \rangle -2 |\underline{1020} \rangle - 2|\underline{0120} \rangle \right) ~, \nonumber \\
|\underline{2100}\rangle_{\tiny
\begin{ytableau}
1& 3& 4  \\
2 & \none & \none \\
\end{ytableau}
} &=& \frac{1}{\sqrt{6}} \left( |\underline{2100} \rangle+|\underline{0120} \rangle+|\underline{0102} \rangle - |\underline{1200} \rangle - |\underline{0210} \rangle - |\underline{0201} \rangle \right)~, \nonumber \\
|\underline{2001}\rangle_{\tiny
\begin{ytableau}
1& 3& 4  \\
2 & \none & \none \\
\end{ytableau}
} &=& \frac{1}{3\sqrt{2}} \left( |\underline{2001} \rangle+|\underline{0021} \rangle+|\underline{1002} \rangle+|\underline{2010}\rangle +|\underline{1020} \rangle+|\underline{0012} \rangle 
\right. \nonumber \\
&& \qquad \qquad \left.
- 2|\underline{0201} \rangle -2 |\underline{1200} \rangle - 2|\underline{0210} \rangle \right) ~; \nonumber \\
|\underline{2001}\rangle_{\tiny
\begin{ytableau}
1& 2 \\
3 & 4 \\
\end{ytableau}
} &=&  \frac{1}{2\sqrt{6}} \left(  |\underline{2001} \rangle + |\underline{0201} \rangle + |\underline{2010} \rangle + |\underline{0210} \rangle + |\underline{0120} \rangle + |\underline{1020}\rangle \right. \nonumber \\
&& \qquad \left.  + |\underline{0102} \rangle + |\underline{1002} \rangle -2|\underline{0021} \rangle -2|\underline{0012} \rangle - 2|\underline{2100} \rangle - 2|\underline{1200}\rangle  \right) ~, \nonumber \\
|\underline{2001}\rangle_{\tiny
\begin{ytableau}
1& 3 \\
2 & 4 \\
\end{ytableau}
} &=& \frac{1}{2\sqrt{6}} \left(  |\underline{2001} \rangle + |\underline{0021} \rangle + |\underline{2100} \rangle + |\underline{0120} \rangle + |\underline{0210} \rangle + |\underline{1200}\rangle \right. \nonumber \\
&& \qquad \left.  + |\underline{0012} \rangle + |\underline{1002} \rangle -2|\underline{0201} \rangle -2|\underline{0102} \rangle - 2|\underline{2010} \rangle - 2|\underline{1020}\rangle  \right) ~; \nonumber \\
|\underline{2010}\rangle_{\tiny
\begin{ytableau}
1& 4  \\
2 & \none \\
3 & \none \\
\end{ytableau}}
&=& \frac{1}{4} \left( |\underline{2010} \rangle+|\underline{1200} \rangle  +|\underline{0012} \rangle +|\underline{0201} \rangle+ 2|\underline{0120}\rangle  
\right. \nonumber \\
 && \qquad \left. - |\underline{1020} \rangle -|\underline{2100} \rangle  - |\underline{0021} \rangle -|\underline{0102} \rangle -2 |\underline{0210} \rangle \right) ~, \nonumber \\
|\underline{2001}\rangle_{\tiny
\begin{ytableau}
1& 3  \\
2 & \none \\
4 & \none \\
\end{ytableau}
} &=& \frac{1}{4} \left( |\underline{2001} \rangle+|\underline{1200} \rangle  +|\underline{0021} \rangle +|\underline{0210} \rangle+ 2|\underline{0102}\rangle \right. \nonumber \\
&& \qquad \left. 
- |\underline{1002} \rangle -|\underline{2100} \rangle  - |\underline{0012} \rangle -|\underline{0120} \rangle -2 |\underline{0201} \rangle \right) ~, \nonumber \\
|\underline{2001}\rangle_{\tiny
\begin{ytableau}
1& 2  \\
3 & \none \\
4 & \none \\
\end{ytableau}
} &=& \frac{1}{4} \left( |\underline{2001} \rangle+|\underline{1020} \rangle  +|\underline{0201} \rangle +|\underline{0120} \rangle+ 2|\underline{0012}\rangle 
\right. \nonumber \\
&& \qquad \left. 
- |\underline{1002} \rangle -|\underline{2010} \rangle  - |\underline{0102} \rangle -|\underline{0210} \rangle -2 |\underline{0021} \rangle \right) ~. \nonumber \\
\end{eqnarray}

\item $|\underline{2011}\rangle$~:
\begin{eqnarray}
 |\underline{2011}\rangle_{\tiny
\begin{ytableau}
1& 2& 3 & 4 \\
\end{ytableau}
} &=& \frac{1}{2\sqrt{3}} \left( |\underline{2011} \rangle+|\underline{0211} \rangle+|\underline{1021} \rangle+|\underline{1012}\rangle +|\underline{2101} \rangle+|\underline{2110} \rangle \right. \nonumber \\
&& \qquad \left. + |\underline{1201} \rangle+|\underline{0121} \rangle+|\underline{0112} \rangle+|\underline{1210}\rangle +|\underline{1120} \rangle+|\underline{1102} \rangle \right) ~; \nonumber \\
|\underline{2011}\rangle_{\tiny
\begin{ytableau}
1& 2& 3  \\
4 & \none & \none \\
\end{ytableau}
} &=& \frac{1}{3\sqrt{2}} \left( |\underline{2011} \rangle+|\underline{0211} \rangle+|\underline{1021} \rangle+|\underline{2101}\rangle +|\underline{1201} \rangle+|\underline{0121} \rangle  
\right. \nonumber \\
&& \qquad \qquad \left.
- 2|\underline{1012} \rangle -2 |\underline{0112} \rangle - 2|\underline{1102} \rangle \right) ~, \nonumber \\
|\underline{2110}\rangle_{\tiny
\begin{ytableau}
1& 2& 3  \\
4 & \none & \none \\
\end{ytableau}
} &=& \frac{1}{\sqrt{6}} \left( |\underline{2110} \rangle +|\underline{1210} \rangle +|\underline{1120} \rangle - |\underline{0112}\rangle - |\underline{1012} \rangle  -|\underline{1102} \rangle   \right) ~, \nonumber \\
|\underline{2011}\rangle_{\tiny
\begin{ytableau}
1& 2& 4  \\
3 & \none & \none \\
\end{ytableau}
} &=& \frac{1}{3\sqrt{2}} \left( |\underline{2011} \rangle+|\underline{0211} \rangle+|\underline{1012} \rangle+|\underline{2110}\rangle +|\underline{1210} \rangle+|\underline{0112} \rangle 
\right. \nonumber \\
&& \qquad \qquad \left.
- 2|\underline{1021} \rangle -2 |\underline{0121} \rangle - 2|\underline{1120} \rangle \right) ~, \nonumber \\
|\underline{2101}\rangle_{\tiny
\begin{ytableau}
1& 2& 4  \\
3 & \none & \none \\
\end{ytableau}
} &=& \frac{1}{\sqrt{6}} \left( |\underline{2101} \rangle +|\underline{1201} \rangle +|\underline{1102} \rangle - |\underline{0121}\rangle - |\underline{1021} \rangle  -|\underline{1120} \rangle   \right) ~, \nonumber \\
|\underline{2101}\rangle_{\tiny
\begin{ytableau}
1& 3& 4  \\
2 & \none & \none \\
\end{ytableau}
} &=& \frac{1}{3\sqrt{2}} \left( |\underline{2101} \rangle+|\underline{0121} \rangle+|\underline{1102} \rangle+|\underline{2110}\rangle +|\underline{1120} \rangle+|\underline{0112} \rangle  
\right. \nonumber \\
&& \qquad \qquad \left.
- 2|\underline{1201} \rangle -2 |\underline{0211} \rangle - 2|\underline{1210} \rangle \right) ~, \nonumber \\
|\underline{2011}\rangle_{\tiny
\begin{ytableau}
1& 3& 4  \\
2 & \none & \none \\
\end{ytableau}
} &=& \frac{1}{\sqrt{6}} \left( |\underline{2011} \rangle +|\underline{1021} \rangle +|\underline{1012} \rangle - |\underline{0211}\rangle - |\underline{1201} \rangle  -|\underline{1210} \rangle   \right) ~; \nonumber \\
|\underline{2011}\rangle_{\tiny
\begin{ytableau}
1& 2 \\
3 & 4 \\
\end{ytableau}
} &=& \frac{1}{2\sqrt{6}} \left( 2|\underline{2011} \rangle+2|\underline{0211} \rangle+ 2|\underline{1120} \rangle + 2|\underline{1102}\rangle -|\underline{1021} \rangle - |\underline{0121} \rangle \right. \nonumber \\
&& \qquad \left.  - |\underline{1012} \rangle - |\underline{0112} \rangle - |\underline{2110} \rangle - |\underline{1210}\rangle - |\underline{2101} \rangle -|\underline{1201} \rangle \right) ~,\nonumber \\
|\underline{2101}\rangle_{\tiny
\begin{ytableau}
1& 3 \\
2 & 4 \\
\end{ytableau}
} &=& \frac{1}{2\sqrt{6}} \left( 2|\underline{2101} \rangle+2|\underline{0121} \rangle+ 2|\underline{1210} \rangle + 2|\underline{1012}\rangle -|\underline{1201} \rangle - |\underline{0211} \rangle \right. \nonumber \\
&& \qquad \left.  - |\underline{1102} \rangle - |\underline{0112} \rangle - |\underline{2110} \rangle - |\underline{1120}\rangle - |\underline{2011} \rangle -|\underline{1021} \rangle \right) ~; \nonumber \\
|\underline{2011}\rangle_{\tiny
\begin{ytableau}
1& 4  \\
2 & \none \\
3 & \none \\
\end{ytableau}
} &=& \frac{1}{4} \left( |\underline{2011} \rangle+|\underline{0121} \rangle  +|\underline{1012} \rangle +|\underline{1120} \rangle+ 2|\underline{1201}\rangle 
\right. \nonumber \\
&& \qquad \left. 
- |\underline{0211} \rangle -|\underline{2101} \rangle  - |\underline{1210} \rangle -|\underline{1102} \rangle -2 |\underline{1021} \rangle \right) ~, \nonumber \\
|\underline{2011}\rangle_{\tiny
\begin{ytableau}
1& 3  \\
2 & \none \\
4 & \none \\
\end{ytableau}
} &=& \frac{1}{4} \left( |\underline{2011} \rangle+|\underline{0112} \rangle  +|\underline{1021} \rangle +|\underline{1102} \rangle+ 2|\underline{1210}\rangle 
\right. \nonumber \\
&& \qquad \left. 
- |\underline{0211} \rangle -|\underline{2110} \rangle  - |\underline{1201} \rangle -|\underline{1120} \rangle -2 |\underline{1012} \rangle \right) ~, \nonumber \\
|\underline{2101}\rangle_{\tiny
\begin{ytableau}
1& 2  \\
3 & \none \\
4 & \none \\
\end{ytableau}
} &=& \frac{1}{4} \left( |\underline{2101} \rangle+|\underline{0112} \rangle  +|\underline{1201} \rangle +|\underline{1012} \rangle+ 2|\underline{1120}\rangle 
\right. \nonumber \\
&& \qquad \left. 
- |\underline{0121} \rangle -|\underline{2110} \rangle  - |\underline{1021} \rangle -|\underline{1210} \rangle -2 |\underline{1102} \rangle \right) ~. \nonumber \\
\end{eqnarray}

\end{itemize}

Below let
\begin{eqnarray}
    |\underline{2201}\rangle_{\tiny \ydiagram{2,1,1}} &=& |\underline{2012}\rangle_{\tiny
\begin{ytableau}
1& 4  \\
2 & \none \\
3 & \none \\
\end{ytableau}}
+
|\underline{2021}\rangle_{\tiny
\begin{ytableau}
1& 3  \\
2 & \none \\
4 & \none \\
\end{ytableau}
}+
|\underline{2201}\rangle_{\tiny
\begin{ytableau}
1& 2  \\
3 & \none \\
4 & \none \\
\end{ytableau}
}~, \nonumber \\
 |\underline{2001}\rangle_{\tiny \ydiagram{2,1,1}} &=& 
 |\underline{2010}\rangle_{\tiny
\begin{ytableau}
1& 4  \\
2 & \none \\
3 & \none \\
\end{ytableau}}
+
|\underline{2001}\rangle_{\tiny
\begin{ytableau}
1& 3  \\
2 & \none \\
4 & \none \\
\end{ytableau}
}+
|\underline{2001}\rangle_{\tiny
\begin{ytableau}
1& 2  \\
3 & \none \\
4 & \none \\
\end{ytableau}
} \nonumber \\
|\underline{2011}\rangle_{\tiny \ydiagram{2,1,1}} &=& 
|\underline{2011}\rangle_{\tiny
\begin{ytableau}
1& 4  \\
2 & \none \\
3 & \none \\
\end{ytableau}}
+
|\underline{2011}\rangle_{\tiny
\begin{ytableau}
1& 3  \\
2 & \none \\
4 & \none \\
\end{ytableau}
}+
|\underline{2101}\rangle_{\tiny
\begin{ytableau}
1& 2  \\
3 & \none \\
4 & \none \\
\end{ytableau}
} 
\end{eqnarray}
the Krylov basis generated from $|Z_2 \rangle$ are therefore 
\begin{eqnarray}
|K_0 \rangle &=& |{10101010} \rangle = |\underline{2222}\rangle  = |\underline{2222}\rangle_{\tiny
\begin{ytableau}
1& 2& 3 & 4 \\
\end{ytableau}
} ~; \\
|K_1 \rangle &=& \frac{1}{2} \left( |10101000 \rangle + |10100010 \rangle + |10001010\rangle+|00101010\rangle \right) \nonumber \\
&=& \frac{1}{2} \left( |\underline{2220} \rangle + |\underline{2202} \rangle + |\underline{2022} \rangle+|\underline{0222} \rangle \right) 
 = |\underline{2220}\rangle_{\tiny
\begin{ytableau}
1& 2& 3 & 4 \\
\end{ytableau}
} ~,
\\
|K_2 \rangle &=& \frac{1}{\sqrt{6}} \left( |10100000\rangle+|10001000\rangle+|10000010\rangle \right. \nonumber \\ 
 && \left.  +|00101000\rangle 
 +|00100010\rangle+|00001010\rangle \right) \nonumber \\
&=& \frac{1}{\sqrt{6}} \left( |\underline{2200} \rangle+|\underline{2020} \rangle+|\underline{2002} \rangle+|\underline{0220}\rangle +|\underline{0202} \rangle+|\underline{0022} \rangle \right) \nonumber \\
&=& |\underline{2200}\rangle_{\tiny
\begin{ytableau}
1& 2& 3 & 4 \\
\end{ytableau}
} ~; \\
|K_3 \rangle & =& 0.158114 \left( |00101001\rangle+|10100100\rangle+|10010010\rangle+|01001010\rangle \right) \nonumber \\
 &&+ 0.474342 \left(|10000000\rangle+|00100000\rangle+|00001000\rangle+|00000010\rangle \right) \nonumber \\
 & =& \frac{1}{2\sqrt{10}} \left( |\underline{0221} \rangle+|\underline{2210} \rangle+|\underline{2102} \rangle+|\underline{1022} \rangle \right) \nonumber \\
 && +\frac{3}{\sqrt{10}} \left(|\underline{2000} \rangle+|\underline{0200} \rangle+|\underline{0020} \rangle+|\underline{0002} \rangle \right) \nonumber \\
&=&\frac{1}{2\sqrt{10}} \left[  \frac{2}{\sqrt{3}}\left( |\underline{2201}\rangle_{\tiny
\begin{ytableau}
1& 2& 3 & 4 \\
\end{ytableau}
} -\frac{1}{\sqrt{2}} 
|\underline{2021}\rangle_{\tiny
\begin{ytableau}
1& 3 \\
2 & 4 \\
\end{ytableau}
}  \right)  -  |\underline{2201}\rangle_{\tiny
\ydiagram{2,1,1}}
\right] \nonumber \\
&& +\frac{6}{\sqrt{10}}  |\underline{2000}\rangle_{\tiny
\begin{ytableau}
1& 2& 3 & 4 \\
\end{ytableau}
} ~,\nonumber \\
\end{eqnarray}

\begin{eqnarray}
|K_4 \rangle &=& 
0.241943\left(|10010000\rangle+|10000100\rangle+|01001000\rangle+|01000010\rangle \right. \nonumber\\
&& \qquad \qquad \qquad \left.+|00100100\rangle+|00100001\rangle+|00010010\rangle+|00001001\rangle \right) \nonumber \\
&&+0.0201619 \left(|10100000\rangle+|10000010\rangle+|00101000\rangle+|00001010\rangle \right) \nonumber\\
&&-0.0403239 \left(|10001000\rangle+|00100010\rangle \right) + 0.72583|00000000\rangle \nonumber \\
&=&  0.241943\left(|\underline{2100} \rangle+|\underline{2010} \rangle+|\underline{1020} \rangle+|\underline{1002} \rangle +|\underline{0210} \rangle+|\underline{0201} \rangle+|\underline{0102} \rangle+|\underline{0021} \rangle \right) \nonumber \\
&&+ 0.0201619 \left(|\underline{2200} \rangle+|\underline{2002} \rangle+|\underline{0220} \rangle+|\underline{0022} \rangle -2|\underline{2020} \rangle -2 |\underline{0202} \rangle \right) \nonumber \\
&&+ 0.72583|\underline{0000} \rangle \nonumber \\
&=& 
 0.241943\left[ \frac{2}{\sqrt{3}}\left( 2|\underline{2001}\rangle_{\tiny
\begin{ytableau}
1& 2& 3 & 4 \\
\end{ytableau}
} -\frac{1}{\sqrt{2}} 
|\underline{2001}\rangle_{\tiny
\begin{ytableau}
1& 3 \\
2 & 4 \\
\end{ytableau}
}  \right)  -  |\underline{2001}\rangle_{\tiny \ydiagram{2,1,1}}
 \right] \nonumber \\
&& - 0.0201619 \cdot 2\sqrt{3} |\underline{2020}\rangle_{\tiny
\begin{ytableau}
1& 3 \\
2 & 4 \\
\end{ytableau}
}
+  0.72583|\underline{0000}\rangle_{\tiny
\begin{ytableau}
1& 2& 3 & 4 \\
\end{ytableau}
} ~;
\nonumber \\
\\
 |K_5\rangle &=& 
 0.0347245 \left( |00101001\rangle+|10100100\rangle+|10010010\rangle+|01001010\rangle \right) \nonumber \\
&& +0.185197 \left( |01001001\rangle+|00100101\rangle+|10010100\rangle+|01010010\rangle \right) \nonumber \\
&&-0.0115748 \left( |10000000\rangle+|00100000\rangle+|00001000\rangle+|00000010\rangle \right) \nonumber \\
&&+0.462993 \left( |00000001\rangle+|01000000\rangle+|00010000\rangle+|00000100\rangle \right) \nonumber\\
&=& 
 0.0347245 \left( |\underline{0221} \rangle+|\underline{2210} \rangle+|\underline{2102} \rangle+|\underline{1022} \rangle \right) \nonumber \\
 && +0.185197 \left( |\underline{1021} \rangle+|\underline{0211} \rangle+|\underline{2110} \rangle+|\underline{1102} \rangle \right)  \nonumber \\
&&  -0.0115748 \left( |\underline{2000} \rangle+|\underline{0200} \rangle+|\underline{0020} \rangle+|\underline{0002} \rangle\right) \nonumber \\
&& +0.462993 \left( |\underline{0001} \rangle+|\underline{1000} \rangle+|\underline{0100} \rangle+|\underline{0010} \rangle \right)  \nonumber \\
&=&  0.0347245
\left[ \frac{2}{\sqrt{3}}\left( |\underline{2201}\rangle_{\tiny
\begin{ytableau}
1& 2& 3 & 4 \\
\end{ytableau}
} -\frac{1}{\sqrt{2}} 
|\underline{2021}\rangle_{\tiny
\begin{ytableau}
1& 3 \\
2 & 4 \\
\end{ytableau}
}  \right)  - 
|\underline{2201}\rangle_{\tiny \ydiagram{2,1,1}}   \right] 
\nonumber \\
&&+0.185197
\left[ \frac{2}{\sqrt{3}}\left( |\underline{2011}\rangle_{\tiny
\begin{ytableau}
1& 2& 3 & 4 \\
\end{ytableau}
} -\frac{1}{\sqrt{2}} 
|\underline{2101}\rangle_{\tiny
\begin{ytableau}
1& 3 \\
2 & 4 \\
\end{ytableau}
}  \right)  -  |\underline{2011}\rangle_{\tiny \ydiagram{2,1,1}}
 \right] 
 \nonumber \\
&&  -0.0115748 \cdot 2  |\underline{2000}\rangle_{\tiny
\begin{ytableau}
1& 2& 3 & 4 \\
\end{ytableau}
} +0.462993 \cdot 2  |\underline{0001}\rangle_{\tiny
\begin{ytableau}
1& 2& 3 & 4 \\
\end{ytableau}
} ~, \nonumber \\
\end{eqnarray}

\begin{eqnarray}
|K_6\rangle &=& 
0.0151446\left(|10010000\rangle+|10000100\rangle+|01001000\rangle+|01000010\rangle \right. \nonumber \\
&& \qquad \qquad \qquad \left.+|00100100\rangle+|00100001\rangle+|00010010\rangle+|00001001\rangle \right) \nonumber \\
&&-0.0158865\left(|10100000\rangle+|10000010\rangle+|00101000\rangle+|00001010\rangle \right)   \nonumber \\
&& \qquad -0.03509|00000000\rangle +0.0317731\left(|10001000\rangle+|00100010\rangle \right) \nonumber \\
&&+0.42946 \left(|01000001\rangle+|00000101\rangle+|01010000\rangle+|00010100\rangle \right) \nonumber \\
&& \qquad +0.357883\left(|00010001\rangle+|01000100\rangle\right)  \nonumber  \\
&=& 
 0.0151446\left(|\underline{2100} \rangle+|\underline{2010} \rangle+|\underline{1020} \rangle+|\underline{1002} \rangle +|\underline{0210} \rangle+|\underline{0201} \rangle+|\underline{0102} \rangle+|\underline{0021} \rangle\right) \nonumber \\
&&-0.0158865\left(|\underline{2200} \rangle+|\underline{2002} \rangle+|\underline{0220} \rangle+|\underline{0022} \rangle -2|\underline{2020} \rangle-2|\underline{0202} \rangle \right)   -0.03509|\underline{0000} \rangle  \nonumber \\
&&+0.405601\left(|\underline{1001} \rangle+|\underline{0011} \rangle+|\underline{1100} \rangle+|\underline{0110} \rangle + |\underline{0101} \rangle+|\underline{1010} \rangle \right) \nonumber \\
&&  +0.023859 \left(|\underline{1001} \rangle+|\underline{0011} \rangle+|\underline{1100} \rangle+|\underline{0110} \rangle  -2|\underline{0101} \rangle-2|\underline{1010} \rangle\right)  \nonumber\\
&=& 
 0.0151446\left[ \frac{2}{\sqrt{3}}\left( 2|\underline{2001}\rangle_{\tiny
\begin{ytableau}
1& 2& 3 & 4 \\
\end{ytableau}
} -\frac{1}{\sqrt{2}} 
|\underline{2001}\rangle_{\tiny
\begin{ytableau}
1& 3 \\
2 & 4 \\
\end{ytableau}
}  \right) - 
|\underline{2001}\rangle_{\tiny \ydiagram{2,1,1}}\right] \nonumber \\
&&+ 0.0158865 \cdot 2 \sqrt{3} |\underline{2020}\rangle_{\tiny
\begin{ytableau}
1& 3 \\
2 & 4 \\
\end{ytableau}
} 
-0.03509|\underline{0000}\rangle_{\tiny
\begin{ytableau}
1& 2& 3 & 4 \\
\end{ytableau}
}
  \nonumber\\
&&  +0.405601 \cdot \sqrt{6}  |\underline{0011}\rangle_{\tiny
\begin{ytableau}
1& 2& 3 & 4 \\
\end{ytableau}
}   - 0.023859\cdot 2 \sqrt{3} |\underline{0101}\rangle_{\tiny
\begin{ytableau}
1& 3 \\
2 & 4 \\
\end{ytableau}
} ~; \\
|K_7\rangle &=& -0.0309286\left(|00101001\rangle+|10100100\rangle+|10010010\rangle+|01001010\rangle \right) \nonumber \\
&&-0.00796488 \left(|01001001\rangle+|00100101\rangle+|10010100\rangle+|01010010\rangle \right) \nonumber \\
&&+0.00576334\left(|00000001\rangle+|01000000\rangle+|00010000\rangle+|00000100\rangle \right) \nonumber \\
&&+0.0103095\left(|10000000\rangle+|00000010\rangle+|00100000\rangle+|00001000\rangle \right) \nonumber \\
&&+0.498839\left( |01010001\rangle+|01000101\rangle+|00010101\rangle+|01010100\rangle \right)  \nonumber \\
&=& -0.0309286\left(|\underline{0221} \rangle+|\underline{2210} \rangle+|\underline{2102} \rangle+|\underline{1022} \rangle \right) 
\nonumber \\
&& -0.00796488 \left(|\underline{1021} \rangle+|\underline{0211} \rangle+|\underline{2110} \rangle+|\underline{1102} \rangle \right) \nonumber \\
&&+0.00576334\left(|\underline{0001} \rangle+|\underline{1000} \rangle+|\underline{0100} \rangle+|\underline{0010} \rangle \right) \nonumber \\
&& +0.0103095\left(|\underline{2000} \rangle+ |\underline{0002} \rangle+|\underline{0200} \rangle+|\underline{0020} \rangle \right) \nonumber \\
&&+0.498839\left( |\underline{1101} \rangle+|\underline{1011} \rangle+|\underline{0111} \rangle+|\underline{1110} \rangle \right)  \nonumber \\
&=& -0.0309286 
\left[
\frac{2}{\sqrt{3}}\left( |\underline{2201}\rangle_{\tiny
\begin{ytableau}
1& 2& 3 & 4 \\
\end{ytableau}
} -\frac{1}{\sqrt{2}} 
|\underline{2021}\rangle_{\tiny
\begin{ytableau}
1& 3 \\
2 & 4 \\
\end{ytableau}
}  \right) - 
|\underline{2201}\rangle_{\tiny
\ydiagram{2,1,1}
}  
 \right]
\nonumber \\
&&-0.00796488 
\left[
\frac{2}{\sqrt{3}}\left( |\underline{2011}\rangle_{\tiny
\begin{ytableau}
1& 2& 3 & 4 \\
\end{ytableau}
} -\frac{1}{\sqrt{2}} 
|\underline{2101}\rangle_{\tiny
\begin{ytableau}
1& 3 \\
2 & 4 \\
\end{ytableau}
}  \right) -  |\underline{2011}\rangle_{\tiny \ydiagram{2,1,1}
}
 \right]
 \nonumber \\
&& +0.00576334 \cdot 2  |\underline{0001}\rangle_{\tiny
\begin{ytableau}
1& 2& 3 & 4 \\
\end{ytableau}
}  +0.0103095 \cdot 2  |\underline{2000}\rangle_{\tiny
\begin{ytableau}
1& 2& 3 & 4 \\
\end{ytableau}
} \nonumber \\
&& +0.498839 \cdot 2  |\underline{0111}\rangle_{\tiny
\begin{ytableau}
1& 2& 3 & 4 \\
\end{ytableau}
} ~, 
\end{eqnarray}

\begin{eqnarray}
|K_8\rangle &=&
-0.0295582\left(|10010000\rangle+|10000100\rangle+|01001000\rangle+|01000010\rangle \right. \nonumber \\
&& \qquad \qquad \qquad 
\left. +|00100100\rangle+|00100001\rangle+|00010010\rangle+|00001001\rangle \right) \nonumber \\
&& +0.0140673\left(|10100000\rangle+|10000010\rangle+|00001010\rangle+|00101000\rangle \right) \nonumber \\
&& \qquad  -0.0281346\left(|10001000\rangle+|00100010\rangle \right) \nonumber \\
&& -0.0229144\left(|01000001\rangle+|00000101\rangle+|01010000\rangle+|00010100\rangle \right)  \nonumber \\
&& \qquad
+0.0673788\left(|00010001\rangle+|01000100\rangle \right) \nonumber\\
&&+0.0741327|00000000\rangle +0.986894|01010101\rangle \nonumber\\
&=& 
-0.0295582\left(|\underline{2100} \rangle+|\underline{2010} \rangle+|\underline{1020} \rangle+|\underline{1002} \rangle +|\underline{0210} \rangle+|\underline{0201} \rangle+|\underline{0102} \rangle+|\underline{0021} \rangle \right) \nonumber \\
&&+0.0140673\left(|\underline{2200} \rangle+|\underline{2002} \rangle+|\underline{0022} \rangle+|\underline{0220} \rangle -2|\underline{2020} \rangle-2|\underline{0202} \rangle \right) \nonumber \\
&& + 0.00718333 \left(|\underline{1001} \rangle+|\underline{0011} \rangle+|\underline{1100} \rangle+|\underline{0110} \rangle + |\underline{0101} \rangle+|\underline{1010} \rangle \right) \nonumber \\
&&  -0.0300977 \left(|\underline{1001} \rangle+|\underline{0011} \rangle+|\underline{1100} \rangle+|\underline{0110} \rangle -2|\underline{0101} \rangle-2|\underline{1010} \rangle\right)  \nonumber\\
&&+0.0741327|\underline{0000} \rangle +0.986894|\underline{1111} \rangle \nonumber \\
&=& 
-0.0295582\left[ \frac{2}{\sqrt{3}}\left( 2|\underline{2001}\rangle_{\tiny
\begin{ytableau}
1& 2& 3 & 4 \\
\end{ytableau}
} -\frac{1}{\sqrt{2}} 
|\underline{2001}\rangle_{\tiny
\begin{ytableau}
1& 3 \\
2 & 4 \\
\end{ytableau}
}  \right) - 
|\underline{2001}\rangle_{\tiny
\ydiagram{2,1,1}
}   \right] \nonumber \\
&& - 0.0140673\cdot 2 \sqrt{3} |\underline{2020}\rangle_{\tiny
\begin{ytableau}
1& 3 \\
2 & 4 \\
\end{ytableau}
}  + 0.00718333 \cdot \sqrt{6}  |\underline{0011}\rangle_{\tiny
\begin{ytableau}
1& 2& 3 & 4 \\
\end{ytableau}
}    + 0.0300977 \cdot 2 \sqrt{3} |\underline{0101}\rangle_{\tiny
\begin{ytableau}
1& 3 \\
2 & 4 \\
\end{ytableau}
} \nonumber\\
&&+0.0741327|\underline{0000}\rangle_{\tiny
\begin{ytableau}
1& 2& 3 & 4 \\
\end{ytableau}
} +0.986894|\underline{1111}\rangle_{\tiny
\begin{ytableau}
1& 2& 3 & 4 \\
\end{ytableau}
} ~; \\
|K_9\rangle &=& +0.120005\left(|00101001\rangle+|10100100\rangle+|10010010\rangle+|01001010\rangle \right) \nonumber\\
&&-0.452501\left( |01001001\rangle+|00100101\rangle+|10010100\rangle+|01010010\rangle \right)  \nonumber\\
&&-0.000933504\left(|01010001\rangle+|01000101\rangle+|00010101\rangle+|01010100\rangle \right) \nonumber\\
&&-0.0400016\left(|00100000\rangle+|00001000\rangle+|10000000\rangle+|00000010\rangle \right) \nonumber\\
&&+0.171\left(|01000000\rangle+|00010000\rangle+|00000001\rangle+|00000100\rangle \right) \nonumber\\
&=& 0.120005\left(|\underline{0221} \rangle+|\underline{2210} \rangle+|\underline{2102} \rangle+|\underline{1022} \rangle \right)
\nonumber \\ &&-0.452501\left( |\underline{1021} \rangle+|\underline{0211} \rangle+|\underline{2110} \rangle+|\underline{1102} \rangle \right) \nonumber\\
&&-0.000933504\left(|\underline{1101} \rangle+|\underline{1011} \rangle+|\underline{0111} \rangle+|\underline{1110} \rangle \right) \nonumber \\ && -0.0400016\left(|\underline{0200} \rangle+|\underline{0020} \rangle+|\underline{2000} \rangle+|\underline{0002} \rangle \right) \nonumber\\
&&+0.171\left(|\underline{1000} \rangle+|\underline{0100} \rangle+|\underline{0001} \rangle+|\underline{0010} \rangle \right) \nonumber \\
&=&
+0.120005 
\left[ \frac{2}{\sqrt{3}}\left( |\underline{2201}\rangle_{\tiny
\begin{ytableau}
1& 2& 3 & 4 \\
\end{ytableau}
} -\frac{1}{\sqrt{2}} 
|\underline{2021}\rangle_{\tiny
\begin{ytableau}
1& 3 \\
2 & 4 \\
\end{ytableau}
}  \right) - 
|\underline{2201}\rangle_{\tiny
\ydiagram{2,1,1}
}   \right] 
\nonumber\\
&&-0.452501
\left[ \frac{2}{\sqrt{3}}\left( |\underline{2011}\rangle_{\tiny
\begin{ytableau}
1& 2& 3 & 4 \\
\end{ytableau}
} -\frac{1}{\sqrt{2}} 
|\underline{2101}\rangle_{\tiny
\begin{ytableau}
1& 3 \\
2 & 4 \\
\end{ytableau}
}  \right) -  |\underline{2011}\rangle_{\tiny \ydiagram{2,1,1}
}
 \right]
 \nonumber\\
&&-0.000933504\cdot 2  |\underline{0111}\rangle_{\tiny
\begin{ytableau}
1& 2& 3 & 4 \\
\end{ytableau}
} -0.0400016 \cdot 2  |\underline{2000}\rangle_{\tiny
\begin{ytableau}
1& 2& 3 & 4 \\
\end{ytableau}
} +0.171\cdot 2  |\underline{0001}\rangle_{\tiny
\begin{ytableau}
1& 2& 3 & 4 \\
\end{ytableau}
} ~, \nonumber \\
\end{eqnarray}
\begin{eqnarray}
|K_{10}\rangle 
&=&
-0.210296 \left(|00100100\rangle+|10000100\rangle+|01001000\rangle+|00100001\rangle \right. \nonumber \\
&& \qquad \qquad \qquad \left. +|00010010\rangle+|00001001\rangle+|10010000\rangle+|01000010\rangle \right) \nonumber\\
&& +0.0404742\left(|00101000\rangle+|10100000\rangle+|00001010\rangle+|10000010\rangle\right) \nonumber\\
&& \qquad -0.0809484 \left(|10001000\rangle+|00100010\rangle\right)  \nonumber\\
&& -0.116275 \left(|01010000\rangle+|01000001\rangle+|00010100\rangle+|00000101\rangle \right) \nonumber\\
&& \qquad + 0.352267\left(|00010001\rangle + |01000100\rangle \right)\nonumber\\
&&+0.547297 |00000000\rangle -0.157323|01010101\rangle  \nonumber\\
&=&
-0.210296 \left(|\underline{0210} \rangle+|\underline{2010} \rangle+|\underline{1020} \rangle+|\underline{0201} \rangle+|\underline{0102} \rangle +|\underline{0021} \rangle+|\underline{2100} \rangle+|\underline{1002} \rangle \right) \nonumber\\
&& +0.0404742\left(|\underline{0220} \rangle+|\underline{2200} \rangle+|\underline{0022} \rangle+|\underline{2002} \rangle -2|\underline{2020} \rangle-2|\underline{0202} \rangle \right) \nonumber\\
 && + 0.0399057 \left(|\underline{1100} \rangle+|\underline{1001} \rangle+|\underline{0110} \rangle+|\underline{0011} \rangle + |\underline{0101} \rangle+|\underline{1010} \rangle \right) \nonumber \\
&&  -0.156181 \left(|\underline{1100} \rangle+|\underline{1001} \rangle+|\underline{0110} \rangle+|\underline{0011} \rangle -2|\underline{0101} \rangle-2|\underline{1010} \rangle\right)  \nonumber\\
&&+0.547297 |\underline{0000} \rangle -0.157323|\underline{1111} \rangle  \nonumber \\
&=&
-0.210296 \left[ \frac{2}{\sqrt{3}}\left( 2|\underline{2001}\rangle_{\tiny
\begin{ytableau}
1& 2& 3 & 4 \\
\end{ytableau}
} -\frac{1}{\sqrt{2}} 
|\underline{2001}\rangle_{\tiny
\begin{ytableau}
1& 3 \\
2 & 4 \\
\end{ytableau}
}  \right) - 
|\underline{2001}\rangle_{\tiny
\ydiagram{2,1,1}
}   \right] \nonumber\\
&& - 0.0404742\cdot 2 \sqrt{3} |\underline{2020}\rangle_{\tiny
\begin{ytableau}
1& 3 \\
2 & 4 \\
\end{ytableau}
}  + 0.0399057 \cdot \sqrt{6}  |\underline{0011}\rangle_{\tiny
\begin{ytableau}
1& 2& 3 & 4 \\
\end{ytableau}
}   + 0.156181 \cdot 2 \sqrt{3} |\underline{0101}\rangle_{\tiny
\begin{ytableau}
1& 3 \\
2 & 4 \\
\end{ytableau}
}  \nonumber\\
&&+0.547297 |\underline{0000}\rangle_{\tiny
\begin{ytableau}
1& 2& 3 & 4 \\
\end{ytableau}
}  -0.157323|\underline{1111}\rangle_{\tiny
\begin{ytableau}
1& 2& 3 & 4 \\
\end{ytableau}
} ~; \\
 |K_{11}\rangle &=&  -0.456548\left(|01001010\rangle+|00101001\rangle+|10100100\rangle+|10010010\rangle \right) \nonumber\\
&&-0.104315\left(|01001001\rangle+|01010010\rangle+|10010100\rangle+|00100101\rangle \right) \nonumber\\
&& -0.0340389 \left(|01010001\rangle+|01010100\rangle+|00010101\rangle+|01000101\rangle \right)\nonumber\\
&&+0.0797719\left(|01000000\rangle+|00010000\rangle+|00000001\rangle+|00000100\rangle \right) \nonumber\\
&&+0.152183\left(|00001000\rangle+|00000010\rangle+|00100000\rangle+|10000000\rangle \right) \nonumber\\
&=&  -0.456548\left(|\underline{1022} \rangle+|\underline{0221} \rangle+|\underline{2210} \rangle+|\underline{2102} \rangle  \right) 
\nonumber \\ && -0.104315\left(|\underline{1021} \rangle+|\underline{1102} \rangle+|\underline{2110} \rangle+|\underline{0211} \rangle  \right) \nonumber\\
&& -0.0340389 \left(|\underline{1101} \rangle+|\underline{1110} \rangle+|\underline{0111} \rangle+|\underline{1011} \rangle  \right)
\nonumber \\ &&  +0.0797719\left(|\underline{1000} \rangle+|\underline{0100} \rangle+|\underline{0001} \rangle+|\underline{0010} \rangle  \right) \nonumber\\
&&+0.152183\left(|\underline{0020} \rangle+|\underline{0002} \rangle+|\underline{0200} \rangle+|\underline{2000} \rangle  \right) \nonumber \\
&=&  -0.456548 
\left[  \frac{2}{\sqrt{3}}\left( |\underline{2201}\rangle_{\tiny
\begin{ytableau}
1& 2& 3 & 4 \\
\end{ytableau}
} -\frac{1}{\sqrt{2}} 
|\underline{2021}\rangle_{\tiny
\begin{ytableau}
1& 3 \\
2 & 4 \\
\end{ytableau}
}  \right) - 
|\underline{2201}\rangle_{\tiny
\ydiagram{2,1,1}
}    \right]
\nonumber \\
&& -0.104315
\left[ \frac{2}{\sqrt{3}}\left( |\underline{2011}\rangle_{\tiny
\begin{ytableau}
1& 2& 3 & 4 \\
\end{ytableau}
} -\frac{1}{\sqrt{2}} 
|\underline{2101}\rangle_{\tiny
\begin{ytableau}
1& 3 \\
2 & 4 \\
\end{ytableau}
}  \right) -  |\underline{2011}\rangle_{\tiny \ydiagram{2,1,1}
}
  \right]
 \nonumber\\
&& -0.0340389 \cdot 2  |\underline{0111}\rangle_{\tiny
\begin{ytableau}
1& 2& 3 & 4 \\
\end{ytableau}
}
+0.0797719 \cdot 2  |\underline{0001}\rangle_{\tiny
\begin{ytableau}
1& 2& 3 & 4 \\
\end{ytableau}
}  
+0.152183 \cdot 2  |\underline{2000}\rangle_{\tiny
\begin{ytableau}
1& 2& 3 & 4 \\
\end{ytableau}
} ~,
\nonumber \\
\end{eqnarray}
\begin{eqnarray}
|K_{12}\rangle &=& 
-0.108981\left(|00100100\rangle+|00100001\rangle+|00001001\rangle+|10000100\rangle \right. \nonumber \\
&& \qquad \qquad \qquad \left.+|01001000\rangle+|10010000\rangle+|01000010\rangle+|00010010\rangle \right) \nonumber\\
&& -0.209288\left(|00001010\rangle+|00101000\rangle+|10000010\rangle+|10100000\rangle \right)  \nonumber \\
&& \qquad +0.418575\left(|00100010\rangle+|10001000\rangle \right) \nonumber\\
&& +0.120341 \left(|01010000\rangle+|01000001\rangle+|00010100\rangle+|00000101\rangle \right) \nonumber \\
&& \qquad  -0.308446\left(|00010001\rangle+|01000100\rangle\right) \nonumber\\
&&+0.0359095 |01010101\rangle +0.360379 |00000000\rangle \nonumber \\
&=&
-0.108981\left(|\underline{0210} \rangle+|\underline{0201} \rangle+|\underline{0021} \rangle+|\underline{2010} \rangle \right. \nonumber \\
&& \qquad \qquad \qquad \left. +|\underline{1020} \rangle+|\underline{2100} \rangle+|\underline{1002} \rangle+|\underline{0102} \rangle \right) \nonumber\\
 && -0.209288\left(|\underline{0022} \rangle+|\underline{0220} \rangle+|\underline{2002} \rangle+|\underline{2200} \rangle -2|\underline{0202} \rangle-2|\underline{2020} \rangle \right) \nonumber\\
 && -0.022588 \left(|\underline{1100} \rangle+|\underline{1001} \rangle+|\underline{0110} \rangle+|\underline{0011} \rangle + |\underline{0101} \rangle+|\underline{1010} \rangle \right) \nonumber \\
&& + 0.142929 \left(|\underline{1100} \rangle+|\underline{1001} \rangle+|\underline{0110} \rangle+|\underline{0011} \rangle -2|\underline{0101} \rangle-2|\underline{1010} \rangle\right)  \nonumber\\
&&+0.0359095 |\underline{1111} \rangle +0.360379 |\underline{0000} \rangle \nonumber\\
&=&
-0.108981\left[  \frac{2}{\sqrt{3}}\left( 2|\underline{2001}\rangle_{\tiny
\begin{ytableau}
1& 2& 3 & 4 \\
\end{ytableau}
} -\frac{1}{\sqrt{2}} 
|\underline{2001}\rangle_{\tiny
\begin{ytableau}
1& 3 \\
2 & 4 \\
\end{ytableau}
}  \right) -
|\underline{2001}\rangle_{\tiny
\ydiagram{2,1,1}
}   \right] \nonumber\\
 && + 0.209288 \cdot 2 \sqrt{3} |\underline{2020}\rangle_{\tiny
\begin{ytableau}
1& 3 \\
2 & 4 \\
\end{ytableau}
} \nonumber \\
&& -0.022588 \cdot \sqrt{6}  |\underline{0011}\rangle_{\tiny
\begin{ytableau}
1& 2& 3 & 4 \\
\end{ytableau}
}   - 0.142929  \cdot 2 \sqrt{3} |\underline{0101}\rangle_{\tiny
\begin{ytableau}
1& 3 \\
2 & 4 \\
\end{ytableau}
}   \nonumber\\
&&+0.0359095|\underline{1111}\rangle_{\tiny
\begin{ytableau}
1& 2& 3 & 4 \\
\end{ytableau}
}
 +0.360379 |\underline{0000}\rangle_{\tiny
\begin{ytableau}
1& 2& 3 & 4 \\
\end{ytableau}
} ~.
\nonumber \\
\end{eqnarray}
The Lanczos coefficients are
\begin{eqnarray}
    &&b_1 = 2~; ~b_2= 2.44949~; ~b_3 = 2.58199~; ~ b_4 = 2.61406~; \nonumber \\ 
    &&b_5 = 2.61282~;~ b_6 = 2.58739~ ;  b_7 = 2.43927~; ~ b_8 = 2.02185~; \nonumber \\
    && b_9 = 0.145695~; ~ b_{10} = 0.937686~; ~b_{11} = 1.07906 ~; ~ b_{12} = 0.935826 ~; \nonumber \\
    && b_{n\ge 13} =0 ~.
\end{eqnarray}

\begin{table}[h!]
\centering
\begin{minipage}[t]{0.45\linewidth}
\centering
\textbf{$|{Z}_2\rangle$}\\[5pt] 
\begin{tabular}{ccc}
\toprule
$E_i$ & $c_i$ & \textbf{Degeneracy} \\
\midrule
0 & 0.44349  & 246 \\
1.3419      & 0.40987  & 1   \\
-1.3419      & 0.40987  & 1   \\
2.6658     & 0.33883  & 1   \\
-2.6658     & 0.33883  & 1   \\
3.9516     & 0.24182  & 1   \\
-3.9516      & 0.24182  & 1   \\
5.1702      & 0.14758  & 1   \\
-5.1702     & 0.14758  & 1   \\
1.1409     & 0.11007  & 1   \\
-1.1409      & 0.11007  & 1   \\
2.238      & 0.090772 & 1   \\
-2.238       & 0.090772 & 1   \\
6.2744     & 0.078847 & 1   \\
-6.2744      & 0.078847 & 1   \\
3.3272      & 0.065032 & 1   \\
-3.3272     & 0.065032 & 1   \\
4.475       & 0.036129 & 1   \\
-4.475      & 0.036129 & 1   \\
0.77907    & 0.031627 & 1   \\
-0.77907     & 0.031627 & 1   \\
0.29667      & 0.031552 & 1   \\
-0.29667     & 0.031552 & 1   \\
\bottomrule
\end{tabular}

~\\
~\\
~\\

\textbf{$|{Z}_4\rangle$}\\[5pt]
\begin{tabular}{ccc}
\toprule
$E_i$ & $c_i$ & \textbf{Degeneracy} \\
\midrule
0 & 0.29504  & 213 \\
0.48614     & 0.27071  & 2   \\
-0.48614    & 0.27071  & 2   \\
1.7439      & 0.25293  & 1   \\
-1.7439     & 0.25293  & 1   \\
3.8946      & 0.2121   & 2   \\
-3.8946     & 0.2121  & 2   \\
\bottomrule
\end{tabular}
\end{minipage}
\quad
\begin{minipage}[t]{0.45\linewidth}
\centering
\textbf{$|{Z}_4\rangle$ (Continued)}\\[5pt]
\begin{tabular}{ccc}
\toprule
$E_i$ & $c_i$ & \textbf{Degeneracy} \\
\midrule
2.4841     & 0.20911  & 1   \\
-2.4841     & 0.20911  & 1   \\
1.7427     & 0.18124  & 1   \\
-1.7427      & 0.18124  & 1   \\
2.9468     & 0.1513   & 2   \\
-2.9468     & 0.1513   & 2   \\
 4.9114     &0.14498          &2 \\    
        -4.9114     &0.14498         & 2    \\
        0.77907    & 0.14326         &1    \\
       -0.77907     &0.14326         & 1    \\
        -2.7902     & 0.1341         & 1    \\  
         2.7902     & 0.1341         & 1    \\
          2.238    & 0.11919         & 1    \\
         -2.238     &0.11919         & 1    \\
         1.1349    & 0.11506         & 2    \\
        -1.1349    & 0.11506         & 1    \\
         1.3772    & 0.11494         & 2    \\
        -1.3772   &  0.11494         & 2    \\
        2.6922     & 0.1067         & 1    \\
         -2.6922     & 0.1067         & 1    \\
        0.70684    & 0.10021         & 2    \\
       -0.70684    & 0.10021         & 2    \\
       0.82595   & 0.099861         & 1    \\
        -0.82595   & 0.099861        &  1    \\
        1.1409   & 0.080521         & 1    \\
         -1.1409   & 0.080521         & 1    \\
        6.2744  &  0.078847         & 1    \\
         -6.2744   & 0.078847         & 1    \\
         3.3272    & 0.07571         & 1    \\
        -3.3272    & 0.07571         & 1    \\
        5.1702   & 0.070433         & 1    \\
         -5.1702   & 0.070433         & 1    \\
        1.3845   & 0.058015         & 1    \\
         -1.3845   & 0.058015         & 1    \\
        7.2429   & 0.047984         & 1    \\
         -7.2429    &0.047984         & 1    \\
\bottomrule
\end{tabular}
\end{minipage}
\caption{The energies $E_i$, component factors $c_i$, and degeneracies of the eigenstates that make up $|Z_k\rangle$ (see \eqref{eigenexpand}), in particular $|Z_2\rangle$ and $|Z_4\rangle$. For practical reasons, the eigenstates contained in $|Z_2\rangle$ with $c_i<0.03$, and the eigenstates contained in $|Z_4\rangle$ with $c_i<0.04$, are not shown. Here the lattice size $L=12$.}
\label{tab:two_tables}
\end{table}

\section{Initial states $|Z_2\rangle$ and $|Z_4\rangle$ in the Eigenbasis of the PXP Hamiltonian} \label{eigenbasisZ2Z4}

\clearpage

\bibliography{Refs}
\bibliographystyle{JHEP-2}


\end{document}